\documentclass[a4paper, 11pt]{Sources/Thesis}  
\graphicspath{{Figures/}}  

\usepackage[square, numbers, comma, sort&compress]{natbib}  
\usepackage{verbatim}  
\usepackage{vector}  
\usepackage{mathrsfs}
\usepackage{Sources/axodraw4j}
\usepackage{epsfig} 
\usepackage{multirow}
\usepackage{hhline}
\usepackage{slashed}

\def\D0{\slash\!\!\!\!\!\!\!\!\!\:D0}

\newcommand{\ltap}{\stackrel{\displaystyle <}{\,_{\! \,_{\displaystyle
\sim}}}}  
\hypersetup{urlcolor=blue, colorlinks=true}  

\begin{document}
\frontmatter	  

\title  {Collider phenomenology of the 4D~composite Higgs model}
\authors  {\texorpdfstring
            {\href{d.barducci@soton.ac.uk}{Daniele Barducci}}
            {Daniele Barducci}
            }
\addresses  {\groupname\\\deptname\\\univname}  
\date       {September 2014}
\subject    {}

\maketitle

\setstretch{1.3}  

\fancyhead{}  
\rhead{\thepage}  
\lhead{}  

\pagestyle{fancy}  


\addtotoc{Abstract}  
\abstract{
\addtocontents{toc}{\vspace{1em}}  

This Thesis is devoted to the phenomenological analysis at the large hadron collider (LHC), as well at a future electron positron collider,
of the 4 dimensional (4D) composite Higgs model (4DCHM), a compelling beyond
the standard model scenario where the Higgs state arises as a pseudo Nambu Goldstone boson.
The motivations and the main characteristics of the model are summarised and then an analysis of the gauge and Higgs
sectors of the 4DCHM is performed.
Finally we propose a general framework for the analysis of models with an extended quark sector that we have applied to
a simplified composite Higgs scenario.

}

\clearpage  


\pagestyle{fancy}  

\lhead{\emph{Contents}}  
\tableofcontents  

\lhead{\emph{List of Figures}}  
\listoffigures  

\lhead{\emph{List of Tables}}  
\listoftables  
\Declaration{

\addtocontents{toc}{\vspace{1em}}  

I, Daniele Barducci, declare that this thesis titled
\emph{Collider phenomenology of the 4D composite Higgs model} and the work
presented in it are my own. I confirm that:

\begin{itemize} 
\item[\tiny{$\blacksquare$}] This work was done wholly or mainly while
  in candidature for a research degree at this University. 
 
\item[\tiny{$\blacksquare$}] Where any part of this thesis has
  previously been submitted for a degree or any other qualification at
  this University or any other institution, this has been clearly
  stated. 
 
\item[\tiny{$\blacksquare$}] Where I have consulted the published work
  of others, this is always clearly attributed. 
 
\item[\tiny{$\blacksquare$}] Where I have quoted from the work of
  others, the source is always given. With the exception of such
  quotations, this thesis is entirely my own work. 
 
\item[\tiny{$\blacksquare$}] I have acknowledged all main sources of
  help. 
 
\item[\tiny{$\blacksquare$}] Where the thesis is based on work done by
  myself jointly with others, I have made clear exactly what was done
  by others and what I have contributed myself. 
\\
\end{itemize}

Signed:\\
\rule[1em]{25em}{0.5pt}  
 
Date:\\
\rule[1em]{25em}{0.5pt}  
}
\clearpage  

\newpage\null\thispagestyle{empty}\newpage

\pagestyle{empty}  

\null\vfill
\begin{flushright}
\textit{``...per aspera sic itur ad astra...''} \\
\textit{Lucio Anneo Seneca}
\end{flushright}

\vfill\vfill\vfill\vfill\vfill\vfill\null
\clearpage  

\setstretch{1.3}  
\newpage\null\thispagestyle{empty}\newpage
\acknowledgements{
\addtocontents{toc}{\vspace{1em}}  
                                
Ringrazio i miei genitori, per il supporto che mi hanno dato durante questa mia prima esperienza fuori casa\\
\emph{
I would like to thank my parents, for the support that they gave me during this first
experience away from home
}

Ringrazio tutti i miei amici e le persone che ho conosciuto in Southampton, che hanno contribuito tutti a farmi crescere come persona durante questa avventura\\
\emph{
I would like to thank all my friends and the people that I met in Southampton, who have all contributed to help me grow as a person during this adventure
}

Ringrazio i miei supervisors, Alexander Belyaev e Stefano Moretti, e tutte le persone con cui ho lavorato negli ultimi tre anni, con cui ho avuto una vita lavorativa stimolante e piacevole\\
\emph{
I would like to thank my supervisors, Alexander Belyaev and Stefano Moretti, and all the people who I worked with in the last three years and with whom I had a stimulating and pleasant working life
}

Grazie a tutti, senza di voi questo non sarebbe stato possibile\\
\emph{
Thank you all, without you this would not have been possible
}

\newpage

}
\clearpage  

\addtocontents{toc}{\vspace{2em}}  

\mainmatter	  
\pagestyle{fancy}  


\chapter{Introduction}
\label{chap:1}
\lhead{Chapter 1. \emph{Introduction}}

\section{The Standard Model of fundamental interactions}

The discovery of the Higgs boson by the ATLAS \cite{Aad:2012tfa} and CMS \cite{Chatrchyan:2012ufa}
collaborations announced at the \emph{Organisation Europ\'eenne pour la Recherche Nucl\'eaire} (CERN) on the 4th of July 2012 has established
the existence of the last missing piece of the Standard Model (SM) of fundamental interactions and has ended a nearly forty years
search for this particle that was theorized in 1964 by Peter Higgs and Fran\c cois Englert.
The two scientists proposed a mechanism, now commonly called the \emph{Higgs mechanism}
\footnote{Even though names with more scientists that have contributed to this idea have been coined,
such as the Brout-Englert-Higgs-Guralnik-Hagen-Kibble mechanism, from now on we will refer to it as the \emph{Higgs mechanism}.},
through which the gauge bosons of the SM acquire mass and for this idea on the 8th of
October 2014 they were awarded with the Nobel Prize for physics
\footnote{Besides the name of Higgs and Englert it should also be mentioned that of Robert Brout,
collaborator of the latter, who would probably have been awarded with the Nobel Prize too if he hadn't passed away in May 2011.}.

The SM represents at the moment the best theoretical description of the fundamental interactions in particle
physics. It unifies the weak, electromagnetic and strong forces, three of the four fundamental forces
of Nature, which are
described by a gauge theory invariant under the $SU(3)_c \otimes SU(2)_L \otimes U(1)_Y$ group.
Since its complete formulation in 1967 this theory has been more and more validated by experimental
evidences such as the discovery
of the bottom quark in 1977, the top quark in 1995, the weak neutral current mediated by the $Z^0$
boson in 1983 and the $\tau$ neutrino in 2000 until the already mentioned discovery of the Higgs boson, that
has been reached thanks to the huge effort of 
the physics community in building the Large Hadron Collider (LHC).

However, despite all its experimental validations, there are theoretical and experimental indications that the SM cannot be the ultimate theory of Nature.
For example, on the experimental side, the evidence of dark matter (DM) and dark energy makes the SM describing only 4\% of
the universe content, while the observation of neutrino oscillation, and therefore the evidence
of at least two massive neutrinos, has no trivial explanation in this theory as well as the matter-antimatter asymmetry observed in the Universe.
Conversely, from the theoretical point of view, the non inclusion of a quantistic description of the gravitational force seems the biggest limitation of the SM.

\begin{figure}[!h]
\centering
\epsfig{file=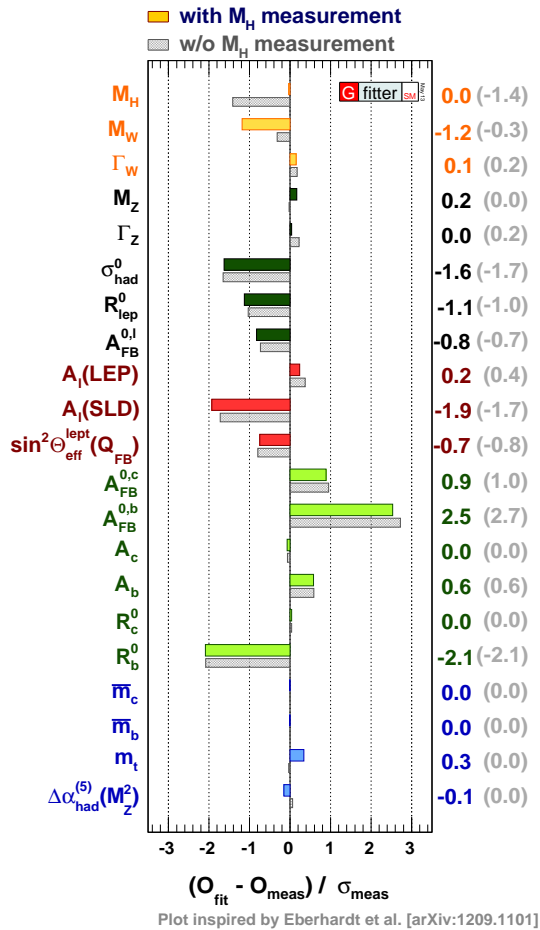, width=.48\textwidth}
\caption[Standard Model pull fit with and without the inclusion of the Higgs boson mass]{Comparison of the
SM fit with direct measurements with (coloured) and without (gray) the inclusion of $m_H$ into the fit.
The figure has been taken from \url{http://gfitter.desy.de/Standard_Model}.}
\label{fig:SM-pull_fit}
\end{figure}

Nevertheless, as already mentioned, the SM extraordinarily agrees with a large number of data collected
so far by various collider experiments (such as LEP, LEP2, Tevatron and LHC) as reported in Fig.\ref{fig:SM-pull_fit}.

Therefore the following questions arise naturally: what is the main reason that makes the physics community think that
we should find new physics at the TeV scale, that is, at the energy scale that is being tested at the LHC?
Could all the problems mentioned before be solvable at a scale that is and will be unreachable by the CERN machine?

\section{Naturalness and fine tuning}

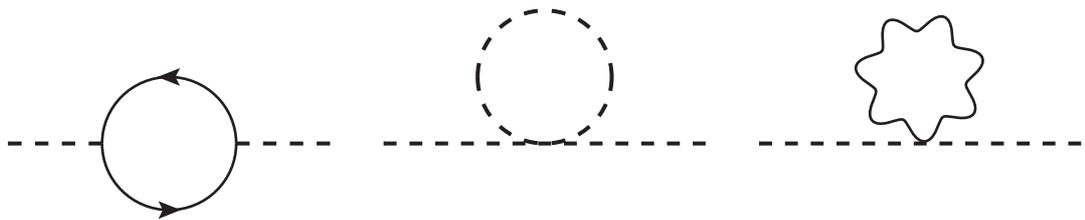
\begin{figure}[!h]
\begin{center}
 \begin{picture}(170,120)
    \SetWidth{1.5}
    \SetColor{Black}
    \Line[dash,dashsize=5](-110,50)(-75,50)
    \SetWidth{1.0}
    \Arc[arrow,arrowpos=0.5](-50,50)(25,0,180)
    \Arc[arrow,arrowpos=0.5](-50,50)(25,180,360)
    \SetWidth{1.5}
    \Line[dash,dashsize=5](-25,50)(10,50)
    
    \SetWidth{1.5}
    \SetColor{Black}
    \Line[dash,dashsize=5](30,50)(150,50)
    \SetWidth{1.5}
    \Arc[dash,dashsize=5](90,75)(25,0,180)
    \Arc[dash,dashsize=5](90,75)(25,180,360)
    
    \SetWidth{1.5}
    \SetColor{Black}
    \Line[dash,dashsize=5](170,50)(290,50)
    \SetWidth{1.0}
    \PhotonArc(230,75)(20.777,4,387){3.5}{7}

  \end{picture}
\end{center}
\caption[Higgs quadratic divergences]{Quadratic divergences to the Higgs mass given by a top quark loop (left), Higgs loop
(centre) and a gauge boson loop (right)}
\label{fig:quad-div}
\end{figure}

A possible answer to this question relies on what the particle physics community calls \emph{naturalness}
or \emph{fine tuning} argument.
In the SM the Higgs mass receives large radiative corrections from loops contributions to its two point function
involving fermions, vector bosons and the Higgs itself (Fig.~\ref{fig:quad-div}), and these corrections grow as $\Lambda^2$, where $\Lambda$ is the theory cut off.
If we assume the SM to be valid up to the Grand Unified Theory (GUT) scale, $10^{16}$ GeV, or Planck scale, $10^{19}$ GeV,
the one loop corrections to the Higgs mass will then be roughly 24 order of magnitude bigger than its measured value,
and this implies that the related counter terms would need to be fine-tuned up to $10^{-24}$ to have the correct cancellations that give the 126 GeV Higgs mass.
Although consistent from a theoretical point of view, since the SM is a renormalizable theory, this huge and precise
cancellation is believed to be unnatural and is taken as a hint of a relative small value of the cut off $\Lambda$, that indicates the
presence of new physics at a scale which, for a reasonable level of fine tuning of the order of a few \%, is expected to be of the order of 1 TeV.
A complementary point of view is to consider that, while in the SM the fermions and vector bosons masses are protected
respectively by the chiral and gauge symmetries, no symmetry is actually protecting the scalar mass from the mentioned
radiative corrections. Conversely the presence of a new symmetry, with the implication of new physics, will stabilize the mass of the scalar.

\begin{figure}[!h]
\begin{center}
 \begin{picture}(70,100)
    \SetWidth{1.5}
    \SetColor{Black}
    \Line[dash,dashsize=5](-110,50)(-75,50)
    \SetWidth{1.0}
    \Arc[arrow,arrowpos=0.5](-50,50)(25,0,180)
    \Arc[arrow,arrowpos=0.5](-50,50)(25,180,360)
    \SetWidth{1.5}
    \Line[dash,dashsize=5](-25,50)(10,50)
    
    \SetWidth{1.5}
    \SetColor{Black}
    \Line[dash,dashsize=5](50,50)(85,50)
    \SetWidth{1.0}
    \Arc[dash,dashsize=5](110,50)(25,0,360)
    \SetWidth{1.5}
    \Line[dash,dashsize=5](135,50)(170,50)    
  \end{picture}
\end{center}
\caption[Top and stop loop contribution to the Higgs mass in Supersymmetry]{Top quark (left) and stop squark (right) loop contributions to
the Higgs mass. In SUSY theories these contributions lead to the cancellation of quadratic divergences. }
\label{fig:susy-canc}
\end{figure}
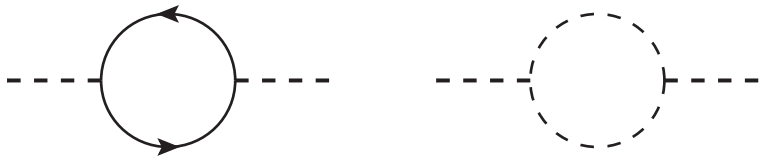

This is exactly what happens in the most common BSM (beyond the Standard Model) scenario for which physicists are searching evidence: Supersymmetry (SUSY). The boson-fermion symmetry present in SUSY theories predicts the existence of partners of the SM particles which have a value of the spin
1/2 smaller, and whose existence provides a precise cancellation of the quadratic divergence arising from the loop of, for example, the top quark with the loop involving its SUSY counter part, the stop (Fig.~\ref{fig:susy-canc}), due to their different statistical behaviour: Fermi-Dirac for the former and Bose-Einstein for the latter.
Despite other fascinating properties of SUSY theories, like a possible explanation of DM, the unification of the gauge
couplings and the fact that a Supersymmetric algebra is the only non trivial extension of the Poincar\'e group, so far we do not have any evidence of Supersymmetric particles at the LHC and bounds on the masses of the SM partners are
being set higher and higher.

For this reason other BSM solutions ought to be investigated and from the point of view of the \emph{fine tuning} argument,
another possibility to prevent large radiative corrections affecting the Higgs mass is to postulate it to be a composite state.

\section{The composite Higgs idea}

The idea of postulating the Higgs boson to be a composite state was first formulated by Georgi and Kaplan \cite{Kaplan:1983fs} in the '80s.
In their theory the Higgs arises as a bound state of a strongly coupled sector at some energy scale higher than the electro-weak (EW) scale.
In order to achieve a light spinless particle, therefore lighter than the other resonances of this strong sector, the Higgs is postulated to be a
Goldstone boson (GB) arising from the spontaneous symmetry breaking (SSB) of a global symmetry $G$ of the strong sector.
An eventual weak and explicit breaking of $G$ can then give rise to a radiative generated scalar potential \cite{Coleman:1973jx} that guarantees a mass
for the Higgs which will then become a pseudo GB.
Being the Higgs a composite state automatically solves the hierarchy problems since all the radiative corrections affecting its mass
will be saturated at the composite scale, that is its mass will not be sensitive to virtual effects above such scale, and also agrees with a historical pattern that has so
far seen all the (pseudo)scalar particles known in Nature to be composite states.

This idea closely resembles the pattern with which it is possible to explain the lightness of the pions, $\pi^\pm$ and $\pi^0$,
with respect to all the other mesons, like for example the $\rho$'s\footnote{In fact $m_{\pi^{\pm/0}}=139.6/135.0$ MeV
and $m_{\rho^{\pm/0}}=775.4/775.5$ MeV.}, that is, postulating them to be pseudo GB arising from the spontaneously
breaking of the $SU(2)_L\otimes SU(2)_R$ chiral symmetry, emerging as bound states of the quantum chromo-dynamic (QCD) strong sector.

A relevant input to the study of the pseudo GB Higgs has been given in a paper of Agashe, Contino and Pomarol
entitled \emph{The Minimal Composite Higgs Model} \cite{Agashe:2004rs}, where the authors introduced the most economical coset of symmetry breaking: $SO(5)/SO(4)$.
This coset develops four GBs, which is the minimum number to be identified with the SM Higgs doublet, and, after three of them have been "eaten" to give mass to the $W^\pm$ and $Z^0$, the spectrum presents just one physical scalar: the Higgs boson. Moreover the authors show that the presence of a $SO(4)$ custodial symmetry is crucial to protect the $\rho$ parameter from dangerous corrections.

The composite Higgs idea could then be a compelling alternative to SUSY scenarios so as to stabilize the EW scale and solve
the \emph{naturalness} issue, and in the last decade we have witnessed a proliferation of models based on this paradigm, among which we also find constructions that can incorporate DM into the composite Higgs framework (see e.g. \cite{Frigerio:2012uc}).
The reason for this, besides a theoretical appeal, is that this kind of theories is actually testable at the LHC.
If the \emph{hierarchy problem} is in fact solved by some strong dynamics, we expect the appearance of new resonances and,
in particular, again for the \emph{fine tuning} argument, the fermionic ones (usually called \emph{top partners}) are expected to be at the TeV scale in order to stabilize the Higgs mass \cite{Matsedonskyi:2012ym,Redi:2012ha,Panico:2012uw}, in the same way that third generation sfermions are expected to be relatively light in SUSY theories, and this is in general an energy scale that is actually explorable at the LHC.


\section{The LHC and future proposed colliders}

The Large Hadron Collider built at CERN between 1998 and 2008 is at the moment the world's largest and most powerful particle collider and four experiments are actually present: ATLAS, CMS, LHCb and ALICE.
Built in the 27 km circumference LEP tunnel, it started operations on the 10th of September 2008.
The centre of mass energy of the beam has been then gradually increased first to the energy of 7 TeV on the 30 of May 2010 
and then to the energy of 8 TeV on the 5th of April 2012, while the collected integrated luminosity has now reached the value
of $\simeq$ 20 fb$^{-1}$.
On the 14th of February 2013 the first long planned shut-down of the machine began, waiting for the 14 TeV run that is scheduled to start in 2015\footnote{This is the centre of mass energy that we have used in considering the forthcoming run of the LHC. However CERN has recently decided to restart the collisions with a centre of mass energy of 13 TeV, which however will not change our conclusions.} and with which the collected integrated luminosity is predicted to reach $\simeq$ 300 fb$^{-1}$ in 10 years of operations.
A planned upgrade of the CERN machine, the high luminosity LHC (HL-LHC), plans to bring the integrated luminosity up to the level of $\simeq$ 3 ab$^{-1}$.

\begin{figure}[!h]
\centering
\epsfig{file=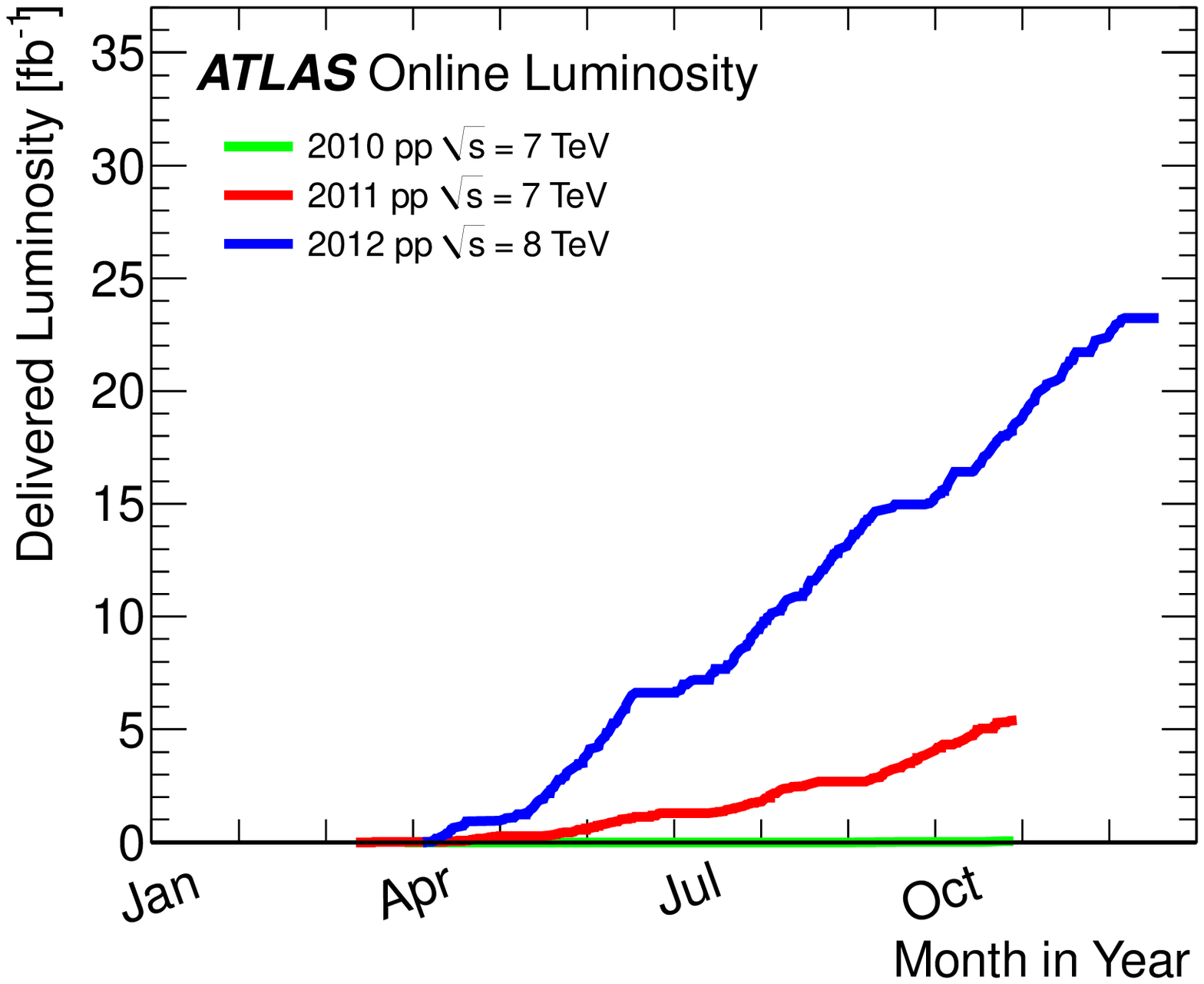, width=.48\textwidth}
\epsfig{file=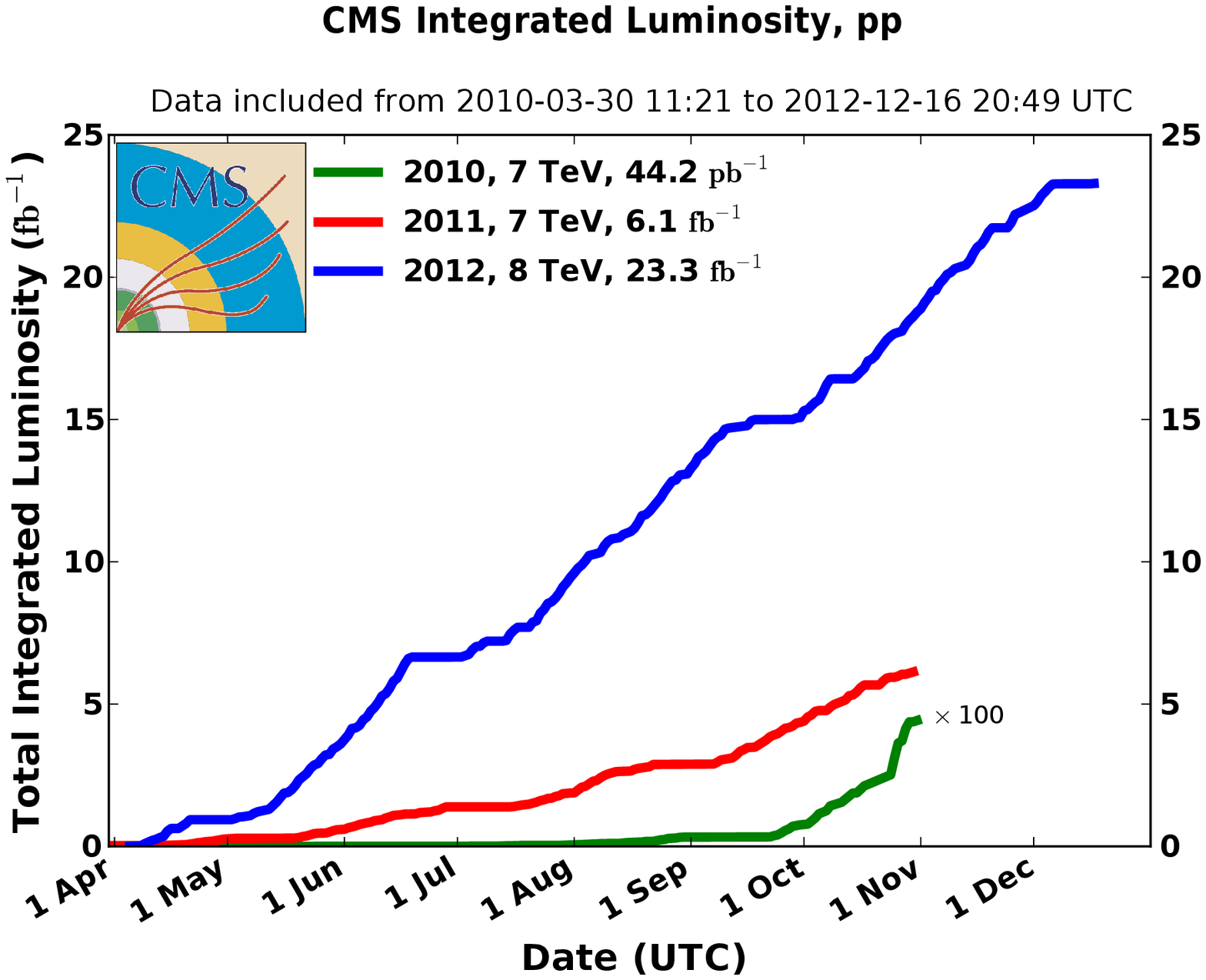, width=.48\textwidth}
\caption[LHC collected luminosity]{ATLAS and CMS collected luminosity in fb$^{-1}$ with the 7 TeV and 8 TeV runs of the LHC during the years
2010, 2011 and 2012.}
\label{fig:LHC-lumi}
\end{figure}

Besides the already mentioned discovery of the Higgs boson\footnote{The LHC has also discovered the bottomonium meson $\chi_b$(3P) (22nd December 2011) and made the first observation of the rare $B_s\to \mu^+\mu^-$ decay (8th November 2012).}, the LHC has achieved important results in testing the Higgs properties and many BSM scenarios.

\begin{figure}[!h]
\centering
\epsfig{file=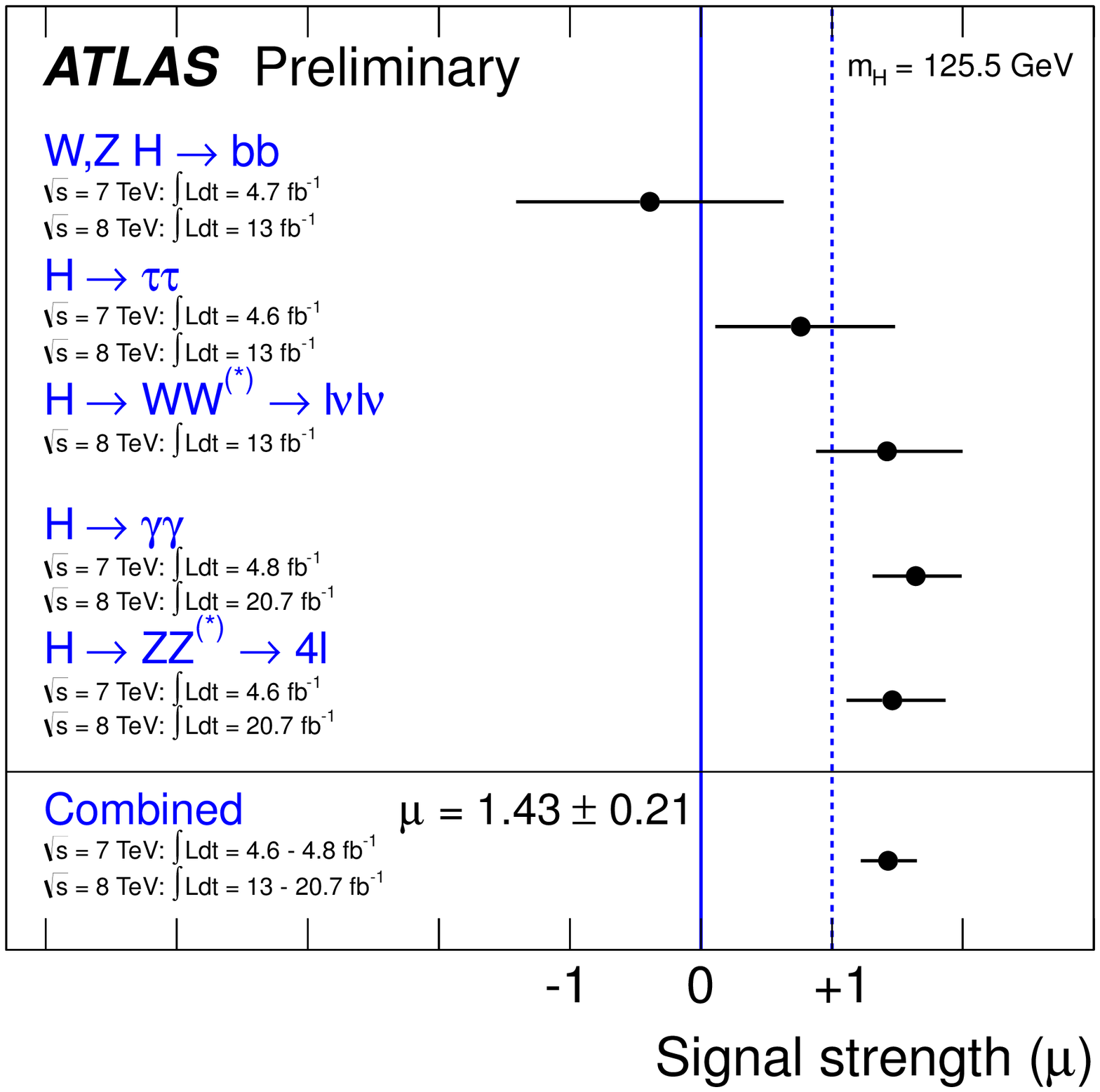, width=.48\textwidth}
\epsfig{file=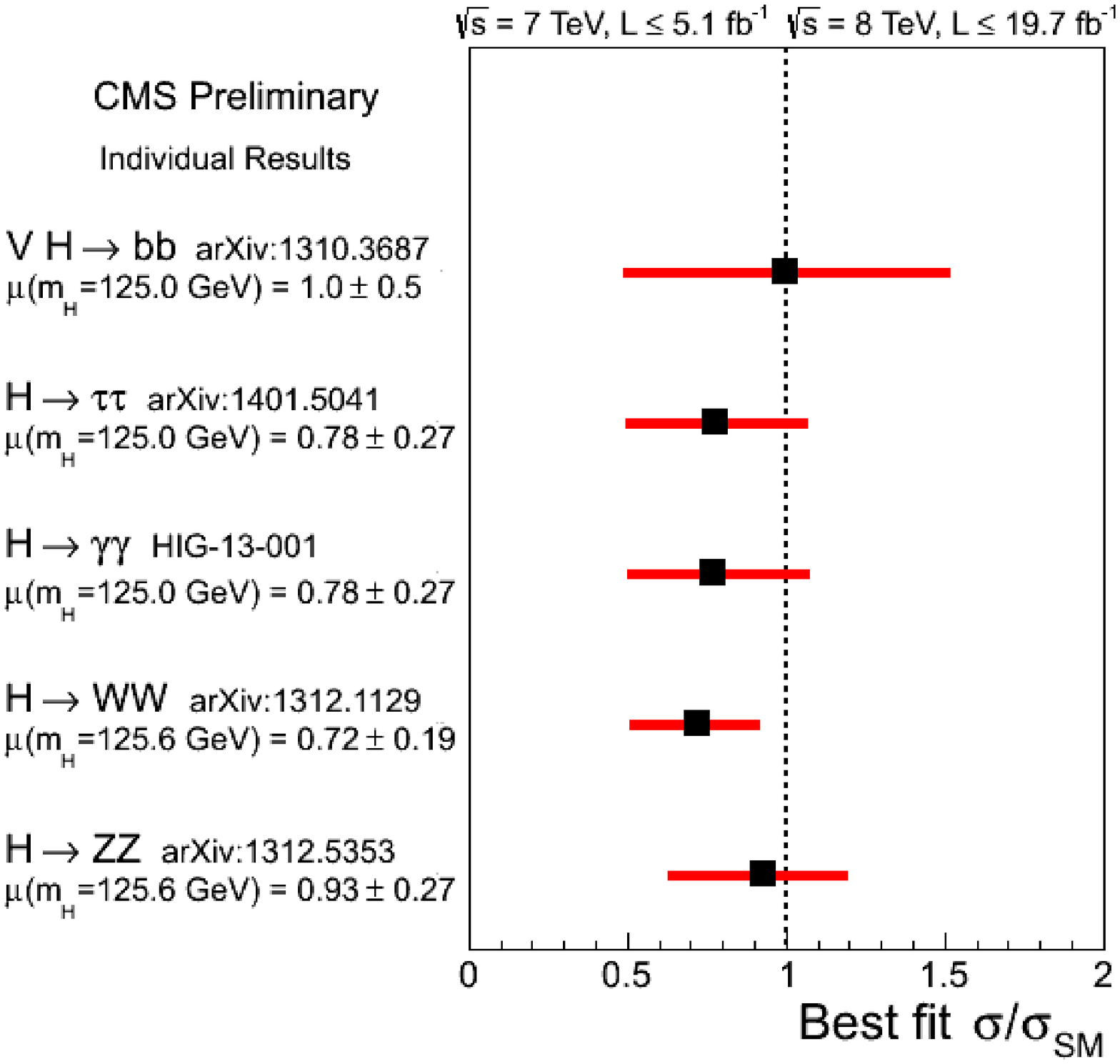, width=.48\textwidth}
\caption[Higgs signal strengths from ATLAS and CMS collaborations]{ATLAS and CMS signal strengths for various Higgs decay channels with the 7 and 8 TeV dataset.}
\label{fig:higgs-mu}
\end{figure}

The measures of the Higgs couplings are usually recast in terms of the so called \emph{signal strengths}, that are the ratio of the 
number of events observed in a given Higgs decay channel per production mode over the SM expectation, and the results
in Fig.~\ref{fig:higgs-mu} for these measurements,
with the 7 and 8 TeV dataset, show that both agreements and
tensions with the SM predictions are present, though nothing definitive can yet be said due do the still big error bars
on these measurements. 

\begin{figure}[!h]
\centering
\epsfig{file=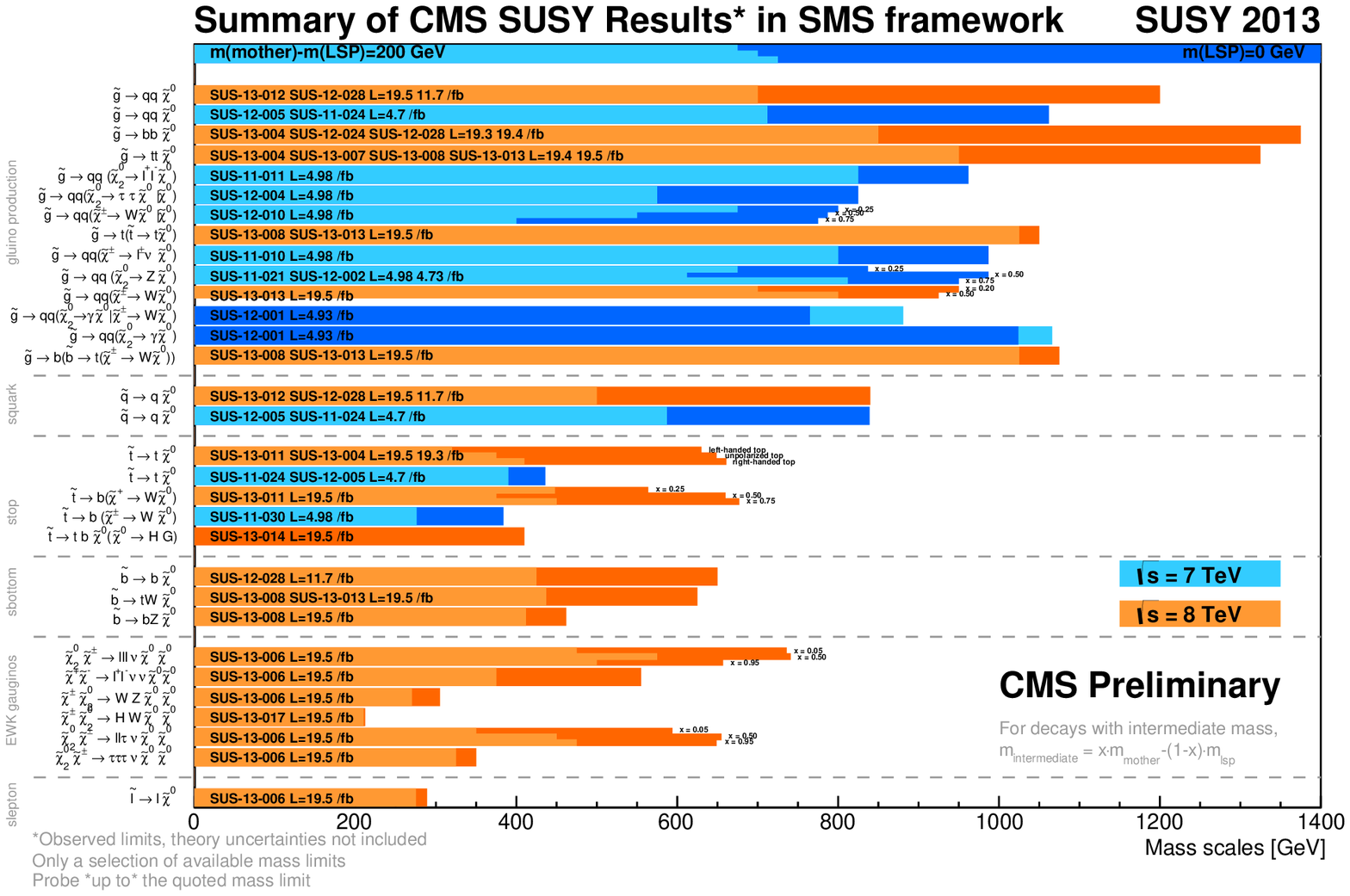, width=.54\textwidth}
\hfill
\epsfig{file=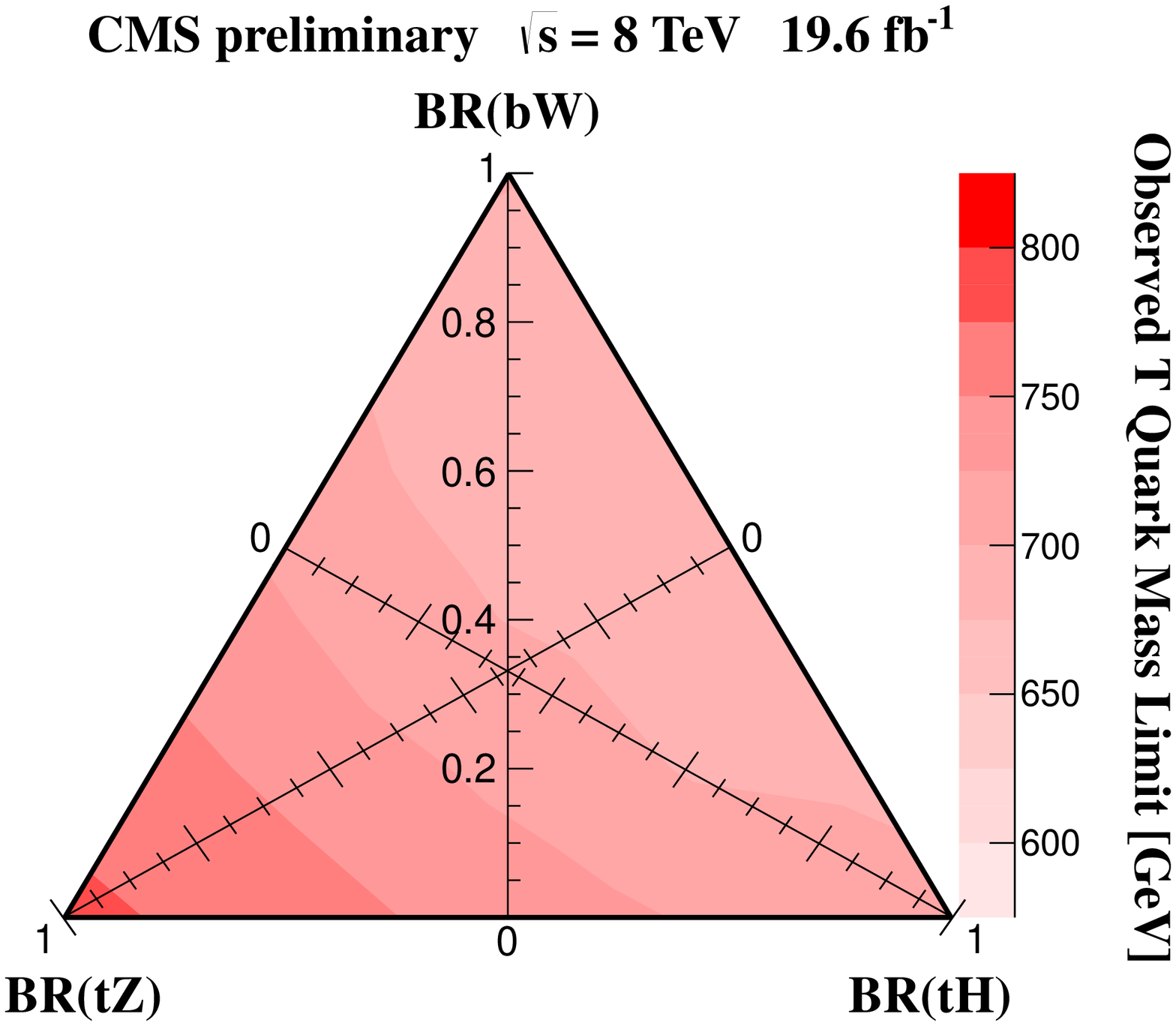, width=.42\textwidth}
\caption[CMS mass bounds on SUSY particles and top partners]{CMS exclusions limits for SUSY particles (left)  with the 7 and 8 TeV dataset.
Limits are set on gluinos, squark, stop, sbottoms, electroweak gauginos and sleptons according to the process considered. CMS mass bound on a single \emph{top partner} (right) in function of its branching ratios into a SM boson and third generation quarks with 19.6 fb$^{-1}$ of integrated luminosity.}
\label{fig:CMS-bounds}
\end{figure}

From the point of view of BSM theories the LHC has so far found no evidence of new particles belonging to any new physics theory
and bounds on the masses of these new states are being set higher and higher, as shown for example in Fig.~\ref{fig:CMS-bounds}.
The left plot shows the CMS limits on the masses of different SUSY particles with the 7 and 8 TeV LHC dataset for different
search channels and integrated luminosity while the right plot shows the CMS mass limit on a generic top partner present in composite Higgs
models (CHMs) in function of its branching ratios into SM bosons and third generation quarks with the 8 TeV dataset and 19.6 fb$^{-1}$ of integrated luminosity.
Though it has to be stressed that these bounds strongly depend on the underlying model's assumptions
(a framework for the reinterpretation of the \emph{top partners} limits will be given in Chapter 5 of this Thesis) it is however
clear that a large part of parameter space of these BSM theories has already been ruled out by the first run of the LHC which
will test the remaining parameter space in the forthcoming run at 13 TeV of centre of mass energy.

We conclude this introduction by pointing out that the study of the Higgs boson properties is one of the main motivations for the proposal of a next generation of colliders. In fact, even without the discovery of any new particle in the second stage of the LHC running, understanding the properties of the Higgs boson is an essential step to shed light on the mechanism of electroweak symmetry breaking (EWSB).
From the point of view of the measurement of the Higgs couplings to SM matter, gauge bosons and the triple and quartic Higgs self couplings, the CERN machine has some limitations due to its hadron-hadron collider nature and in particular the latter two measurements 
seem respectively challenging and impossible to achieve with sufficient precision.
In conclusion, at the end of the LHC operations, it is not so unlikely that there will be a situation where the measurements of the Higgs properties
are achieved with an error not small enough to disentangle the BSM nature of this state.
The physics community must then make an important decision regarding a new generation of colliders, and at present both
hadron colliders, like the already mentioned HL-LHC and a 100 TeV LHC, and electron-positron colliders, such as the Compact Linear Collider (CLIC), the International Linear Collider (ILC) and new circular electron positron collider (TLEP), are being considered.
For this reason it is important to study the capability of such machines in testing particular BSM scenarios and in this
Thesis we will also focus on the potential of the ILC on testing the Higgs sector of a CHM.

\section{Plan of the Thesis}
The plan of the Thesis is the following.

In Chapter~\ref{chap-2} we will discuss the general properties of CHMs and present the particular model used as a framework
for most of the phenomenological analysis of this Thesis, the 4DCHM, discussing its main characteristics
and its implementation in Monte Carlo (MC) generators.

In Chapter 3 we will analyse the gauge sector of the 4DCHM presenting the LHC analysis of Drell-Yan (DY), diboson and $t\bar t$ production at the LHC and the discovery potential of the new gauge bosons present in the model of such processes.

In Chapter 4 we will discuss the Higgs sector of the 4DCHM showing the compatibility of our model with the 7 and 8 TeV LHC data and the prospects of a future $e^+e^-$ collider, namely the ILC,
in testing the scalar sector of the model.

Finally in Chapter 5 we present an analysis strategy and a dedicated  program that allows the recasting of the results of the experimental direct searches for \emph{top partners} in models with a complete and
extended fermionic sector, as typical in CHMs, for which a general analysis would otherwise be challenging.
We will apply our tool to simplified as well as physically motivated models, highlighting also its potentiality in setting bounds
on scenario not yet covered  by the experimental analyses.


\chapter{Composite Higgs and the 4DCHM}
\label{chap-2}
\lhead{Chapter 2. \emph{Composite Higgs and the 4DCHM}}

In this Chapter we discuss the idea and general properties of CHMs, starting from their origins up
to their more recent developments.
We introduce the formalism for construction of an effective Lagrangian thorough the Coleman, Callan, Wess and Zumino (CCWZ)
method specializing it then to the framework that has been chosen for our analysis, the 4DCHM.
After describing the properties of this concrete realization, whose most important feature is having an Higgs potential fully calculable making then the Higgs mass a dependent parameter of the model, we will then discuss its implementation
 into automated tools that allow a phenomenological study of the composite pseudo Nambu GB (pNGB) scenario for the Higgs scalar
in a complete way up to event generation level.

\section{The Higgs as a composite pNGB}

The idea of postulating the Higgs as a composite pNGB goes back to the 80's and was introduced in a
paper by Georgi and Kaplan titled \emph{$SU(2) \otimes U(1)_Y$ breaking by vacuum misalignment} \cite{Kaplan:1983fs}.
Their work was written shortly after the discover of the $W^\pm$ and $Z^0$ boson that confirmed that a spontaneously
broken $SU(2)_L \otimes U(1)_Y$ gauge group was describing the electroweak interactions, but at that time it was not
clear how this symmetry was broken.
The authors postulated that besides the Higgs mechanism of the SM theory and the Standard Hypercolour scheme
(now called Technicolour) a third possibility exists.

They described the Higgs boson as a bound state of a strongly interacting sector arising from a SSB at an energy scale $f \gg v_{SM}=246$ GeV, where $v_{SM}$ is the SM vacuum expectation value (VEV).
This breaking describes a certain number of GBs one of which should be identified with the Higgs boson,
that will then be a massless particle. It is the explicit breaking of the global symmetry of the strong sector that
can give rise to a potential
generated by radiative corrections which will then trigger EWSB and make then the Higgs a pseudo GB.
The EWSB is then realized at a scale different from that of the strong sector.
The Higgs boson created in this way will then be a composite state, therefore insensitive to radiative corrections
above its compositness scale, and lighter than the other resonances of the strong sector due to its pseudo GB nature.

As mentioned in Chapter~\ref{chap:1} this idea resembles quite closely what happens for pions in massless QCD, where it can be postulated that they are GBs arising from the spontaneous symmetry breaking of the $SU(2)_L\otimes SU(2)_R\to SU(2)_V$ chiral symmetry of the strong sector. If this symmetry was exact the pions would be massless GB, however, the coupling of the charged pions with the photon, that can be introduced by gauging the $U(1)_{em}\in SU(2)_V$ group, breaks this symmetry explicitly and, in the same way as for the composite Higgs case, a radiatively induced potential can be generated and give rise to a mass terms for $\pi^\pm$, leaving $\pi^0$ massless, thus explaining the mass difference between the charged and neutral pions.

Coming back to the composite Higgs idea, let's take a strongly interacting theory that possesses a global
symmetry $\mathcal{G}$ broken to a subgroup $\mathcal{H}$ at a scale $f$ that then develops, in virtue of the Goldstone theorem, $n$ GBs with
$n=dim(\mathcal{G})-dim(\mathcal{H})$.
If we assume now to have an external sector with an associated gauge symmetry $\mathcal{H}_g\in \mathcal{G}$
we then have that $n_0=dim(\mathcal{H}_0)-dim(\mathcal{H}\cap \mathcal{H}_g)$ GBs will be eaten to give mass to an equal
number of gauge bosons, leaving then $n-n_0$ GBs in the spectrum.
Let's now identify  $\mathcal{H}_g=G_{SM}$, where $G_{SM}$ is the SM gauge group, in order to have the minimum number of external (not belonging
to the strong sector) fields. In order to have a correct pNGB composite Higgs two conditions need to be
satisfied \cite{Contino:2010rs}:

\begin{enumerate}
\item $G_{SM}\in \mathcal{H}$ so that the EW gauge group is not broken at tree-level
\item Four of the GBs arising from $\mathcal{G}/\mathcal{H}$ should transform under $G_{SM}$ as the SM Higgs doublet.
\end{enumerate}

The most economical SSB pattern that respects these conditions is the one described in \cite{Agashe:2004rs} with the assignments $\mathcal{G}=SO(5)\otimes U(1)_X$ and $\mathcal{H}=SO(4)\otimes U(1)_X$. 
The breaking $\mathcal{G}/\mathcal{H}$ has the following characteristics

\begin{itemize}
\item It develops $dim(\mathcal{G})-dim(\mathcal{H})$=11-7=4 GBs, the minimum number to be identified with the SM Higgs doublet. They transform according to the fundamental representation of $SO(4)$ and they can then be identified with the SM Higgs doublet since $SO(4)\simeq SU(2)_L \otimes SU(2)_R$.
\item $\mathcal{G}_{SM}\in \mathcal{H}$ thanks to the identification $T^Y=T^{3R}+T^X$ and so the EW gauge group is unbroken at tree-level.
\end{itemize}

As already mentioned, it will be the weak but explicit breaking of the global symmetry of the strong sector, due to the gauging of a subgroup of $\mathcal{H}$ identified with $\mathcal{G}_{SM}$, that will be able to generate a 
potential for the Higgs at loop level and then, if this potential has the correct \emph{Mexican hat} shape,
deliver a VEV that can then break the EW symmetry.
The ratio between the electroweak scale, $v_{SM}$, and the scale of breaking of the strong sector, $f$,
is an important parameter in CHMs and the parameter $\xi=v/f$ substantially describes the misalignment 
between the vacuum of the theory before and after the gauging of $SU(2)_L\otimes U(1)_Y$ and functions of it
will also give the deviations of the composite Higgs couplings with respect to the SM theory.

\section{The 4DCHM}

We want to describe now the framework that we chose for our phenomenological analysis, the 4DCHM of \cite{DeCurtis:2011yx}, which is a concrete realization of the composite Higgs paradigm based on the $SO(5)/SO(4)$ breaking
pattern described in the previous Section.

\subsection{The gauge sector}

The starting point is the construction of a non linear $\sigma$ model which describes the breaking
of $SO(5)/SO(4)$ in terms of the low energy dynamics of the associated GBs, following the CCWZ prescription for which
we refer to Appendix~\ref{chap:App-A}.

We first introduce the pion matrix 
\begin{equation}
U(\Pi)=e^{i \sqrt{2} \pi^{\hat a} T^{\hat a}/f} \quad \Pi=\sqrt{2}\pi^{\hat a}T^{\hat a} \quad \hat a=1,2,3,4
\end{equation}
where $T^{\hat a}$ are the broken generators of $SO(5)/SO(4)$ whose algebra is described in Appendix B.
This matrix transforms under the symmetry $\mathcal{G}=SO(5)$ as 
\begin{equation}
U(\Pi)\to g U(\Pi) h^\dag(\Pi,g)\quad g\in \mathcal{G}, h(\Pi,g)\in \mathcal{H}
\end{equation}
from which we can construct the Maurer-Cartan form
\begin{equation}
U^\dag \partial_\mu U=i d_\mu^{\hat a} T^{\hat a} + i e_\mu^a T^a = i d_\mu + i e_\mu
\end{equation}
that transforms under $\mathcal{G}$ as
\begin{equation}
\begin{split}
& e_\mu \to h^\dag(\Pi,g)e_\mu h(\Pi,g)-i h(\Pi,g)\partial_\mu h^\dag(\Pi,g),\\
& d_\mu \to h^\dag(\Pi,g)d_\mu h(\Pi,g).
\end{split}
\end{equation}

At the lowest derivative order the generic $\mathcal{G}$ invariant Lagrangian written just in terms of the broken
generators is then
\begin{equation}
\mathcal{L}=\frac{f^2}{4}tr[d_\mu d^\mu]
\label{eq:ccwz}
\end{equation}
which, for the case $SO(5)/SO(4)$, can explicitly be written as (see again Appendix A)
\begin{equation}
\mathcal{L}=\frac{f^2}{2}(\partial_\mu \Phi)^T(\partial_\mu \Phi)
\label{eq:pion-lag}
\end{equation}
where
\begin{equation}
\Phi=U(\Pi)\phi_0=e^{i \sqrt{2} \pi^{\hat a} T^{\hat a}/f}, \quad \phi_0 =\delta_{i5}
\label{eq:phi-expr}
\end{equation}

To introduce spin-1 resonances into this construction the authors chose to add a second non-linear $\sigma$ model which describes the breaking $SO(5)_L \otimes SO(5)_R / SO(5)_V$.
This approach follows that of \cite{Son:2003et} (see also \cite{Becciolini:2009fu}) where the authors show that with a similar construction it is possible to describe the pions and the heaviest mesons like the $\rho$s and the $a_1$s in the context of QCD.
Following again the Maurer-Cartan formalism it is possible to show that the Lagrangian of eq.(\ref{eq:ccwz})
can be written in the case of this breaking pattern as
\begin{equation}
\mathcal{L}=\frac{f^2}{4}Tr[(\partial_\mu \Omega)^\dag(\partial^\mu \Omega)], \quad \Omega=\exp(i \theta^a T^a), \quad T^a \in SO(5).
\end{equation}

The complete Lagrangian then now reads
\begin{equation}
\mathcal{L}=\frac{f^2}{2}(\partial_\mu \Phi)^T(\partial^\mu \Phi)+\frac{f^2}{4}Tr[(\partial_\mu \Omega)^\dag(\partial^\mu \Omega)]
\end{equation}
and describes 4+10=14 scalars arising from the two $\sigma$ models that for now do not interact\footnote{
We are neglecting for the moment the $U(1)_X$ group since it is irrelevant to the present discussion, while it will be introduced again in describing the fermionic sector of the model.}.
We now gauge the diagonal subgroup of $SO(5)_R$ and $SO(5)$ adding a complete $SO(5)$ multiplet of resonances $\rho$
living in the $Adj(SO(5))$ obtaining then the following Lagrangian
\begin{equation}
\mathcal{L}=\frac{f^2}{2}(D^\mu \Phi)^T(D_\mu \Phi)+\frac{f^2}{4}Tr[(D_\mu \Omega)^\dag(D^\mu \Omega)] -\frac{1}{4}Tr[\rho^{\mu\nu}\rho_{\mu\nu}]-\frac{1}{4}Tr[A^{\mu\nu}A_{\mu\nu}]
\label{eq:lag-gauge}
\end{equation}
with
\begin{equation}
\begin{split}
& D^\mu \Phi = \partial^\mu \Phi - g_\rho \rho^\mu \Phi, \\
& D^\mu \Omega = \partial^\mu \Omega - g_0 i A^\mu \Omega + i g_\rho \rho^\mu \Omega,
\end{split}
\end{equation}
and where $\rho^{\mu\nu}$ and $A^{\mu\nu}$ are the field strengths for the new spin-1 resonances and the SM gauge fields.
The $A^\mu$ fields that have been introduced are responsible for the explicit breaking of the $SO(5)_L$ symmetry
and will contribute then to the generation of a loop induced Higgs potential.
Finally with $g_0$ we indicate in a compact way both the gauge couplings of $SU(2)_L\otimes U(1)_Y$, namely $\{g_0,g_{0y} \}$,
while the gauge couplings of $SO(5)\otimes U(1)_X$ have been fixed at a common value $g_\rho$.
Another possible construction for the pNGB Higgs, based on the $SO(5)/SO(4)$ breaking pattern, is the one formulated in \cite{Panico:2011pw} which adopts a different construction from ours in introducing the resonances.

Let's now pause and recap the particle content that has been introduced so far in the model.

\begin{itemize}
\item The two non linear $\sigma$ models parametrising the breaking of $SO(5)/SO(4)$ and $SO(5)_L\otimes SO(5)_R/SO(5)_V$ give 14 GBs in the spectrum.
\item The gauging of the diagonal subgroup of $SO(5)_R$ and $SO(5)$ causes, via a sort of Higgs like mechanism, the reabsorbing of 10 GBs into the longitudinal components of the $\rho$ resonances, leaving the spectrum with 10 massive vector bosons and 4 massless GBs.
\item The EWSB caused by a radiative generated Higgs potential will cause the disappearance, in the unitary gauge, of 3 of the GBs that will give a longitudinal component to the SM gauge bosons, leaving the spectrum with just one physical scalar: the (composite) Higgs boson.
\item The extension of the symmetry breaking pattern $SO(5)\to SO(5)\otimes U(1)_X$ has the only consequence of having 11 massive vector bosons instead of 10.
\end{itemize}

As a last step in the description of the gauge sector we introduce a convenient parametrization, called unitary gauge, in which the mixing terms between the $\pi$s and the vector resonances disappear.
This gauge is defined
\begin{equation}
\Omega=\exp(\frac{i \Pi}{2 f} )=U(\Pi)
\end{equation}
from which we can write an explicit expression for $\Omega$ as
\begin{equation}
\Omega=\textbf{1}+i \frac{\sin(h/(2f))}{h}\Pi+\frac{\cos{h/(2f)}-1}{h^2}\Pi^2, \quad h=\sqrt{h^{\hat a} h^{\hat a}}.
\label{eq:omega-expr}
\end{equation}

\subsection{The fermionic sector}

As explained in \cite{Agashe:2004rs} the gauge bosons contribution is not enough to cause a correct misalignment
of the vacuum that can provide
EWSB and for this reason CHMs usually present also an extended fermionic sector in their spectrum.
While the choice of the gauge sector of the model depends only on the pattern of symmetry breaking that develops
the pseudo GB Higgs, the fermionic sector has quite a large freedom of choice which will then cause a strong model
dependency.

In the model under consideration the new fermions are embedded in fundamental
(vector) representations of $SO(5)\otimes U(1)_X$ and the authors considered four copies of 
this representation in order to ensure the finiteness of the Higgs potential and
a dynamical generated shape with the correct VEV in order to trigger EWSB \cite{DeCurtis:2011yx}.
It is worth stressing, however, that other choices of fermionic embedding
are possible \footnote{Among the most common ones are the adjoint, \textbf{10}, and the spinorial, \textbf{4}, of $SO(5)$.}
and that, in order to have a simpler spectrum
for a phenomenological analysis, the finiteness of the Higgs potential is not always required
in literature.
However, a dynamical generated potential
is still one of the most attractive features of CHMs and  efforts have been made
in the calculation of it in other patterns of SBB, see for example \cite{Redi:2012ha} for the case of $SO(6)/SO(5)$ breaking.

Coming back to the 4DCHM, the authors considered the following $SO(5)\otimes U(1)_X$
quantum number assignments for the embedding of the new fermions
\begin{equation}
\begin{split}
& \Psi_T,\Psi_{\tilde T} \in (\textbf{5},2/3),\\
& \Psi_B,\Psi_{\tilde B} \in (\textbf{5},-1/3).
\end{split}
\label{eq:ferm1}
\end{equation}

Under $SO(4)$, the fundamental representation of $SO(5)$ decomposes as \textbf{4}+\textbf{1} which,
under $SU(2)_L\otimes SU(2)_R$, reads (\textbf{2},\textbf{2})+\textbf{1},
so that each representation of eq.(\ref{eq:ferm1}) is made up of a bidoublet
and a singlet of $SU(2)_L \otimes SU(2)_R$. Then, thanks to the identification $T^Y=T^{3R}+T^X$,
it is possible to assign to these extra quarks the SM quantum numbers and also embed the SM quarks,
the left handed third generation doublet and the two right handed third generation singlets, into incomplete
representations of $SO(5) \otimes U(1)_X$.
The four representations of eq.(\ref{eq:ferm1}) can then be written in terms of the $SU(2)_L$ doublets
$(X_{5/3},X_{2/3})$, $(T_{2/3},B_{-1/3})$ and $(Y_{-1/3},Y_{-4/3})$ as
\begin{equation}
\Psi_T=\Psi_{\tilde T}=\left(
\begin{array}{c c c c}
\begin{multirow}{2}{*}{\Big(}\end{multirow} & X_{5/3} & T_{2/3} & \begin{multirow}{2}{*}{\Big)}\end{multirow}\\
					    & X_{2/3} & B_{-1/3}\\
 & \multicolumn{2}{c}{\psi_{2/3}} & \\
\end{array}
\right)
\qquad
\Psi_B= \Psi_{\tilde B}=\left(
\begin{array}{c c c c}
\begin{multirow}{2}{*}{\Big(}\end{multirow} & T_{2/3} & Y_{-1/3} & \begin{multirow}{2}{*}{\Big)}\end{multirow}\\
					    & B_{-1/3} & Y_{-4/3}\\
 & \multicolumn{2}{c}{\psi_{-1/3}} & \\
\end{array}
\right)
\label{eq:ferm2}
\end{equation}
where with $X$ and $Y$ we have labelled the quarks belonging to the $SU(2)_L$ doublet with exotic (not present in the SM) hypercharge and the subscript indicates the electric charge of the quark.
We can then see from eq.(\ref{eq:ferm2}) that the SM doublet $(t_L, b_R)$ can be embedded in both the vector representations with charge $X=2/3$ and $X=-1/3$ respectively, with a proper assignment of the $SU(2)_R$ charge, while the SM singlets $t_R$ and $b_R$ needs to be identified with $\psi_{2/3}$ and $\psi_{1/3}$ respectively.

Differently from the SM fermions these new quarks have the properties that their left
and right handed component transforms in the same way under the gauge group (for this reason they are called vector-like quarks), so we haven't labelled them in eq.(\ref{eq:ferm2}) with a chirality index, that needs to be done in writing explicitly the embedding of the SM quarks into incomplete representations of $SO(5)\otimes U(1)_X$.

Before writing the fermionic Lagrangian of the 4DCHM we want to discuss briefly the motivations for
extending just the third generation of quarks with a new sector and the motivation is twofold.
Firstly, since the SM top quark is the main cause of the \emph{hierarchy problem}, it is natural to extend
as a first instance just the heavy quark generation in order to provide cancellations so as to
restore the naturalness of the EW scale and for this reason the new quarks that appear in CHMs are usually called 
\emph{top partners}, to emphasize the role that they have in relieving the \emph{hierarchy problem}.
Secondly, CHMs also present a mechanism called \emph{partial compositness} \cite{Kaplan:1991dc} that tries to explain the mass
hierarchy between the masses of the SM fermions (especially the top-bottom hierarchy) postulating
the existence of linear mixing between them and the new quarks.
This mixing allows the SM quarks to interact directly with the composite sector in which the Higgs
boson lives, and then generate  Yukawa interactions for the SM quarks.
The idea is that the larger the mixing angle the more the SM fermions are massive and, since the top quark is the heaviest of them, the first natural extension of a quark sector
will again be related to the third generation of quarks. These two motivations are of course not
separated and usually one speaks of \emph{CHMs with partial compositness}.

With this in mind, and dictated by the symmetry of the model chosen, the fermionic Lagrangian of the 4DCHM is

\begin{equation}
\begin{split}
\mathcal{L}_{fermions}&=\mathcal{L}_{fermions}^{el}+ (\Delta_{t_L}\bar{q}^{el}_L\Omega\Psi_T+\Delta_{t_R}\bar{t}^{el}_R\Omega\Psi_{\tilde{T}}+h.c.)\\
&+\bar{\Psi}_T(i\slashed{D}-m_*)\Psi_T+\bar{\Psi}_{\tilde{T}}(i\slashed{D}-m_*)\Psi_{\tilde{T}}\\
&-(Y_T\bar{\Psi}_{T,L}\Phi^T\Phi\Psi_{\tilde{T},R}+m_{Y_T}\bar{\Psi}_{T,L}\Psi_{\tilde{T},R}+h.c.)\\
&+(T\rightarrow B)
\end{split}
\label{eq:lag-ferm}
\end{equation}

where $\mathcal{L}_{fermions}^{el}$ indicates the kinetic Lagrangian of the third generation SM fermions, $\slashed D$ are the 
covariant derivatives with respect to the $SO(5)\otimes U(1)_X$ fields and, contrary to the general choice present
in the original work \cite{DeCurtis:2011yx}, we have chosen a common mass parameter, $m_*$, for all four 
representations of $SO(5)\otimes U(1)_X$ present in the model.

We have so far introduced the composite Higgs boson via a low energy effective theory, through the CCWZ construction, with the addition of extra spin-1 and spin-1/2 resonances.
This is the most popular approach used in literature for considering the pNGB Higgs state and studying its properties at colliders, but it should be noticed that there is no guarantee that there is an UV theory from which these constructions, and so the 4DCHM, can arise.
Attempts have been made in order to provide a UV completion of models with a composite Higgs based on the partial compositness paradigm, based on both SUSY and non-SUSY approaches \cite{Caracciolo:2012je,Barnard:2013zea}.
While a discussion on this issue is beyond the scope of this Thesis, it is however important to stress that our calculations based on a low energy approach remains valid for phenomenological analyses even though the general theory is not known.
To take into account effects induced by strong interactions, naive dimensional analysis power counting \cite{Manohar:1983md,Georgi:1986kr} can be used and, without wanting to be exhaustive on this subject, we refer to \cite{Georgi:1992dw,Antola:2011at}  where strongly interacting QCD-like theories are considered.
Though not rigorously motivated, we can then state that the low energy effective theory approach commonly adopted, and also limited to the lightest set of extra resonances that can be introduced in these constructions, will provide the necessary ingredients for a phenomenological collider study of the composite Higgs idea, although a construction of a UV completion of this paradigm remains an interesting, and challenging, question.

Finally, even though it has been shown in a recent work \cite{DeCurtis:2014oza} that this is not the most general
Lagrangian dictated by the symmetries, it will be, together with the gauge sector Lagrangian
in eq.(\ref{eq:lag-gauge}), the framework for our phenomenological analysis of the composite Higgs scenario.

\subsection{The particle spectrum and the parameter space of the 4DCHM}

The extended gauge symmetry and the new fermionic sector bring new particles into the model and 
in particular in the 4DCHM there are 11 degrees of freedom associated with vector bosons and 20 associated 
with spin 1/2 Dirac fermions.
The spin 1 resonances can be identified according to their $SO(5)$ generators (see Appendix \ref{chap:App-B}) 
and it is possible to recast their 11 degrees of freedom in terms of 3 charged and 5 neutral gauge bosons, 
that will be from now on called in a generalized way as $W_i^\prime$ and $Z_j^\prime$ with $i=1,2,3$
and $j=1,\dots,5$, where an increase in the subscript number will indicate a higher mass for 
the corresponding particle.
From eq.(\ref{eq:ferm2}) we see that the fermionic spectrum of the 4DCHM also contains
 fermions with an exotic electric charge, besides the usual ones with the SM $U(1)_{em}$ quantum number.
In particular we have 8 new quarks with charge 2/3 and 8 with charge $-1/3$, called in a 
generalized way $t^\prime_i$ and $b^\prime_i$, and 2 quarks with charge 5/3 and $-4/3$,
called $X_j$ and $Y_j$ respectively\footnote{With the same mass ordering convention as for the gauge boson, $i=1,\dots,8$ 
and $j=1,2$.}.
In the spirit of the \emph{partial compositness} these fields mix with their SM counterparts (except for $X$ and
$Y$ for which no SM partner is present)
in such a way that the physical states will be a superposition of the original fields
\begin{equation}
\begin{split}
& | {\textrm{SM}}_{phys}\rangle = \cos{\psi} | {\textrm{SM}} \rangle + \sin{\psi} | {\textrm{4DCHM}} \rangle\\
& | {\textrm{4DCHM}}_{phys}\rangle = -\sin{\psi} | {\textrm{SM}} \rangle + \cos{\psi} | {\textrm{4DCHM}} \rangle\\
\label{eq:partialcompositness}
\end{split}
\end{equation}
where in principle the mass eigenstate could be a superposition of more than just two fields 
and where the mixing angles are dependent on the model parameter.
The remainder of the spectrum is made up of the rest of the SM fermions, the three generations of leptons 
and the two lightest generations of quarks which will be treated as 
massless particles belonging to the elementary sector, and the Higgs boson.

From the Lagrangian of eq.(\ref{eq:lag-gauge}) and eq.(\ref{eq:lag-ferm})
we notice that the 4DCHM presents quite a large parameter space.
In particular the gauge sector parameters are the model scale $f$, the couplings of the
SM gauge group $g_0$ and $g_{0Y}$, the $SO(5)\otimes U(1)_X$ gauge
coupling called $g_{\rho}$ and, after the Higgs field develops a VEV, $\langle h \rangle$.
The parameters of the fermionic sector are $m_*$, the mass parameters of the new quarks,
the mixing parameters $\Delta_{t/L,R}$ and $\Delta_{b/L,R}$ responsible for the elementary-composite mixing of the top and bottom sectors, and $Y_{T,B}$, $m_{Y_{T,B}}$ responsible for 
the interactions of the extra quarks with the Higgs boson.
Not all these parameters are independent and in the next subsection, while describing the 
implementation of the 4DCHM into dedicated automated tools, we will also explain the
constraints on the model parameters due to physical observables.

\subsection{The implementation of the 4DCHM into automated tools}
\label{subsec:2-implement}

It is clear that, due to the large number of particles and parameters
of the model chosen for our analysis, a detailed study of the latter, considering
the full particle spectrum without any expansion approximation in function of 
the model parameters, would have been a really hard task to achieve.
However in order to perform a rigorous phenomenological analysis of this framework at 
the LHC it is necessary to take into account all the possible contributions that can modify the 
physical observables and it is therefore important to keep the full particle spectrum of the
model and to use the least approximation possible.
For this reason we chose to implement the 4DCHM into automated tools to allow for an exact calculation
of the model spectrum and a fast phenomenological analysis up to event generation.
In this respect the publicly available tools that we have used are LanHEP \cite{Semenov:2010qt},
which is a package for the automatic generations of Feynman rules
in a format that can be read by several MC generators, among which there is CalcHEP \cite{Belyaev:2012qa} 
which has been our main MC generator for all of the analyses that we will discuss in this Thesis.
The model files in CalcHEP format have been uploaded for public use onto the \emph{High Energy Physics Model Database} (HEPMD)
\cite{hepmdb} under the name \emph{4DCHM}\footnote{That can be found at the following URL: 
\url{http://hepmdb.soton.ac.uk/hepmdb:1212.0120}.}.

Nevertheless the parameter space of the model ought to be constrained to be 
given as an input to our MC generator, and in order to do so we have written a 
stand-alone Mathematica \cite{mathematica} program that is able to perform a parameter
scan accounting for various constraints.
We firstly wrote the gauge and the fermionic mass matrices, so after the shifting
$h\to \langle h \rangle$, in function of the following parameters
\begin{equation}
 f,g_\rho,g_0,g_{0y},\langle h \rangle, m_*,\Delta_{t/L,R}, \Delta_{b/L,R}, Y_{T,B}, m_{Y_{T,B}} 
 \label{eq:param}
\end{equation}
where the neutral and charged spin-1 mass matrices just depends on the first five.
These $7\times7$ and $4\times4$ matrices, containing the $\gamma$, $Z$ and $W$ fields beside the new resonances, have been diagonalised in order to find the corresponding eigenvalues (the masses of the physical
states) and eigenvectors (for the determinations of the angles in eq.(\ref{eq:partialcompositness})) constraining 
$g_0$, $g_{0y}$ and $\langle h \rangle$ with respect to the EW precision observables
\begin{equation}
\begin{split}
& \frac{1}{\alpha} = 128.88, \\
& M_Z = 91.1876 \textrm{ GeV}, \\
& G_F = 1.16639 \cdot 10^{-5} \textrm{ GeV}^{-2},
\end{split}
\end{equation}
and leaving $f$ and $g_\rho$ as input parameters.
In the fermionic sector, due to the larger number of parameters and particles, we have chosen to perform a random scan on 
$m_*$, $\Delta_{t/L,R}$, $\Delta_{b/L,R}$, $Y_{T,B}$, $m_{Y_{T,B}}$ after fixing the model scale $f$ and the coupling $g_\rho$.
The scan has been performed requiring that the lowest eigenvalues of the 9$\times$9 matrices for the charge 2/3 and $-1/3$ quarks 
 correspond with the top and bottom mass that, taking into account the data from LEP, SLC, Tevatron and LHC, has been chosen
to lie in the conservative interval $165$ GeV $<m_{top}<175$ GeV and $2$ GeV $<m_{bot}<6$ GeV, and requiring
the generation of a Higgs mass $m_H$ compliant with the latest LHC results\footnote{In some parts of this
Thesis these bounds could have been made more or less stringent. In that case we will explicitly mention the range of values 
used in a specific analysis.}.

We conclude this Chapter by mentioning that, being CalcHEP a default tree-level MC generator, it has been necessary in the 
analysis of the Higgs sector of the 4DCHM to manually implement into it loop-induced interactions such as $hgg$ and $h\gamma\gamma$, in order to be able to study the related phenomenology.
The results of the calculations of these vertices will be presented in the related Chapter, when we will discuss the analysis of the 4DCHM Higgs boson.


\chapter{Phenomenology of the gauge sector of the 4DCHM}
\label{chap-3}
\lhead{Chapter 3. \emph{Phenomenology of the gauge sector of the 4DCHM}}

In this Chapter we analyse the LHC phenomenology of the $Z^\prime$s and $W^\prime$s of the 4DCHM,
focusing on the study of the DY and diboson production modes of such states.
After describing in detail the properties of the gauge sector of the model, we discuss the potentiality of such processes in discovering the new gauge bosons belonging to the 4DCHM, both in cross sections and in asymmetries distributions, also paying  attention to the presence of multiple degenerate resonances that can be distinguishable in certain regions of the parameter space.
Finally, since these new resonances have sizeable couplings to the third generation of SM quarks, we also study top-antitop pair production at the LHC as a test-bed for discovering $Z^\prime$s states again both in cross section as well as in various asymmetries.
However, since the EW precision data generally disfavour extra gauge bosons with mass below the TeV range and with large couplings to light fermions, we will just focus on the 14 TeV stage of the CERN machine.

\section{Masses and couplings of the gauge bosons of the 4DCHM}

From the Lagrangian of eq.(\ref{eq:lag-gauge}), using the explicit expression of $\Phi$ and $\Omega$ given in
eq.(\ref{eq:phi-expr}) and eq.(\ref{eq:omega-expr}) and expanding the Higgs field $h$ onto its VEV, it is possible to write an exact expression for the neutral and charged spin-1 mass matrices that can be diagonalised in order to obtain masses and rotation matrices of the model.
Plugging the rotations into the fermionic Lagrangian of eq.(\ref{eq:lag-ferm}) and in the Higgs field dependent part of eq.(\ref{eq:lag-gauge}) we can then find the couplings of the SM gauge bosons and extra resonances to the SM and extra fermions, and also the coupling of the spin 1 states to the Higgs boson.
As mentioned, in all our phenomenological analysis we have performed an exact numerical diagonalisation of the relevant matrices, however we will present here some useful analytical expressions obtained at the leading order in the expansion parameter $\xi=v^2/f^2$ that, we recall, measures the misalignment between the vacuum of the theory before and after the $SU(2)_L\otimes U(1)_Y$ gauging or, in other words, measures the degree of compositness of the theory.

\subsection{Masses of the gauge bosons}

At the leading order in the expansion parameter $\xi=v^2/f^2$ the expressions for the neutral gauge boson masses, $M_{Z^\prime_{i+1}}>M_{Z^\prime_i}$, are
\begin{equation}
\begin{split}
& M^2_{\gamma}= 0,\\
& M^2_{Z}\simeq  \frac{f ^2}{4} g_\rho^2(s^2_\theta+\frac{s^2_\psi}{2})  \xi,\\
& M^2_{Z^\prime_1}= f ^2g_\rho^2,\\
& M^2_{Z^\prime_2}\simeq \frac{f ^2g_\rho^2}{ c_\psi^2}(1-\frac{s^2_\psi c^4_\psi }{4 c_{2\psi}}\xi),\\
& M^2_{Z^\prime_3}\simeq \frac{f ^2g_\rho^2}{ c_\theta^2}(1-\frac{s^2_\theta c^4_\theta }{4 c_{2\theta}}\xi),\\
& M^2_{Z^\prime_4}=2 f ^2g_\rho^2,\\
& M^2_{Z^\prime_5}\simeq 2 f ^2g_\rho^2\left[1+\frac 1 {16} (\frac 1 {c_{2\theta}}+\frac 1{2 c_{2 \psi}})\xi\right]
\end{split}
\label{eq:neutral-gauge-mass}
\end{equation}
where
\begin{equation}
\tan\theta=\frac{s_\theta}{c_\theta}=\frac{g_0}{g_\rho}, \qquad  \tan\psi=\frac{s_\psi}{c_\psi}=\sqrt{2} \frac{g_{0Y}}{g_\rho}
\end{equation}
and where the masses of $Z^\prime_1$ and $Z^\prime_4$\footnote{And of course of the $\gamma$.} are exact at all orders in the expansion parameter and are therefore completely determined by the composite sector, so that they do not receive any contribution from EWSB.
For the charged sector we have, $M_{W^\prime_{i+1}}>M_{W^\prime_i}$ ,
\begin{equation}
\begin{split}
& M^2_{W}\simeq  \frac{f ^2}{4} g_\rho^2s^2_\theta  \xi,\\
& M^2_{W^\prime_1}= f ^2g_\rho^2,\\
& M^2_{W^\prime_2}\simeq \frac{f ^2g_\rho^2}{ c_\theta^2}(1-\frac{s^2_\theta c^4_\theta}{2 c_{2\theta}}\xi),\\
& M^2_{W^\prime_3}\simeq2 f ^2g_\rho^2 (1- \frac{s^2_\theta}{4c_{2\theta}}\xi)
\end{split}
\label{eq:charged-gauge-mass}
\end{equation}
where $W^\prime_1$ does not receive any mass correction after EWSB.
From eq.(\ref{eq:neutral-gauge-mass}) and eq.(\ref{eq:charged-gauge-mass}) we see that the neutral spectrum is made up of three resonances with mass $\simeq f g_\rho$, called the vector resonances, and two with mass $\simeq \sqrt{2} f g_\rho$, called the axial resonances.
The former mainly correspond to the gauge bosons associated with the $T^{3R}-T^X$, $T^{3L}$ and $T^{3R}+T^X$ generators while the latter with $T^{\hat 3}$ and $T^{\hat 4}$.
The charged partners of the vector resonances are $W^\prime_1$ and $W^\prime_2$, associated with $T^{R\pm}\propto T^{R1}\pm i T^{R2}$ and $T^{L\pm}\propto T^{L1}\pm i T^{L2}$, while the axial partner 
is $W^\prime_3$, associated with the $T^{\hat \pm}\propto T^{\hat 1}\pm T^{\hat 2}$ combination.

\subsection{Couplings of the gauge bosons to SM light fermions}

The couplings of the neutral gauge bosons to the light SM fermions can be expressed by the following Lagrangian
\begin{equation}
{\mathcal L}\supset\sum_f\big[ e\bar\psi^f \gamma_\mu Q^f \psi^f A^\mu+ \sum_{i=0}^5   (\bar\psi^f_L  g_{Z^\prime_i}^L(f) \gamma_\mu  \psi^f_L+\bar\psi^f_R  g_{Z^\prime_i}^R(f) \gamma_\mu  \psi^f_R ) Z^{\prime\mu}_i \big]
\label{eq:neutral-gauge-coupl-light}
\end{equation}
where $\psi_{L,R}=[(1\pm\gamma_5)/2]\psi$ and where $Z^\prime_0$ and $A$ corresponds to the neutral SM gauge bosons $Z$ and $\gamma$.
The photon field is coupled to the electromagnetic current in the standard way with the electric charge which is defined as
\begin{equation}
e=\frac{g_Lg_Y}{\sqrt{g_L^2+g_Y^2}}, \qquad g_L=g_0 c_\theta, \qquad g_Y = g_{0Y} c_\psi.
\label{eq:electric-charge}
\end{equation}
The $g^{L,R}_{Z^\prime_i}$ couplings have the following expression
\begin{equation}
g_{Z^\prime_i}^L(f)= A_{Z^\prime_i}T^3_L(f)+ B_{Z^\prime_i} Q^f, \quad\quad
g_{Z^\prime_i}^R(f)=  B_{Z^\prime_i}Q^f,
\end{equation}
where, at the leading order in the expansion parameter $\xi$, $A_{Z^\prime_i}$ and $B_{Z^\prime_i}$ read
\begin{equation}
\begin{alignedat}{2}
& A_{Z}\simeq \frac{e}{s_\omega c_\omega}\big[1+(c^2_\omega a_Z+ s^2_\omega b_Z)\xi\big], \qquad
&& B_{Z}\simeq -e\frac{s_\omega}{c_\omega}(1+b_Z\xi), \\
& A_{Z^\prime_1}=0, \qquad && B_{Z^\prime_1}=0,\\
& A_{Z^\prime_2}\simeq  -\frac{e}{c_\omega} \frac{s_\psi}{c_\psi} \Big[1+(\frac{c_\omega}{s_\omega} a_{Z^\prime_2}-b_{Z^\prime_2})\xi\Big], 
\qquad &&  B^\prime_{Z^\prime_2}\simeq \frac{e}{c_\omega} \frac{s_\psi}{c_\psi} \Big[1-b_{Z^\prime_2}\xi\Big],\\
& A_{Z^\prime_3}\simeq  -\frac{e}{s_\omega}\frac{s_\theta}{c_\theta}\big[1+(a_{Z^\prime_3}+\frac{s_\omega}{c_\omega} b_{Z^\prime_3})\xi\big]   
,\qquad &&  B_{Z^\prime_3}\simeq   \frac{e}{c_\omega} \frac{s_\theta}{c_\theta} b_{Z^\prime_3}\xi,\\
& A_{Z^\prime_4}=0, \qquad && B_{Z^\prime_4}=0,\\
& A_{Z^\prime_5}\simeq  e(\frac{1}{s_\omega} a_{Z^\prime_5}-\frac{1}{c_\omega} b_{Z^\prime_5})\sqrt{\xi},  
\qquad && B_{Z^\prime_5}\simeq  \frac{e}{c_\omega} b_{Z^\prime_5}\sqrt{\xi},
\label{eq:coup-neu-me}
\end{alignedat}
\end{equation}
with
\begin{equation}
\tan{\omega}=\frac{g_Y}{g_L}, \qquad e=g_L s_\omega = g_Y c_\omega, \qquad \frac{e}{s_\omega c_\omega}=\sqrt{g_L^2+g_Y^2},
\end{equation}
and
\begin{equation}
\begin{alignedat}{2}
& a_{Z}=  (2 s_\theta^2+s_\psi^2)(4 cˆ_\theta^2-1)/32, \qquad 
&& b_{Z}=  (2 s_\theta^2+s_\psi^2)(4 cˆ_\psi^2-1)/32, \\
& a_{Z^\prime_2}= \frac{\sqrt{2} s_\theta s_\psi c_\psi^6}{4(c_\psi^2-c_\theta^2)(2 c_\psi^2-1)}, 
\qquad && b_{Z^\prime_2}= \frac{c_\psi^4(2-7c_\psi^2+9 c_\psi^4-4 c_\psi^6)}{8s_\psi^2 (1-2 c_\psi^2)^2},\\
& a_{Z^\prime_3}=   \frac{-2 c_\theta^4+5 c_\theta^6-4 c_\theta^8}{4(1-2 c_\theta^2)^2},
\qquad && b_{Z^\prime_3}=  \frac{\sqrt{2} s_\theta s_\psi c_\theta^6}{4 (2 c_\theta^2-1)(c_\theta^2-c_\psi^2) },\\
& a_{Z^\prime_5}=   \frac{s_\theta}{2 \sqrt{2}(1-2 c_\theta^2)},
\qquad && b_{Z^\prime_5}= - \frac{s_\psi}{4(1-2 c_\psi^2)}.
\end{alignedat}
\end{equation}

The $Z^\prime_1$ and $Z^\prime_4$ couplings to the light quarks are zero at all orders in the expansion parameter $\xi$ and these gauge bosons will therefore be inert for the processes under consideration in this Chapter, as they don't couple to the proton constituents. 

In the same way we can work out the expressions for the charged sector, that are
\begin{equation}
\mathcal{L}_{CC}=\sum_{i=0}^4 g^+_{W^\prime_i}W^{\prime+}_iJ^-+h.c.
\end{equation}
with $J^\pm=(J^1\pm i J^2)/2$, $ J^i_\mu=\bar\psi T^i_L \gamma_\mu[(1-\gamma_5)/2]\psi$, with $W^{\prime +}_0$ corresponding to the SM $W^+$ boson and
\begin{equation}
\begin{split}
& g_W^\pm\simeq-\frac{g_\rho s_\theta}{\sqrt 2}(1+\frac  {s_\theta}{4 c_\theta}  a_{12}  \xi)\label{Wff},\\
& g_{W^\prime_1}^\pm= 0,\\
& g_{W^\prime_2}^\pm\simeq \frac{g_\rho s_\theta^2}{\sqrt{2}c_\theta}    (1+\frac 1 4 (a_{22}-\frac{ c_\theta}{s_\theta}a_{12})\xi ),\\
& g_{W^\prime_3}^\pm\simeq \frac{g_\rho s_\theta^2}{2\sqrt{2}c_\theta} a_{24}\sqrt{\xi},
\end{split}
\end{equation}
where 
\begin{equation}
a_{12}= -\frac 1 4 c_\theta (1-4 c_\theta^2) s_\theta,  \quad \quad
a_{22}=-\frac{ c_\theta^2}{4(1-2 c_\theta^2)^2}, \quad \quad
a_{24}=-\frac{c_\theta }{\sqrt{2}(1-2c_\theta^2)},
\end{equation}
and where in this case it is  $W^\prime_1$ that will be inert with respect to the processes under consideration.
The masses of the gauge bosons and their couplings to light generations of quarks and all the three generations of leptons are independent of the parameters of the fermionic sector at all orders in the expansion parameter $\xi$ since the partial compositness mechanism is implemented in the 4DCHM just for the third generation of quarks.

Finally, as mentioned, the parameters $f,g_\rho,g_0,g_{0Y}$ and $\xi$ (recast from $\langle h \rangle$) are not all independent and three of them can be constrained from $M_Z$, eq.(\ref{eq:neutral-gauge-mass}), the electric charge $e$, eq.(\ref{eq:electric-charge}),
and the Fermi constant $G_F$ obtained from the charged current processes at $q^2\ll M^2$,
\begin{equation}
G_F=\sqrt{2}\frac{g_0^2}{8}\sum_{i=0}^4\frac{g^2_{W^\prime_i}}{M^2_{W^\prime_i}},
\end{equation}
leaving us with just two free parameters in the gauge sector that have been chosen to be $f$ and~$g_\rho$.
Again, these constraints have been imposed in an exact numerical way in implementing the model for our forthcoming
phenomenological analysis.


\section{Drell-Yan signals at the LHC}

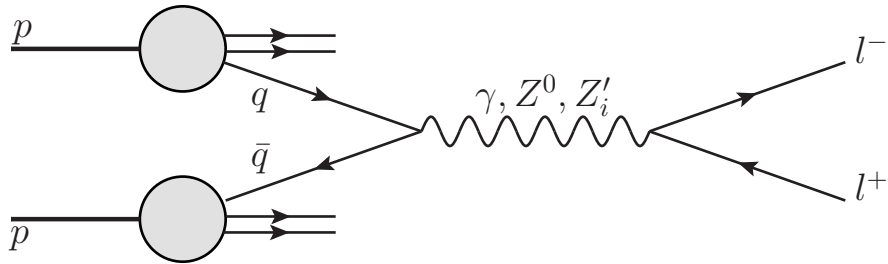
\begin{figure}[!h]
  \begin{picture}(292,118) (55,-47)
    \SetWidth{1.0}
    \SetColor{Black}
    \GOval(160,34)(16,16)(0){0.882}
    \GOval(160,-30)(16,16)(0){0.882}
    \SetWidth{1.7}
    \Text(97,36)[lb]{\Large{\Black{$p$}}}    
    \Line[](96,34)(144,34)
    \SetWidth{1.7}
    \Text(96,-42)[lb]{\Large{\Black{$p$}}}      
    \Line[](96,-30)(144,-30)
    \SetWidth{1.0}   
    \Line[arrow,arrowpos=0.5,arrowlength=5,arrowwidth=2,arrowinset=0.2](175,39)(217,39)
    \Line[arrow,arrowpos=0.5,arrowlength=5,arrowwidth=2,arrowinset=0.2](176,33)(217,33)
    
    \Line[arrow,arrowpos=0.5,arrowlength=5,arrowwidth=2,arrowinset=0.2](176,-29)(217,-29)
    \Line[arrow,arrowpos=0.5,arrowlength=5,arrowwidth=2,arrowinset=0.2](175,-35)(217,-35)
    
    \Text(186,10)[lb]{\Large{\Black{$q$}}}  
    \Line[arrow,arrowpos=0.5,arrowlength=5,arrowwidth=2,arrowinset=0.2](175,29)(249,3)
    \Text(186,-13)[lb]{\Large{\Black{$\bar q$}}}      
    \Line[arrow,arrowpos=0.5,arrowlength=5,arrowwidth=2,arrowinset=0.2](249,3)(176,-23)    
    \Text(270,9)[lb]{\Large{\Black{$\gamma,Z^0,Z^\prime_i$}}}      
    \Photon(249,3)(334,3){5.5}{6}
    \Text(412,29)[lb]{\Large{\Black{$l^-$}}}     
    \Line[arrow,arrowpos=0.5,arrowlength=5,arrowwidth=2,arrowinset=0.2](334,3)(407,29)
    \Text(412,-23)[lb]{\Large{\Black{$l^+$}}}      
    \Line[arrow,arrowpos=0.5,arrowlength=5,arrowwidth=2,arrowinset=0.2](407,-23)(334,3)
  \end{picture}
  \caption[Feynman diagram for the Drell-Yan process]{Neutral current DY processes via SM and extra gauge bosons. The corresponding charged current channel
  is mediated by a charged SM or extra gauge boson, $W^-$ or $W^{\prime-}_i$, and leads to a $l^-\bar \nu_l$ final state ($l^+ \nu_l$ for the charged conjugated process).}
  \label{fig:dy-proc}
\end{figure}

The DY mechanism is one of the most important probes in the search for new vector boson resonances associated with possible BSM scenarios. It consists of di lepton production from hadron-hadron scattering via neutral current (NC) or charged current (CC) processes, as shown in Fig.~\ref{fig:dy-proc}.

The importance of this process arises from the fact that from the theoretical point of view such a mechanism is well under control as higher order effects from both EW and 
QCD interactions are known up to one and two loop contributions respectively (see, e.g., \cite{Campbell:2006wx} for a review), while
from the experimental point of view the directions and energies of the particles of such final state are well reconstructed in a generic detector at an hadron collider, consisting in fact of electrons or muons ($e$ or $\mu$) and their related neutrinos.
Therefore, for all such reasons, this class of processes is ideal for identifying the mass of the intermediate bosons being produced and studying their properties, and nowadays the LHC offers us the chance to test DY phenomenology at high energy proton proton scattering.
DY processes for analogous scenarios have already been studied in literature (see  \cite{Agashe:2007ki} and \cite{Agashe:2008jb} for a general review), however, the purpose of this Thesis is to perform a detailed phenomenological study of a specific CHM, the 4DCHM, that implies also taking into account all the possible model dependent effects that can escape from a general analysis.
In this respect we will stress the importance of:
\begin{itemize}
\item[-] the impact of the fermionic parameter of the model onto the line shape of the emerging gauge bosons,
\item[-] the possibility of being able to resolve, albeit limited to certain regions of the parameter space, the two lightest (quasi) degenerate neutral active (non-inert) resonances
\end{itemize}

\subsection{Parameter space and benchmark points}
\label{subsec:2-bench}

Before tackling the analysis of the DY processes we ought to comment on the bounds from the EWPT that affect the 4DCHM.
As is well known, extra gauge bosons give a positive contribution to the Peskin-Takeuchi $S$ parameter and the requirement
of consistency with the EWPT generally gives a bound on the masses of these resonances in CHMs around few TeV \cite{Marzocca:2012zn}, while the fermionic 
sector is quite irrelevant for $S$ since the extra fermions are weakly coupled to the SM gauge boson.
Either way, as stressed in \cite{Contino:2006qr}, when dealing with effective theories one can only parametrize $S$ rather than calculating it.
In other words, since the construction of the 4DCHM just deals with the lowest lying resonances neglecting those that, allowed by the symmetries, may exist and that are not introduced in the construction under our consideration\footnote{While they arise naturally in the context of CHMs formulated in 5 dimensions, see for example \cite{Agashe:2004rs}.}, we need to invoke an ultra violet (UV) completion for the physics effects that we are not including.
These effects can in general make the bound stronger, but not significantly. For example in order to satisfy $S\le0.3$ in the minimal composite Higgs model of Ref.\cite{Agashe:2004rs}, the lowest vectors have to be heavier that $1.6-2.3$ TeV, depending on the choice of some coupling constants. Higher order operators in the chiral expansions can however compensate for $S$, albeit with some tuning, and one example is illustrated in \cite{Contino:2006qr}, while in \cite{DeCurtis:2011yx} it is shown that the inclusion of non-minimal interactions in the picture of the 4DCHM can lead to the reduction of the $S$ parameter.
For all of these reasons in our phenomenological analysis we will choose, omitting a systematic and detailed study of the EWPT, values of the spin-1 resonances masses around 2 TeV in order to avoid big contributions to the $S$ parameter, that will reflect on the choice of values of $f\simeq$~1~TeV and $g_\rho\simeq$~2, according to eq.(\ref{eq:neutral-gauge-mass}) and eq.(\ref{eq:charged-gauge-mass}).
Furthermore we have checked that our 4DCHM parameter choices, for any sorted point, are compatible with  LHC direct searches for heavy gauge bosons \cite{Aad:2011fe,Chatrchyan:2012meb,Hayden:2012np,Chatrchyan:2012it} while, for the moment, we are neglecting the bounds arising from the direct searches of extra fermions.
In particular, we will present our results for the following combinations of the parameter $f$ and $g_\rho$
\begin{equation}
\begin{split}
&a)\quad (f,g_\rho)=(750\textrm{ GeV},2)\\
&b)\quad (f,g_\rho)=(800\textrm{ GeV},2.5)\\
&c)\quad (f,g_\rho)=(1000\textrm{ GeV},2)\\
&d)\quad (f,g_\rho)=(1200\textrm{ GeV},1.8)
\end{split}
\label{eq:bench-fvar}
\end{equation} for each of which we have scanned on the parameters of the fermionic sector with the following constraints: 500 GeV $<\Delta_{t/L,R},Y_{T,B}<$ 5000 GeV, $-5000$ GeV $< m_{Y_{T/B}}<$ $-500$ GeV and, in the spirit of partial compositness, 50 GeV $<\Delta_{b/L,R},Y_{T,B}<$ 500 GeV, requiring the values of $m_{top}$, $m_{bot}$ and $m_{H}$ described in Section~\ref{subsec:2-implement}.
These choices of $f$ and $g_\rho$ have been made with the purpose of illustrating the salient features of the model for a quasi constant mass of the gauge resonances, varying however the model scale $f$.

\begin{figure}[!h]
\centering
\epsfig{file=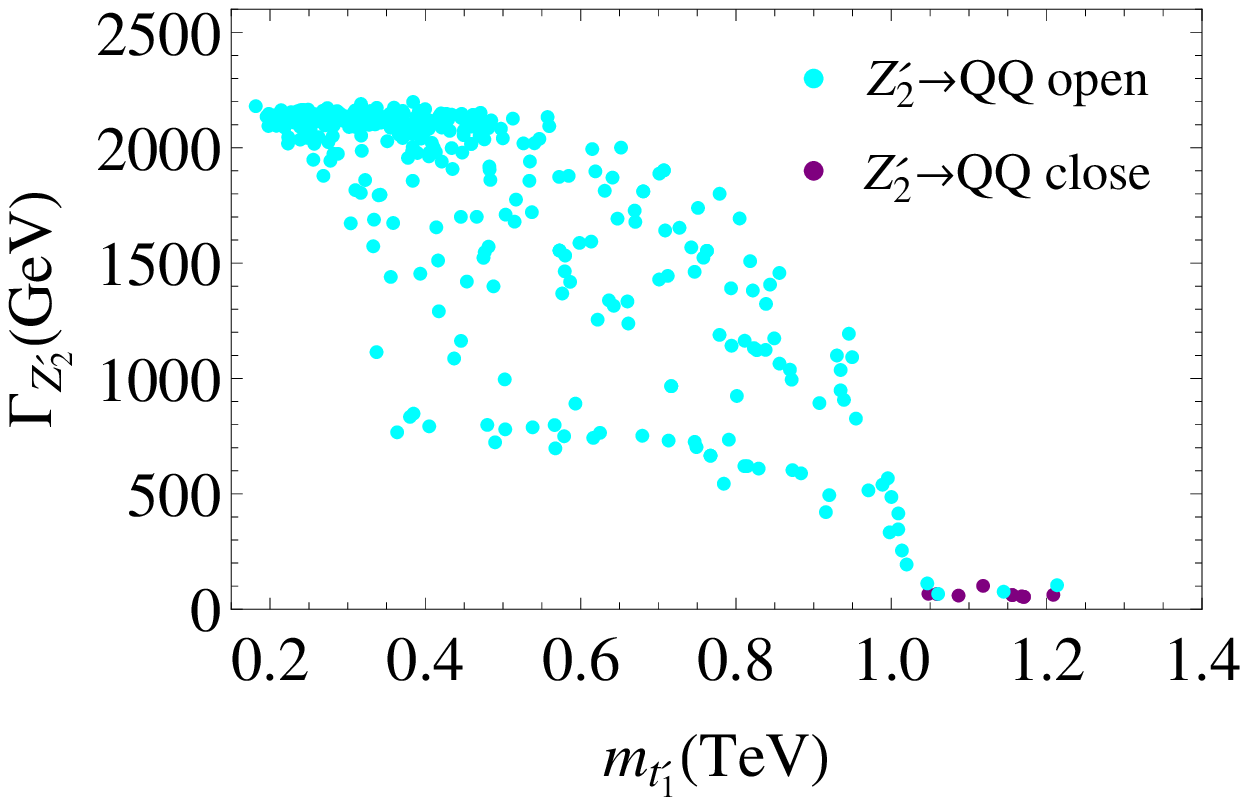, width=.48\textwidth}
\hfill
\epsfig{file=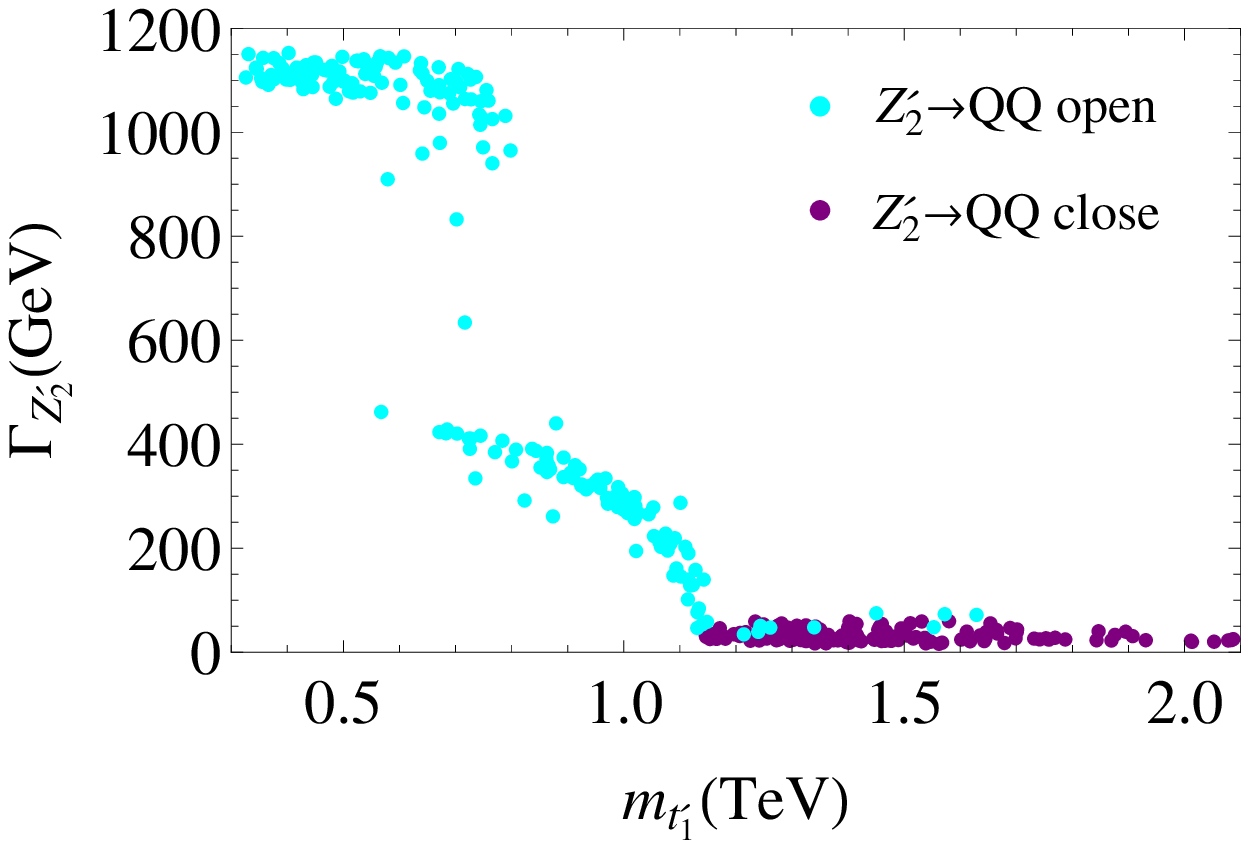, width=.48\textwidth}\\
\epsfig{file=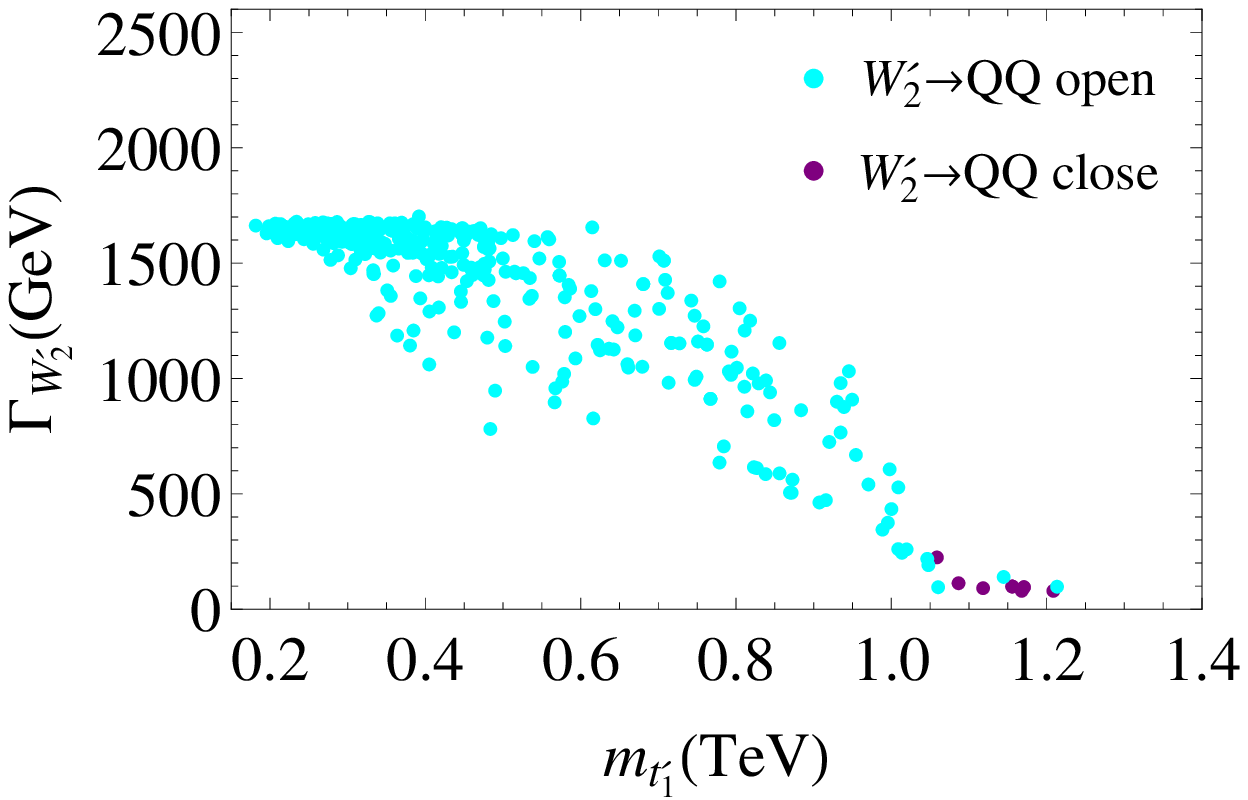, width=.48\textwidth}
\hfill
\epsfig{file=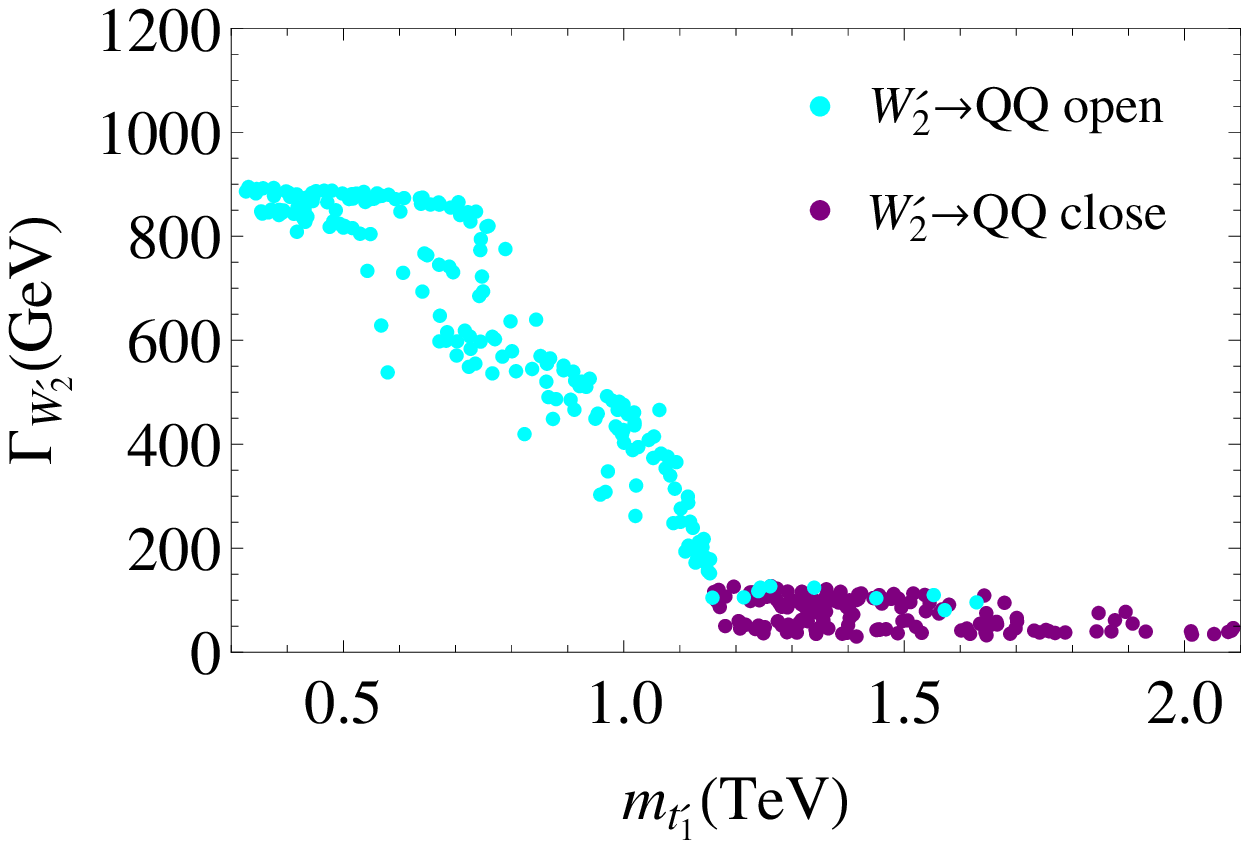, width=.48\textwidth}
\caption[Lightest gauge resonance widths]{Widths of the lightest non-inert neutral (upper row) and charged (lower row) resonances in function of the mass of the lightest charge 2/3 extra fermion for the choices $f=800$ GeV, $g_\rho=2.5$ (left) and $f=1200$ GeV, $g_\rho=1.8$ (right). The purple points are the ones where the decay of the resonance in a pair of extra fermions is forbidden, the cyan points where this process is allowed. }
\label{fig:DY-light-width-mt1}
\end{figure}

\begin{figure}[!h]
\centering
\epsfig{file=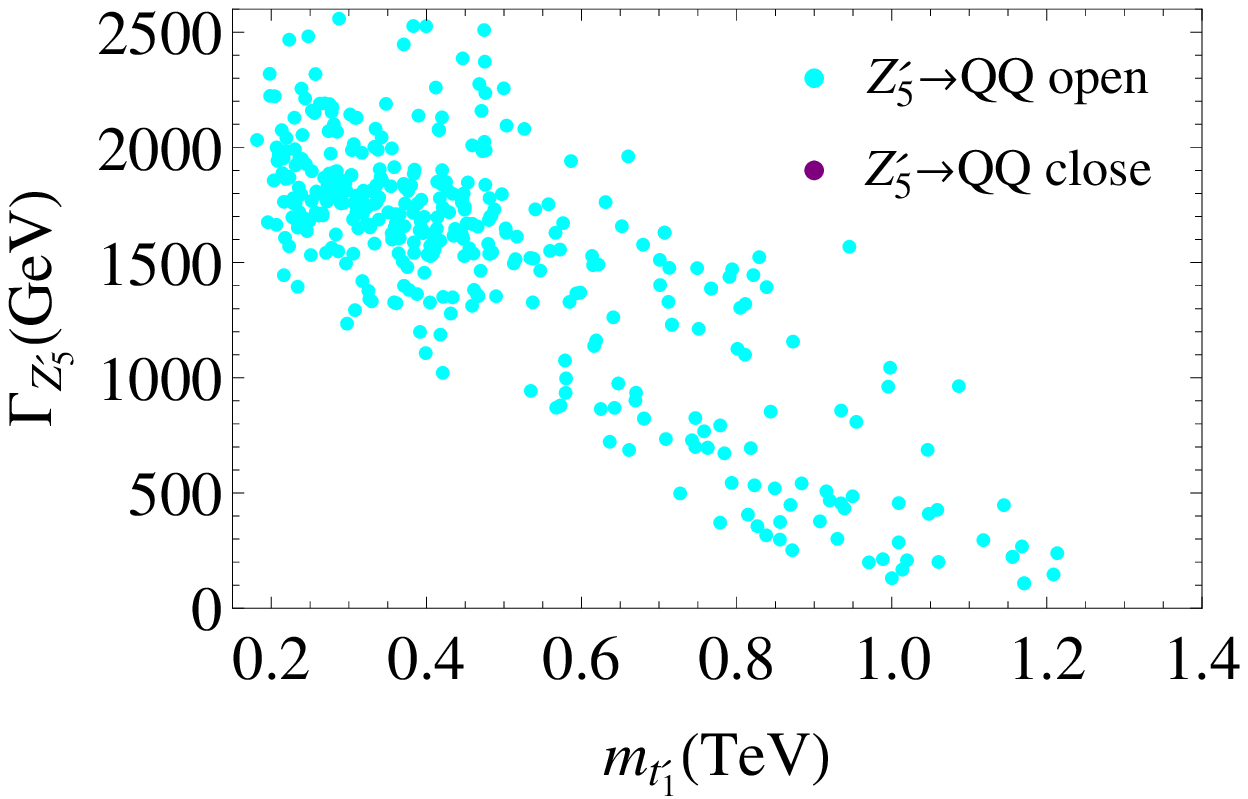, width=.48\textwidth}
\hfill
\epsfig{file=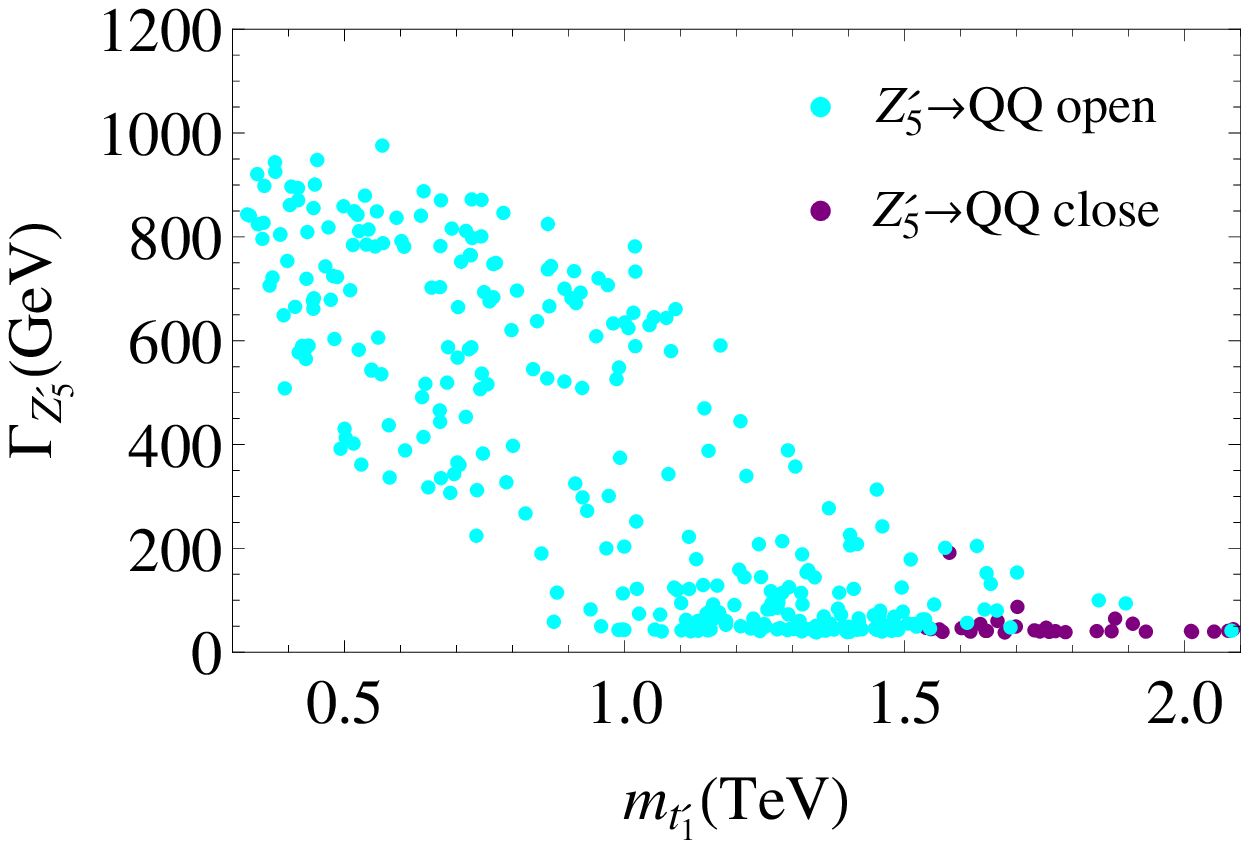, width=.48\textwidth}\\
\epsfig{file=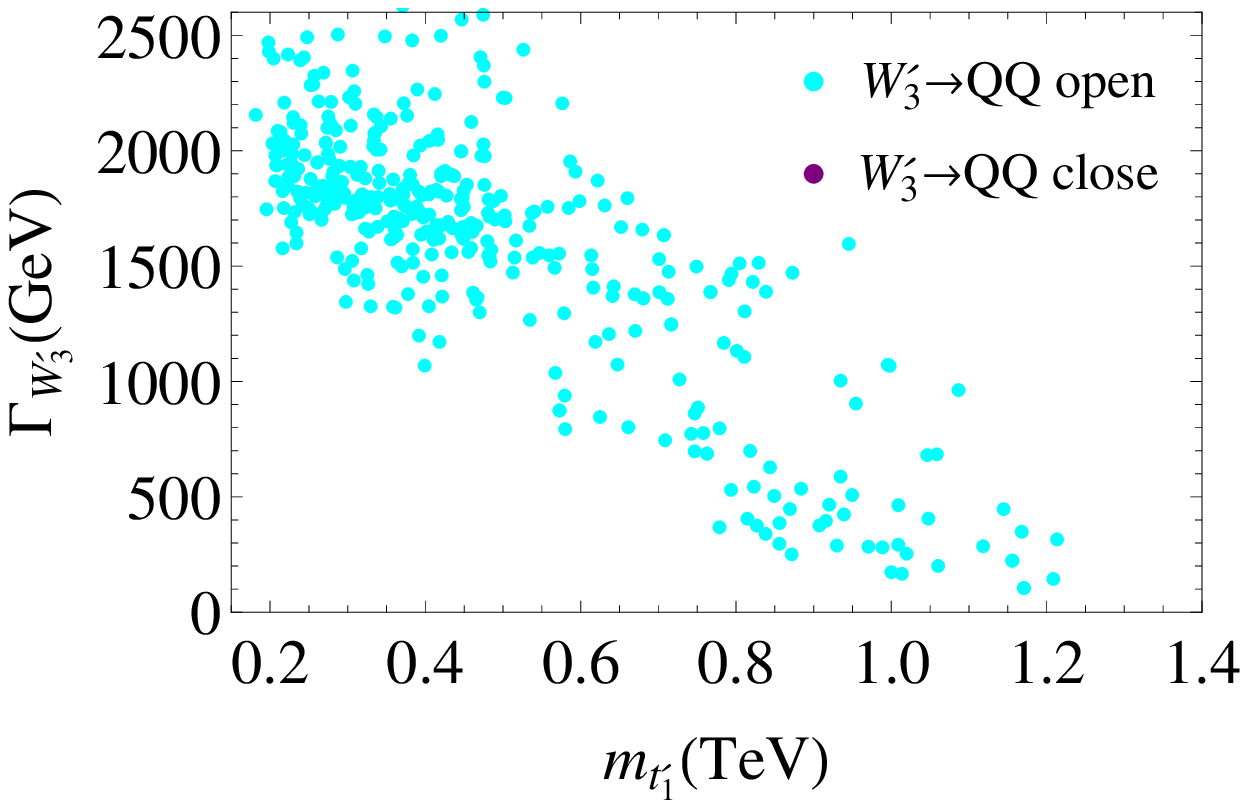, width=.48\textwidth}
\hfill
\epsfig{file=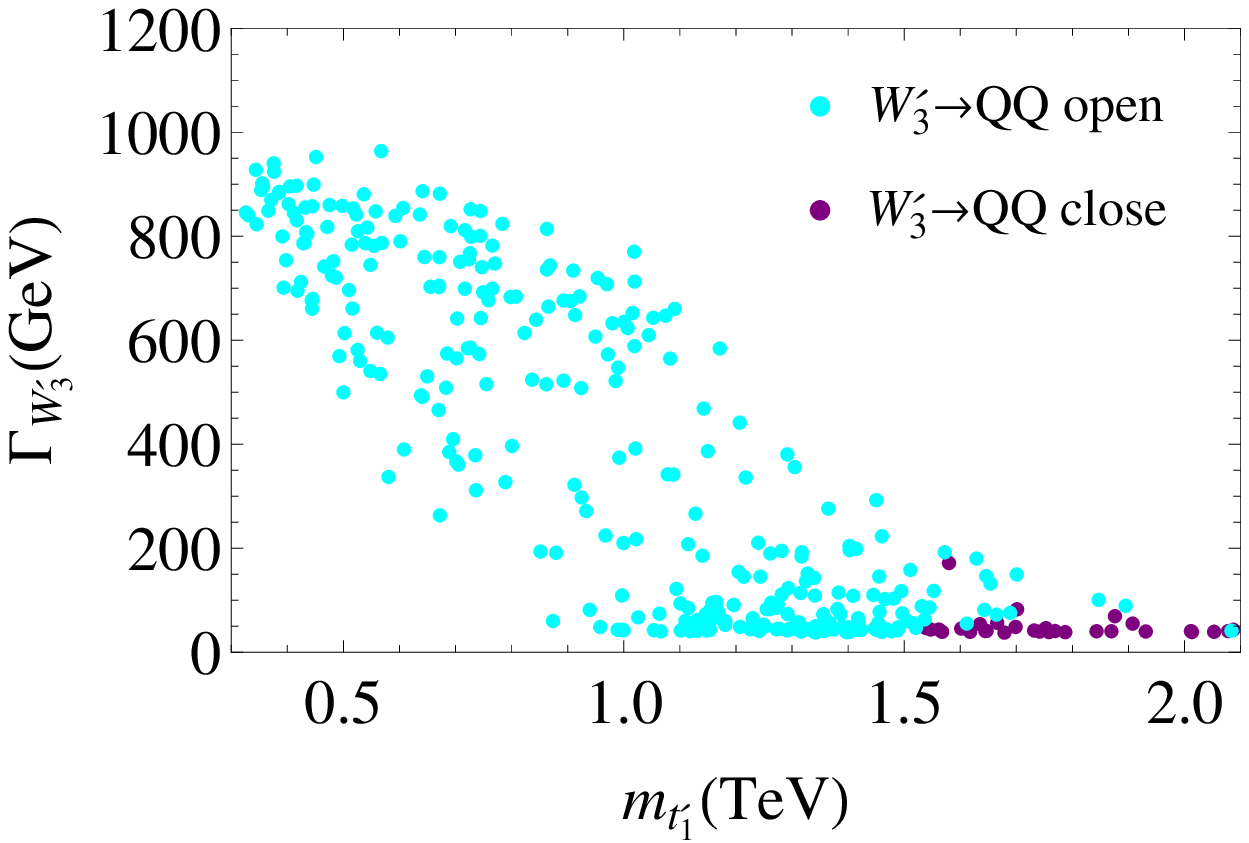, width=.48\textwidth}
\caption[Heaviest gauge resonance widths]{Widths of the heaviest neutral (upper row) and charged (lower row) resonances in function of the mass of the lightest charge 2/3 extra fermion for the choices $f=800$ GeV, $g_\rho=2.5$ (left) and $f=1200$ GeV, $g_\rho=1.8$ (right). The purple points  (where present) are the ones where the decay of the resonance in a pair of extra fermions is forbidden, the cyan points where this process is allowed.}
\label{fig:DY-heavy-width-mt1}
\end{figure}

 In varying the parameters of the fermionic sector we found that one of the extreme model dependent effects, that will have a dramatic influence on the effectiveness of the DY analysis, is the dependence of the widths of the resonances with respect to the fermionic spectrum of the model.
From eq.(\ref{eq:lag-ferm}) we observe in fact that the coupling constant of the extra gauge bosons with the extra fermions of the model is proportional to $g_\rho$ modulus a combination of $\cos{\psi_i}$, where $\psi_i$ are the rotation angles of the transformation of eq.(\ref{eq:partialcompositness}).
Therefore, when the mass of a spin-1 resonances is such that the decay in a pair of extra fermions is allowed, its decay widths will become much larger with respect to the region of parameter space where this decay is not allowed.
We then chose to divide these configurations in various \emph{regimes} of the extra resonances widths that we can present as follows:
\begin{itemize}
\item[-] \emph{Small width regime}: the threshold for the decay in a pair of extra fermions has not been reached and the dominant decay channels are the ones in SM gauge bosons and fermions. In this situation the widths of the extra gauge bosons are well below 100 GeV for a mass of $\simeq$ 2 TeV.
\item[-] \emph{Medium width regime}: the threshold for the decay in a pair of extra fermions has just been reached. In this situation the width over mass ratio can be of order $\simeq 25\%$
\item[-] \emph{Large width regime}: the threshold for the decay in a pair of extra fermions has been abundantly surpassed and the widths of the extra gauge bosons can become comparable with their mass.
\end{itemize}
These situations are illustrated in Fig.~\ref{fig:DY-light-width-mt1} and Fig.~\ref{fig:DY-heavy-width-mt1} for the choices of $f=800$ GeV, $g_\rho=2.5$ and $f=1200$ GeV, $g_\rho=1.8$, respectively, where we plot the widths, $\Gamma$s, of the lightest non-inert and heaviest neutral and charged spin-1 resonances in function of the mass of the lightest charge 2/3 extra quark. The masses of the non-inert gauge bosons for these two configurations are reported in Tab.~\ref{tab:mass-f08g25-f12g18}.
We notice that the \emph{small} and \emph{medium/large} width regime corresponds to almost two different regions, which are clearly
distinguishable for the case of $Z_2^\prime$ and $W_2^\prime$ and less so in the case of $Z_5^\prime$ and $W_3^\prime$, and separated roughly at the point $2 m_{t_1^\prime}=m_{Z^\prime,W^\prime}$. Already this aspect is of interest from a phenomenological point of view since it relates the size of the width of the extra gauge boson with the presence of an extra quark under or above the threshold $\simeq~f~g_\rho/2$.
These different configurations in the parameter space might lead to interestingly phenomenological possibilities, but in order to assess whether they are accessible at the LHC via DY processes we first need to define some specific benchmark points representative of the regimes already described.
We will do so by investigating both the different regimes for a fixed choice of the parameters $f$ and $g_\rho$, namely $f=1200$ GeV and $g_\rho=1.8$, that correspond to fixed masses of the extra gauge bosons and fixed couplings of them to SM light quarks, by just varying the parameters affecting the fermionic sector of the model, and then analysing benchmark points for the other combinations of the model scale and the extra gauge coupling of eq.(\ref{eq:bench-fvar}).

\begin{table}[htb]
\begin{center}

\begin{tabular}{|c| c c|| c| c |}
\hline 
	&	M (GeV)			& &		& M (GeV)			\\
\hline
$Z^\prime_2$	&	2041			& & $W^\prime_2$		&	2067	\\
$Z^\prime_3$	&	2068			& & $W^\prime_3$		&	2830	\\
\cline{4-5}
$Z^\prime_5$	&	2830			& 			\\
\cline{1-3}
\end{tabular}
{(a)}
\begin{tabular}{|c| c c|| c| c |}
\hline
	&	M (GeV)			& &		& M (GeV)			\\
\hline
$Z^\prime_2$	&	2249			& & $W^\prime_2$		&	2312	\\
$Z^\prime_3$	&	2312			& & $W^\prime_3$		&	3056	\\
\cline{4-5}
$Z^\prime_5$	&	3056			& 			\\
\cline{1-3}
\end{tabular}
{(b)}
\end{center}
\caption[Masses of the extra gauge bosons for $f=800$ GeV, $g_\rho=2.5$ and $f=1200$ GeV, $g_\rho=1.8$]{Masses of the non-inert additional gauge boson in the 4DCHM for the parameter point $f=800$ GeV, $g_\rho=2.5$ (a) and $f=1200$ GeV, $g_\rho=1.8$ (b).}
\label{tab:mass-f08g25-f12g18}
\end{table}

For the choice $f=1200$ GeV and $g_\rho$=1.8 the masses of the spin-1 resonances are reported in Tab.~\ref{tab:mass-f08g25-f12g18} (b) and with the choice of model parameters reported in Tab.~\ref{tab:input_bench-fgfixed}~(a), (e) and (f) their widths are the ones reported in Tab.~\ref{tab:f12g18widthsmallmediumlargeregime}.
These three configurations, corresponding to the aforementioned regimes, correspond to a situation where the mass of the lightest spin-1/2 resonance is roughly 1600, 980 and 500 GeV, respectively, that is above, below and well below the threshold for the decay of the gauge boson in a pair of them, and this will clearly reflect on the branching ratios of the states.
While in the small width regime they are dominated by decays into SM particles, mainly third generation quarks and gauge bosons, in the medium and large width regime the main decay modes are the ones in a pair of extra fermions, and rates in SM particles are reduced well below the percent level as we can see in Tab.~\ref{tab:br-z2-f12g18-smallmediumlarge} where we report the branching ratios of the $Z^\prime_2$ for the choice of parameters related to the widths of Tab. \ref{tab:f12g18widthsmallmediumlargeregime}.
For a matter of space we do not report the decay modes of the other gauge bosons for these three regimes, and we refer to \cite{Barducci:2012kk} for a more complete list.

\begin{table}[h!]
\begin{center}
\begin{tabular}{|c| c c|| c| c |}

\hline 
	&	$\Gamma$ (GeV)			& &		& $\Gamma$ (GeV)			\\
\hline
$Z^\prime_2$	&	32			& & $W^\prime_2$		&	55	\\
$Z^\prime_3$	&	55			& & $W^\prime_3$		&	54	\\
\cline{4-5}
$Z^\prime_5$	&	54			& 			\\
\cline{1-3}
\end{tabular}
{(a)}
\begin{tabular}{|c| c c|| c| c |}
\hline
	&	$\Gamma$ (GeV)			& &		& $\Gamma$ (GeV)			\\
\hline
$Z^\prime_2$	&	301			& & $W^\prime_2$		&	434	\\
$Z^\prime_3$	&	434			& & $W^\prime_3$		&	522	\\
\cline{4-5}
$Z^\prime_5$	&	526			& 			\\
\cline{1-3}
\end{tabular}
{(b)}
\vskip 5pt
\begin{tabular}{|c| c c|| c| c |}
\hline
	&	$\Gamma$ (GeV)			& &		& $\Gamma$ (GeV)			\\
\hline
$Z^\prime_2$	&	1099			& & $W^\prime_2$		&	820	\\
$Z^\prime_3$	&	827			& & $W^\prime_3$		&	614	\\
\cline{4-5}
$Z^\prime_5$	&	413			& 			\\
\cline{1-3}
\end{tabular}
{(c)}
\end{center}
\caption[Widths of the extra gauge bosons for $f=1200$ GeV and $g_\rho$=1.8 in the small, medium and large width regime.]{Widths of the non-inert additional gauge bosons of the 4DCHM in the small (a), medium (b) and large (c) width regime for the choice $f=1200$ GeV, $g_\rho$=1.8 and with the other parameters fixed to the values reported in Tab.~\ref{tab:input_bench-fgfixed} (a), (e) and (f) in Appendix \ref{chap:App-C}.}
\label{tab:f12g18widthsmallmediumlargeregime}
\end{table}

\begin{table}[h!]
\begin{tabular}{|l|}
\hline
$Z^\prime_2$				\\
\hline
68\% in $t\bar{t}$			\\
9\% in $W^+W^-$/$Z h$			\\
5\% in $b\bar{b}$			\\
1\% in $u \bar u, d \bar d$	\\
1\% in $l_i\bar l_i$	\\
0.5\% in $d\bar d,s\bar s$	\\
0.2\% in $b \bar b^\prime_3$ and c.c.	\\
\hline
\end{tabular}
{(a)}
\begin{tabular}{|l|}
\hline
$Z^\prime_2$			  	\\
\hline
46\% in $b^\prime_1 \bar b^\prime_1$				\\
40\%  in $Y_1 \bar Y_1$	\\
8\% in $t\bar{t}$					\\
2\% in $b^\prime_2{\bar b^\prime_2}$				\\
1\% in $t^\prime_1{\bar t^\prime_1}$		  		\\
1\% in $W^+W^-$ and $Z h$			\\
$<$1\% in $b\bar{b}$, $u\bar u, d\bar d$ and $l_i\bar l_i$\\
\hline
\end{tabular}
{(b)}
\begin{tabular}{|l|}
\hline
$Z^\prime_2$			  			\\
\hline
31\% in $t^\prime_2{\bar t^\prime_2}$				\\
29\% in $Y_1 \bar Y_1$	\\
16\% in $X_1 \bar X_1$	\\
5\% in $t\bar{t}$					\\
4\% in $b^\prime_1{\bar b^\prime_2}$ and c.c. 		\\
4\% in $b^\prime_1{\bar b^\prime_1}$				\\
2\% in $b^\prime_{2,3}{{\bar b^\prime}_{2,3}}$		\\
\hline
\end{tabular}
{(c)}
\caption[Branching ratios of the extra gauge bosons for $f=1200$ GeV and $g_\rho$=1.8 in the small, medium and large width regime.]{Branching ratios of the $Z^\prime_2$ for the choice of $f=1200$ GeV, $g_\rho$=1.8 and the other parameters fixed to the values reported in  Tab.~\ref{tab:input_bench-fgfixed} (a), (e) and (f) in Appendix~\ref{chap:App-C}.}
\label{tab:br-z2-f12g18-smallmediumlarge}
\end{table}

Other combinations of fermionic parameters that will be used in the analysis, corresponding then to other widths of the gauge bosons, can be found in Tab.~\ref{tab:input_bench-fgfixed} in Appendix \ref{chap:App-C} together with the input parameters corresponding to the different choices of the parameters $f$ and $g_\rho$ of eq.(\ref{eq:bench-fvar}), see Tab.~\ref{tab:input_bench-fgvar}, with the caveat that in those cases we will just analyse  regimes corresponding to small widths, which we will see to be a necessary condition for the extraction of the 4DCHM signal over the SM background.

\subsection{Simulation}

Our numerical MC simulations have been performed by means of a code based on helicity amplitudes, defined through the HELAS subroutine \cite{Murayama:1992gi} and assembled with MadGraph \cite{Stelzer:1994ta}, which has been validated against CalcHEP \cite{Belyaev:2012qa}.
The matrix elements (MEs) generated account for all off-shellness effects of the particles involved. Two different phase space implementations were used, an \emph{ad hoc one}, based on Metropolis \cite{Kharraziha:1999iw}, and a \emph{blind one} based on RAMBO \cite{Kleiss:1985gy}, checked against each other. The latter was adopted eventually, as it proved the most unbiased one in sampling the multiple resonances existing in the model.
Further, VEGAS~\cite{Lepage:1977sw,Lepage:1980dq} was finally used for the multi dimensional numerical integrations. All these additional subroutines were also validated against CalcHEP outputs. 

The MEs have been computed at leading order (LO). Clearly, in the LHC environment, QCD corrections are not
negligible and associated scale uncertainties may have an impact on the dynamics of $Z'$s and $W'$s production and decay
(see Refs. \cite{Adam:2008pc,Adam:2008ge} for the case of the SM $Z$ and $W$ channels). In fact, EW corrections
may also be relevant \cite{Balossini:2007zzb,Balossini:2009sa}. However, the treatment we are adopting here
of the two DY channels is such that real radiation of gluons and photons would be treated inclusively (i.e., no selection is
enforced here that relies on the gluon and photon dynamics), so that we do not expect such QCD effects to have an impact on the distributions
that we will be considering, neither those of the cross sections nor those of the asymmetries, apart from an
overall rescaling. The latter, in particular, when implemented at large invariant and transverse mass, is affected by a 
residual uncertainty of 5\% at the most \cite{Balossini:2007zzb,Balossini:2009sa}.
The parton distribution functions (PDFs) used were CTEQ5L~\cite{Lai:1999wy}, with factorisation and renormalisation scale set at $Q=\mu=\sqrt{\hat{s}}$.
(we have verified that later PDF sets do not generate any significant difference in the results we are going to 
present\footnote{Furthermore, we have estimated the theoretical uncertainty (at next to leading order (NLO)) due to the PDFs by adopting 
NNPDF sets \cite{Ball:2011uy}, which yielded a 10\% effect at the most, rather independent of the $Z'$ and $W'$ masses involved and with negligible impact onto the shape of the differential distributions presented.}).
 Initial state quarks have been taken as massless, just like the final state leptons and neutrinos. 

\subsection{Results}

We now consider the two tree-level processes of Fig.~\ref{fig:dy-proc}
\begin{equation}
p p \to l^- l^+ \; \textrm{ (NC)}, \qquad p p \to l^- \bar{\nu}_l + c.c\textrm{ (CC)}, \qquad l=e,\mu,
\end{equation}
presenting our results for either of the flavours and not summing on the two, in order to be able to discuss separately the different mass reconstruction resolution, that can change significantly from electrons to muons.
Considering for example the neutral case, where the mass involved in reconstructing the $Z^\prime$ resonance is the invariant one
\begin{equation}
M_{l^+l^-}=\sqrt{(p_{l^-}+p_{l^+})^2}, \quad p_{l}=(p^0, \vec p),
\label{eq:inv-mass}
\end{equation}
and considering a typical resonance of our benchmarks around 2 TeV, the mass resolutions are about 1\% and 10\% for $e^+e^-$ and $\mu^+\mu^-$, respectively.
In the CC channel the mass concerned with the reconstruction of the $W^\prime$ resonance is the transverse mass
\begin{equation}
M_T=\sqrt{(E^T_l+E^T_{miss})^2-\sum_{i=x,y}(p^i_l+p^i_{miss})^2}
\label{eq:tran-mass}
\end{equation}
with
\begin{equation}
E^T_l=\sqrt{(p^x_l)^2+(p^y_l)^2} \quad i=e,\mu
\end{equation}
and $E^T_{miss}$ being the lepton transverse energy and missing transverse energy respectively,
where $p^{x,y}$ are the momenta component in $x$-$y$ plane assuming that the proton beams are directed along the $z$ direction.
For $M_T$ the mass resolution is about 20\%, being dominated by the uncertainty in reconstructing the $E^T_{miss}$, due to the neutrino escaping the detector.
We then impose standard acceptance cuts on the lepton transverse momentum, $p^T_l$, and pseudorapidity, $\eta_l$
\begin{equation}
\begin{alignedat}{2}
& p^T_l > 20\textrm{ GeV}, \quad && p^T_l=\sqrt{{p^x_l}^2+{p^y_l}^2}\\
& |\eta_l|<2.5, \quad &&  \eta_l=-\log[\tan(\frac{\theta_l}{2})]
\label{eq:dy-cuts}
\end{alignedat}
\end{equation}
where $\theta_l$ is the angle between the particle momenta and the beam axis.
In order to reduce the SM background we will apply a cut on the invariant and transverse mass for the NC and CC channel, respectively, that will depend on the choice of the $(f$, $g_\rho)$ combination and that will be listed when we present the results.

\begin{figure}[!h]                                      
\centering
\epsfig{file=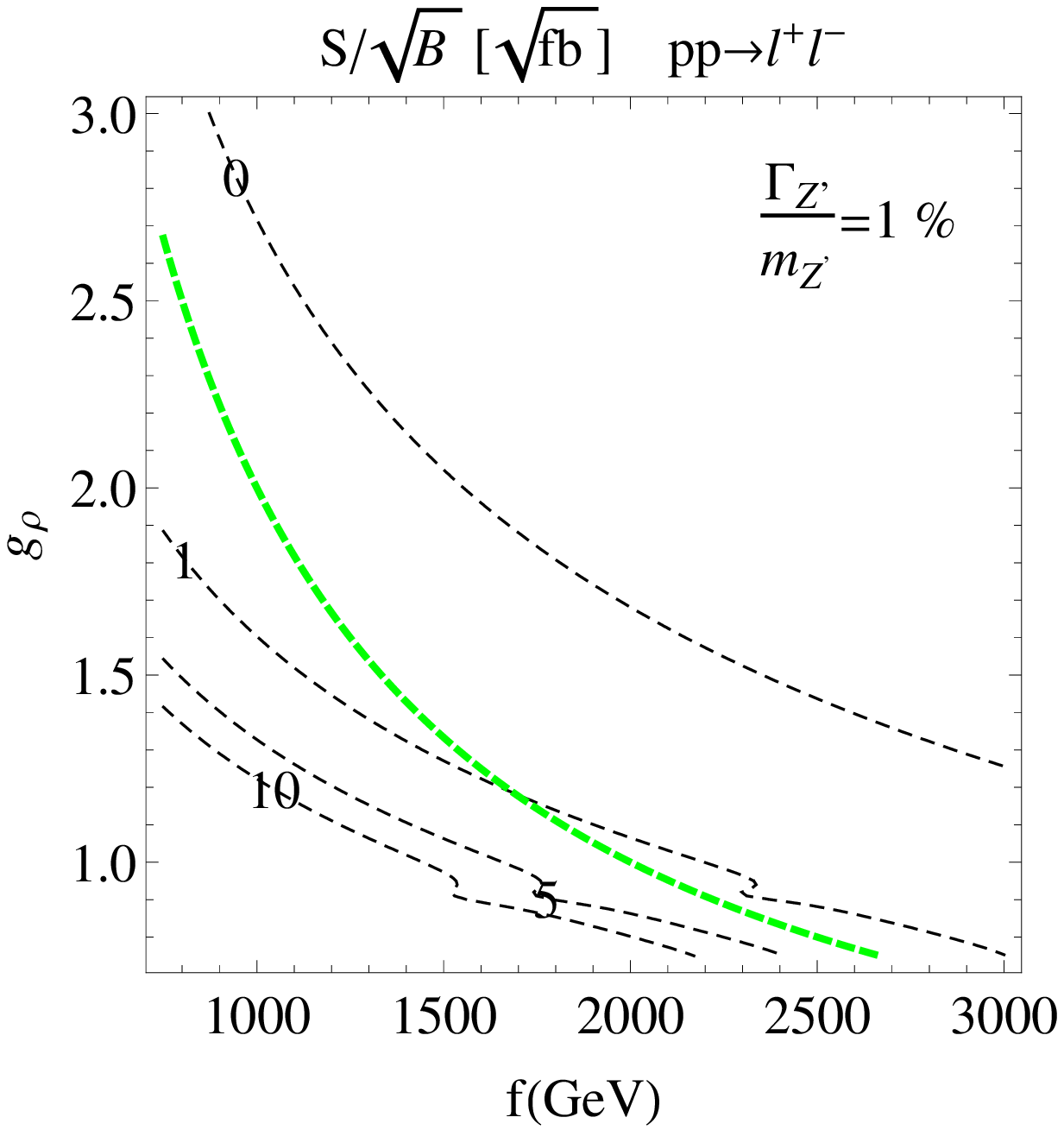, width=.46\textwidth}{(a)}\hfill
\epsfig{file=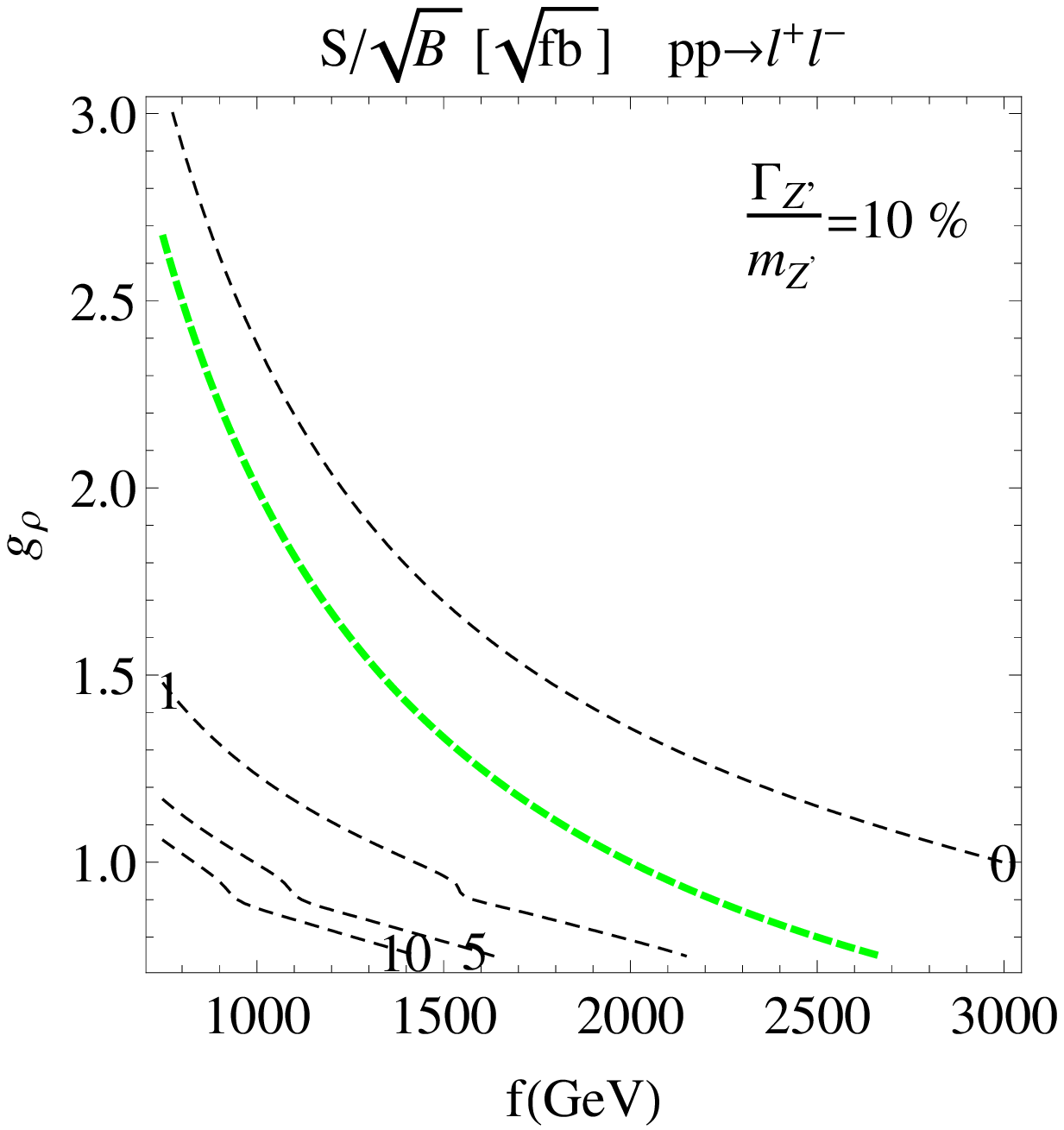, width=.46\textwidth}{(b)}
\caption[$S/\sqrt{B}$ contours in the $f$ $g_\rho$ plane for the neutral current DY process at the LHC]{$S/\sqrt{B}$ contours in the $f$ $g_\rho$ plane for the NC process of eq.(\ref{fig:dy-proc}) at the 14 TeV LHC for the choice $\Gamma_{Z^\prime}/m_{Z^\prime}$=1\% (a) and 10\% (b). Beside the cuts of eq.(\ref{eq:dy-cuts}) we have applied $M_{l^+l^-}>fg_{\rho}$. The green dashed line represent the contour $f g_\rho=M_{Z^\prime_1}=2$~TeV.}
\label{fig:dy-sign-fg}
\end{figure}

As a first result we show in Fig.~\ref{fig:dy-sign-fg} the contours for the $S/\sqrt{B}$ values, from which it is possible to compute the statistical significance $\alpha$
\begin{equation}
\alpha=\frac{S}{\sqrt{B}}\sqrt{\mathcal L}\sqrt{\epsilon}
\label{eq:signif}
\end{equation}
where $\mathcal L$ is the integrated luminosity and $\epsilon$ the reconstruction efficiency of a given final state.
We show the results for the NC process at the 14 TeV LHC for two arbitrary configurations of $\Gamma_{Z^\prime}/m_{Z^\prime}$, 1\% (a) and 10 \% (b).
From the plots we already observe that in order to have a high $S/\sqrt{B}$ value, and so a high exclusion or discovery power, a small value of the $\Gamma_{Z^\prime}/m_{Z^\prime}$ ratio is compulsory, due to the otherwise decrease of the branching ratio into leptonic final state, no matter the choice of $f$ and $g_\rho$. We do not show the results for a higher value of the width over mass ratio, for which the $S/\sqrt{B}$ value becomes even smaller.

\begin{figure}[!h]                                      
\centering
\epsfig{file=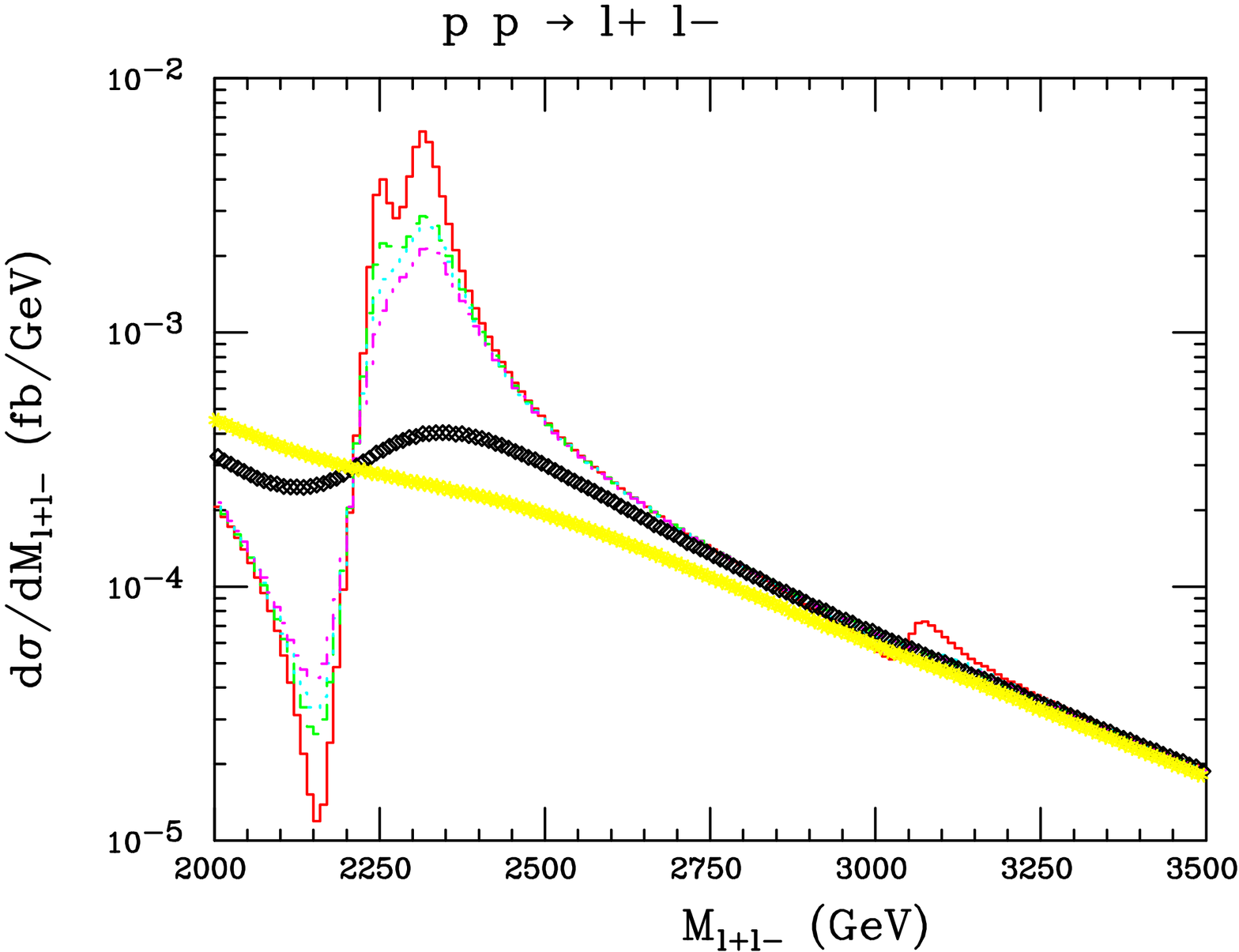,angle=90, width=.46\textwidth}{(a)}\hfill
\epsfig{file=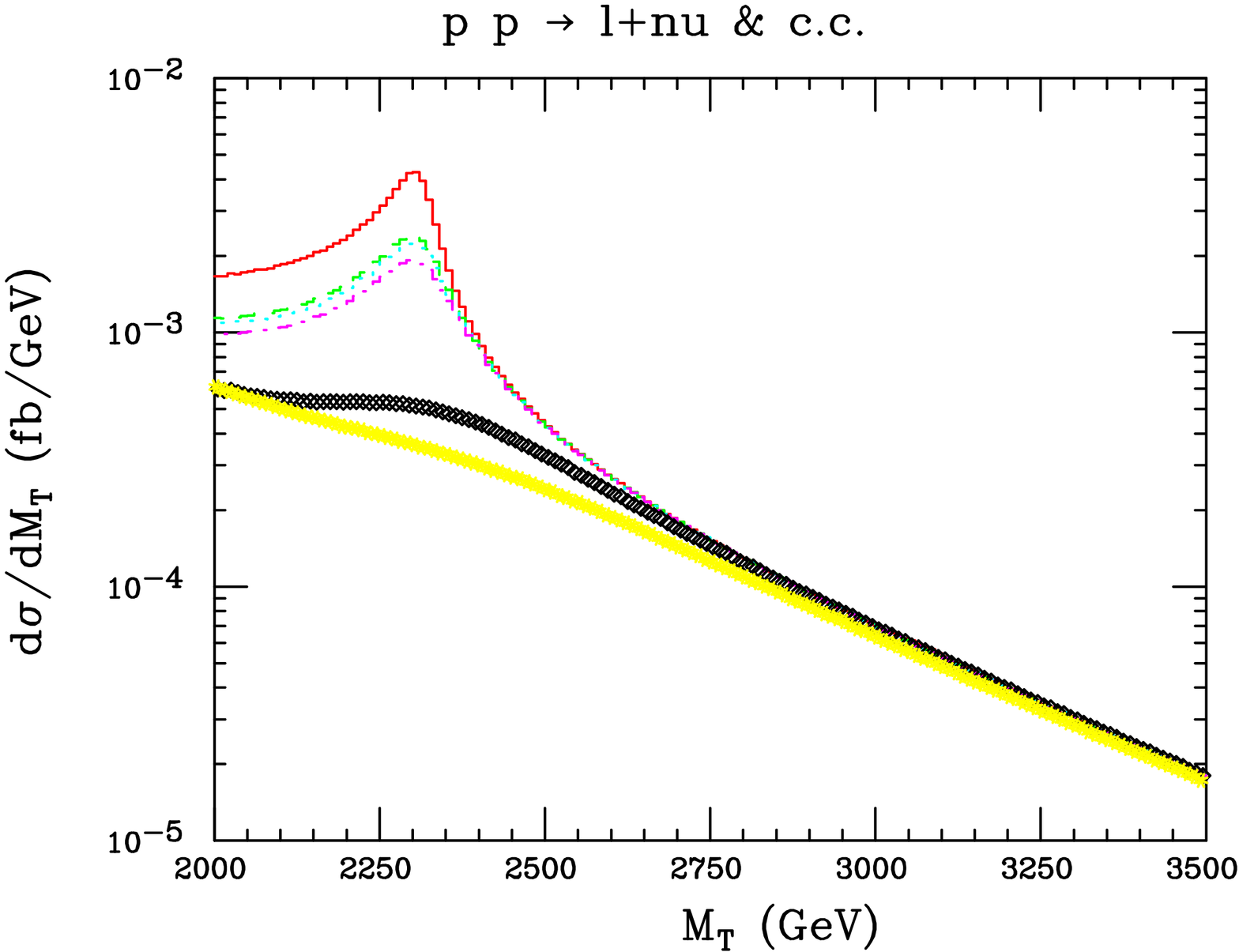,angle=90, width=.46\textwidth}{(b)}
\caption[Invariant and transverse mass distributions for the cross section for DY processes at the LHC with variable extra gauge bosons widths]{Invariant (a) and transverse (b) mass differential distribution for the cross section at the 14 TeV LHC for the NC (a) and CC (b) DY process for the choice of benchmark points with $f$=1200 GeV and $g_\rho=1.8$ of Tab.~\ref{tab:input_bench-fgfixed} where the  colours correspond to: red (a), green (b), cyan (c), magenta (d), black (e) and yellow (f). Red, black and yellow lines correspond to the small, medium and large width regime previously discussed.
Cuts on the invariant and transverse mass are 2 TeV. Integrated cross sections are 0.78, 0.56, 0.52, 0.47, 0.26 and 0.23 fb with a SM background of 0.21 fb in the NC case and 1.11, 0.79, 0.76, 0.70, 0.36 and 0.30 fb with a SM background of 0.23 fb in the CC case.}
\label{fig:DY-CCNC-ffixed-width}
\end{figure}

We now focus on realistic points in the 4DCHM parameter space, that is, we choose benchmarks arising from the constrained parameter scan described in Chapter~\ref{chap-2}, and we first show the results for the cross section distributions in invariant mass for the choices of the benchmark points with fixed
$f=$ 1200 GeV and $g_\rho=1.8$ of Tab.~\ref{tab:input_bench-fgfixed}, that correspond to different decreasing values of the masses of the extra fermions and therefore increasing values of the gauge bosons widths (see again \cite{Barducci:2012kk} for the values of the widths of all the gauge bosons in these configurations).
As we can see from Fig.~\ref{fig:DY-CCNC-ffixed-width}, when the values of the $Z^\prime s$ and $W^\prime s$ widths are small enough, say for the red, green cyan and magenta lines, a clear excess over the SM background, that is covered by the yellow line, is visible, while as the values of $\Gamma_{Z^\prime s}$ and $\Gamma_{W^\prime s}$ grow (black and yellow lines) the signal shape flattens down with the ultimate tendency, for very large widths, that the extraction of a narrow resonance is no longer possible, both in the neutral and in the charged sector.
For this reason we will now present results for benchmark points with different values of $f$ and $g_\rho$ (Tab.~\ref{tab:input_bench-fgvar}), focusing however only on the case where the widths are small enough, $\Gamma_{Z^\prime_i,W^\prime_i}/M_{Z^\prime_i,W^\prime_i}\simeq 1-5\%$, to test this sector of the model with DY processes.

\begin{figure}[!h]                                      
\centering
\epsfig{file=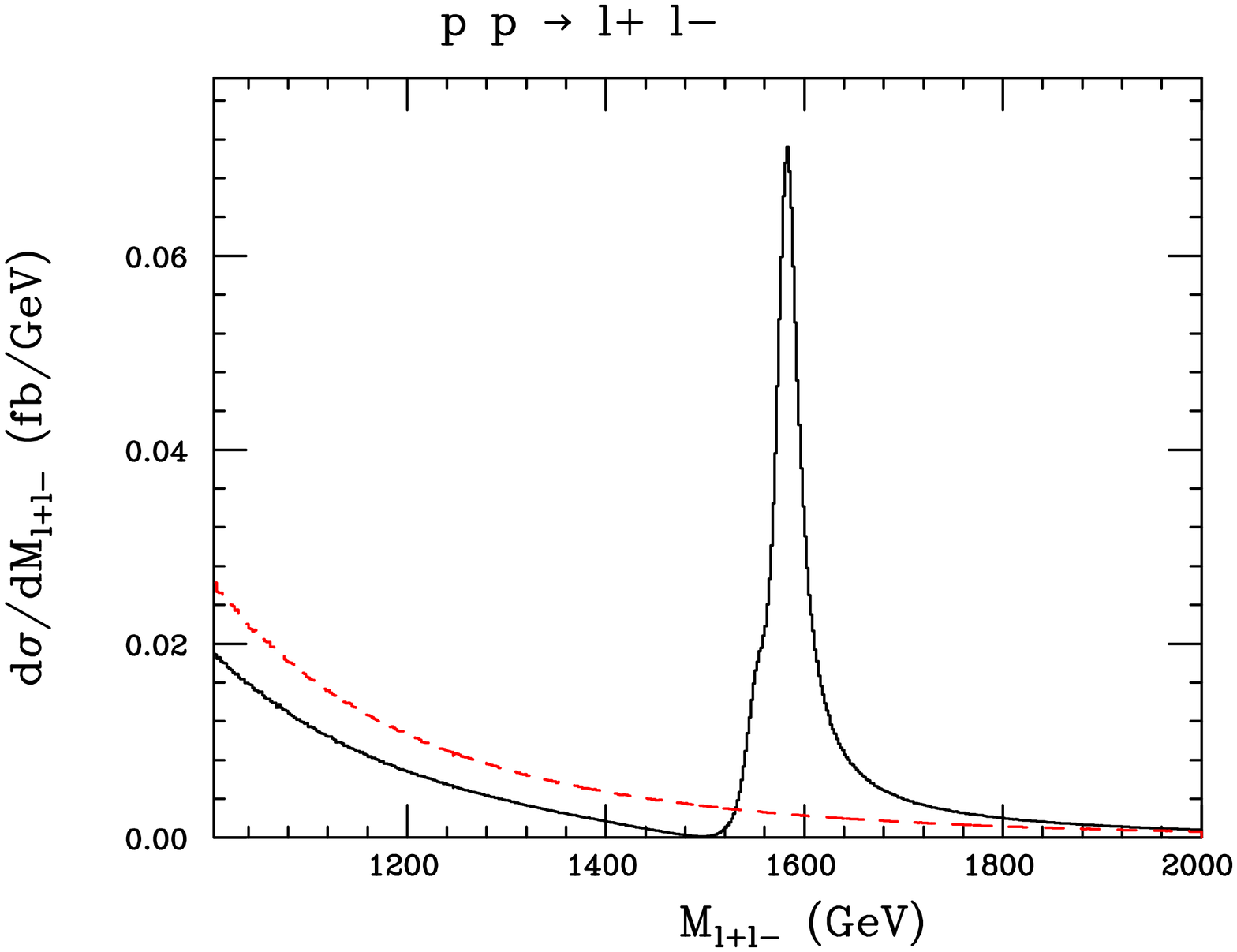,angle=90, width=.46\textwidth}{(a)}
\epsfig{file=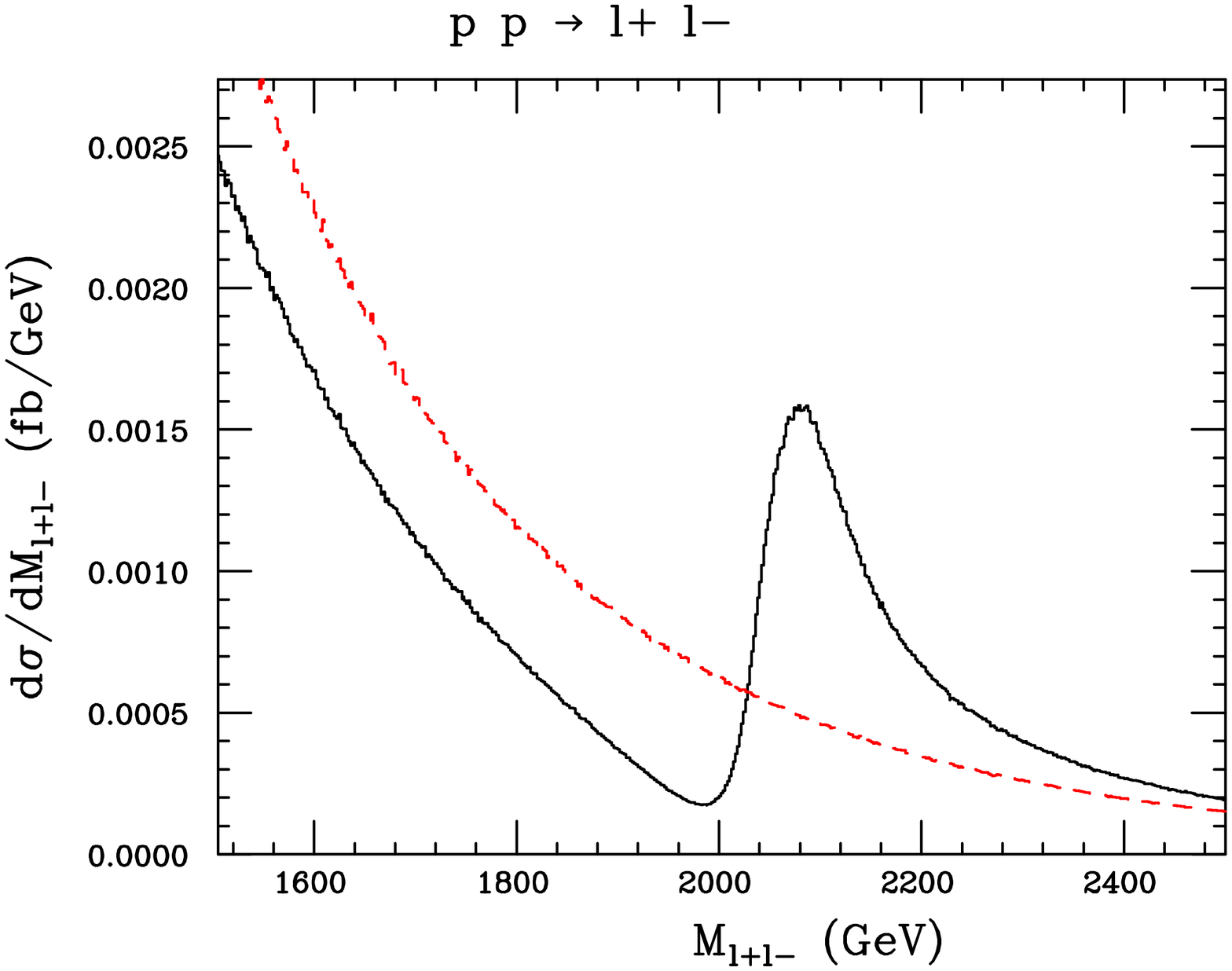,angle=90, width=.46\textwidth}{(b)}
\epsfig{file=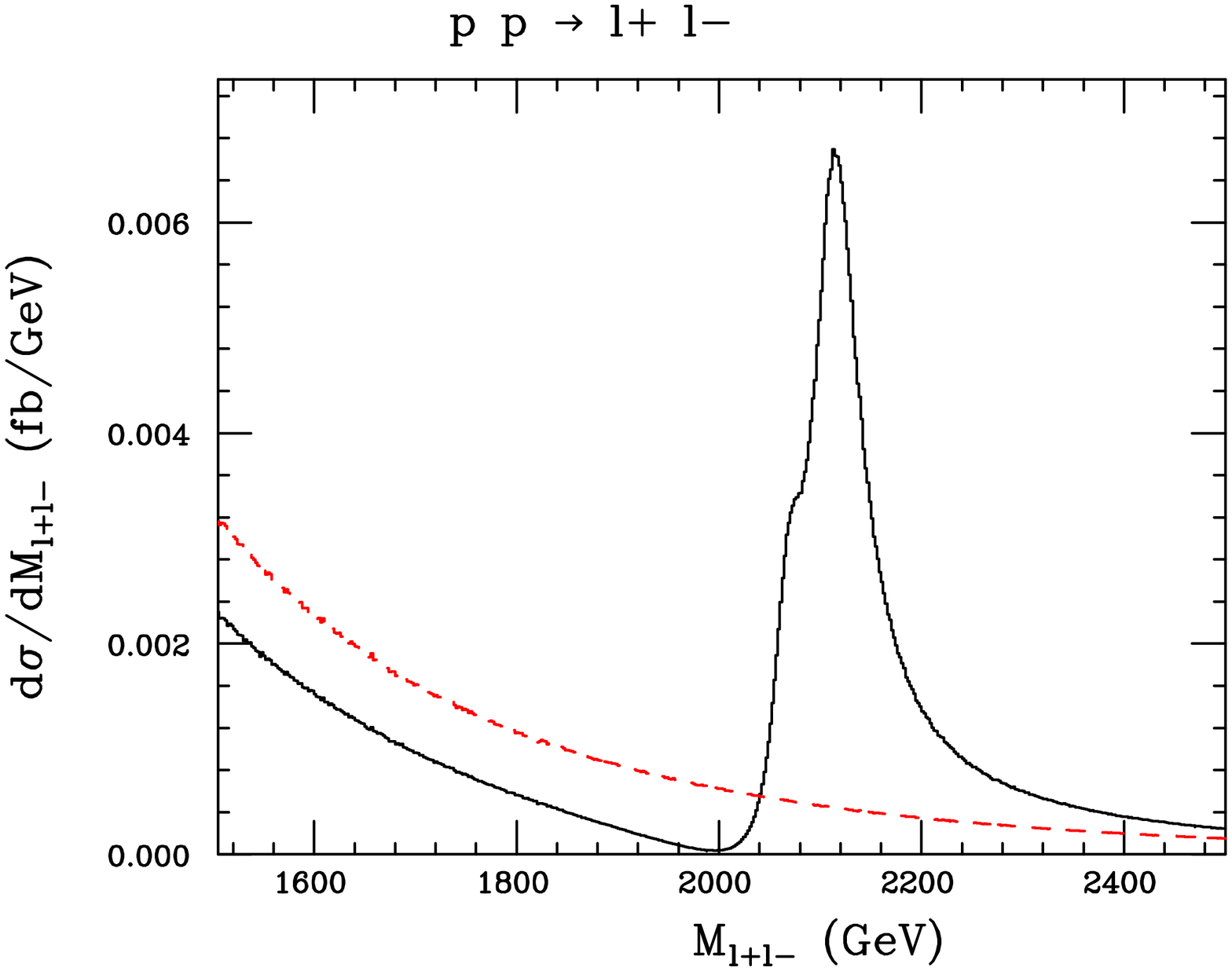,angle=90, width=.46\textwidth}{(c)}
\epsfig{file=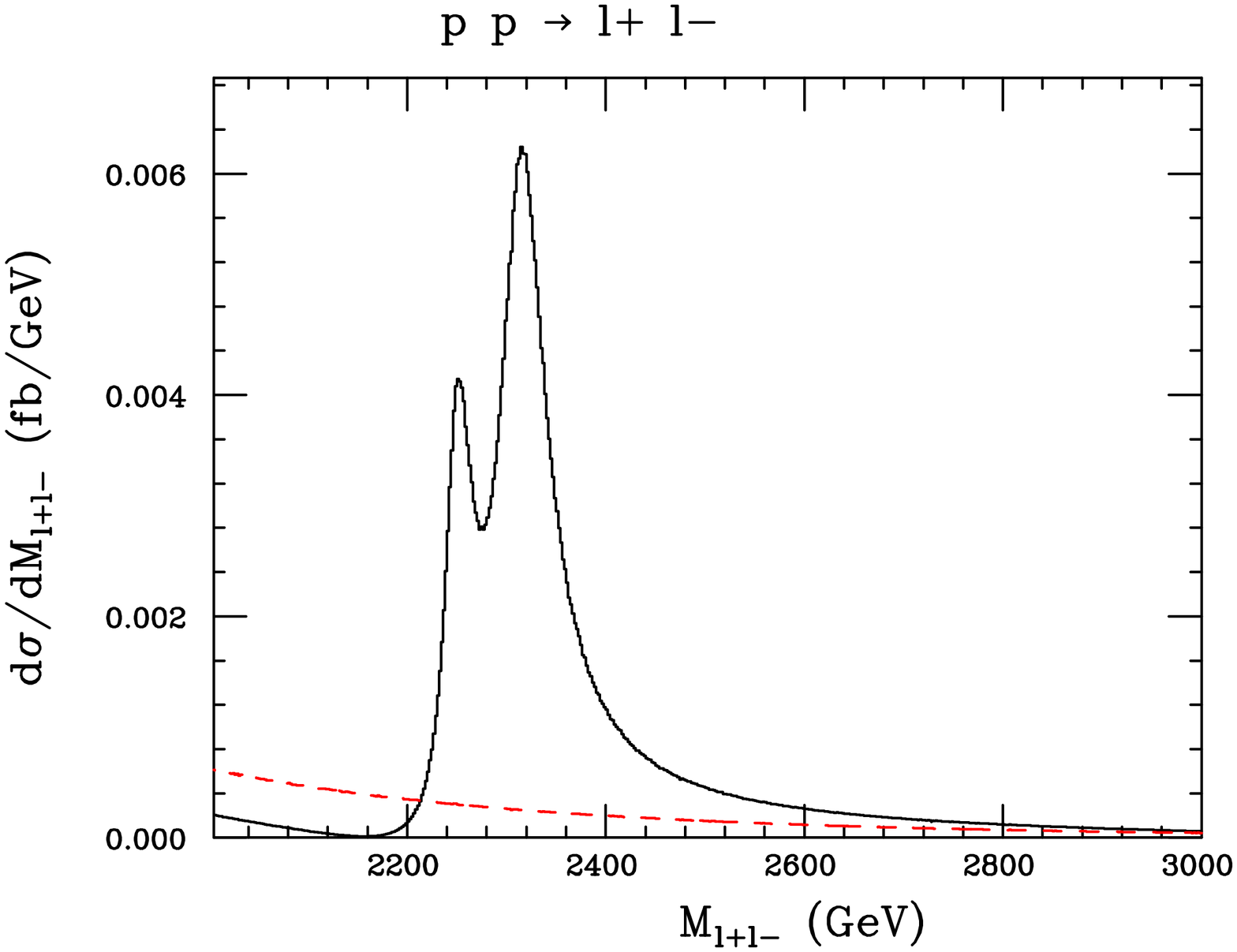,angle=90, width=.46\textwidth}{(d)}
\caption[Invariant mass distributions for the  cross section for DY processes at the LHC with variable model scale $f$ and coupling constant $g_\rho$]{Invariant mass differential distribution for the cross section at the 14 TeV LHC for the NC DY process for the choice of benchmark points $f=750\textrm{ GeV},g_\rho=2$ (a), $f=800\textrm{ GeV},g_\rho=2.5$ (b), $f=1000\textrm{ GeV},g_\rho=2$ (c) and $f=1200\textrm{ GeV},g_\rho=1.8$ (d) for which the complete set of input parameter for the small width regime is reported in Tab.~\ref{tab:input_bench-fgvar}. The solid black line represent the total signal while the dashed red the SM background. For integrated cross sections and cuts see Tab.~\ref{tab:cs-DY-NC-CC-fvar}.}
\label{fig:DY-NC-fvar}
\end{figure}

\begin{figure}[!h]                                      
\centering
\epsfig{file=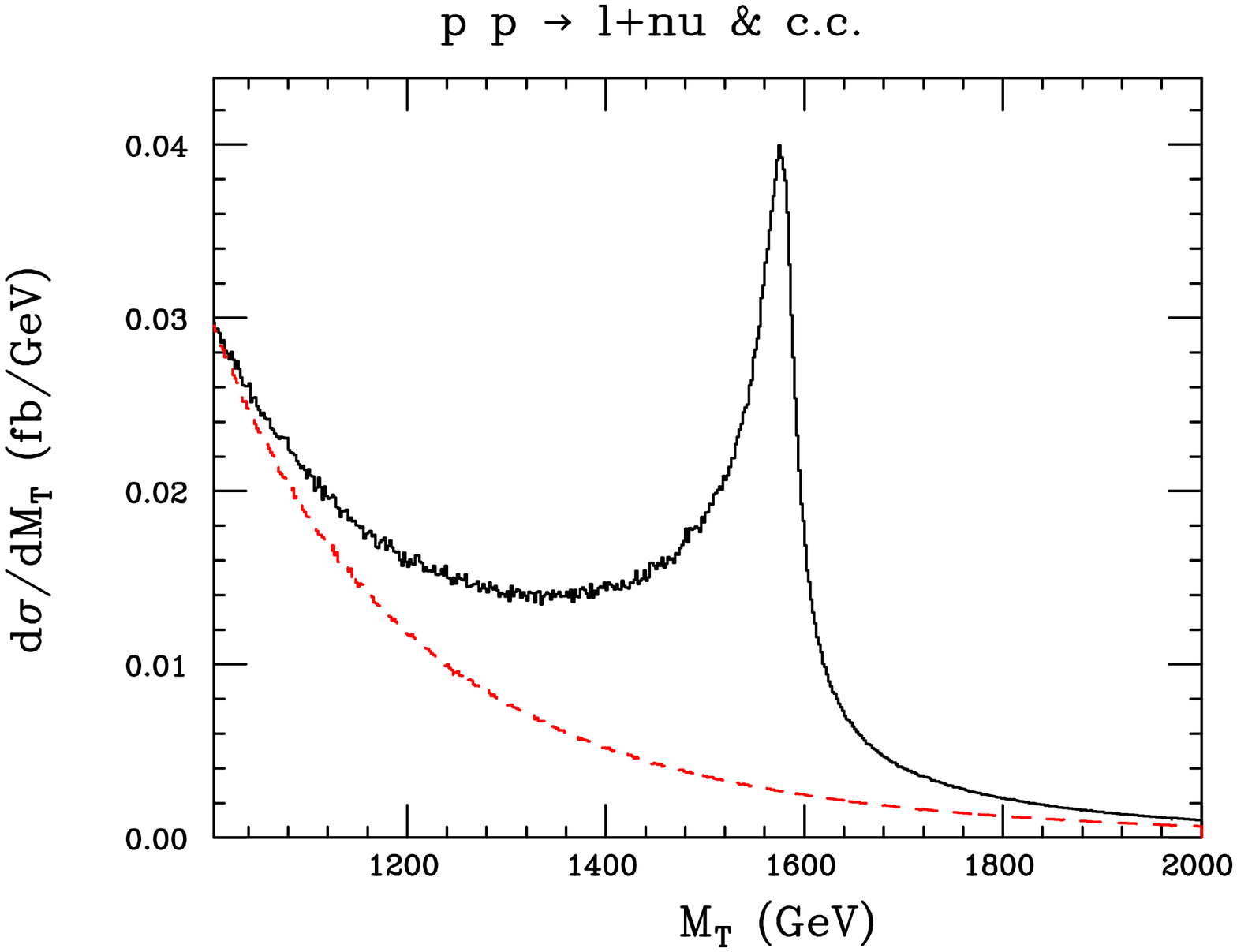,angle=90, width=.46\textwidth}
\epsfig{file=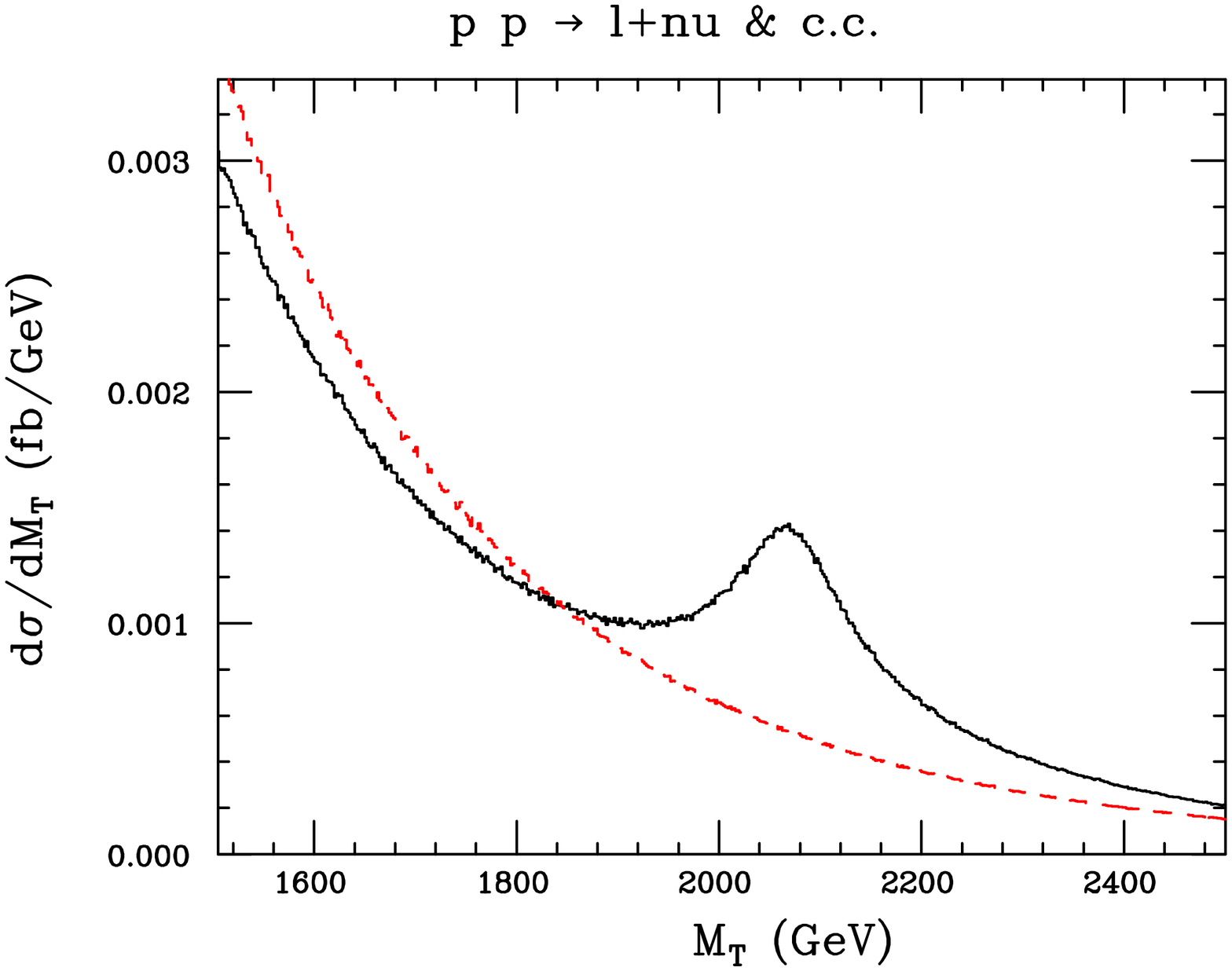,angle=90, width=.46\textwidth}
\epsfig{file=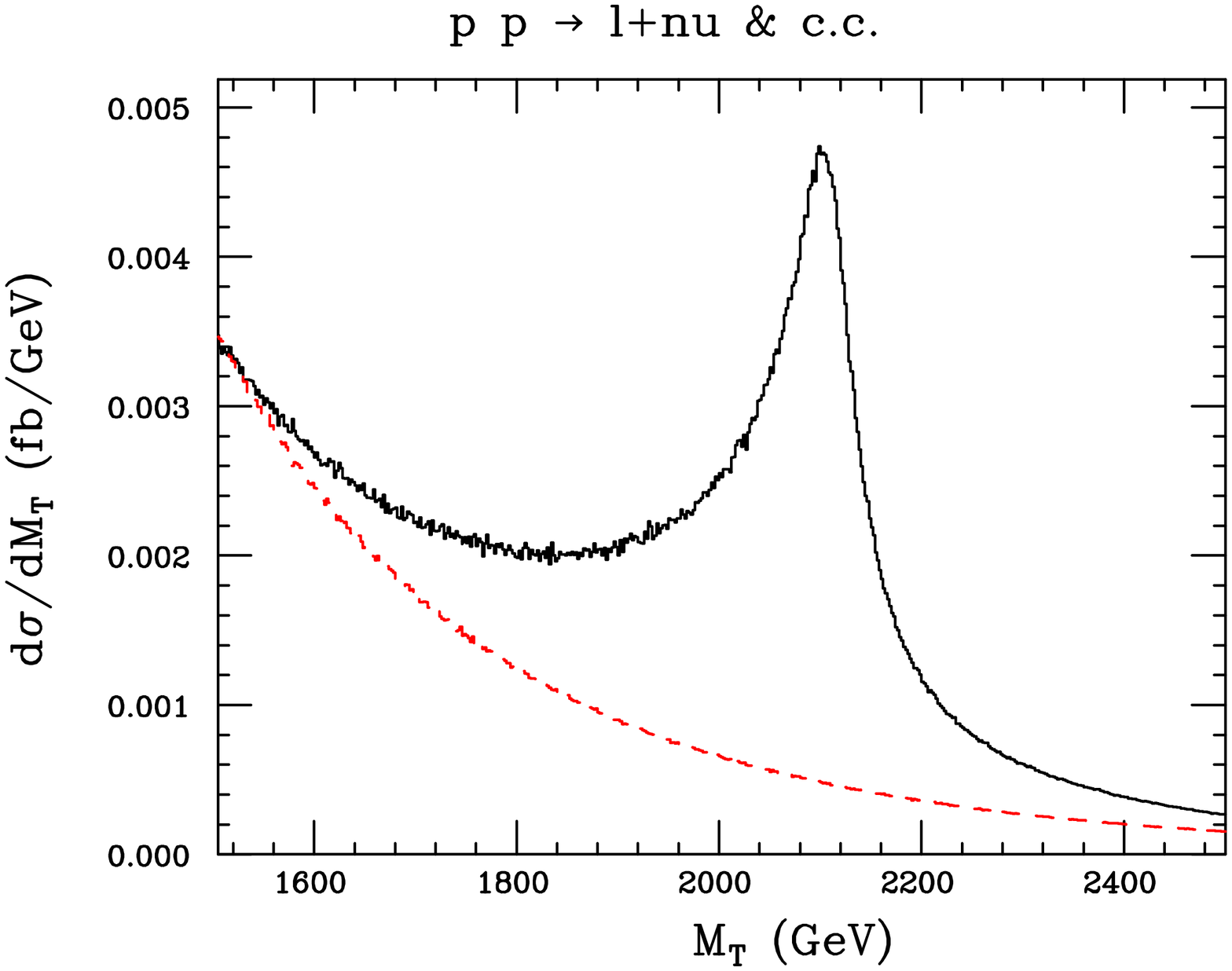,angle=90, width=.46\textwidth}
\epsfig{file=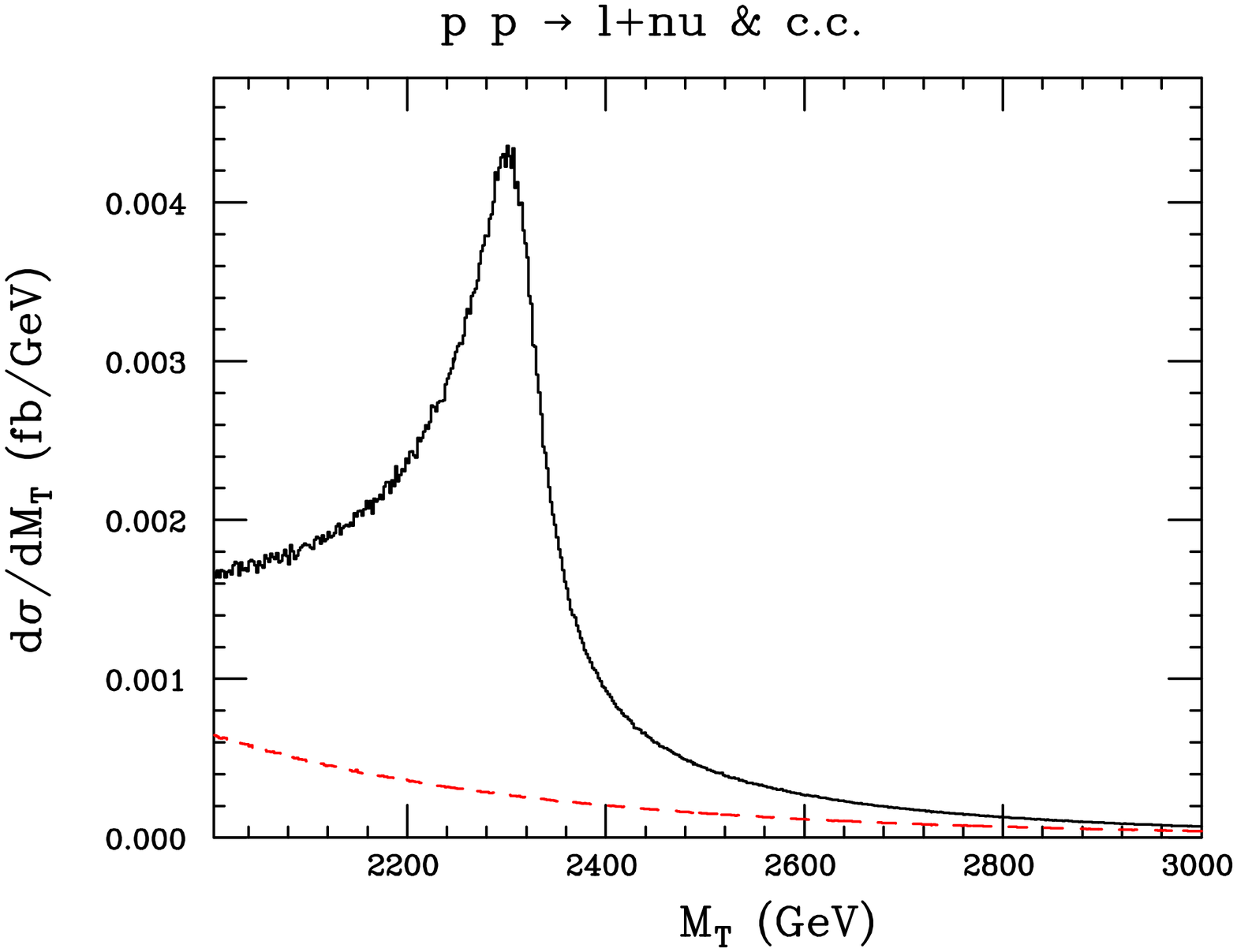,angle=90, width=.46\textwidth}
\caption[Transverse mass distributions for the cross section for DY processes at the LHC with variable model scale $f$ and coupling constant $g_\rho$]{Transverse mass differential distribution for the cross section at the 14 TeV LHC for the CC DY process for the choice of benchmark points $f=750\textrm{ GeV},g_\rho=2$ (a), $f=800\textrm{ GeV},g_\rho=2.5$ (b), $f=1000\textrm{ GeV},g_\rho=2$ (c) and $f=1200\textrm{ GeV},g_\rho=1.8$ (d) for which the complete set of input parameter for the small width regime is reported in Tab.~\ref{tab:input_bench-fgvar}. The solid black line represent the total signal while the dashed red the SM background. For integrated cross sections and cuts see Tab.~\ref{tab:cs-DY-NC-CC-fvar}.}
\label{fig:DY-CC-fvar}
\end{figure}

The results of the invariant mass distributions for these choices are shown in Fig.~\ref{fig:DY-NC-fvar} for the NC and Fig.~\ref{fig:DY-CC-fvar} for the CC case.
From Fig.~\ref{fig:DY-NC-fvar} we note that, while the two lightest $Z^\prime$ resonances are clearly accessible, this is not the case for the heaviest one, $Z^\prime_5$, which is much suppressed due to its higher mass, $\simeq \sqrt{2}f g_\rho$, and smaller couplings which have a leading order contribution proportional to $\sqrt{\xi}$, eq.(\ref{eq:coup-neu-me}), and for which we have not shown the corresponding mass range in the plots except, marginally, in Fig.~\ref{fig:DY-NC-fvar} (d).
A clear excess is always seen near the quasi degenerate $Z^\prime_2$ and $Z^\prime_3$ masses and, while in some cases they are not resolvable, in others they can be separated, certainly in $e^+e^-$ but not in $\mu^+\mu^-$. Notice in fact that the bin width in the plots is 2 GeV, so that even assembling ten of them in the case
of an electron pair final state will not spoil the ability to establish the two resonances separately, while this is not the case for a muon pair final state since, due to the mentioned 10\% mass resolution, this will imply integrating over a hundred bins.
The overall cross sections are $\mathcal{O}(1-10\; \rm{fb})$ which render each of these parameter configurations accessible at the 14 TeV LHC, albeit limited to the lowest resonances only.
From the CC process of Fig.~\ref{fig:DY-CC-fvar} we can see that only the lightest of the two resonances involved, $W^\prime_2$ and $W^\prime_3$, can be detected since for the heaviest one the same consideration applies as in the case of $Z^\prime_5$. Having to use the transverse mass variable, since in this case is not possible to reconstruct the invariant mass of the decaying resonances, which is less correlated with the mass of the decaying $W^\prime$, the visible peaks are broader compared to the NC case. However even in this case the total signal is clearly visible over the SM background, although limited to the lightest resonance, and event rates are somewhat larger with respect to the neutral current case.

\begin{table}[htb]
\centering
 {\begin{tabular}{|l|l|l|}
\hline
(a) & $M$ (GeV) & $\Gamma$ (GeV) \\
\hline
$Z^\prime_2$ &1549 &28 \\
$Z^\prime_3$ &1581 &26 \\
$Z^\prime_5$ &2124 &34 \\
\hline
$W^\prime_2$ &1581 &26 \\
$W^\prime_3$ &2123 &33 \\
\hline
\hline
\multicolumn{1}{|c|}{} & $l^+ l^-$& $l^+\nu$ + c.c. \\
\hline
\multicolumn{1}{|c|}{$\sigma$ (fb)} & 7.44[5.46]& 13.22[6.96] \\
\hline
\multicolumn{1}{|c|}{$p_l^T$(GeV)} &$>20$ &$>20$ \\
\multicolumn{1}{|c|}{$|\eta_l|$} &$<2.5$ & $<2.5$ \\
\multicolumn{1}{|c|}{$M_{l^+l^-/T}$(TeV)} & $>1$ &$>1$  \\
\cline{1-3}
\end{tabular}}\hfill
  {\begin{tabular}{|l|l|l|}
\hline
(b) & $M$ (GeV) & $\Gamma$ (GeV) \\
\hline
$Z^\prime_2$ &2041 &61 \\
$Z^\prime_3$ &2068 &98 \\
$Z^\prime_5$ &2830 &223 \\
\hline
$W^\prime_2$ &2067 &98 \\
$W^\prime_3$ &2830 &221 \\
\hline
\hline
\multicolumn{1}{|c|}{} & $l^+ l^-$& $l^+\nu$ + c.c. \\
\hline
\multicolumn{1}{|c|}{$\sigma$ (fb)} & 0.90[0.91]& 1.19[1.06]\\
\hline
\multicolumn{1}{|c|}{$p_l^T$(GeV)} &$>20$ &$>20$ \\
\multicolumn{1}{|c|}{$|\eta_l|$} &$<2.5$ & $<2.5$ \\
\multicolumn{1}{|c|}{$M_{l^+l^-/T}$(TeV)} & $>1.5$ &$>1.5$  \\
\cline{1-3}
\end{tabular}}
\\
\vskip 7pt
  {\begin{tabular}{|l|l|l|}
\hline
(c) & $M$ (GeV) & $\Gamma$ (GeV) \\
\hline
$Z^\prime_2$ &2066 &39 \\
$Z^\prime_3$ &2111 &52 \\
$Z^\prime_5$ &2830 &71 \\
\hline
$W^\prime_2$ &2111 &52 \\
$W^\prime_3$ &2830 &50 \\
\hline
\hline
\multicolumn{1}{|c|}{} & $l^+ l^-$& $l^+\nu$ + c.c. \\
\hline
\multicolumn{1}{|c|}{$\sigma$ (fb)} & 1.24[0.91]& 2.04[1.06] \\
\hline
\multicolumn{1}{|c|}{$p_l^T$(GeV)} &$>20$ &$>20$ \\
\multicolumn{1}{|c|}{$|\eta_l|$} &$<2.5$ & $<2.5$ \\
\multicolumn{1}{|c|}{$M_{l^+l^-/T}$(TeV)} & $>1.5$ &$>1.5$  \\
\cline{1-3}
\end{tabular}}\hfill
  {\begin{tabular}{|l|l|l|}
\hline
(d) & $M$ (GeV) & $\Gamma$ (GeV) \\
\hline
$Z^\prime_2$ &2249 &32 \\
$Z^\prime_3$ &2312 &55 \\
$Z^\prime_5$ &3056 &54 \\
\hline
$W^\prime_2$ &2312 &55 \\
$W^\prime_3$ &3056 &54 \\
\hline
\hline
\multicolumn{1}{|c|}{} & $l^+ l^-$& $l^+\nu$ + c.c. \\
\hline
\multicolumn{1}{|c|}{$\sigma$ (fb)} & 0.78[0.21]& 1.11[0.23] \\
\hline
\multicolumn{1}{|c|}{$p_l^T$(GeV)} &$>20$ &$>20$ \\
\multicolumn{1}{|c|}{$|\eta_l|$} &$<2.5$ & $<2.5$ \\
\multicolumn{1}{|c|}{$M_{l^+l^-/T}$(TeV)} & $>2$ &$>2$  \\
\cline{1-3}
\end{tabular}}
\caption[Masses and widths of the extra gauge bosons for various model scales $f$ and coupling constants $g_\rho$]{Extra gauge bosons masses and widths for the 
gauge boson arising from the benchmarks of eq.(\ref{eq:bench-fvar}) for which the complete set of input parameter for the small width regime is reported in Tab.~(\ref{tab:input_bench-fgvar}) and corresponding integrated event rates for the NC and CC after the applications of the indicated selection cuts. In square brackets the SM background. $f$=750 GeV, $g_\rho=2$ (a), $f$=800 GeV, $g_\rho=2.5$ (b), $f$=1000 GeV, $g_\rho=2$ (c), $f$=1200 GeV, $g_\rho=1.8$ (d).}
\label{tab:cs-DY-NC-CC-fvar}
\end{table}

\begin{figure}[!h]                                      
\centering
\epsfig{file=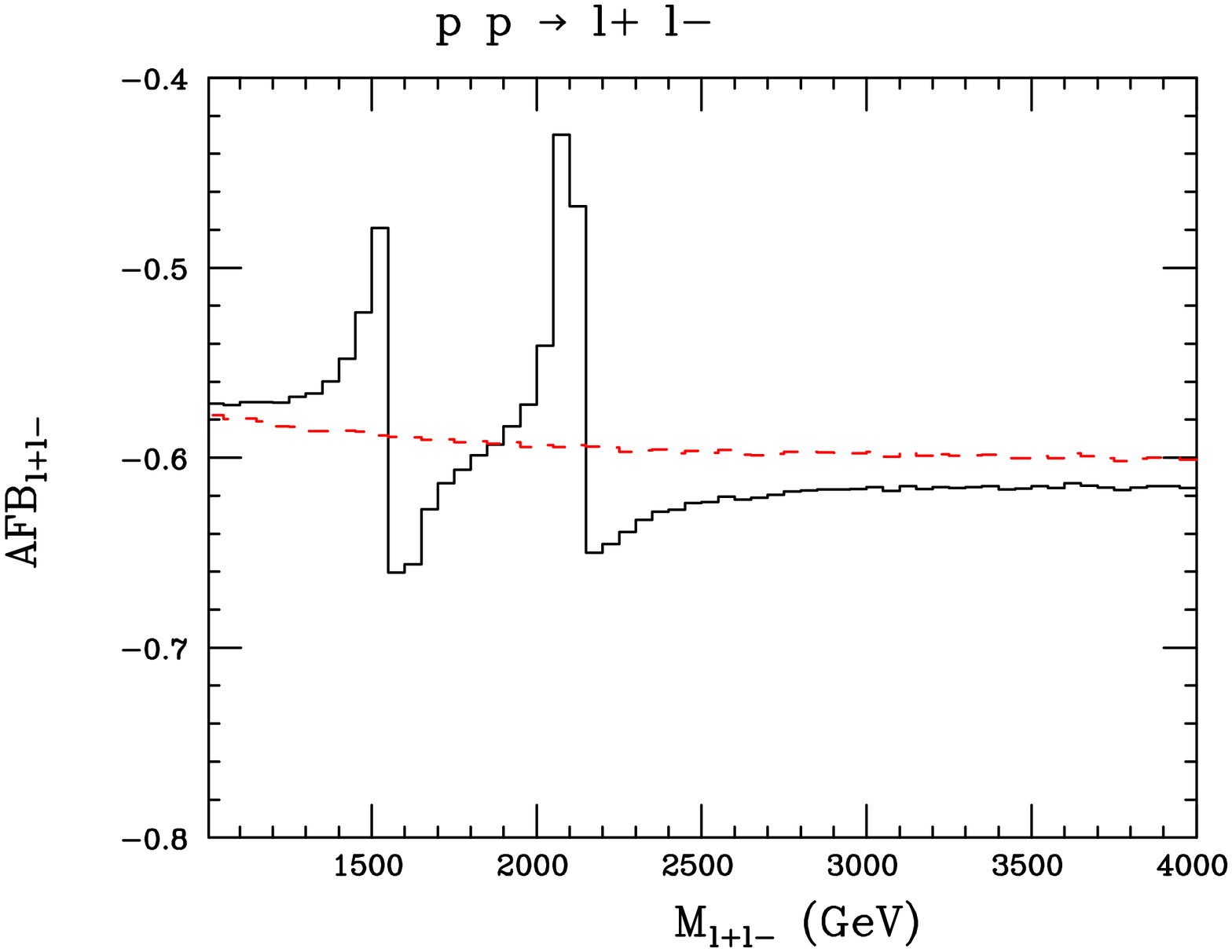,angle=90, width=.46\textwidth}
\epsfig{file=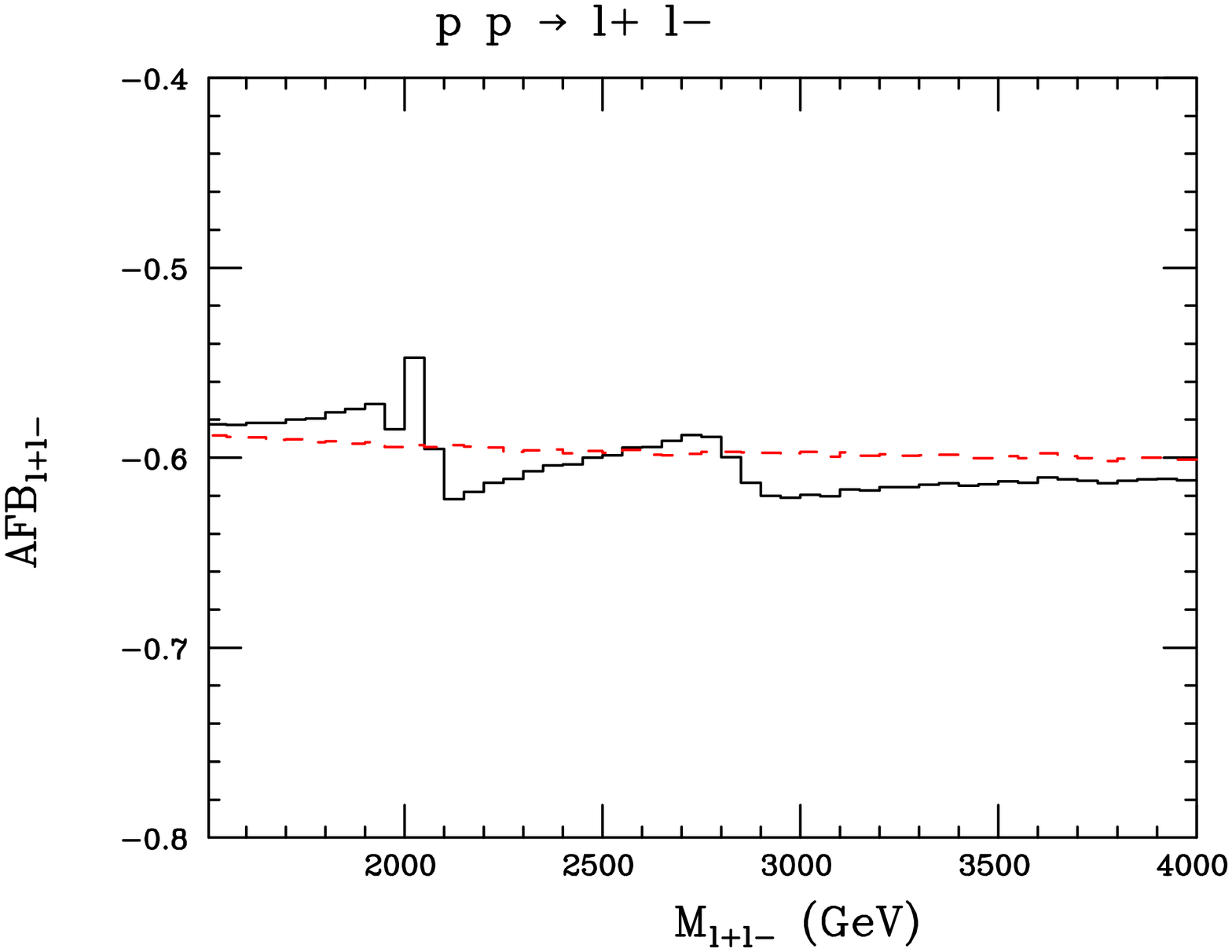,angle=90, width=.46\textwidth}
\epsfig{file=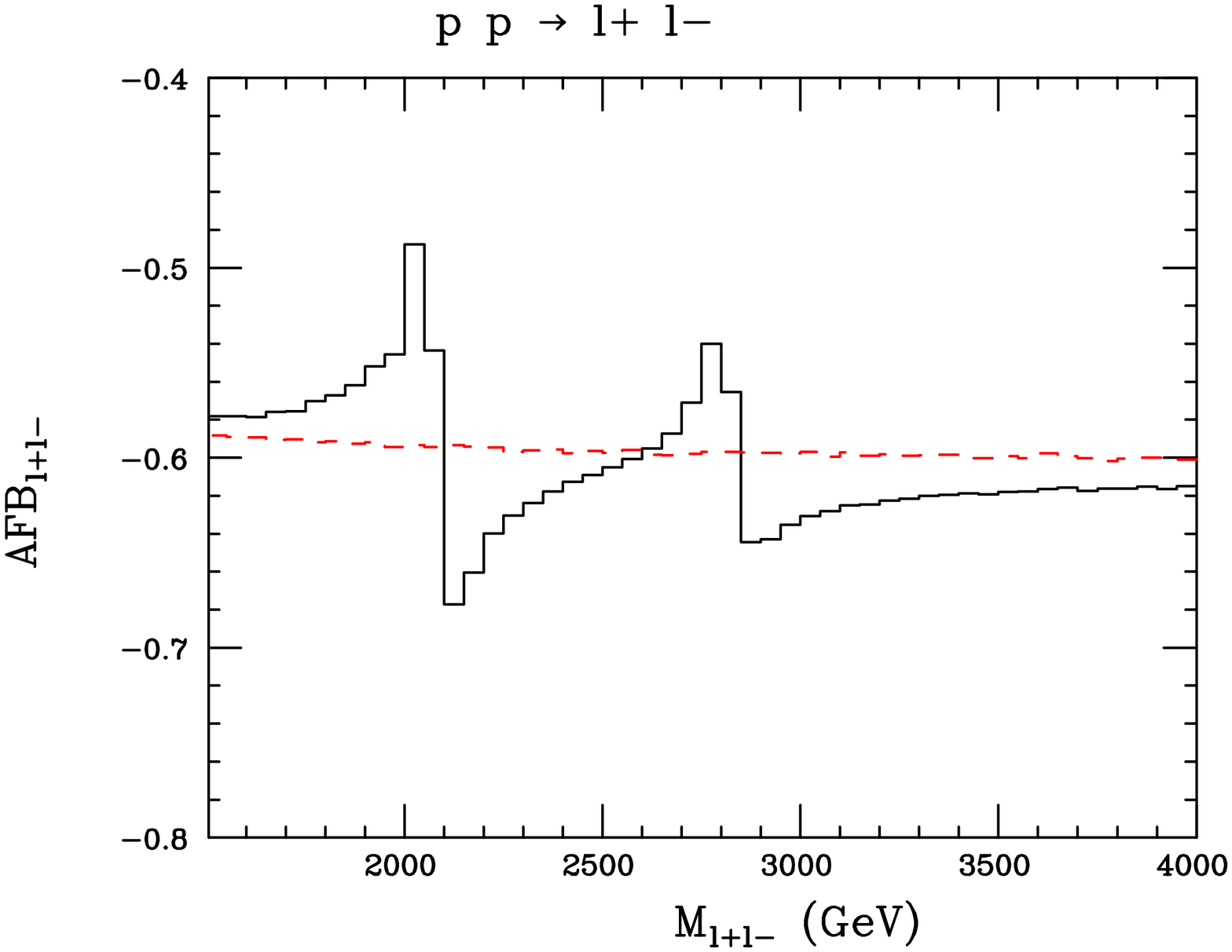,angle=90, width=.46\textwidth}
\epsfig{file=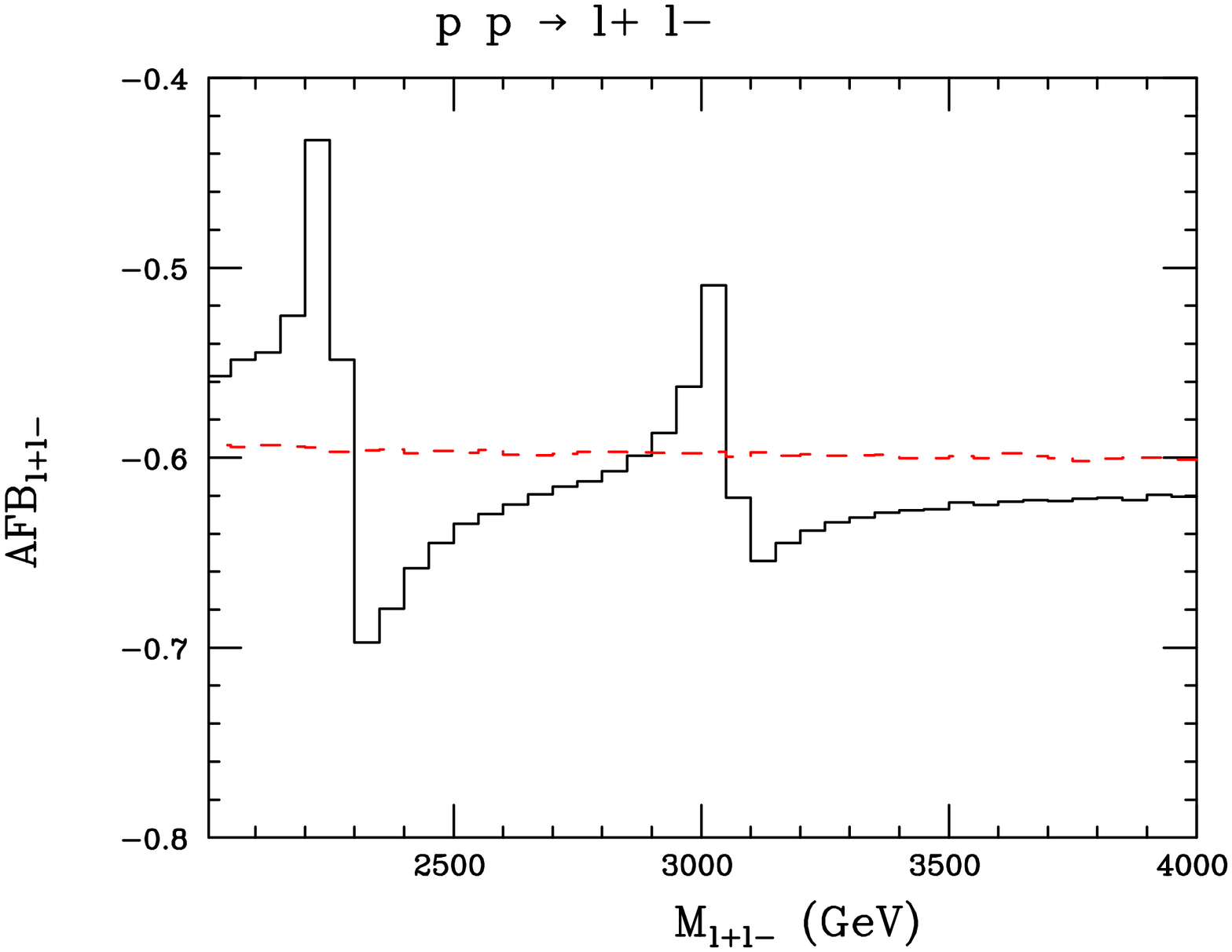,angle=90, width=.46\textwidth}
\caption[Invariant mass distributions for the AFB for DY processes at the LHC with variable model scale $f$ and coupling constant $g_\rho$]{Invariant mass differential distribution for the AFB at the 14 TeV LHC for the CC DY process for the choice of benchmark points $f=750\textrm{ GeV},g_\rho=2$ (a), $f=800\textrm{ GeV},g_\rho=2.5$ (b), $f=1000\textrm{ GeV},g_\rho=2$ (c) and $f=1200\textrm{ GeV},g_\rho=1.8$ (d) for which the complete set of input parameters for the small width regime is reported in Tab.~\ref{tab:input_bench-fgvar}. The solid black line represent the total signal while the dashed red the SM background. For integrated cross sections and cuts see Tab.~\ref{tab:cs-DY-NC-CC-fvar}.}
\label{fig:DY-AFB-NC-fgvar}
\end{figure}

\begin{figure}[!h]                                      
\centering
\epsfig{file=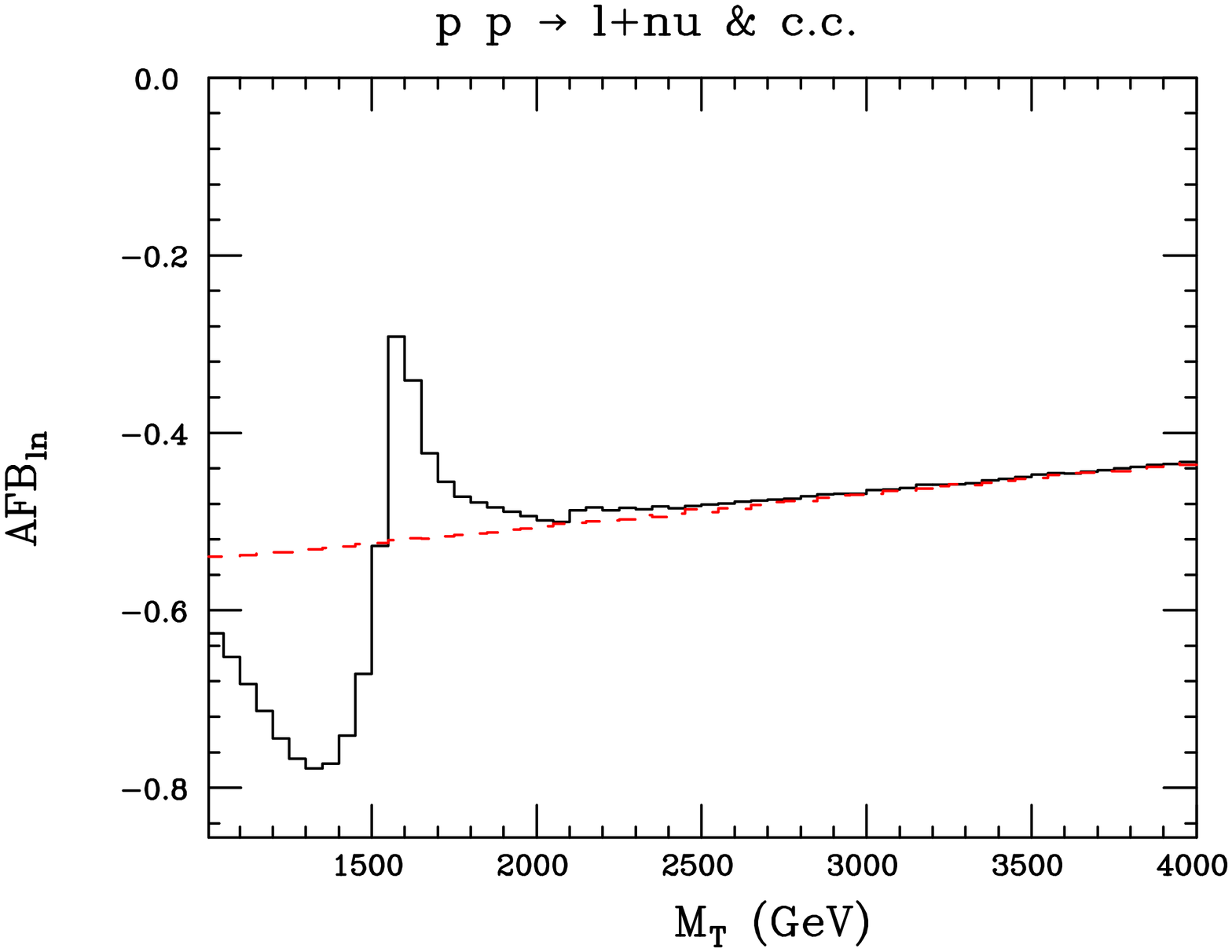,angle=90, width=.46\textwidth}
\epsfig{file=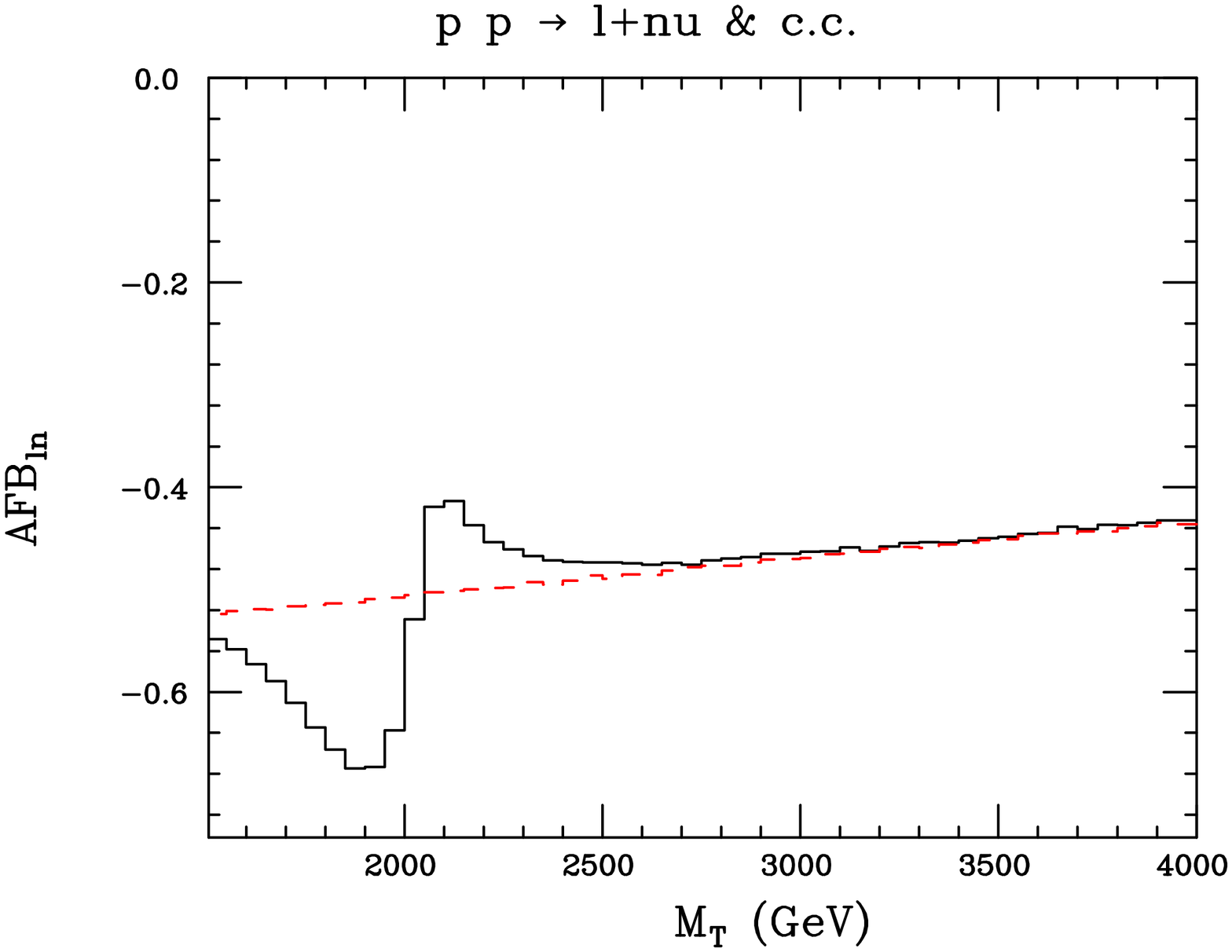,angle=90, width=.46\textwidth}
\epsfig{file=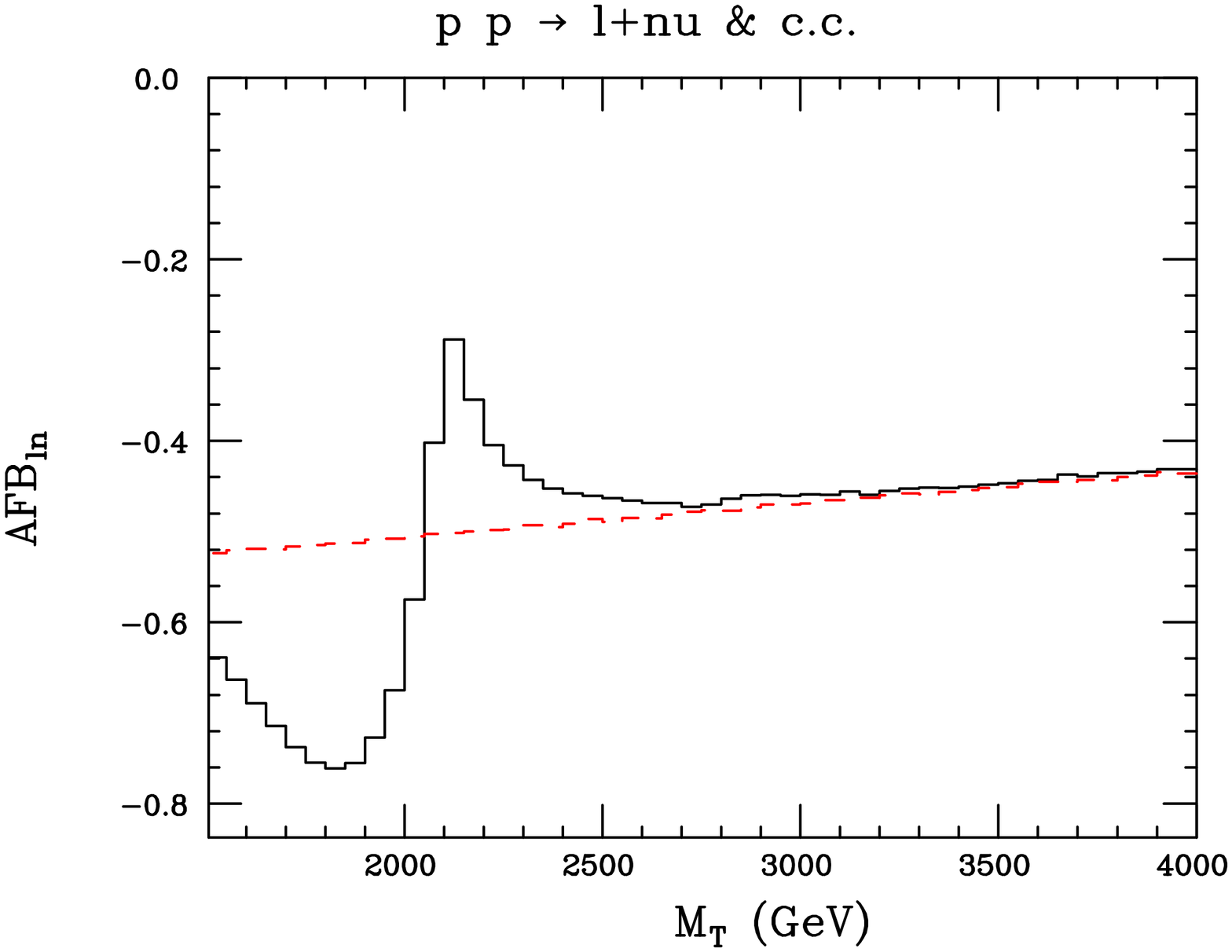,angle=90, width=.46\textwidth}
\epsfig{file=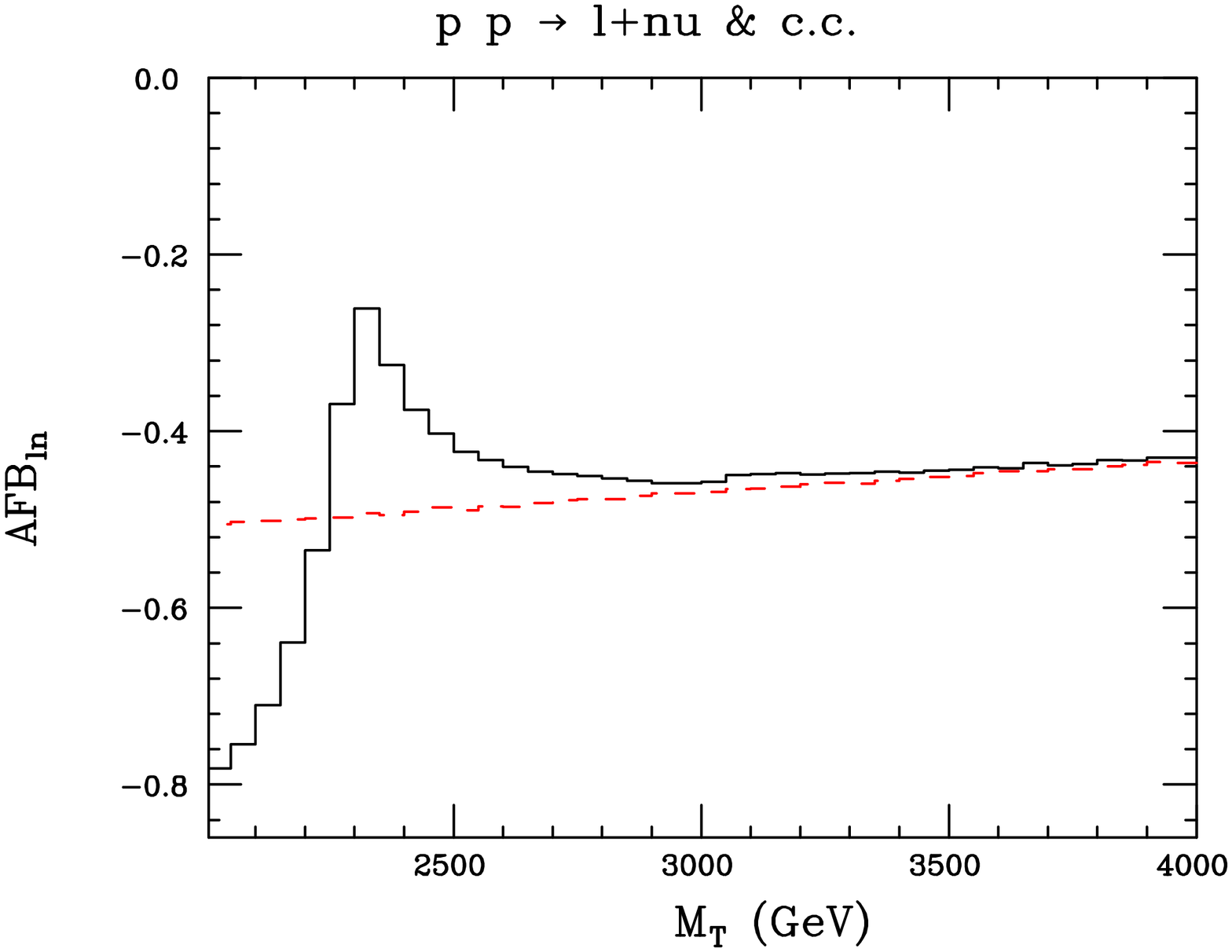,angle=90, width=.46\textwidth}
\caption[Transverse mass distributions for the AFB for DY processes at the LHC with variable model scale $f$ and coupling constant $g_\rho$]{Transverse mass differential distribution for the AFB at the 14 TeV LHC for the CC DY process for the choice of benchmark points $f=750\textrm{ GeV},g_\rho=2$ (a), $f=800\textrm{ GeV},g_\rho=2.5$ (b), $f=1000\textrm{ GeV},g_\rho=2$ (c) and $f=1200\textrm{ GeV},g_\rho=1.8$ (d) for which the complete set of input parameters for the small width regime is reported in Tab.~\ref{tab:input_bench-fgvar}. The solid black line represent the total signal while the dashed red the SM background. For integrated cross sections and cuts see Tab.~\ref{tab:cs-DY-NC-CC-fvar}.}
\label{fig:DY-AFB-CC-fgvar}
\end{figure}

In both the NC and the CC case it is possible to define the forward backward asymmetry (AFB) of the cross section, as the direction of the reference incoming quark or antiquark can be inferred from the direction of the boost onto the final state in the laboratory frame.
The AFB can be sampled in invariant (NC) or transverse (CC) mass bins by defining
\begin{equation}
\frac{dAFB}{dM}=\frac{d\sigma(\cos \theta >0)/dM-d\sigma(\cos \theta <0)/dM}{d\sigma(\cos \theta >0)/dM+d\sigma(\cos \theta <0)/dM}
\label{eq:DY-AFB}
\end{equation}
where $\theta$ is the polar angle of the reference outgoing lepton or anti lepton relative to the direction of the reference incoming quark or antiquark and $M=M_{l^+l^-}$, $M_T$ for the NC and CC processes respectively.
In the CC case however one cannot reconstruct unambiguously the longitudinal component of the neutrino momentum in order to define the asymmetry, because of the twofold solution from the mass equation. We therefore take both solutions and plot them with a half weight each, which somewhat dilutes the asymmetry, in order to individuate the direction of the boost. This is done by assuming that the invariant mass of the final state corresponds to the SM $W$
mass in case of the SM hypothesis and with the $W^\prime_{2,3}$ in case of the 4DCHM one, where an indicative value of the latter can be obtained from the neutral gauge bosons mass, given the correlation between the $Z^\prime s$ and $W^\prime s$ masses.
We show the results for the NC and CC cases in Fig.~\ref{fig:DY-AFB-NC-fgvar} and Fig.~\ref{fig:DY-AFB-CC-fgvar} respectively, where the bins width are now 50 GeV, for the same benchmark points used for Fig.~\ref{fig:DY-NC-fvar} and Fig.~\ref{fig:DY-CC-fvar}.
From the NC case we see that such an observable displays a peculiar dependence in the vicinity of the $Z^\prime s$ masses, including also the heaviest one, $Z^\prime_5$, for all the chosen combinations of $f$ and $g_\rho$, while in the CC case we notice that the resolving power of the resonance in the AFB is diminished, as the presence of the heaviest $W^\prime$ is hardly visible. Also, while in the NC case we can claim that these effects should be resolvable, no matter what the final state, since the bin width here is 50 GeV, the same is not true for the CC case, due to the worst resolution for the transverse mass.

\begin{figure}[!h]                                      
\centering
\epsfig{file=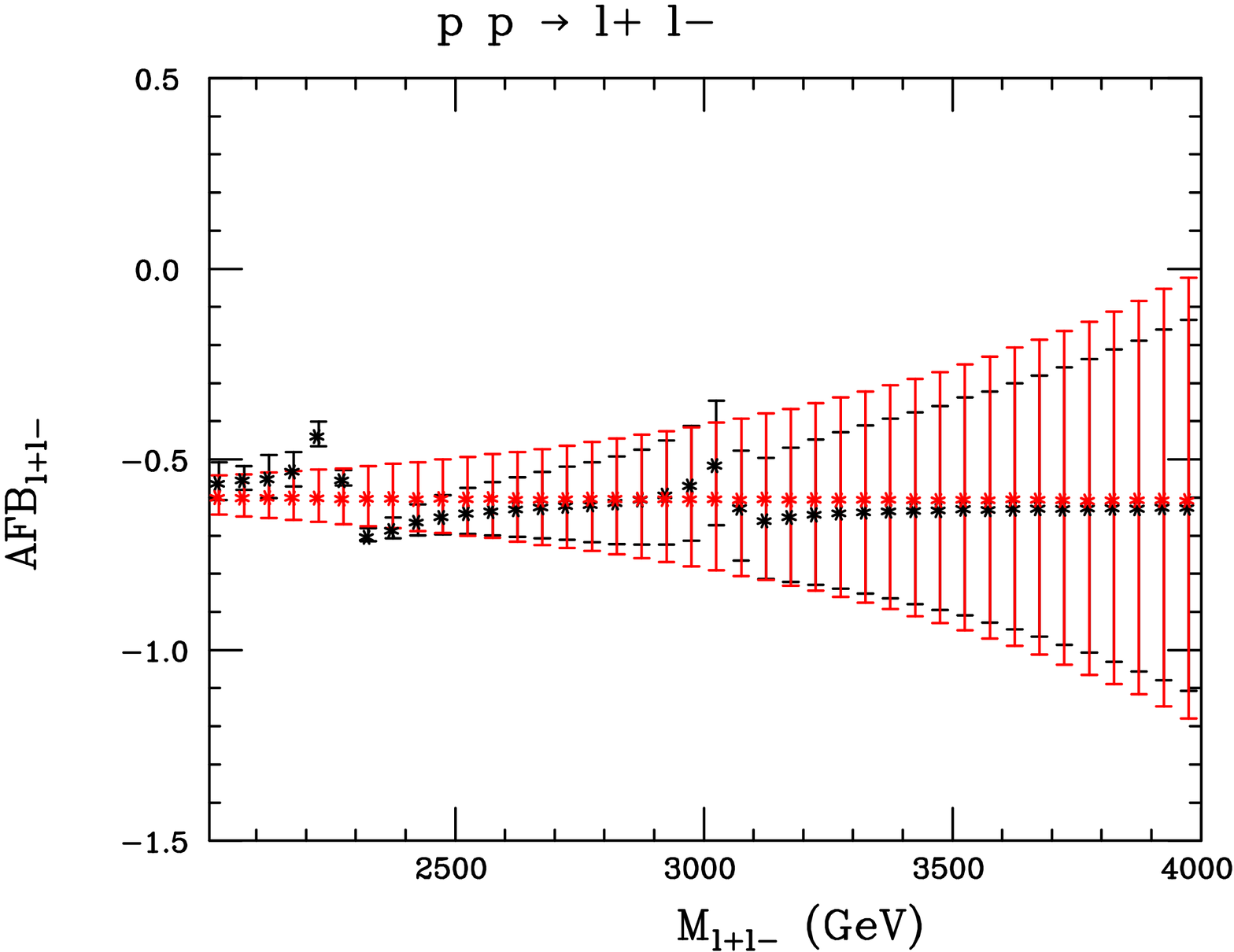,angle=90, width=.46\textwidth}{(a)}
\epsfig{file=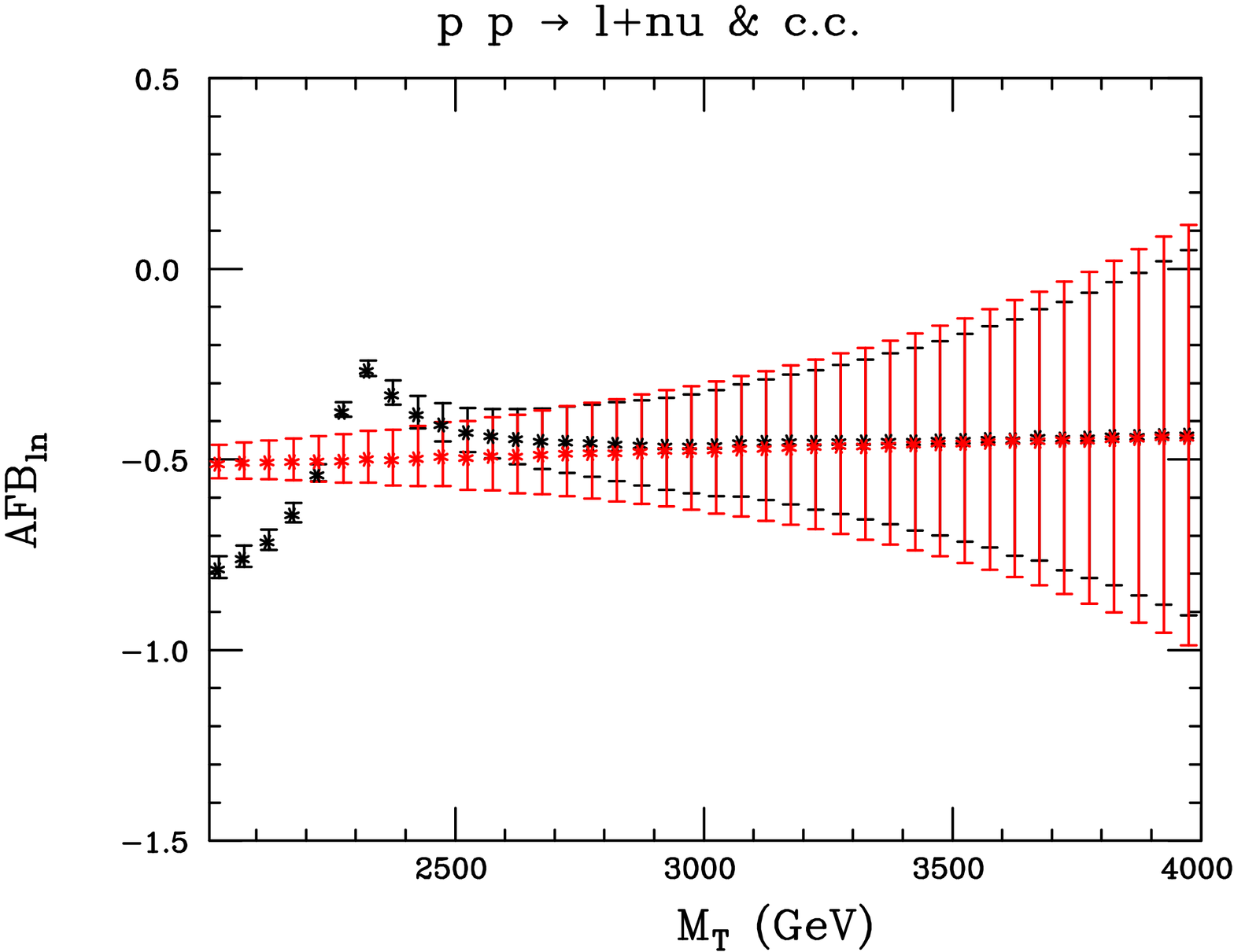,angle=90, width=.46\textwidth}{(b)}
\caption[Invariant and transverse mass distributions for the AFB at the LHC for DY processes with statistical error bars]{Invariant (a) and transverse (b) mass differential distribution for the AFB with corresponding error bars at the 14 TeV LHC for the CC DY process for the choice of benchmark point $f=1200\textrm{ GeV},g_\rho=1.8$ for which the complete set of input parameter for the small width regime is reported in Tab.~\ref{tab:input_bench-fgvar}. Error bars assume $\mathcal L=$1500 fb$^{-1}$ and $\epsilon$=90\%. In red the total signal and in black the SM background. For integrated cross sections and cuts see Tab.~\ref{tab:cs-DY-NC-CC-fvar}.}
\label{fig:DY-AFB-NC-CC-Err}
\end{figure}

However, in order to be able to quantitatively address the distinguishability between the 4DCHM and the SM, the statistical errors on the predictions ought to be calculated. 
While they are quite irrelevant in the case of the cross section, we need to pay particular attention to the AFB distributions.
Given that the AFB is defined in terms of the number of events measured in a forward ($N_F$) and backward ($N_B$) direction, the statistical error is evaluated by propagating the Poisson error on each measured quantity, that is $\delta N_{F,B}=\sqrt{N_{F,B}}$. For a given integrated luminosity $\mathcal L$ the measured number of events will be
$N_{F,B}=\epsilon \mathcal L \sigma_{F,B}$, where $\sigma_{F,B}$ is the integrated or differential forward, backward cross section and $\epsilon$ is the assumed reconstruction efficiency of the $l^+l^-$ and $l^- \bar \nu + c.c.$ system, yielding an uncertainty on AFB of
\begin{equation}
\delta(AFB)=\delta(\frac{N_F-N_B}{N_F+N_B})=\sqrt{\frac{4}{\mathcal L \epsilon}\frac{\sigma_F \sigma_B}{(\sigma_F+\sigma_B)^3}}
\end{equation}
In Fig.~\ref{fig:DY-AFB-NC-CC-Err} we then show  the results for the NC (a) and CC (b) cases with the inclusion of the relative error bars, that are calculated for $\mathcal L=$1500 fb$^{-1}$ and $\epsilon$=90\%, for the choice of $f=1200$ GeV and $g_\rho=1.8$ in the small width regime assumption.
We see that the resonant effects in AFB can still  be discernible with respect to the SM noise in both the NC and CC case, albeit limited to the lowest lying resonances only in both cases, so long as very high luminosity can be attained,
and that in the neutral case, for a di electron final state,  $Z^\prime_5$ also remains however an open possibility.

Before closing this Section we want to point out how the mass correlation between the neutral and charged gauge resonances, dictated by the group structure of the model, can be exploited to improve searches in either of the neutral or charged channels.
Suppose that (justified by the fact that the events rates in the CC case are somewhat bigger than in the NC case) a $W^\prime$ resonance has been seen in the transverse mass spectrum and nothing appears, after standard acceptance cuts, in the invariant mass distribution, the knowledge of the $W^\prime$ mass also implies the knowledge of the $Z^\prime$ mass, so that, to exalt the resonance associated with the latter, one may impose onto the di lepton NC sample a cut, for example $p^T_l>M_{W^\prime}/2$, extracted from the analysis of the CC sample. Clearly the opposite exercise can also be carried out if it is a $Z^\prime$ the first one to have been seen, which could occur, for example, in the case of a very narrow width, helping to improve the $W^\prime$ signal selection also in this case.

\clearpage


\section{Diboson signals at the LHC}

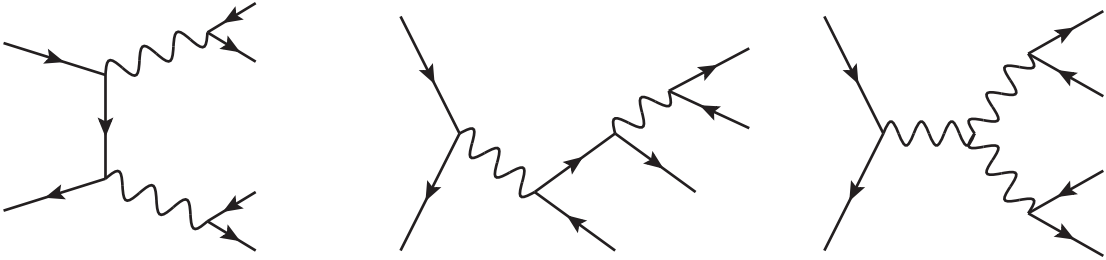
\begin{figure}[!h]
  \begin{picture}(514,130) (79,-63)
    \SetWidth{1.0}
    \SetColor{Black}
    \Line[arrow,arrowpos=0.5,arrowlength=5,arrowwidth=2,arrowinset=0.2](80,46)(118,34)
    \Line[arrow,arrowpos=0.5,arrowlength=5,arrowwidth=2,arrowinset=0.2](118,34)(118,-5)    
    \Line[arrow,arrowpos=0.5,arrowlength=5,arrowwidth=2,arrowinset=0.2](118,-5)(80,-17)   
    \Photon(118,34)(156,50){4.5}{3}   
    \Photon(118,-5)(156,-21){4.5}{3}    
    \Line[arrow,arrowpos=0.5,arrowlength=5,arrowwidth=2,arrowinset=0.2](174,61)(156,50)
    \Line[arrow,arrowpos=0.5,arrowlength=5,arrowwidth=2,arrowinset=0.2](156,50)(174,39)
    \Line[arrow,arrowpos=0.5,arrowlength=5,arrowwidth=2,arrowinset=0.2](174,-10)(156,-21)
    \Line[arrow,arrowpos=0.5,arrowlength=5,arrowwidth=2,arrowinset=0.2](156,-21)(174,-32)
    \Line[arrow,arrowpos=0.5,arrowlength=5,arrowwidth=2,arrowinset=0.2](228,56)(250,12)
    \Line[arrow,arrowpos=0.5,arrowlength=5,arrowwidth=2,arrowinset=0.2](250,12)(228,-32)
    \Photon(250,12)(278,-10){4.5}{3}
    \Line[arrow,arrowpos=0.5,arrowlength=5,arrowwidth=2,arrowinset=0.2](308,-32)(278,-10)
    \Line[arrow,arrowpos=0.5,arrowlength=5,arrowwidth=2,arrowinset=0.2](278,-10)(308,12)
    \Line[arrow,arrowpos=0.5,arrowlength=5,arrowwidth=2,arrowinset=0.2](308,12)(338,-10)
    \Photon(308,12)(328,28){4.5}{2}
    \Line[arrow,arrowpos=0.5,arrowlength=5,arrowwidth=2,arrowinset=0.2](328,28)(358,44)
    \Line[arrow,arrowpos=0.5,arrowlength=5,arrowwidth=2,arrowinset=0.2](358,14)(328,28)
    \Line[arrow,arrowpos=0.5,arrowlength=5,arrowwidth=2,arrowinset=0.2](386,56)(408,12)
    \Line[arrow,arrowpos=0.5,arrowlength=5,arrowwidth=2,arrowinset=0.2](408,12)(386,-32)
    \Photon(408,12)(442,12){4.5}{3}
    \Photon(442,12)(462,42){4.5}{3}
    \Photon(442,12)(462,-18){-4.5}{3}
    \Line[arrow,arrowpos=0.5,arrowlength=5,arrowwidth=2,arrowinset=0.2](462,42)(490,58)
    \Line[arrow,arrowpos=0.5,arrowlength=5,arrowwidth=2,arrowinset=0.2](490,26)(462,42)
    \Line[arrow,arrowpos=0.5,arrowlength=5,arrowwidth=2,arrowinset=0.2](490,-2)(462,-18)
    \Line[arrow,arrowpos=0.5,arrowlength=5,arrowwidth=2,arrowinset=0.2](462,-18)(490,-34)
  \end{picture}
  \caption[Feynman diagrams for diboson processes]{Feynman diagrams topologies contributing to the 2$l$ and 3$l$ processes of eq.(\ref{eq:diboson-proc}). The wavy lines correspond to any of the EW gauge bosons of the 4DCHM.}
  \label{fig:diboson-top}
\end{figure}

In this Section we will present the analysis of some diboson production processes at the LHC in order to complete the analysis of its gauge sector started with the study of the DY channels.
The charged and mixed diboson productions at the LHC
\begin{equation}
\begin{split}
pp\to W^+ W^-\to e^+ \nu_e \mu^- \bar \nu_\mu +c.c \to e^\pm \mu^\mp E^T_{miss}\\
pp\to W^\pm Z\to l^+ \nu_l l^{\prime+}l^{\prime -}+c.c \to l^\pm l^{\prime +}l^{\prime -}E^T_{miss}
\label{eq:diboson-proc}
\end{split}
\end{equation}
yielding opposite charge different flavour, for the former, and all flavour charged tri lepton, for the latter, plus missing transverse energy arise from the topologies of Fig.~\ref{fig:diboson-top}; we will refer to them as the $2l$ and $3l$ signatures respectively.
The importance of such processes lies in the fact that, besides the couplings also involved  in the DY processes, these channels allow us to access the triple gauge self couplings of a model, a crucial ingredient in order to pin down the underlying EW gauge structure.
From an experimental point of view it is again the cleanliness of this channel that renders it a favourite from an experimental point of view, while high order QCD and EW corrections are well under control (see again \cite{Campbell:2006wx} for a review).

For our phenomenological analysis we relied on the same subroutines used for the DY case, with the addition that both PHACT \cite{Ballestrero:1999md} and HELAS \cite{Murayama:1992gi} 
have been used as a code based on the helicity amplitude method, and cross checked against each other.
Again the MEs account for all the off shellness effects of the particles involved and were constructed starting from the topologies in Fig.~\ref{fig:diboson-top}, where the wavy lines refer to any of the gauge bosons belonging to the 4DCHM.

We now define some other kinematic variables, apart from the one already used for the DY study, that will be used to define acceptance and selection criteria of the final states in the processes of eq.(\ref{eq:diboson-proc})
\begin{itemize}
\item $p^T_M=\max^n_{i=1}p^T_i$ is the maximum amongst the transverse momenta of $n$ particles
\item $p^T_{ij}=\sqrt{(p^x_i+p^x_j)^2+(p^y_i+p^y_j)^2}$ is the transverse momentum of a pair of particles
\item $M_{ijk}=\sqrt{(p_i+p_j+p_k)^\mu(p_i+p_j+p_k)_\mu}$ is the invariant mass of a tern of particles
\item $\theta_{ij}$ is the angle between two particles in the longitudinal plane
\item $\cos \phi^T_{ij}=\frac{p^x_i p^x_y+p^y_i p^y_j}{p^T_i p^T_j}$ is the cosine of the relative azimuthal angle between two particles in the plane transverse to the beam
\end{itemize}

\subsection{The $2l$ signature}
\label{subsec:diboson-2l}

The acceptance and selection cuts that maximise the sensitivity to the $2l$ signature of eq.(\ref{eq:diboson-proc}) that have been imposed are
\begin{equation}
\begin{split}
& |\eta_{e,\mu}|<2\\
& p^T_{e,\mu}>20 \textrm{ GeV}\\
& E^T_{miss}>50 \textrm{ GeV}\\
& M_{e,\mu}>180 \textrm{ GeV}\\
&p^T_M>300 \textrm{ GeV}\\ 
& \cos\phi^T_{e,\mu}<-0.9\\
&\cos\theta_{e\mu}<0.5
\end{split}
\label{eq:diboson-2l-cuts}
\end{equation}
where, besides the first two, that are standard acceptance cuts, they have been applied in order to suppress the SM background leaving however a not too small signal cross section.

\begin{figure}[!h]                                      
\centering
\epsfig{file=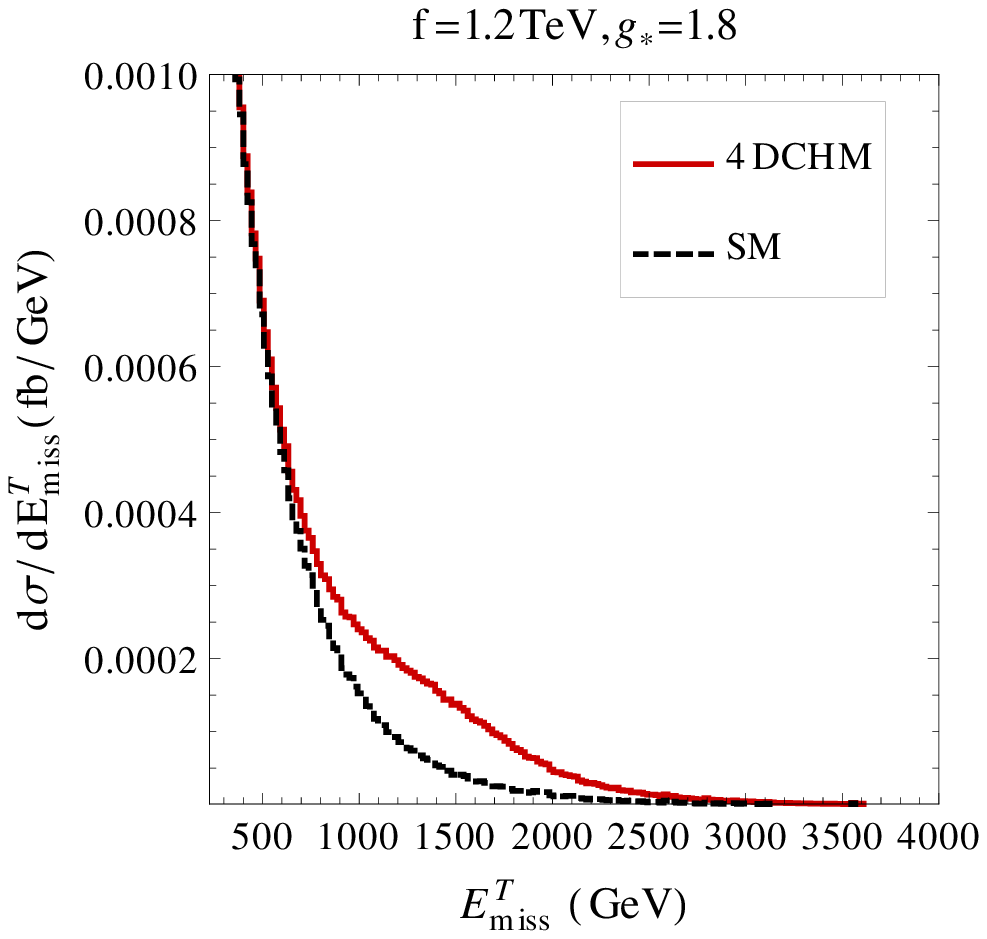, width=.46\textwidth}
\epsfig{file=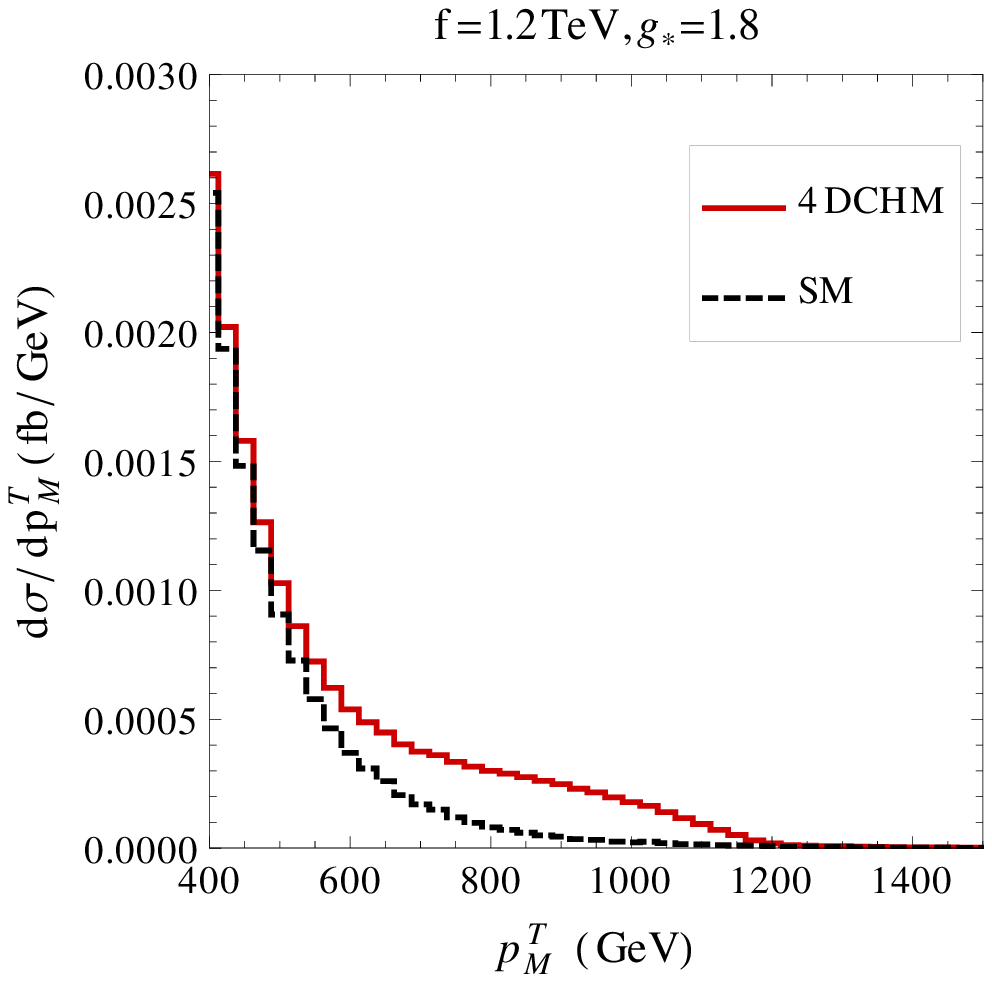, width=.44\textwidth}
\caption[$E^T_{miss}$ and $p^T_M$ distributions for the benchmark with $f$=1200 GeV and $g_\rho$=1.8 for the 2$l$ signature]{Differential cross sections in $E^T_{miss}$ (left) and $p^T_M$ (right) for the benchmark with $f$=1200 GeV and $g_\rho$=1.8 for which the complete set of input parameters can be found in Tab.~\ref{tab:input_bench-fgvar} for the 2$l$ process of eq.(\ref{eq:diboson-proc}). The red solid curve represents the full 4DCHM, while the black dashed the SM background.}
\label{fig:diboson-2l-etmiss-mt-f12g18}
\end{figure}

We then show in Fig.~\ref{fig:diboson-2l-etmiss-mt-f12g18} the cross section distributions in $E^T_{miss}$ and $p^T_M$  for the benchmark point with $f=1200$ GeV and $g_\rho=1.8$ of Tab.~\ref{tab:input_bench-fgvar}.
We notice that the fact that it is not possible to detect all the final state particles, due to the presence of the two neutrinos, makes it very hard to achieve a clear identification of the intermediate vector bosons. However the effect of the extra neutral gauge bosons appears as an excess of events in some energy measure below the value corresponding to the new gauge boson mass, while unfortunately there is no observable that allows us to have a signature of the new charged gauge bosons involved.
Since in the 4DCHM the $Z_2^\prime$ and $Z^\prime_3$ are close in mass, and the $Z^\prime_5$ state is heavier and very weakly coupled to SM fermions, the results in this figure essentially highlight the masses of the quasi degenerate $Z^\prime_{2,3}$ at the end point of the excess region extending to the left of their mass value in the case of $E^T_{miss}$ or to the left of a half of it in the case of $p^T_M$. 
We then present  in Tab.~\ref{tab:diboson-2l-events-fvar} the cross sections for the benchmark points of Tab.~\ref{tab:input_bench-fgvar} after the application of the cuts given in eq.(\ref{eq:diboson-2l-cuts}) for the 14 TeV LHC where we have defined the signal $(S)$ as the difference between the 4DCHM results $(T)$ and the SM one $(B)$.
In the last column we have reported the ratio $S/\sqrt{B}$ that can be used to get an estimate of the statistical significance $\alpha$ from eq.(\ref{eq:signif}).
Assuming for example $\epsilon=0.55$ we see that in order to have an exclusion or discovery, respectively $\alpha=2$ and $\alpha=5$, for the benchmark with $f=1200$ GeV and $g_\rho$=2, it is necessary to require an integrated luminosity for the 14 TeV LHC of $\simeq$ 62 fb$^{-1}$ and 400 fb$^{-1}$. Analysing also the other benchmarks it is generally clear that the lightest $Z^\prime s$ can usually be extracted, if not with the  standard luminosity LHC, surely with a high luminosity upgrade of the CERN machine.
We do not report here the results for the benchmark points with fixed $f$ and $g_\rho$ and variable widths, for which we refer to \cite{Barducci:2012as}, where they have been analysed in detail, but we mention that the extraction of the lightest neutral resonances is possible so long as their widths are small enough with respect to their mass, on the same footing as for the DY analysis.

\begin{table}[!t]
\begin{center}
\begin{tabular}{|c|c|c|c|c|}
\hline 
& $S$~(fb) & $S/\sqrt{B}$~($\sqrt{\rm fb}$) & $\alpha$ & $L_m$ \\
\hline 
(a)   &1.8   &  2.6 & 61 & 7\\
(b) &0.22  &  0.32  & 8 & 444\\
(c)   &0.36  &  0.52 & 12 & 168\\
(d) &0.24  &  0.34 & 8 & 393\\
\hline
\end{tabular}
\end{center}
\caption[Cross sections for the benchmark with variable $f$ and $g_\rho$ for the 2$l$ signature]{
Signal (S), defined as the total cross section (T) minus the SM background (B), $S/\sqrt{B}$ values, statistical significance with $\epsilon=0.55$ and $\mathcal L=$1000 fb$^{-1}$ and minimum luminosity in order to achieve a significance $\alpha=5$ for the benchmarks of eq.(\ref{eq:bench-fvar}) for which the complete set of input parameters can be found in Tab.~\ref{tab:input_bench-fgvar} for the 2$l$ process of eq.(\ref{eq:diboson-proc}).}
\label{tab:diboson-2l-events-fvar}
\end{table}

\subsection{The 3$l$ signature}

Before proceeding with the study of this signature, a subtlety should be noted.
Some of the kinematical variables previously defined implicitly assume the capability to identify, in the final state of process of eq.(\ref{eq:diboson-proc}),
the two leptons coming from the neutral current (propagated by the $\gamma,Z,Z^\prime_{2},Z^\prime_3,Z^\prime_5$ states). In the 
$l=e$ and $l'=\mu$ (or vice versa) case this is trivial, since the pair of leptons
with identical flavour are necessarily those emerging from such a current. In the case $l=l'=e$ or $\mu$ the identification
is in principle ambiguous. However, in Ref.~\cite{Fedeli:2012cn}, an efficient method was devised to overcome this problem, by noting that the maximum amongst the transverse momenta of all pair of 
particles is generally the one induced by the pair of leptons emerging from the $\gamma,Z,Z_{2},Z_3,Z_5$ current, so that this enables us to enforce the same cuts onto the final 
state of the  process of eq.(\ref{eq:diboson-proc}), irrespectively of the actual $l,l'$ combination being generated.

The acceptance and selection criteria that maximise the sensitivity to the $3l$ signature that have been imposed are
\begin{equation}
\begin{split}
&|\eta_{l^\pm,l^{'+},l^{'-}}|<2,\qquad p^T_{l^\pm,l^{'+},l^{'-}}>20~{\rm GeV}, \qquad E^{T}_{\rm miss}>50~{\rm GeV},\\
&M_{ l^\pm l^{'+},~l^\pm l^{'-},~l^{'+}l^{'-}}>20~{\rm GeV},\qquad p^T_M>150~{\rm GeV},\qquad \cos\phi_{l^{'+}l^{'-}}^T<-0.5,\\
&\cos\theta_{l^\pm l^{'+},l^\pm l^{'-},l^{'+}l^{'-}}<0.9, \qquad  p^T_{l^\pm l^{'+},~l^\pm l^{'-},~l^{'+}l^{'-}}>150~{\rm GeV},\\
& M_{l^\pm l^{'+} l^{'-}}>0.9 ~M_{W^\prime_2}.
\end{split}
\label{eq:diboson-3l-cuts}
\end{equation}

The last cut, which unlike the others depends on a 4DCHM parameter, is actually justified by the results of the previous Section, where we have showed that the extraction of a value for $M_{W_2^\prime}$ can be made through the DY channel.
We want to show now that the process 3$l$ of eq.(\ref{eq:diboson-proc}) can act as an effective means to extract part of the mass spectrum of the gauge sector of the 4DCHM that is not accessible elsewhere.
It is crucial however, in order to accomplish this, that the 3$l$ process affords one the possibility to reconstruct the longitudinal momentum of the neutrino, with the algorithm described in \cite{Bach:2011jy}. Therefore, alongside $M_{l^{'+}l^{'-}}$, sensitive to the neutral gauge boson resonances, we can also plot the
reconstructed CM energy at the partonic level, $\sqrt{\hat s}=E^R_{\rm cm}$, which is sensitive to the charged gauge
 boson resonance.
 
\begin{figure}[!h]                                      
\centering
\epsfig{file=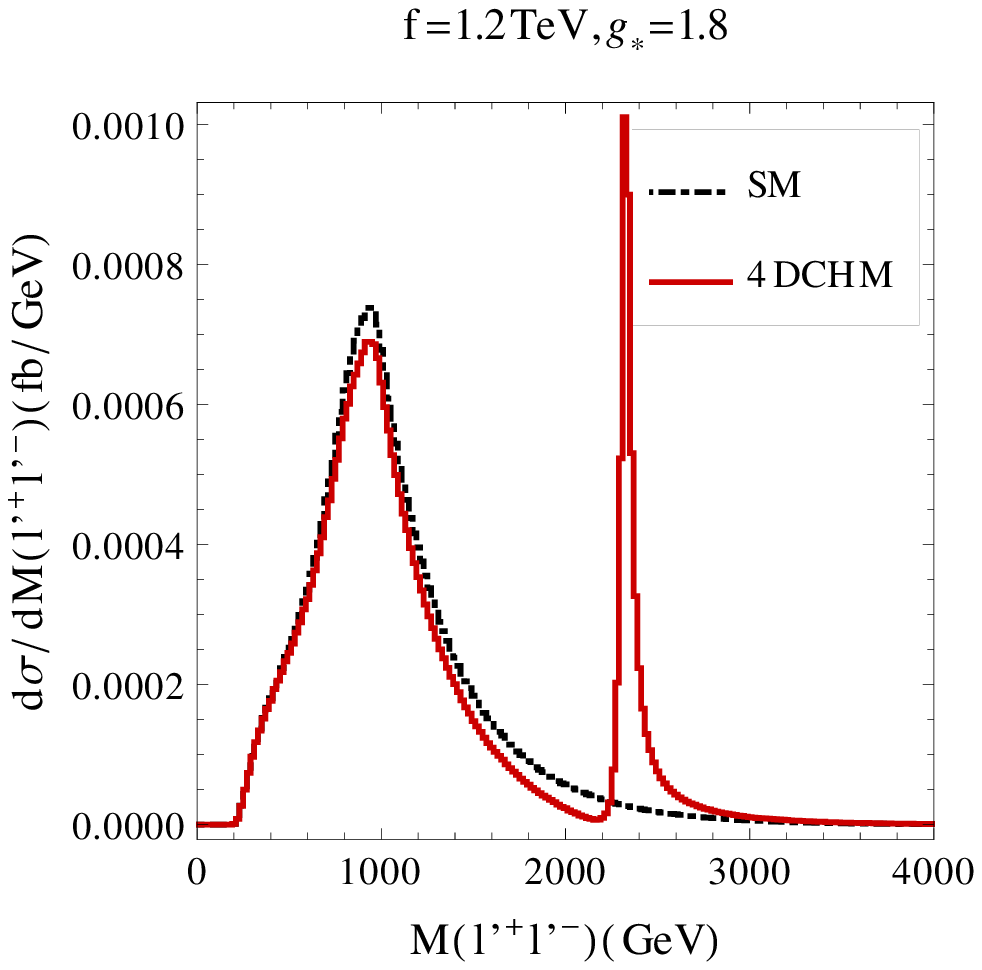 , width=.46\textwidth}
\epsfig{file=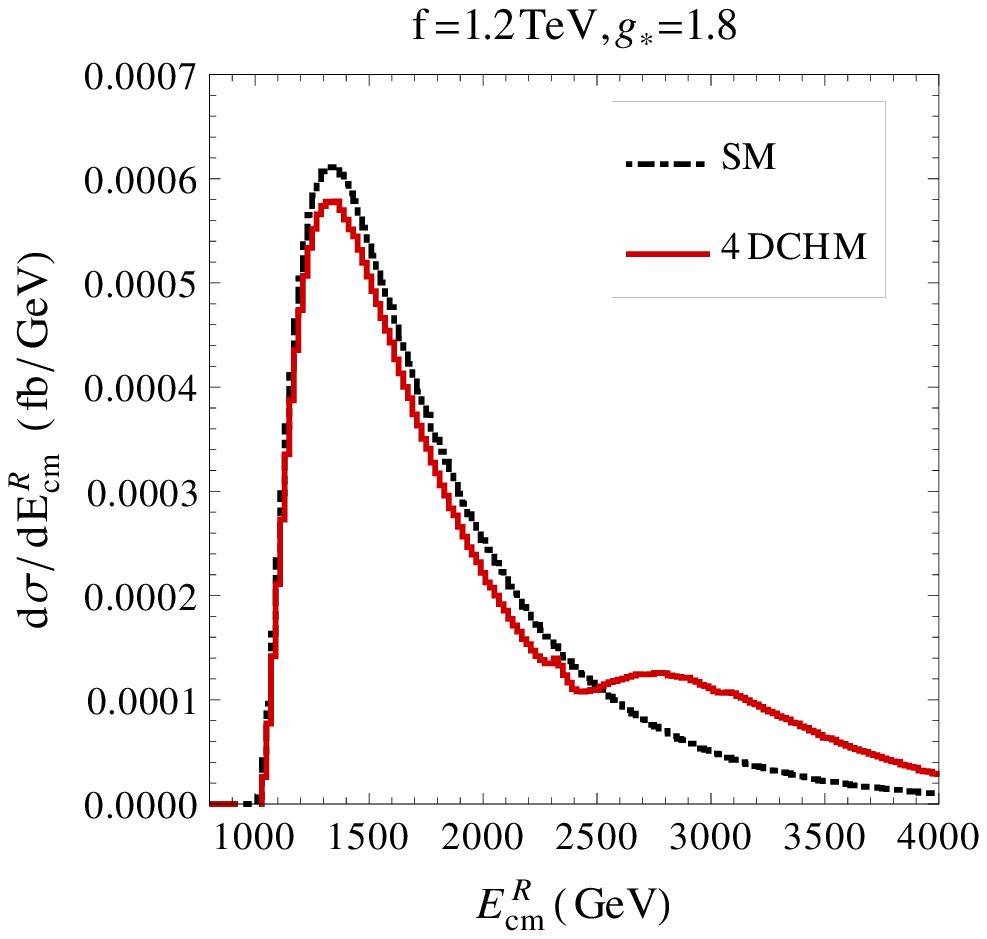, width=.46\textwidth}
\caption[$M_{l^{\prime -}l^{\prime +}}$ and $E^R_{cm}$ distributions for the benchmark with $f$=1200 GeV and $g_\rho$=1.8 for the 3$l$ signature]{Differential cross sections in $M_{l^{\prime -}l^{\prime +}}$ and $E^R_{cm}$ for the benchmark with $f$=1200 GeV and $g_\rho$=1.8 for which the complete set of input parameters can be found in Tab.~\ref{tab:input_bench-fgvar} for the 3$l$ process of eq.(\ref{eq:diboson-proc}). The red solid curve represents the full 4DCHM, while the black dashed the SM background.}
\label{fig:diboson-3l-M-Mr-f12g18}
\end{figure}

We then show  in Fig.~\ref{fig:diboson-3l-M-Mr-f12g18}, again for the choice $f$=1200 GeV and $g_\rho=1.8$, the differential cross section distributions in terms of these two kinematic variables, from which we recognise again the presence of the non resolvable $Z^\prime_{2,3}$ resonances as well as, for the first time, that of the $W^\prime_3$ in the $E_{\rm cm}^R$ distribution, which didn't occur either in the DY analysis or in the 2$l$ channel; in contrast $Z^\prime_5$ does not emerge again over the background.
Finally the $W^\prime_2$, whilst evident in the reconstructed CM energy at partonic level, is clearly mimicked by the 
SM background, in view of the last cut
in eq.(\ref{eq:diboson-3l-cuts}), which renders the signal and background very similar.
 
One peculiar feature of the 3$l$ process is that, unlike the 2$l$ case, it is at times characterized by a large negative interference between the 4DCHM and the SM contribution, induced by the signs of the fermion gauge boson as well as trilinear couplings of the 4DCHM with respect to those of the SM, which can onset precisely where the masses of the new gauge bosons are. This is ultimately responsible for the generic smallness of the cross sections for the 3$l$ process with respect to the 2$l$ ones, up to the point that the actual signal could be a depletion of the expected SM rate, in the relevant mass region.

We then show  in Tab.~\ref{tab:3l-tab1} and Tab.~\ref{tab:3l-tab2} the signal and background rates integrated over the $M_{l^{'+}l^{l'-}}$ range between 2 and 3 TeV (except for benchmark (a) for which we integrate from 1.5 to 2 TeV), i.e., the $Z^\prime_{2,3}$ peak region,
and  over the $E_{cm}^R$  range between 2.5 TeV and 4 TeV (except for benchmark (a) for which we integrate from 2 to 4 TeV), i.e., 
the $W_3$ peak region, respectively.
In the same tables, assuming now an overall tagging efficiency of 50\%, we present the significance $\alpha$ of the signal for 1000 fb$^{-1}$ of integrated luminosity as well as its minimal value required to claim detection, $\alpha=5$.
While the extraction of the signal is possible in a few instances with standard luminosities expected at the 14 TeV stage of the LHC, in 
others this requires much larger data samples, probably obtainable only at the currently considered HL-LHC stage.
Again we do not show the results for the benchmarks with fixed $f$ and variable gauge bosons widths, for which we refer to \cite{Barducci:2012as}, just mentioning here that signal rates dramatically drop down with the growth of the widths of the extra gauge boson.

Finally, if one computes the event rates for process 3$l$ of (\ref{eq:diboson-proc}) after the cuts in eq.(\ref{eq:diboson-3l-cuts}) without any selection and extraction around the resonance peaks, as described before, the effects of the aforementioned negative interferences are apparent, to the extent that most of the chosen benchmark points become inaccessible at the standard LHC and the HL-LHC would become the only available option, as we show in Tab.~\ref{tab:3l-tab-nosel}.
Is is then clear that a high resolution sampling in various kinematic distributions is of fundamental importance to establish a signal in this channel at the LHC and this is clearly impossible at the 7 and 8 TeV energy stages, given the limited data samples collected therein.


\begin{table}[!ht]
\begin{center}
\begin{tabular}{|c|c|c|c|c|}
\hline
~~~~ & $S$~(fb)&$S/\sqrt{B}$~($\sqrt{\rm fb}$)& $\alpha$&$L_{m}$~(fb$^{-1}$)\\
\hline 
(a)&1.1   &2.2  &49  &10\\
(b)&0.067 &0.23 &5 &945\\
(c)&0.25  &0.85 &19  &69\\
(f)&0.3  &1.0  &22  &50\\
\hline
\end{tabular}
\end{center}
\caption[Cross sections for the benchmark with variable $f$ and $g_\rho$ for the 3$l$ signature]{Signal (S), defined as the total cross section (T) minus the SM background (B), $S/\sqrt{B}$ values, statistical significance with $\epsilon$=0.5 and $\mathcal L=$1000 fb$^{-1}$ and minimum integrated luminosity in order to achieve a significance $\alpha$=5, for the benchmarks of eq.(\ref{eq:bench-fvar}) for which the complete set of input parameters can be found in Tab.~\ref{tab:input_bench-fgvar} for the 3$l$ process of eq.(\ref{eq:diboson-proc}) after having imposed the cuts of eq.(\ref{eq:diboson-3l-cuts}) and having integrated the cross sections around the $Z^\prime_{2,3}$ peak.}
\label{tab:3l-tab1}
\end{table}
\begin{table}[!b]
\begin{center}
\begin{tabular}{|c|c|c|c|c|}
\hline 
~~~~ & $S$~(fb)&$S/\sqrt{B}$~($\sqrt{\rm fb}$)& $\alpha$&$L_{m}$~(fb$^{-1}$)\\
\hline 
(a)&0.81   &1.1   &25 &41\\
(b)&0.031  &0.064 &1& NA \\
(c)&0.19   &0.39  &9&328\\
(d)&0.24   &0.48  &11 &217\\
\hline
\end{tabular}
\end{center}
\caption[Cross sections for the benchmark with variable $f$ and $g_\rho$ for the 3$l$ signature]{Signal (S), defined as the total cross section (T) minus the SM background (B), $S/\sqrt{B}$ values, statistical significance with $\epsilon$=0.5 and $\mathcal L=$1000 fb$^{-1}$ and minimum integrated luminosity in order to achieve a significance $\alpha$=5, for the benchmarks of eq.(\ref{eq:bench-fvar}) for which the complete set of input parameters can be found in Tab.~\ref{tab:input_bench-fgvar} for the 3$l$ process of eq.(\ref{eq:diboson-proc}) after having imposed the cuts of eq.(\ref{eq:diboson-3l-cuts}) and having integrated the cross sections around the $W^\prime_{3}$ peak. NA is related to a luminosity value not accessible at present and future proposed colliders.}
\label{tab:3l-tab2}
\end{table}


\begin{table}[!t]
\begin{center}
\begin{tabular}{|c|c|c|c|}
\hline
 & $S=T-B$~(fb) & $S/\sqrt{B}$~($\sqrt{\rm fb}$)\\
\hline
(a) & 0.78   & 0.48\\
(b) & $-7.8\times10^{-2}$ & $-4.8\times10^{-2}$\\
(c) & $7.4\times10^{-2}$ & $3.6\times10^{-2}$\\
(d) & 0.11 &  $6.9\times10^{-2}$\\
\hline
\end{tabular}
\end{center}
\caption[Cross sections for the benchmark with variable $f$ and $g_\rho$ for the 3$l$ signature]{Signal (S), defined as the total cross section (T) minus the SM background (B) values for the benchmarks of eq.(\ref{eq:bench-fvar}) for which the complete set of input parameters can be found in Tab.~\ref{tab:input_bench-fgvar} for the 3$l$ process of eq.(\ref{eq:diboson-proc}) after imposing the cuts of eq.(\ref{eq:diboson-3l-cuts}) without any integration around the resonances peaks.}
\label{tab:3l-tab-nosel}
\end{table} 

\clearpage


\section{$Z^\prime \to t \bar t$ signals at the LHC}

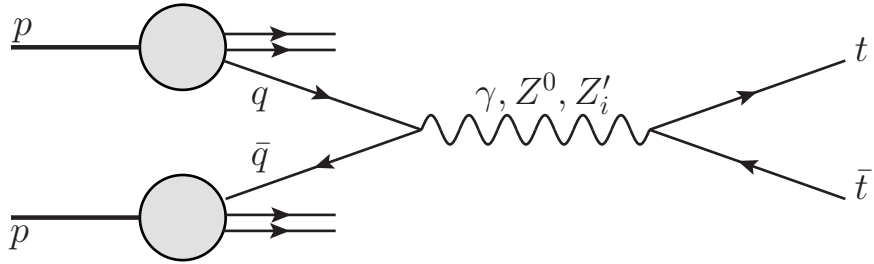
\begin{figure}[!h]
  \begin{picture}(292,118) (55,-47)
    \SetWidth{1.0}
    \SetColor{Black}
    \GOval(160,34)(16,16)(0){0.882}
    \GOval(160,-30)(16,16)(0){0.882}
    \SetWidth{1.7}
    \Text(97,36)[lb]{\Large{\Black{$p$}}}    
    \Line[](96,34)(144,34)
    \SetWidth{1.7}
    \Text(96,-42)[lb]{\Large{\Black{$p$}}}      
    \Line[](96,-30)(144,-30)
    \SetWidth{1.0}   
    \Line[arrow,arrowpos=0.5,arrowlength=5,arrowwidth=2,arrowinset=0.2](175,39)(217,39)
    \Line[arrow,arrowpos=0.5,arrowlength=5,arrowwidth=2,arrowinset=0.2](176,33)(217,33)
    
    \Line[arrow,arrowpos=0.5,arrowlength=5,arrowwidth=2,arrowinset=0.2](176,-29)(217,-29)
    \Line[arrow,arrowpos=0.5,arrowlength=5,arrowwidth=2,arrowinset=0.2](175,-35)(217,-35)
    
    \Text(186,10)[lb]{\Large{\Black{$q$}}}  
    \Line[arrow,arrowpos=0.5,arrowlength=5,arrowwidth=2,arrowinset=0.2](175,29)(249,3)
    \Text(186,-13)[lb]{\Large{\Black{$\bar q$}}}      
    \Line[arrow,arrowpos=0.5,arrowlength=5,arrowwidth=2,arrowinset=0.2](249,3)(176,-23)    
    \Text(270,9)[lb]{\Large{\Black{$\gamma,Z^0,Z^\prime_i$}}}      
    \Photon(249,3)(334,3){5.5}{6}
    \Text(412,29)[lb]{\Large{\Black{$t$}}}     
    \Line[arrow,arrowpos=0.5,arrowlength=5,arrowwidth=2,arrowinset=0.2](334,3)(407,29)
    \Text(412,-23)[lb]{\Large{\Black{$\bar t$}}}      
    \Line[arrow,arrowpos=0.5,arrowlength=5,arrowwidth=2,arrowinset=0.2](407,-23)(334,3)
  \end{picture}
  \caption[Feynman diagram for the production of $t\bar t$ via neutral EW gauge bosons]{Feynman diagram for the production of $t\bar t$ pair via neutral SM and extra EW gauge bosons.}
  \label{fig:ttbar-ew-proc}
\end{figure}

In order to establish the $Z^\prime$ and $W^\prime$ sector of a BSM model it would be sufficient in principle to consider just DY and diboson processes that allow us to test all the couplings of the gauge bosons to fermions, assuming universality across generations, and the triple self interactions of the gauge bosons, further recalling that the quartic couplings are not gauge independent per se. Clearly this statement is not valid any more if we dismiss the universality hypothesis, which is precisely what happens in the 4DCHM where the extended fermionic sector of the third generation of quarks reflects on their different couplings to the gauge bosons with respect the  ones of the light generations.
In particular in our CHM scenario the extra gauge bosons present in the spectrum can have sizeable couplings to the third quark generation, being in fact the $t \bar t $ final state one of the main branching ratio of such states if the heavy fermionic decay channels are not open, see Tab.~\ref{tab:f12g18widthsmallmediumlargeregime}.
We therefore  now study the production of top antitop pairs, Fig.~\ref{fig:ttbar-ew-proc}, at the LHC as a test bed for discovering the heavy $Z^\prime$s present in the 4DCHM.

In such a scenario however various complications, with respect to the DY and diboson processes, must be overcome if one intends to probe the $Z^\prime$s states.
These complications arise from a large background due to the QCD top pair production cross section and a more involved final state, yielding at least six objects in the detector after the decay of the top quarks, with an associated poor efficiency in reconstructing the two heavy quarks.
Although this is an arduous challenge, it reveals its rewards since the spin properties of the top quark are transmitted to the decay products, in view of the fact that it decays before hadronising, and that the electromagnetic charge of the top can be tagged via a lepton or a $b$ jet \cite{Stelzer:1995gc,Mahlon:1995zn}.
This can be extremely useful in profiling the $Z^\prime$s states as various asymmetries, effective in pinning down the couplings of these states, can be defined theoretically and measured experimentally \cite{Fajfer:2012si,Berger:2011hn}.
These asymmetries can be defined in a wider variety, thanks to the decay chain $Z^\prime\to t\bar t \to b\bar b W^+ W^-\to b\bar b X$, where with $X$ we label the possible decay products of the $W$ bosons, and moreover the large top mass induces non trivial spin correlations, again sensitive to the nature of the intermediate $Z^\prime$s states.
Guided by these considerations, experimental collaborations both at Tevatron \cite{Aaltonen:2011ts,Abazov:2008ny} and LHC \cite{ATLAS-CONF-2013-052,Chatrchyan:2012yca} have been studying the $t \bar t$ data samples, while from the theoretical point of view in this process higher order effects from both QCD \cite{Cacciari:2008zb,Kidonakis:2008mu,Bredenstein:2010rs} and EW \cite{Moretti:2006nf,Kuhn:2006vh,Hollik:2007sw,Bernreuther:2008md} interactions are also well known.

\subsection{Couplings of the neutral gauge bosons to $t \bar t$}
\label{sec:ztt}

The couplings of the neutral gauge bosons to the top quark can be derived in a similar way to the one used for deriving the couplings to light quarks and leptons. However, due to the partial compositness mechanism, these couplings also depend  on the parameters of the fermionic sector
and this reflects in more complicated analytical expressions also because we have to diagonalise a nine dimensional matrix for the charge 2/3 quarks in order to find corresponding eigenvalues and eigenvectors.
For this reason we will present the expression for the couplings at the order $\xi=0$, that is, neglecting any contribution arising from EWSB, and, in this approximation, the neutral current Lagrangian for the top sector reads
\begin{equation}
\mathcal{L}\supset\frac{2}{3} e \bar \psi^t \gamma_{\mu} \psi^t A^{\mu}+\sum_{i=0}^5(g^L_{Z_i}(t) \bar\psi^t_L\gamma_{\mu}\psi^t_L+g^R_{Z_i}(t)\bar\psi^t_R\gamma^{\mu}\psi^t_R)Z^{\prime i}_{\mu}
\label{eq:lag_nc_top}
\end{equation}
with
\begin{equation}
\begin{alignedat}{2}
& g_{Z^\prime_0}^L(t)=\frac{e}{s_\omega c_\omega}(\frac 1 2 -\frac 2 3 s^2_\omega),\quad 
&& g_{Z^\prime_0}^R(t)=\frac{e}{s_\omega c_\omega}( -\frac 2 3 s^2_\omega),\\
& g_{Z^\prime_1}^L(t)= 0,\quad \quad\quad\quad\quad&&g_{Z^\prime_1}^R(t)= 0,\\
& g_{Z^\prime_2}^L(t)=\frac{e}{6 c_\omega}\frac{s_\psi}{c_\psi}\frac{1}{(1+F_L)}(1-\frac{c_\psi^2}{s_\psi^2} F_L),\quad && g_{Z^\prime_2}^R(t)=\frac{2 e}{3 c_\omega}\frac{s_\psi}{c_\psi}\frac{1}{(1+F_R)}(1-\frac{c_\psi^2}{s_\psi^2} F_R),\\
& g_{Z^\prime_3}^L(t)=-\frac{e}{2 s_\omega}\frac{s_\theta}{c_\theta}\frac{1}{(1+F_L)}(1-\frac{c_\theta^2}{s_\theta^2} F_L),\quad && g_{Z^\prime_3}^R(t)= 0,\\
& g_{Z^\prime_4}^L(t)=0, && g_{Z^\prime_4}^R(t)=0,\\
& g_{Z^\prime_5}^L(t)= 0, && g_{Z^\prime_5}^R(t)= 0,
\end{alignedat}
\end{equation}
where
\begin{equation}
F_L=\tilde\Delta_{t_L}^2(1+\tilde M^2_{Y_T}),\quad\quad F_R=\tilde\Delta_{t_R}^2(1+(\tilde M_{Y_T}+\tilde Y_T)^2)
\label{FLR}
\end{equation}
and where we have defined
\begin{equation}
\begin{split}
& \tilde \Delta_{t,b/L,R}=\Delta_{t,b/L,R}/m_*,\\
& \tilde Y_{T,B}= Y_{T,B}/m_*,\\
& \tilde m_{Y_{T,B}}=m_{Y_{T,B}}/m_*.
\end{split}
\end{equation}
Notice that, in the $\xi=0$ approximation, $\omega$ is equal to the Weinberg angle defined by: 
\begin{equation}
\label{weinberg}
s^2_W c^2_W=\frac{\sqrt{2}e^2}{8 M^2_Z G_F}.
\end{equation} 
In fact, the following relation holds in the 4DCHM:
\begin{equation}
s_\omega c_\omega \simeq s_W c_W(1-g(\theta,\psi)\xi),\quad\quad g(\theta,\psi)=\frac 1 {32} (-6 s^2_\theta+4 s^4_\theta-3 s^2_\psi+2 s^4_\psi).
\end{equation}
The $Z^\prime_4$ resonance has zero coupling also to $t \bar t$ at all orders in perturbation theory and, together with $Z^\prime_1$, is inert to the process under consideration since it doesn't couple to the proton constituents. Moreover, since we found that the $Z^\prime_5$ state is not accessible even in this process, we will refrain from presenting results related to it.
Finally, the expression for the top mass at the leading order in $\xi$ is the following
\begin{equation}
m_{top}^2 \simeq \xi \frac{m_*^2} 2   \frac{ \tilde \Delta_{t_L} ^2\tilde \Delta_{t_R} ^2\tilde Y_T^2}{(1+F_L)(1+F_R) }.
\label{eq:topmass}
\end{equation}

Besides taking into account the constraints due to the EWPT and the direct searches of extra gauge bosons, we now also  consider the bounds on the extra fermions present in the model, that can be produced in hadron hadron collision, and that, due to the mixing with the top quark, can affect the coupling of the latter to the extra neutral gauge bosons.
The most stringent bounds on the production of $t^\prime$s come from the LHC and
rely on the QCD pair production mechanism.
To compare with the experimental results that assume an exclusive branching ratio in a given final state ($W^+b$ and $Zt$ mainly\footnote{Note that more recent searches, assuming variable branching ratios for the decay channels of extra fermions, have been produced by ATLAS and CMS: we will discuss on this in Chapter \ref{chap-5}.}) we have then rescaled the production cross section for the process $\sigma(pp(q\bar q,gg)\rightarrow t^\prime_i \bar t^\prime_i)$, calculated with the code described in \cite{Cacciari:2011hy}, in order to take into account the non 100\% branching ratios in the 4DCHM of the $t^\prime$s in the decay channels searched for by the experimental collaborations.
As we will show in Chapter \ref{chap-5}, in order to recast in a consistent way the experimental limits, one has to deal also with the possible enhancement of the signal due to the presence of more than one extra quark
in the spectrum. However we will postpone this discussion until the end of this Thesis, using now the simplified results that we have obtained that limits the heavy quark masses to values above 500 GeV
or so, and therefore we have excluded such masses in the forthcoming parameter scans for which we will present our results.

Finally we have used for this analysis the same tools as for the DY and diboson cases, with the only difference that the PDFs set chosen has been CTEQ6L1 \cite{Pumplin:2002vw} with factorisation and renormalisation scale set to $Q=\mu\simeq M_{Z^\prime_{2,3}}$.

\subsection{Asymmetries}
We describe now some asymmetries for which we will present our results highlighting their main characteristics.
\subsubsection{Charge asymmetry}

The charge asymmetry, already defined for the DY case, is a measure of the symmetry of a process under charge conjugation.
In this process it can only be generated from the $q\bar{q}$ initial state due to the symmetry of the gluon gluon system and 
it can occur via both subtle NLO QCD effects, as described in detail in~\cite{Brown:1979dd,Kuhn:1998kw}, as well as more standard EW ones.
We repeat here the definition of this asymmetry given in eq.(\ref{eq:DY-AFB})
\begin{equation}
    AFB^{\ast}=\frac{N(\cos\theta > 0)-N(\cos\theta < 0)}{N(\cos\theta > 0)+N(\cos\theta < 0)}
\label{eq:asy_AFB}    
\end{equation}
where $\cos\theta$ is defined  with the $z$ axis in the direction of the rapidity of the top antitop system $y_{t\bar{t}}$, and
$\theta$ is the polar angle in the $t\bar t$ rest frame, i.e. the CM system at parton level, 
to which the entire event can generally be boosted to, whatever the actual final state produced by the $t\bar t$ pair after it decays~\cite{Chatrchyan:2012saa,Chatrchyan:2012yca,ATLAS:2012sla}.
This observable can only be generated by a $Z^\prime$ boson if its vector and axial couplings to both the initial and final state fermions are non-vanishing. 

\subsubsection{Spin asymmetry}
The spin asymmetry focuses on the helicity structure of the final state fermions and, when such properties are measurable, displays interesting dependences on the chiral structure of the $Z^\prime$s couplings.
The helicity of a final state can only be experimentally determined for a decaying final state, where the asymmetries are extracted as coefficients in the angular distribution of its decay products. This is well described for the case of top quarks in \cite{Stelzer:1995gc,Bernreuther:2001rq}.
As such, since our implementation will be only at the parton level, our analysis does not represent the full reconstruction and extraction chain of such observables but it is still able to highlight their potential strength if we estimate the reconstruction efficiencies from recent experimental publications. We will elaborate more on this point in Section~\ref{subsect:ztt-res}.

The first observable that we define is the polarisation, or single spin asymmetry, $A_L$
\begin{equation}
 A_{L}=\frac{N(-,-) + N(-,+) - N(+,+) - N(+,-)}{N_{\rm Total}},
 \label{eq:ztt-asy-AL}
\end{equation}
where  $N$ denotes the number of observed events and its first and second argument corresponds to the helicity of the final state particle and antiparticle respectively, whereas ${N_{\rm Total}}$ is the total number of events.
It singles out one of the two final state particles, comparing the number of its positive and negative helicities, while summing over the ones of the other antiparticle, or vice versa.
This observable is proportional to the product of the vector and axial couplings of the final state only and is 
therefore additionally sensitive to their relative sign, a unique feature among asymmetries and cross section observables being, in other words, a measure of the relative handedness of the $Z^\prime$ couplings to the final state.

In the case of a highly massive final state, such as the top quark, the spin correlation $A_{LL}$, or double spin asymmetry, is accessible. 
This observable relies on the helicity flipping of either of the final state particles, whose amplitude is proportional to $m_{t}/\sqrt{\hat{s}}$, 
where $\sqrt{\hat{s}}$ is the partonic CM energy, and it gives the proportion of like sign final states against the opposite sign ones
\begin{equation}\label{eqn:asy_ALL}
      A_{LL}=\frac{N(+,+) + N(-,-) - N(+,-) - N(-,+)}{\rm N_{Total}}.
\end{equation} 
This observable depends only on the square of the couplings in a similar way to the total cross section. 
In the massless limit $A_{LL}$ becomes maximal, making it a relevant quantity to measure only in the $t\bar{t}$ final state.

\subsection{Results}
\label{subsect:ztt-res}
Before presenting our results for the process
\begin{equation}
pp\to \gamma,Z^0,Z^\prime_i\to t \bar t
\end{equation}
we ought to make a consideration about the validity of our analysis in terms of reconstruction efficiencies of the $t \bar t $ final state.
As mentioned, one of the primary complications of such a final state is the difficulty in reconstructing a six body final state.
Ideally one would perform a full chain of event generation, parton showering and hadronisation with a final detector simulation
in order to get an accurate representation of the reconstruction process for observables of interest, and the associated
efficiencies will depend on the information required for the 
observable and the particular decay channel of the $t\bar{t}$ system.
However since our analysis is limited to a parton level study, without subsequent decay of the $t \bar t$ final state, we need to give
a reasonable estimate of the reconstruction efficiencies so that our predictions correspond better to the reality of a detector environment.
We have estimated these quantities in a conservative manner by gauging the selection efficiencies from the requirements of each observable in  each decay channel, taking the sum of these and defining a net efficiency to reconstruct a $t\bar{t}$ event coming
from a high mass, $\simeq 2$ TeV, $Z^{\prime}$, weighing by the branching ratios: from this we will take the average efficiencies to be 10\%.
However we want to point out that for some observables, like the top polarisation asymmetry, this assumption might be too optimistic stressing then that our results remain of illustrative nature and are given in order to show that this model has the potentiality to be investigated further in the $t \bar t$ channel.

\begin{figure}[!h]
\centering
\epsfig{file=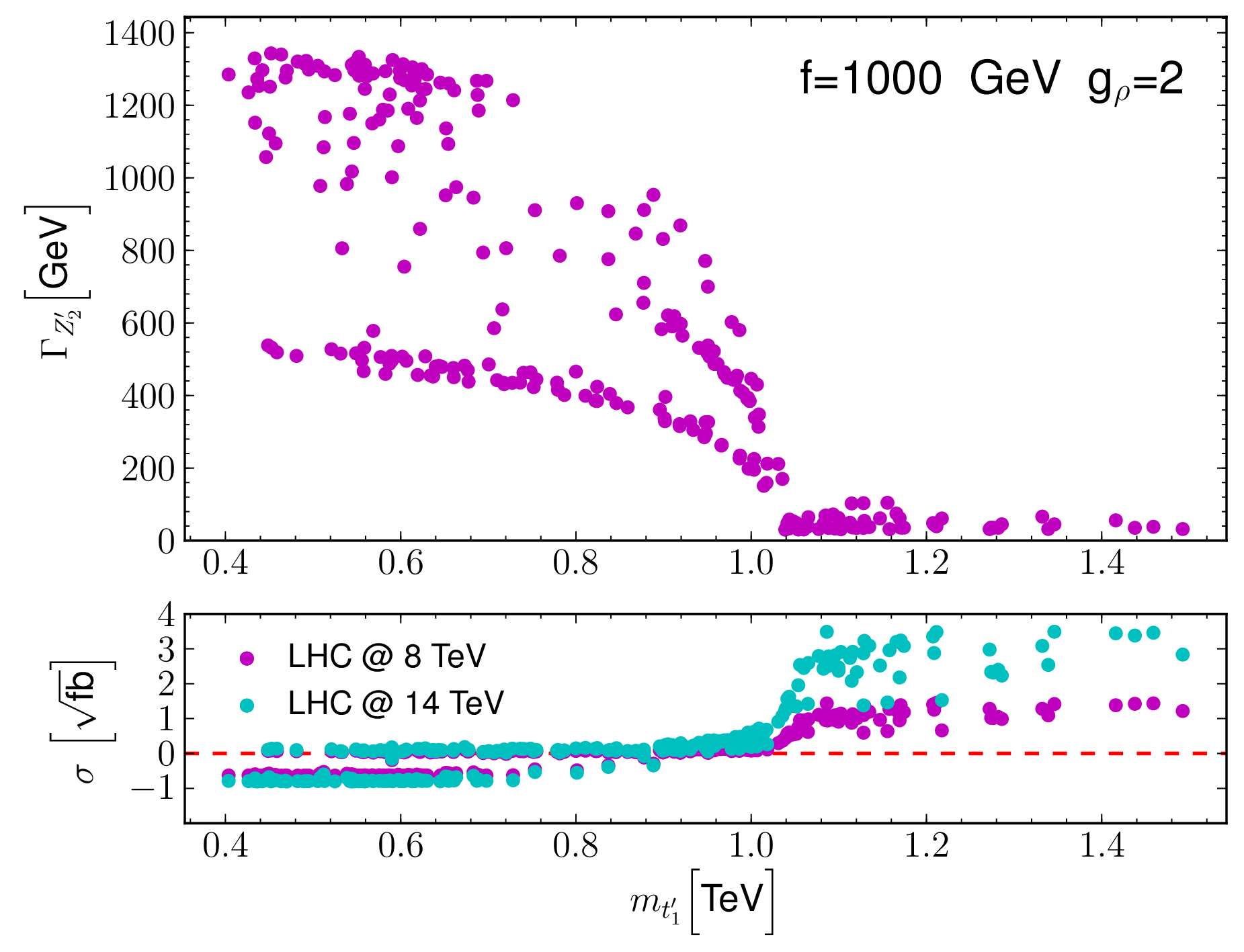, width=.44\textwidth}{(a)}
\hfill
\epsfig{file=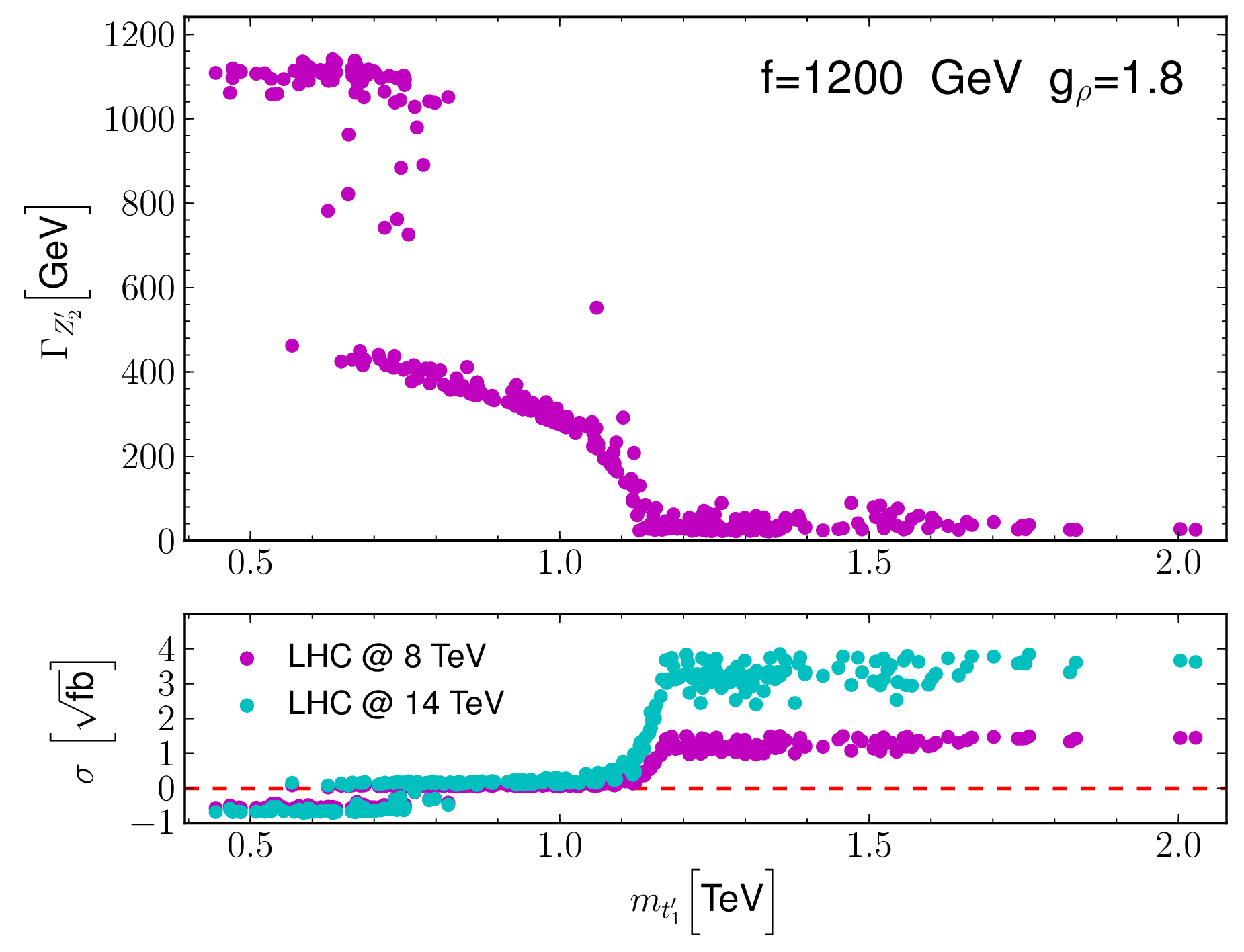, width=.44\textwidth}{(b)}
\caption[8 and 14 TeV LHC significance for the process $pp\to t \bar t$ for $f$=1000 GeV, $g_\rho$=2 and $f$=1200 GeV, $g_\rho$=1.8]{Scatter plots in the plane $m_{t_1^\prime}/\Gamma_{Z^\prime_2}$ for the choice $f$=1000 GeV, $g_\rho$=2 (a) and  $f$=1200 GeV, $g_\rho$=1.8 (b) for which the
complete set of input parameters can be found in Tab.~\ref{tab:input_bench-fgvar}. In the lower frame we show the relative dimensional significance, defined as $S/\sqrt{B}$ for the 8 (purple) and 14 (cyan) TeV LHC.}
\label{fig:ztt-scan}
\end{figure}

In order to establish the regions of the parameter space of the 4DCHM where at least one $Z^\prime \to t \bar t$ signal can be established, we have performed a scan over the fermionic parameters of the model in the same range as declared in Section \ref{subsec:2-bench}, with the only difference that the top mass has now been constrained to be in the 170-175 GeV range.
We have used both CalcHEP \cite{Belyaev:2012qa} and MadGraph \cite{Stelzer:1994ta} to compute cross sections for the 8 and 14 TeV LHC in the presence of the following selection cuts
\begin{equation}
\frac{M_{Z^\prime_2}+M_{Z^\prime_3}}{2}-3 \frac{\Gamma_{Z^\prime_2}+\Gamma_{Z^\prime_3}}{2}<M_{t\bar t}< \frac{M_{Z^\prime_2}+M_{Z^\prime_3}}{2}+3 \frac{\Gamma_{Z^\prime_2}+\Gamma_{Z^\prime_3}}{2}
\label{eq:ztt-cut1}
\end{equation}
that is selecting a window for the lightest $Z^\prime$s bosons, where $M_{t \bar t}=\sqrt{(p_t+p_{\bar t})^2}$ is the invariant mass of the $t\bar t$ final 
state\footnote{Due to the large widths of these $Z^\prime$s in certain region of the parameter space lower and upper bounds on the selection cut have been imposed to be the maximum and minimum between the ones of eq.(\ref{eq:ztt-cut1})
 and $2m_{t}$ and $\sqrt s$.}.
 The signal $S$ has again been defined as the difference between the total 4DCHM rate, $T$, and the SM background, $B$, and the dimensional significance has been computed as $S/\sqrt{B}$ from which we can calculate the statistical significance $\alpha$ with the mean of eq.(\ref{eq:signif}).
 The results for the scans over two benchmark points, $f$=1000 GeV, $g_\rho$=2 and $f$=1200 GeV, $g_\rho$=1.8 defined in Tab.~\ref{tab:input_bench-fgvar}, are presented in Fig.~\ref{fig:ztt-scan}.

We can  clearly see the relationship between the mass scale of the first charge 2/3 quark resonance and the visibility of the extra gauge bosons, connected to the already mentioned increase of their widths with the opening of the extra fermionic decay channels, which prevents any significant deviation from the SM background.
We notice that in the case of large width the value $S/\sqrt{B}$ can become negative and this is due again, as in the diboson case, to interference effects which yield a negative total signal $S$. However we do not associate any physical meaning with this negative dimensional significance, also in view of the fact that these off peak effects may have consequences down to very low invariant masses, perhaps even near the $t\bar{t}$ threshold, and this may not only already be constrainable with current LHC data, but would certainly require analyses
with background estimates beyond leading order to have a more precise prediction of the overall shape and normalisation of the invariant mass spectrum, without which it is difficult to make meaningful statements 
about such deficits in the production cross section over a large $M_{t\bar{t}}$ range.
It is however clear from the plots that once again our intended resonant analysis becomes difficult beyond the limit in which the $Z^\prime s$ are narrow and cannot decay into heavy fermions and, with this in mind, we will now analyse some specific benchmark points, namely the ones of eq.(\ref{eq:bench-fvar}), focusing again on the small width regime, that is using again the values of the input parameters of Tab.~\ref{tab:input_bench-fgvar}, considering only the 14 TeV stage of the LHC with an integrated luminosity of 300 fb$^{-1}$ as the scope at smaller value of $\sqrt{s}$ and $\mathcal L$ is limited.

In order to get a feel for the strength of the asymmetry observables studied, we also define an illustrative measure of theoretical significance of an asymmetry prediction for the signal $A_{S}$ as the number of standard deviations it lies away from the background prediction, $A_{B}$,
\begin{equation}
\alpha_A=\frac{|A_{S}-A_{B}|}{\sqrt{\delta A^{2}_{S}+\delta A^{2}_{B}}},
\end{equation}
which, again, remains within the scope of our parton level analysis. As such, they should not be interpreted as true LHC significances, but be indicative of the strength of a particular observable.

\begin{figure}[!h]
\centering
\epsfig{file=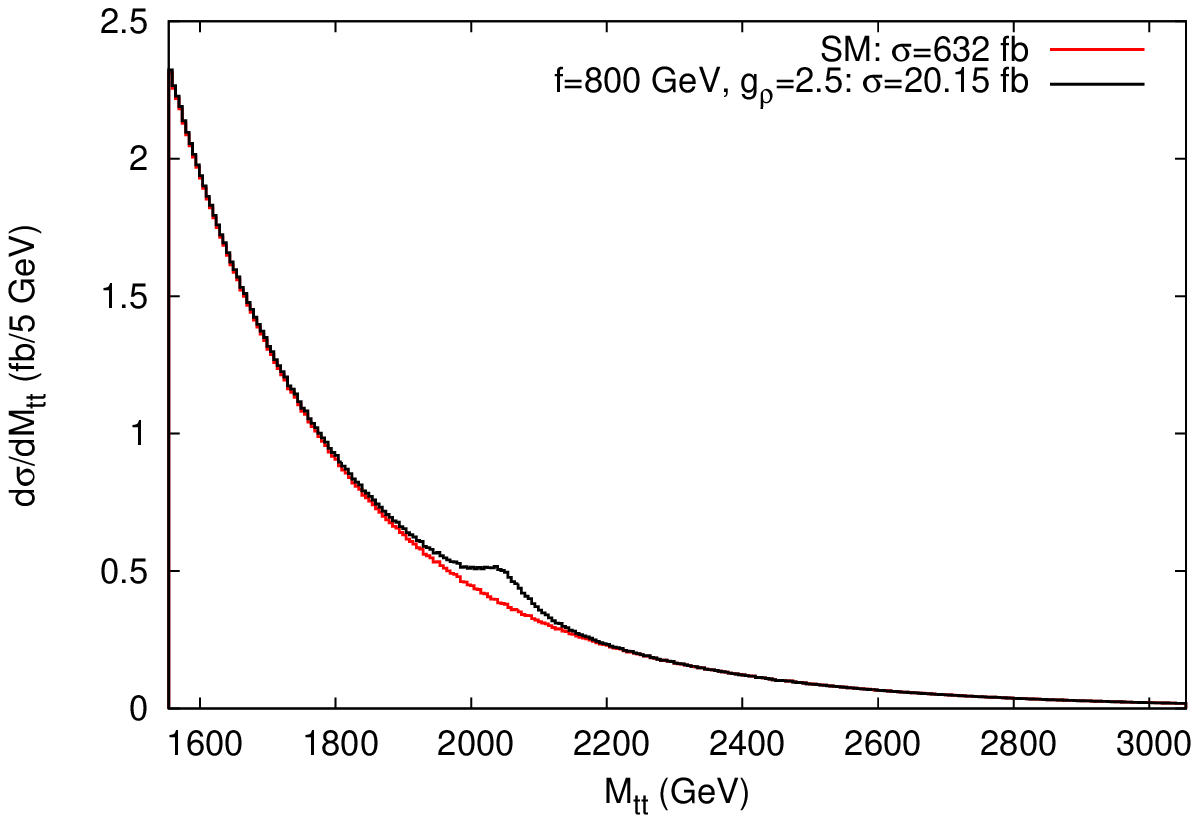, width=.44\textwidth}{(a)}
\hfill
\epsfig{file=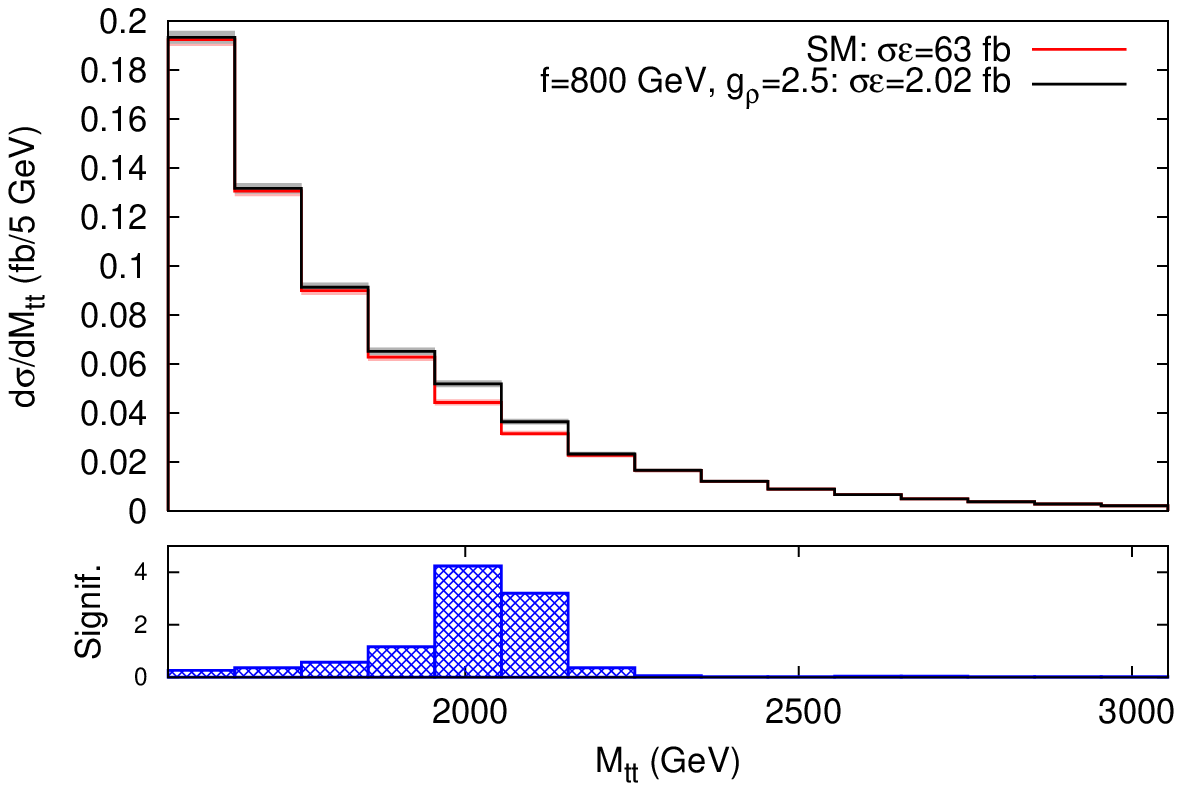, width=.44\textwidth}{(b)}
\epsfig{file=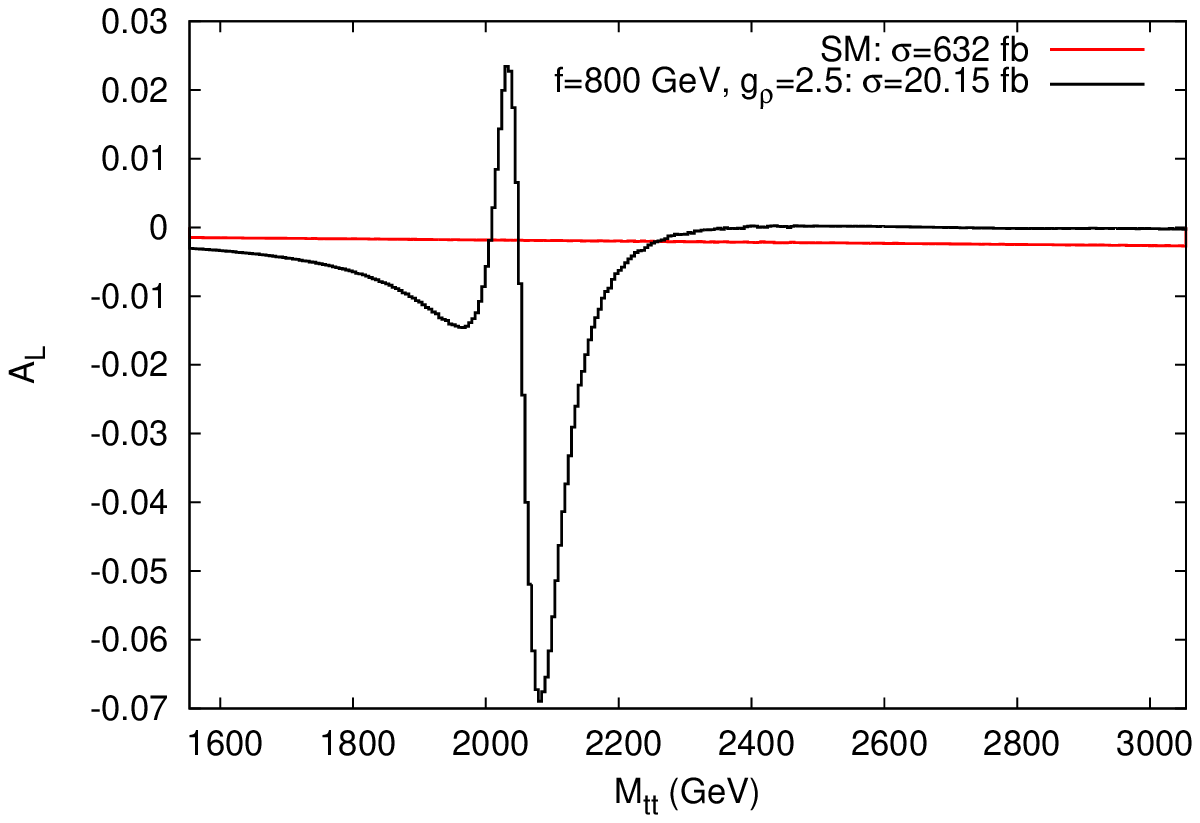, width=.44\textwidth}{(c)}
\hfill
\epsfig{file=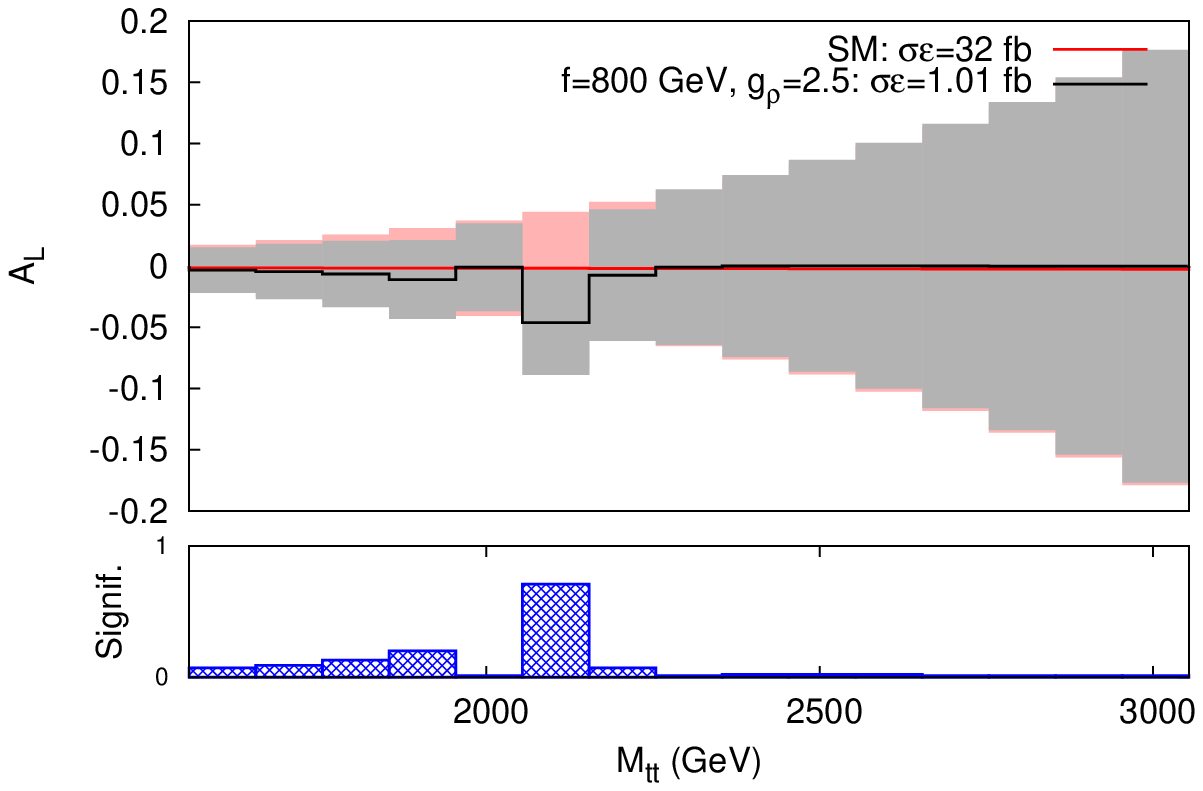, width=.44\textwidth}{(d)}
\epsfig{file=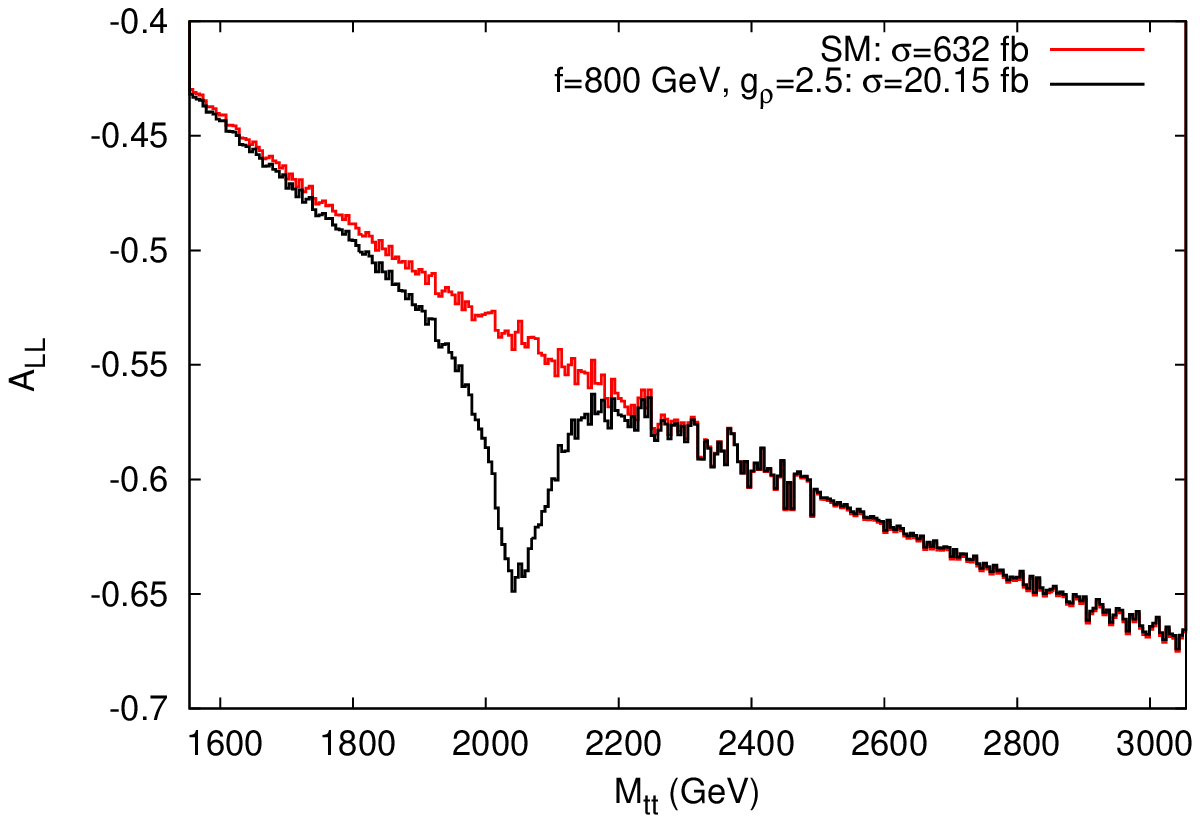, width=.44\textwidth}{(e)}
\hfill
\epsfig{file=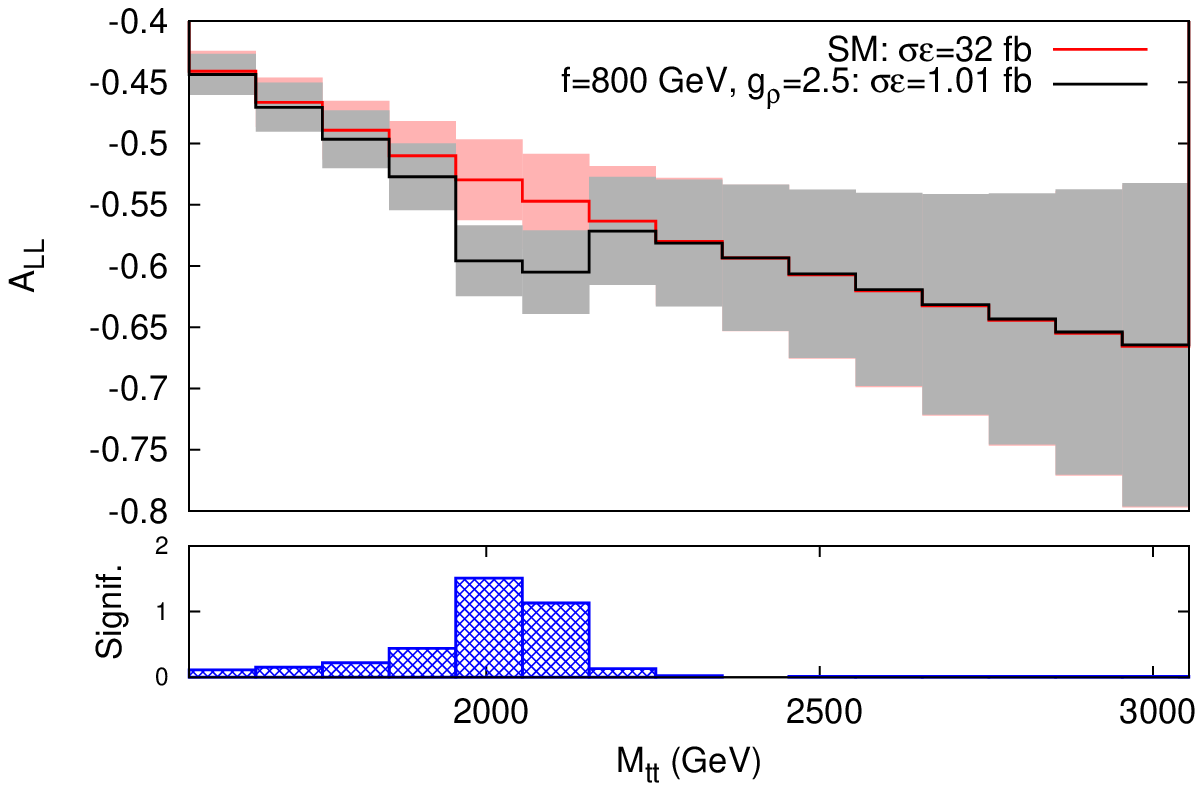, width=.44\textwidth}{(f)}
\epsfig{file=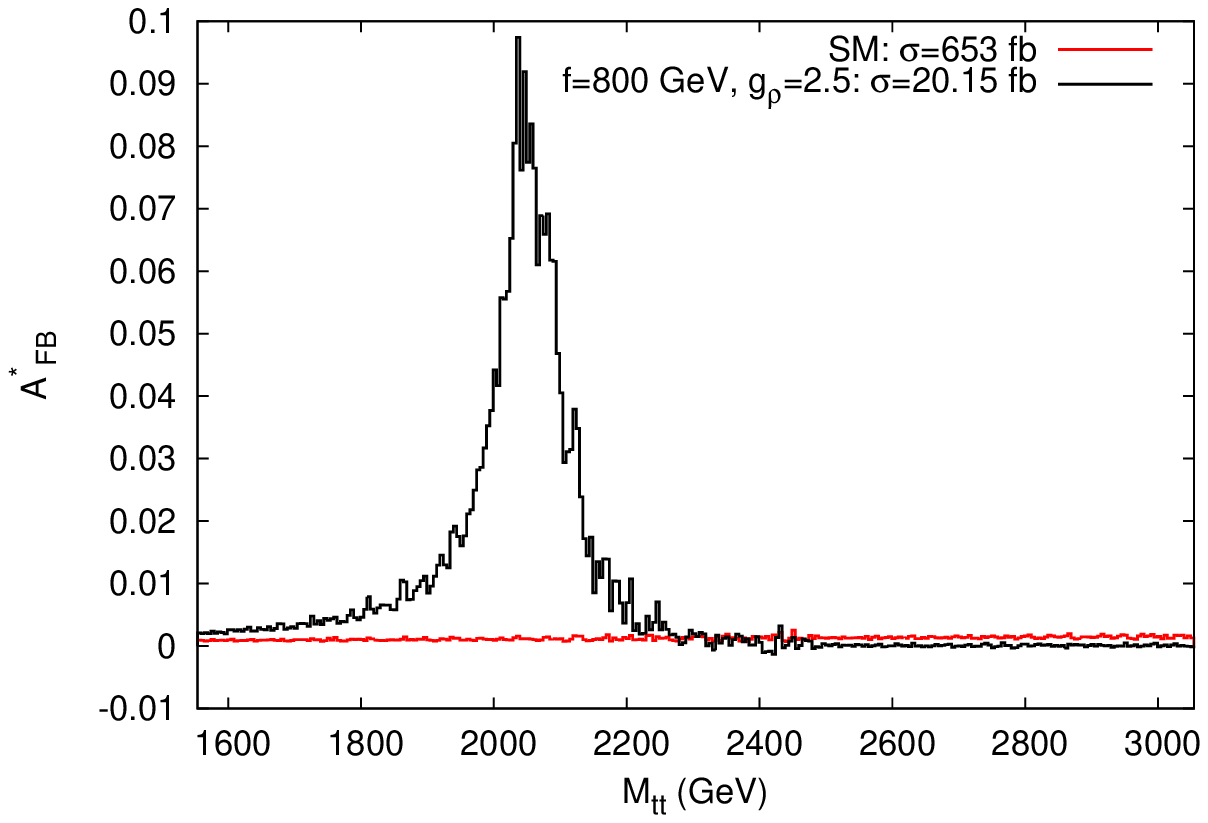, width=.44\textwidth}{(g)}
\hfill
\epsfig{file=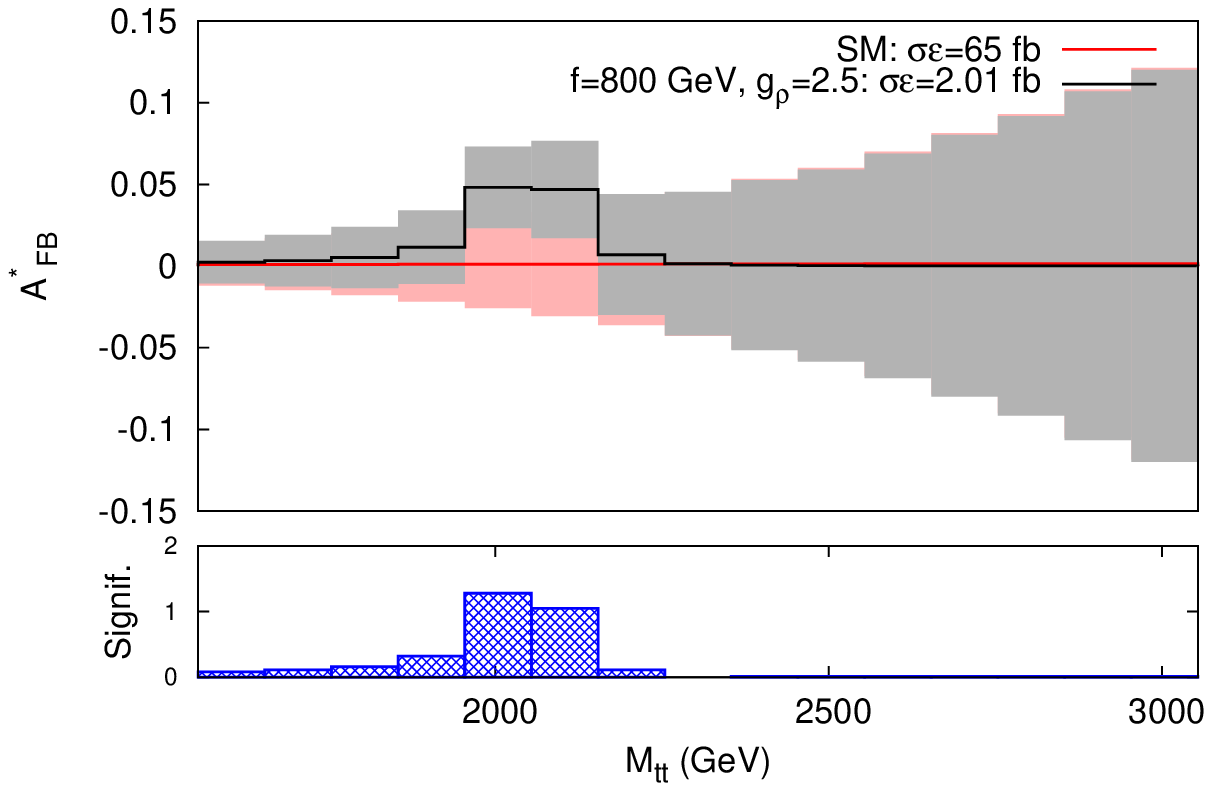, width=.44\textwidth}{(h)}
\caption[Invariant mass differential distributions for the cross section and asymmetries for the process $pp\to t \bar t$ for $f=800$ GeV and $g_\rho$=2.5 at the LHC]{Cross section
and asymmetries distributions as a function of the $t\bar t$ invariant mass  for the choice $f=$800 GeV, $g_\rho=$2.5 for which the
complete set of input parameters can be found in Tab.~\ref{tab:input_bench-fgvar} at the 14 TeV LHC with 300 fb$^{-1}$. The left column shows the fully differential observables. 
Right plots (upper frames) include estimates of statistical uncertainties assuming a realistic 
100 GeV mass resolution and also display (lower frames) the theoretical significances assuming a 10\% reconstruction efficiency. Grey and pink shading represent the statistical error on the 4DCHM and SM rates, shown in black and red solid lines respectively.}
\label{fig:ztt-f08g25}
\end{figure}

\begin{figure}[!h]
\centering
\epsfig{file=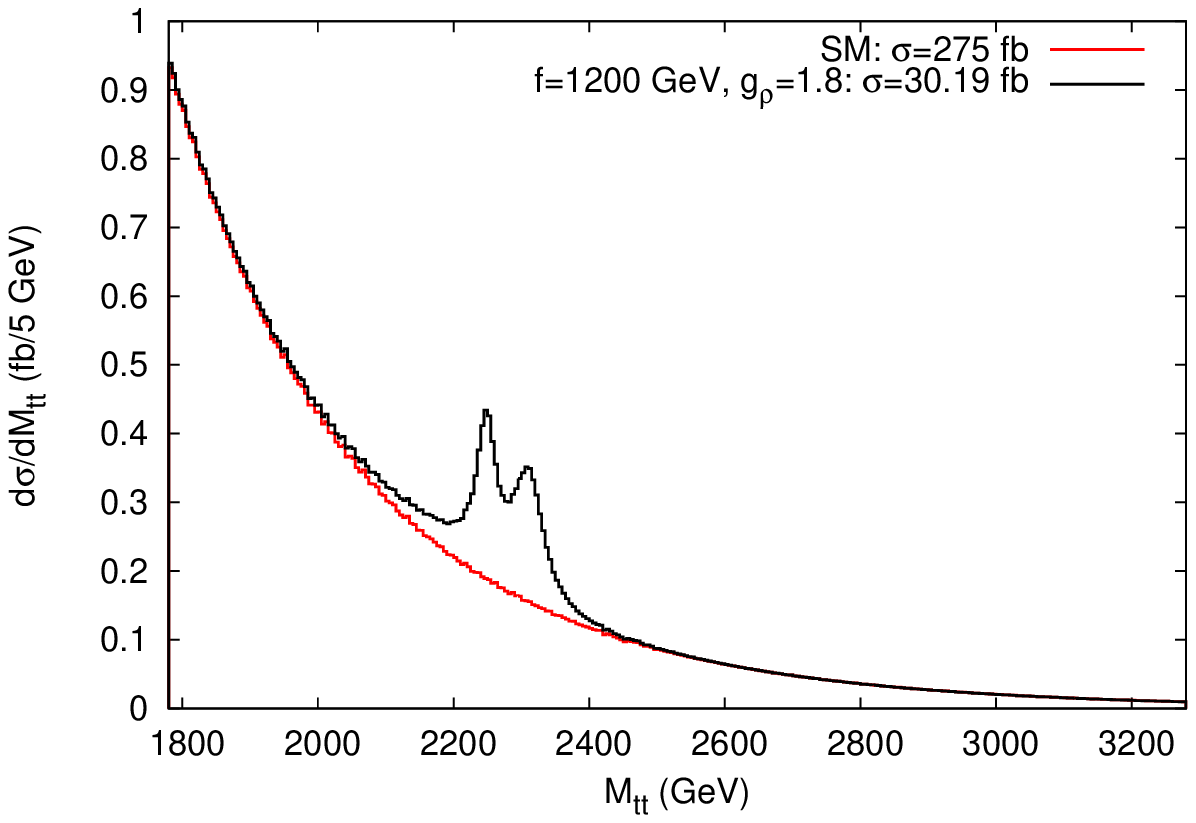, width=.44\textwidth}{(a)}
\hfill
\epsfig{file=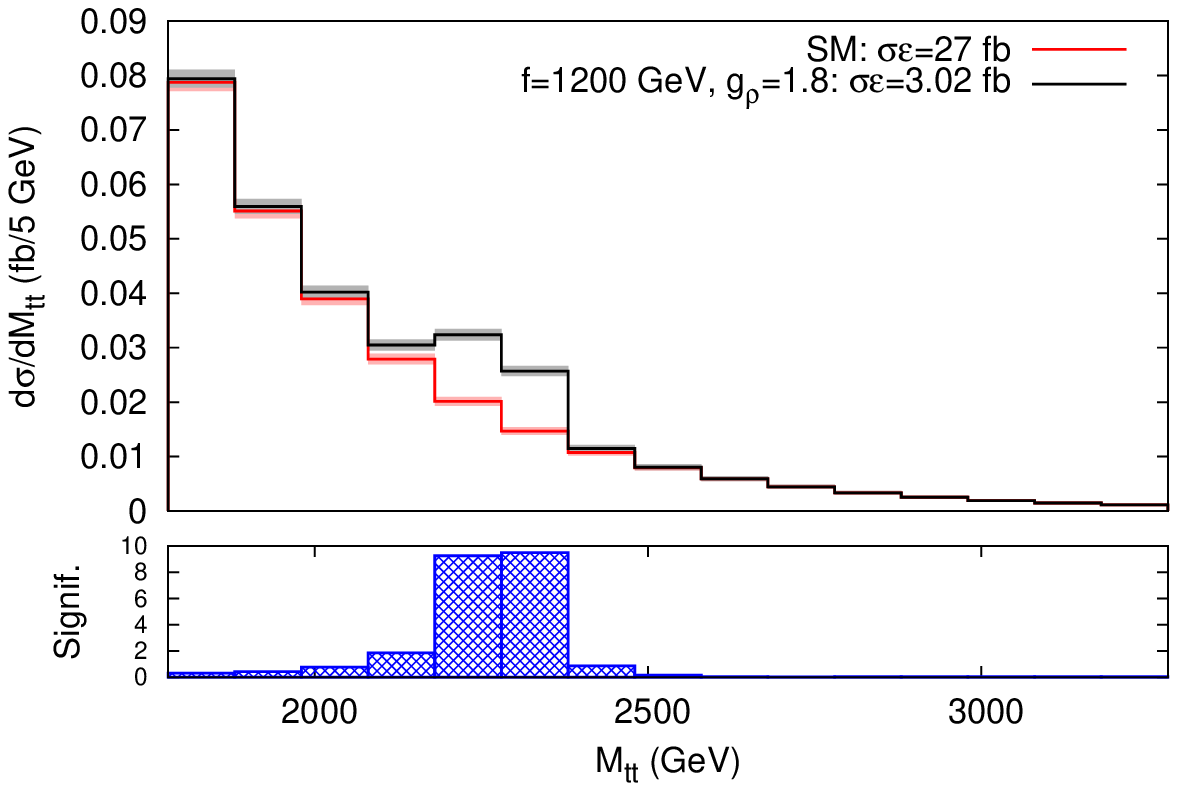, width=.44\textwidth}{(b)}
\epsfig{file=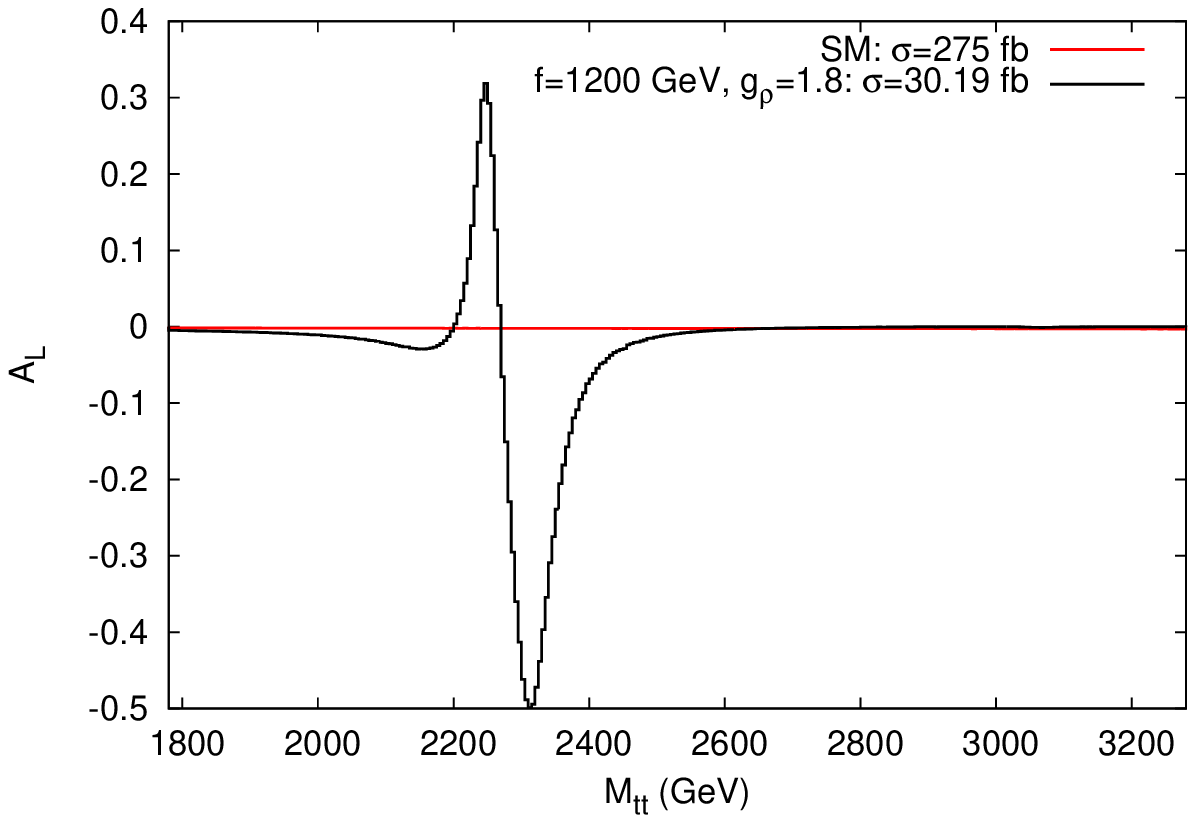, width=.44\textwidth}{(c)}
\hfill
\epsfig{file=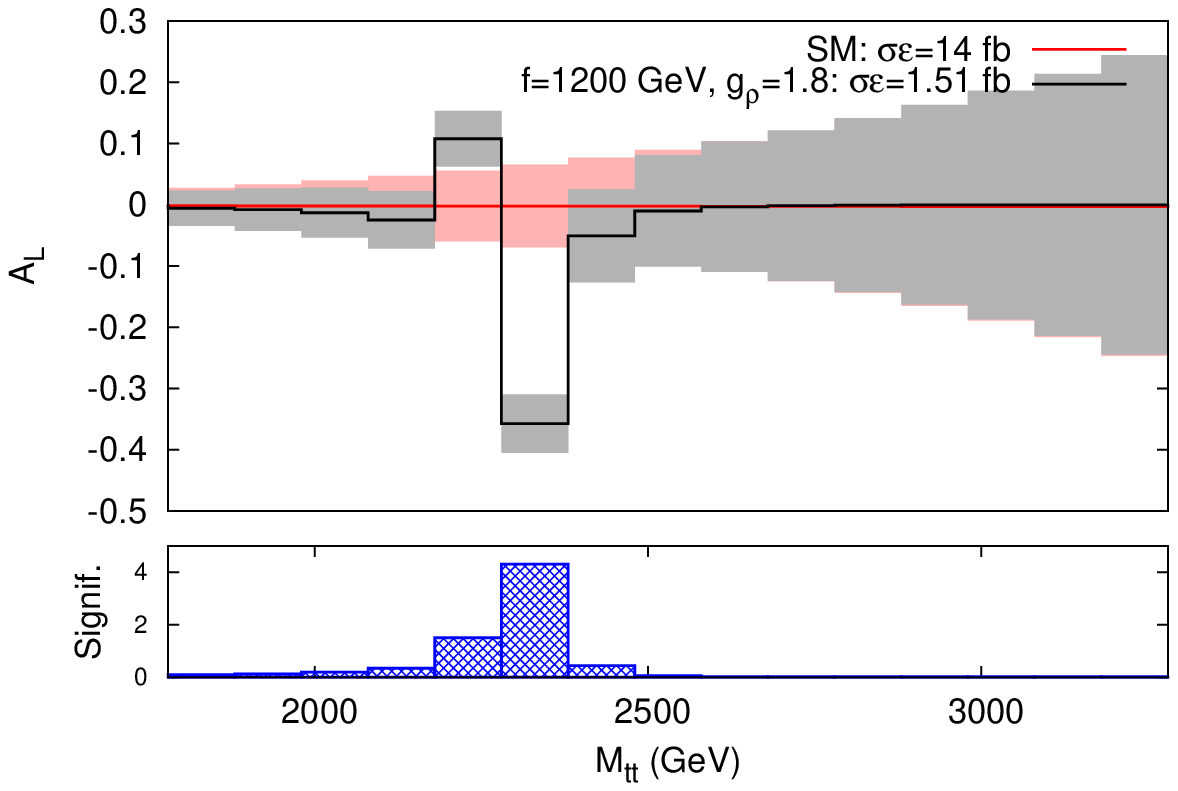, width=.44\textwidth}{(d)}
\epsfig{file=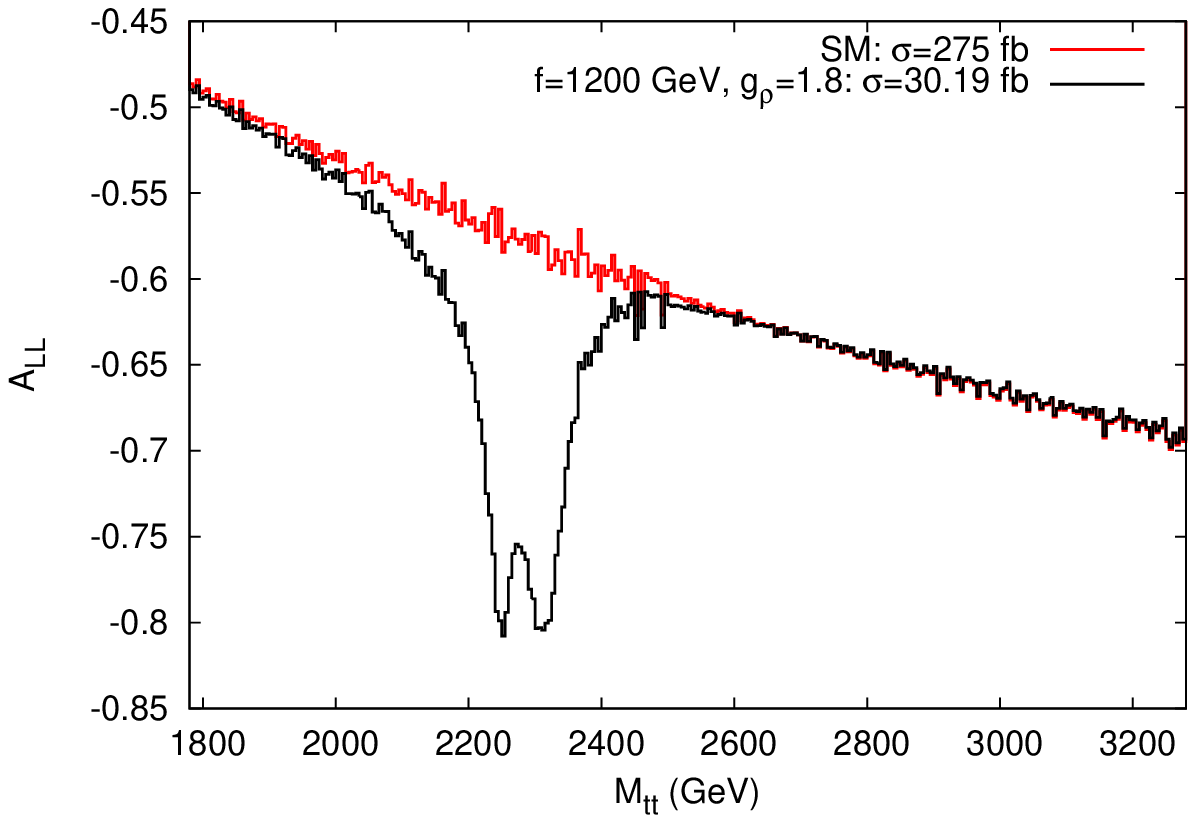, width=.44\textwidth}{(e)}
\hfill
\epsfig{file=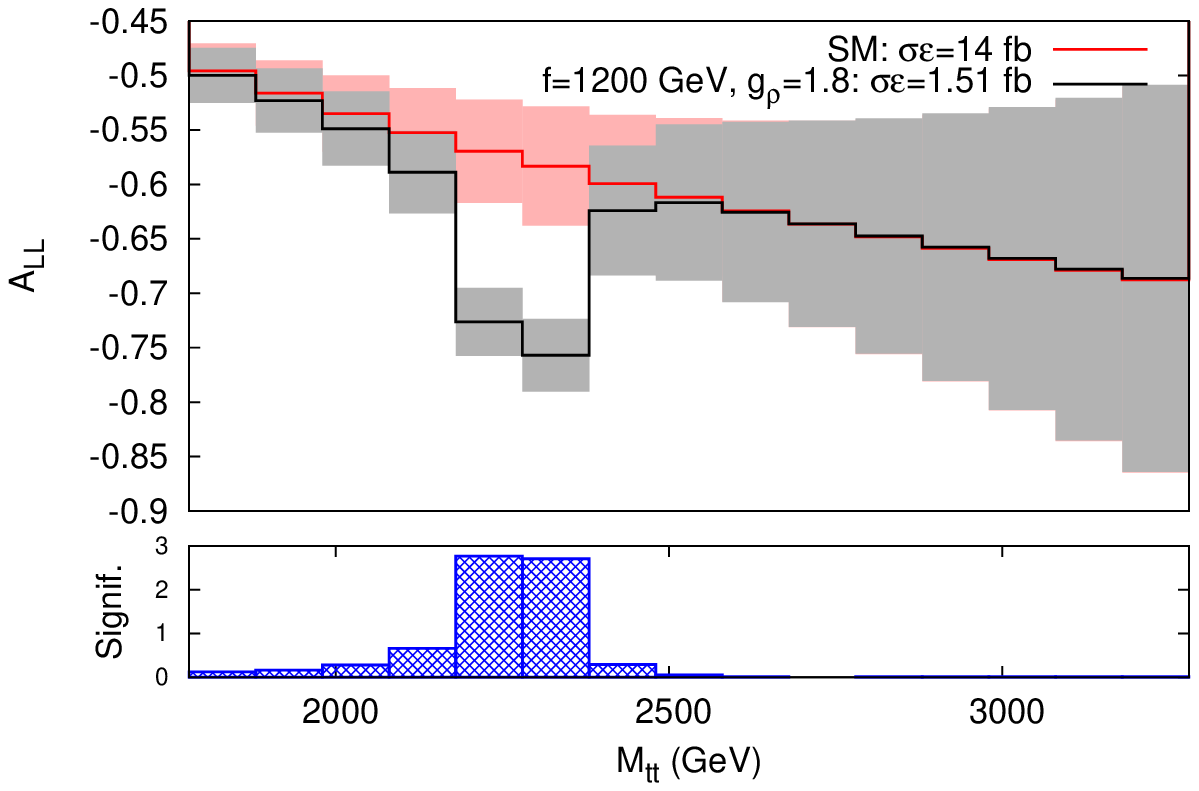, width=.44\textwidth}{(f)}
\epsfig{file=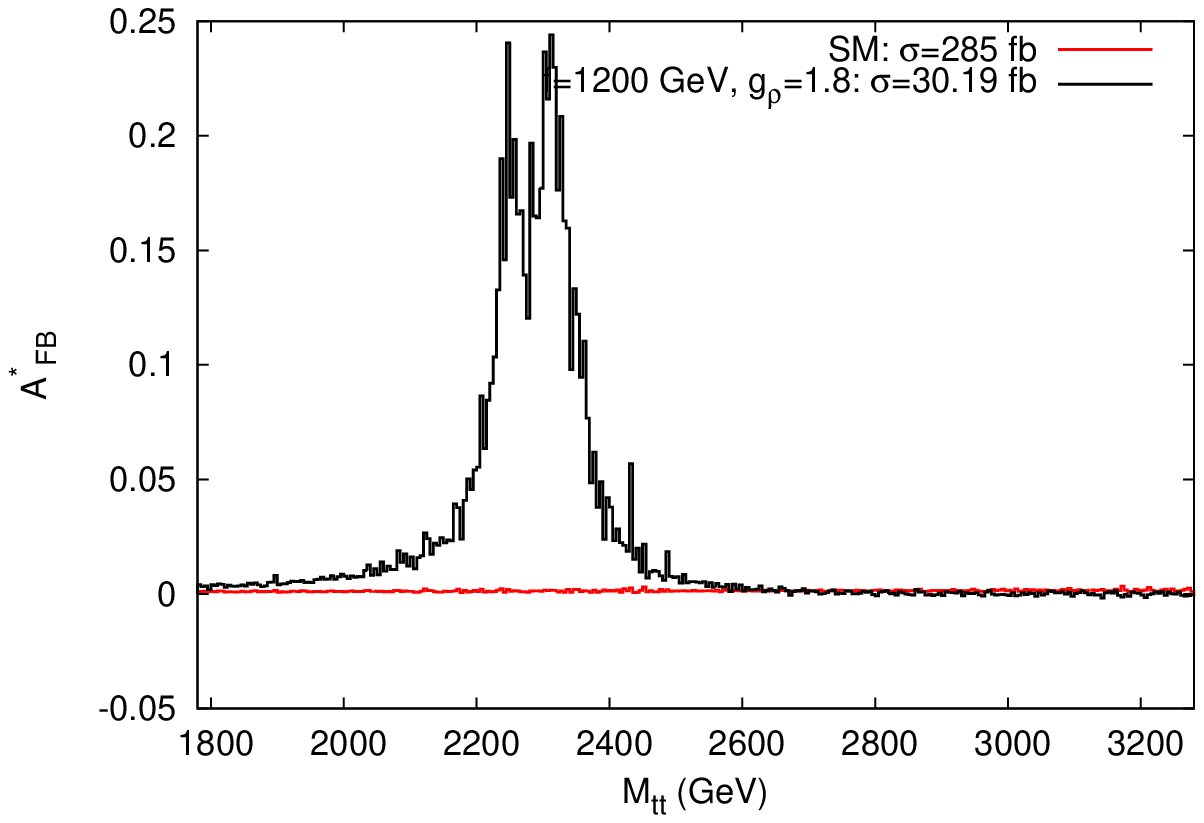, width=.44\textwidth}{(g)}
\hfill
\epsfig{file=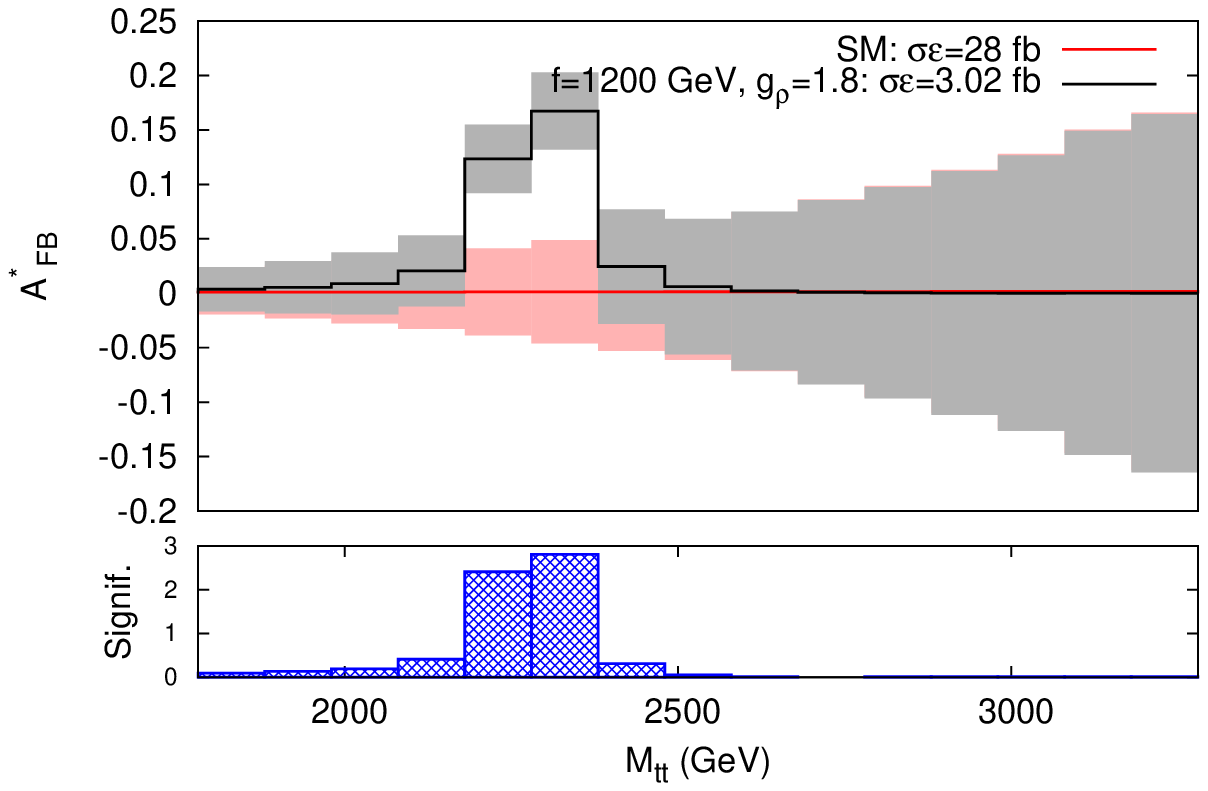, width=.44\textwidth}{(h)}
\caption[Invariant mass differential distributions for the cross section and asymmetries for the process $pp\to t \bar t$ for $f=1200$ GeV and $g_\rho$=1.8 at the LHC]{Cross section and asymmetries distributions
as a function of the $t\bar t$ invariant mass  for the choice $f=$1200 GeV, $g_\rho=$2.5 for which the complete set of input parameters can
be found in Tab.~\ref{tab:input_bench-fgvar} at the 14 TeV LHC with 300 fb$^{-1}$. The left column shows the fully differential observables. 
Right plots (upper frames) include estimates of statistical uncertainties assuming a realistic 
100 GeV mass resolution and also display (lower frames) the theoretical significances assuming a 10\% reconstruction efficiency. Grey and pink shading represent the statistical error on the 4DCHM and SM rates, shown in black and red solid lines respectively.}
\label{fig:ztt-f12g18}
\end{figure}

We then show in Fig.~\ref{fig:ztt-f08g25} and Fig.~\ref{fig:ztt-f12g18} the differential values of the cross section and of the asymmetries $A_L$, $A_{LL}$ and $AFB^*$ as a function of the invariant mass of the top pair for the choice (b) and (d) of Tab.~\ref{tab:input_bench-fgvar}.
As a first result we see that, contrary to the DY case, here it is no longer possible to separate the two light $Z^\prime$s peaks, due to the mass resolution of the $t \bar t$ pairs being of the order of 100 GeV\footnote{Somewhat better for semi leptonic
decay channels and somewhat worst for fully hadronic and leptonic ones.} and that, again, the $Z^\prime_5$ never emerges from the background, and this is shown in frames (a) where the results for the differential cross section are binned over artificially narrow
$M_{t \bar t}$ bins of 5 GeV width and in frames (b) where we adopt more realistic 100 GeV bins. Despite this it is  possible to achieve a significance greater than 5 with $\mathcal L=300$ fb$^{-1}$ and $\epsilon=10$\% for the choice $f=1200$ GeV and $g_\rho=1.8$ and somewhat
smaller for $f=800$ GeV and $g_\rho=2.5$.
In the other panels we show, again with a 5 GeV (left) and 100 GeV (right) binning, the results for the asymmetries that can complement the scope of the cross section as in all the cases they offer a similar level of significance, so that the contributions of the former and the latter can be combined, when needed, to increase significance.
Finally if we allow the extra fermions to have a lighter mass, increasing the widths of the extra gauge bosons, the ability of both the cross section and asymmetries diminish, though for these cases we refer to \cite{Barducci:2012sk} for a detailed analysis and related results.

\clearpage
\section{Conclusions}

In this Chapter we have analysed the $Z^\prime$ and $W^\prime$ sector of the 4DCHM in three various production processes, two yielding leptonic final states and one in which the extra gauge bosons produce a top antitop quark pair.
By means of the first two we have shown that the LHC, limited to its 14 TeV stage and with standard and high luminosity options, has the potentiality to probe such a model, albeit limited to the non-inert lowest lying mass resonances up to a mass of $\sim$2-3 TeV, which are two in the neutral sector and one in the charged sector, with the exception that in the diboson production modes, after the applications of dedicated cuts, the extraction of the heaviest charged resonance can also be possible.
Furthermore, motivated by the enhanced couplings of these extra states to the third generation of SM quarks, we have shown the potentiality of the $t\bar t $ production mode in detecting the extra gauge bosons of such a model. 
We have however proved that all these conclusions are strongly correlated to the condition of the resonances being sufficiently narrow ($\Gamma_{Z^\prime,W^\prime}/m_{Z^\prime,W^\prime}\ltap$10 \%) since, with the opening of extra possible decay modes and the quick growth of their width, the extraction of the signal over the SM background is no longer possible, no matter the final state.
These, together with the possible separation of the two lightest quasi degenerate $Z^\prime$s, again in a regime where the width over mass ratio is below 10\%, and the use of asymmetries, other than cross section, distributions, both in the leptonic and in the top antitop final state, are the main results for the analysis of the gauge sector of this particular CHM framework, for which the extraction of the $Z_{2,3}^\prime$ and $W_2^\prime$, all degenerate in mass, would represent the hallmark signature.
Moreover the use of $t\bar t$ samples to define charge and spin asymmetries, sensitive to the chirality of the couplings of the new states, and the analysis of the line shapes of the resonances, that would reveal, or otherwise, the presence of light additional fermions, will shed further light on the spectrum of this model.

\chapter{Phenomenology of the Higgs sector of the 4DCHM}
\label{chap-4}
\lhead{Chapter 4. \emph{Phenomenology of the Higgs sector of the 4DCHM}}

This Chapter is devoted to the analysis of the phenomenology of the Higgs sector of the 4DCHM.
After presenting the experimental status of the Higgs searches currently ongoing at the LHC and describing in general the main properties of a composite Higgs arising as a pNGB, we will analyse the characteristics of the 4DCHM Higgs state, showing the compatibility of our framework with the Higgs data arising from the 7 and 8 TeV runs of the LHC.
We will then move onto the analysis of the capabilities of the forthcoming 14 TeV run of the CERN machine and of a future proposed electron positron collider in disentangling the nature of this scalar state in the context of our model.

\section{The status of Higgs searches}

\begin{figure}[!h]
\centering
\epsfig{file=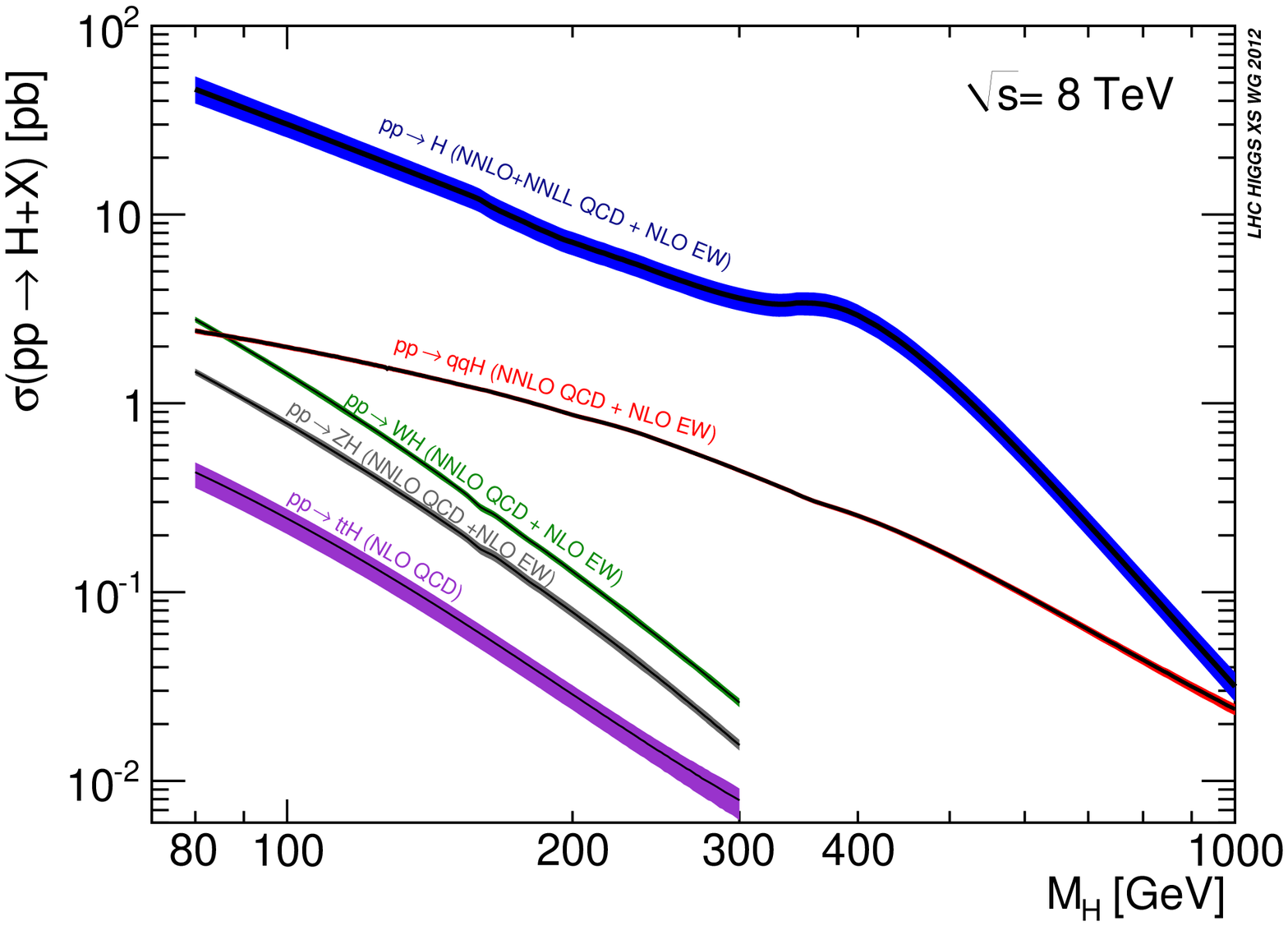, width=.46\textwidth}\hfill
\epsfig{file=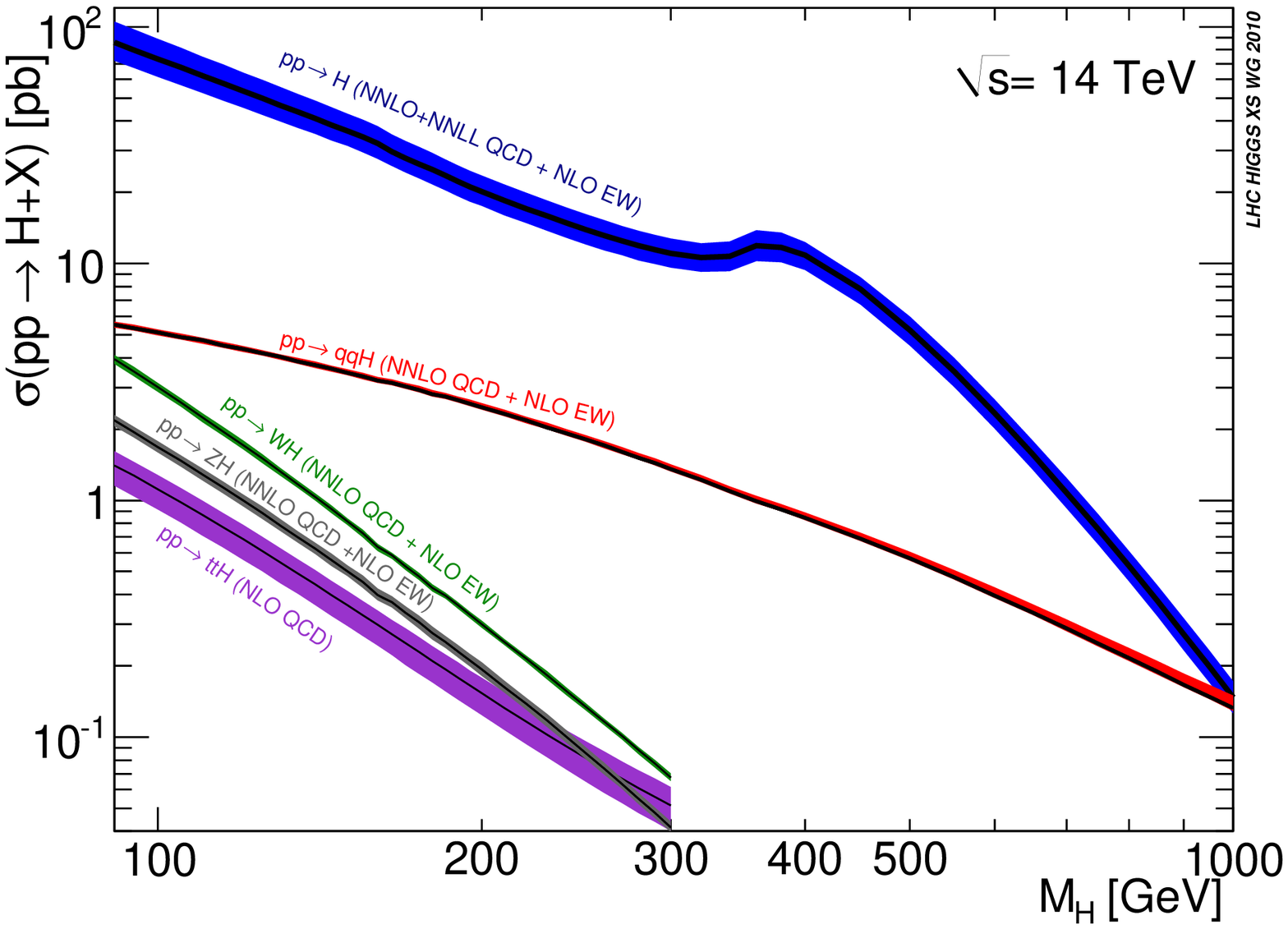, width=.46\textwidth}
\caption[SM Higgs boson production cross sections at the 8 and 14 TeV LHC]{SM Higgs boson production cross sections at the 8 and 14 TeV LHC for various production modes in function of its mass.
Figures are taken from the Higgs cross section working group \cite{Dittmaier:2011ti,Dittmaier:2012vm,Heinemeyer:2013tqa} web page \url{https://twiki.cern.ch/twiki/bin/view/LHCPhysics/CrossSections}.}
\label{fig:higgs-prod-8}
\end{figure}

\begin{figure}[!h]
\centering
\epsfig{file=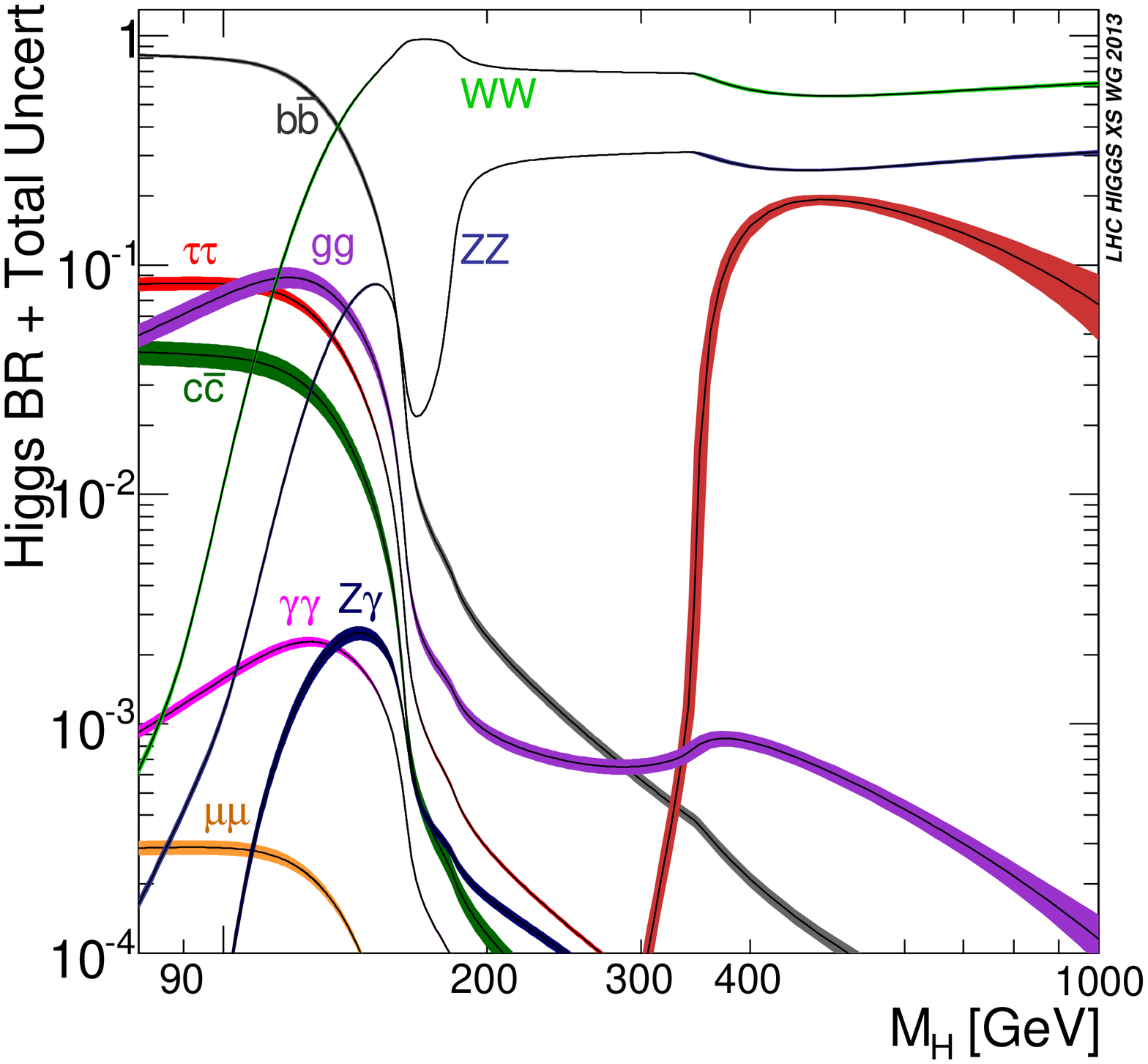, width=.46\textwidth}{(a)}
\hfill
\epsfig{file=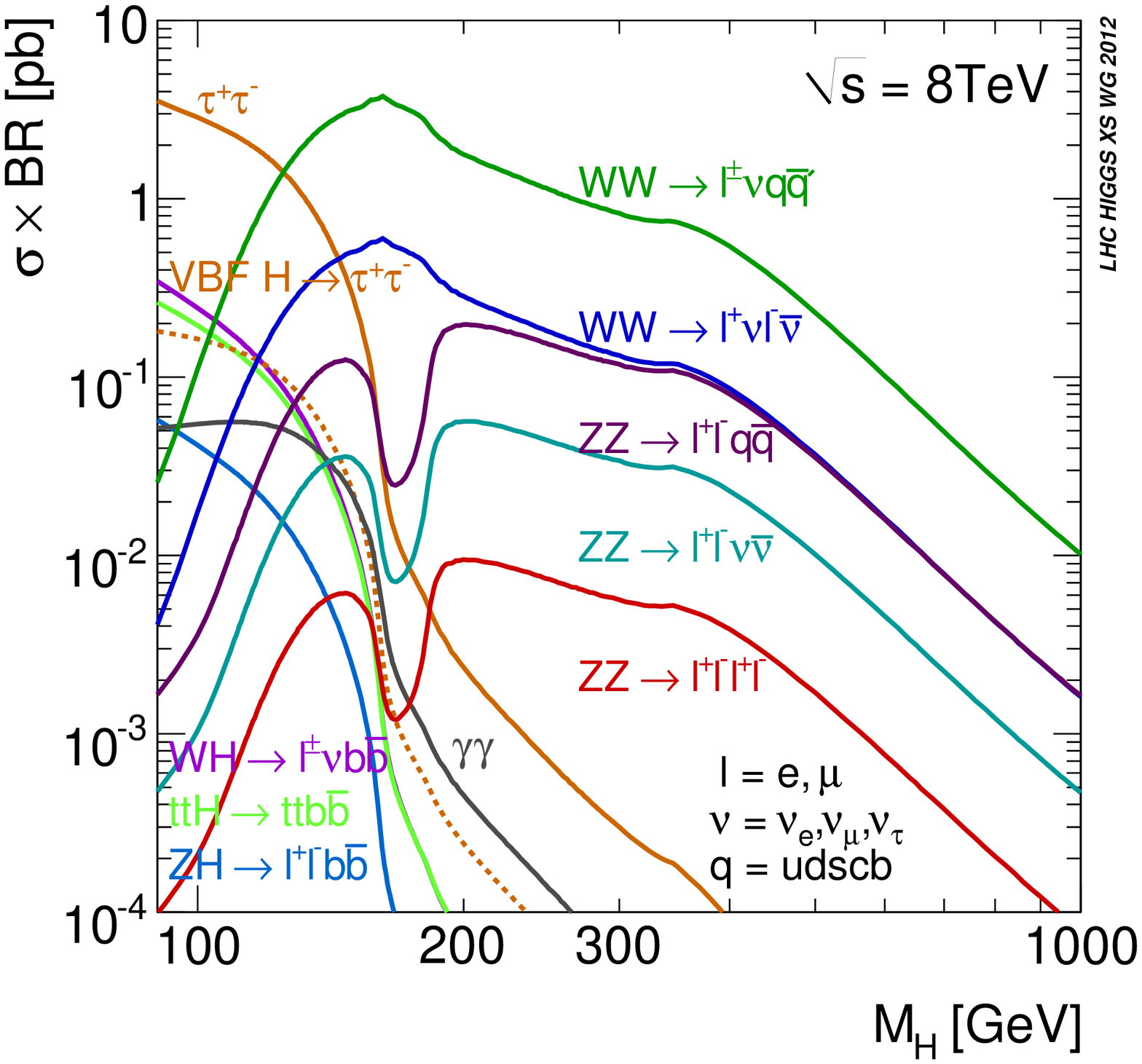, width=.46\textwidth}{(b)}
\caption[Higgs branching ratios and production cross sections times branching ratios for the 8 TeV LHC]{Higgs boson branching ratios (a) and production cross sections for the 8 TeV LHC combined with its branching ratios (b) in function of the mass of a SM Higgs boson. Figures are taken from the Higgs Cross Section Working Group \cite{Dittmaier:2011ti,Dittmaier:2012vm,Heinemeyer:2013tqa} web page \url{https://twiki.cern.ch/twiki/bin/view/LHCPhysics/CrossSections}.}
\label{fig:higgs-br-xsbr}
\end{figure}

After the discovery of a Higgs-like particle at the LHC, one of the primary questions is now to identify whether this state corresponds to the Higgs boson predicted by the SM or if it could belong to other BSM scenarios, and presently the efforts of the ATLAS and CMS collaborations are mainly devoted to giving an answer to this question.
While its spin and parity properties seem quite clear and consistent with the $0^+$ hypothesis \cite{Aad:2013xqa,Chatrchyan:2012jja}, the same is not true for its couplings to the SM gauge bosons and fermions, which are far from being determined with a great accuracy.
The way in which the experimentalists usually present their results concerning these properties of the new scalar state is by defining the so called \emph{signal strengths}, which are the observed number of events for a given Higgs decay channel and production process over the SM expectation
\begin{equation}
\mu_i=\frac{\sigma(pp\to h X)_i Br(h\to YY)_{\phantom{SM}}}{[\sigma(pp\to h X)_i Br(h\to YY)]_{SM}}
\label{eq:mupar_exp}
\end{equation}
where with the subscript $i$ we are indicating a Higgs boson specific production mode, with $X$ any particle produced in association with it and with $YY$ a given final state for its decay.
The final states can be given by tree-level direct decays, such as $h\to b\bar b$, tree-level decays mediated by one of shell gauge boson, such as $h\to WW^*$ and $h\to ZZ^*$ with the subsequent decay of the $W$ and $Z$ bosons yielding a four body final state, as well as loop induced decay process, such as $h\to \gamma\gamma$.
At the LHC the main production processes for the SM Higgs boson are through gluon-gluon-fusion, vector-boson-fusion (VBF) and Higgs-strahlung while the contributions given by associated production with heavy quarks ($t$ and $b$) are sub-dominant, as shown in Fig.~\ref{fig:higgs-prod-8}.
Luckily a Higgs boson of 125 GeV has many decay modes with a significant branching ratios as shown in Fig.~\ref{fig:higgs-br-xsbr} (a)
and this gives us the chance to test its properties at colliders in many different final states, as shown in Fig.~\ref{fig:higgs-br-xsbr} (b) where the decay and the production rates of the Higgs have been combined for the 8 TeV stage of the LHC.

Recently, possible hints of deviations from the SM expectations observed in the $\gamma\gamma$ channel by the CMS collaboration have been downsized by their latest updates while they are still present in the results provided by ATLAS. For other channels both agreement and tensions with the SM are present as shown in Tab.~\ref{tab:higgs-mu-premor2013-postmor2013} where we summarize the LHC measurements of the $\mu$ parameters by decay modes, provided by the ATLAS and CMS collaborations before and after the Moriond winter conference 2013. 
Although more data, that will be collected with the forthcoming high energy run of the CERN machine, are necessary to understand the nature of this state, it is clear that the first two runs of the LHC are pointing towards a scalar particle with properties in good agreement with the ones predicted by the SM, making then important to test the compatibility of our model with the available data.

\begin{table}[!h]
\begin{center}
\begin{tabular}{|l|l|l|}
\hline
 & ~~ATLAS & ~~~~CMS\\
\hline
\hline
$\mu_{\gamma\gamma}$ & $\phantom{-} 1.8 \pm 0.4\phantom{-}$ & $\phantom{-}1.564_{-0.419}^{+0.460}\phantom{-}$ \\
$\mu_{WW}$           & $\phantom{-} 1.5 \pm 0.6\phantom{-}$ & $\phantom{-}0.699_{-0.232}^{+0.245}\phantom{-}$ \\
$\mu_{ZZ}$           & $\phantom{-} 1.0 \pm 0.4\phantom{-}$ & $\phantom{-}0.807_{-0.280}^{+0.349}\phantom{-}$ \\
$\mu_{b\bar b}$           & $- 0.4 \pm 1.0\phantom{-}$ & $\phantom{-}1.075_{-0.566}^{+0.593}\phantom{-}$ \\
\hline
\end{tabular}{(a)}
\hfill
\begin{tabular}{|l|l|l|}
\hline
 & ~~ATLAS & ~~~~CMS\\
\hline
\hline
$\mu_{\gamma\gamma}$ & $\phantom{-} 1.6 \pm 0.3\phantom{-}$ & $\phantom{-}0.77\pm0.27\phantom{-}$ \\
$\mu_{WW}$           & $\phantom{-} 1.4 \pm 0.6\phantom{-}$ & $\phantom{-}0.68\pm0.20\phantom{-}$ \\
$\mu_{ZZ}$           & $\phantom{-} 1.5 \pm 0.4\phantom{-}$ & $\phantom{-}0.92\pm0.28\phantom{-}$ \\
$\mu_{b\bar b}$           & $- 0.4 \pm 1.0\phantom{-}$ & $\phantom{-}1.15\pm0.62\phantom{-}$ \\
\hline
\end{tabular}{(b)}
\end{center}
\caption[$\mu$ parameters from the ATLAS and CMS collaborations before and after the Moriond 2013 winter conference]{Summary of the LHC measurements of the $\mu$ parameters from the ATLAS  and CMS data before the Moriond 2013 winter conference \cite{ATLAS:2012klq,CMS:aya,CMS-twiki-hig12045} (a) and after \cite{ATLAS:2013mma,CMS-PAS-HIG-13-005} (b).
\label{tab:higgs-mu-premor2013-postmor2013}}
\end{table}

\section{The properties of the composite Higgs boson}

In general a composite Higgs arising as a pNGB from a spontaneous breaking of a global symmetry presents couplings to vector bosons and fermions that differ from the ones predicted by the SM, due to the non linear realization of the Goldstone symmetry.
In the context of a low energy effective theory, expressed just in terms of the SM fields and of the composite Higgs and neglecting all the resonances belonging to the strong sector, we have shown in Chapter~\ref{chap-2} that the Lagrangian describing the $SO(5)/SO(4)$ breaking that gives rise to our scalar state can be expressed by the non-linear sigma model of eq.(\ref{eq:pion-lag})
\begin{equation}
\mathcal{L}=\frac{f^2}{2}(\partial_\mu \Phi)^T(\partial_\mu \Phi)
\label{eq:pion-lag-2}
\end{equation}
where, by using the expressions of the $SO(5)$ generators given in Appendix \ref{chap:App-B}, the vector $\Phi$ can be explicitly written as
\begin{equation}
\Phi^T=\frac{\sin h/f}{h}(\textrm{{\bf{\emph{h}}}}^T, h \cot h/f), \quad \textrm{{\bf{\emph{h}}}}^T=(h_1,h_2,h_3,h_4), \quad h=\sqrt{h^{\hat a}h^{\hat a}}, \quad \hat a=1,2,3,4,
\label{eq:leef-phit}
\end{equation}
with \textrm{{\bf{\emph{h}}}} being the Higgs bi-doublet expressed as a vector in the fundamental representation of $SO(4)$.
By means of this Lagrangian we now want to show the modifications of the Higgs couplings to the $W^\pm$ and $Z$ boson with respect to the SM values.
In order to do this we need to gauge the SM gauge group $SU(2)_L\otimes U(1)_Y\subset SO(4)$ by introducing the corresponding covariant derivative and, since we are just interested in the gauge sector of the model, we can identify $T^Y=T^{3R}$, neglecting the $U(1)_X$  gauge group present in the 4DCHM which, as explained in Chapter~\ref{chap-2}, is relevant only to the fermionic sector.
We then make the substitution
\begin{equation}
\partial_\mu \to D_\mu=\partial_\mu - i g W_\mu^i T^i_L-i g^\prime Y_\mu T^3_R
\label{eq:leef-dercov}
\end{equation}
where with $g$ and $g^\prime$ we indicated the gauge couplings of $SU(2)_L$ and $U(1)_Y$ respectively.
Let's assume now that a potential for the Higgs fields exists and that it has the correct \emph{Mexican hat} shape so as to trigger EWSB by making the Higgs acquiring a VEV, as happens in the 4DCHM, choosing for example
\begin{equation}
\langle {\bf{h}} \rangle = (0,0,0,\langle h \rangle)^T.
\label{eq:leef-vev}
\end{equation}
Then, using eq.(\ref{eq:pion-lag-2}-\ref{eq:leef-vev}), we can calculate the part of Lagrangian quadratic in the $W_\mu^i$ fields
\begin{equation}
\mathcal L = \frac{f^2}{2}(\partial_\mu \Phi)^T(\partial_\mu \Phi) \supset \frac{g^2 f^2}{8}\sin^2\frac{\langle h \rangle}{f} W_\mu^i W^{\mu,i}.
\label{eq:leef-wwh}
\end{equation}
Now, expanding the Higgs field onto his VEV, $\langle h \rangle \to \langle h \rangle +h$ and plugging this into eq.(\ref{eq:leef-wwh}), we obtain at the second order in the Higgs field $h$
\begin{equation}
\begin{split}
& \mathcal L   \supset \frac{g^2 f^2}{8}W^\mu W_\mu [\sin^2\frac{\langle h \rangle}{f}+ 2 \sin\frac{\langle h \rangle} {f}\cos\frac{\langle h \rangle}{f}\frac{h}{f}+(1-2 \sin^2\frac{\langle h \rangle}{f})\frac{h^2}{f^2}]=\\
& =\frac{1}{2} W^\mu W_\mu [\frac{g^2 v^2}{4}+\frac{g^2}{2} v \sqrt{1-\xi}h+\frac{g^2}{4}(1-2\xi)h^2]
\label{leef-hvv}
\end{split}
\end{equation}
where we have used the relation
\begin{equation}
v = f \sin{\frac{\langle h \rangle}{f}}
\end{equation}
given by the expression for the $W^\pm$ mass
\begin{equation}
m^2_W=\frac{g^2 f^2}{4}\sin^2{\frac{\langle h \rangle}{f}}
\end{equation}
so that
\begin{equation}
\xi\equiv\frac{v^2}{f^2}=\sin^2{\frac{\langle h \rangle}{f}}.
\end{equation}

From eq.(\ref{leef-hvv}) we see that the couplings of the SM charged vector bosons to one and two Higgs are modified, with respect to the SM, by a factor
\begin{equation}
g_{VVh}=g^{SM}_{VVh}\sqrt{1-\xi} \qquad g_{VVhh}=g^{SM}_{VVhh}(1-2 \xi)
\label{eq:leef-vvh-vvhh-mod}
\end{equation}
which is only dictated by the symmetry breaking pattern of the model. Conversely the modification of the Higgs coupling to the SM fermions also depends on the choice of the fermion embedding and it is possible to prove, see for example \cite{Espinosa:2010vn}, that for the 4DCHM, that is embedding the fermions in fundamental representation of $SO(5)$, the modified coupling is
\begin{equation}
g_{ffh}=g^{SM}_{ffh}\frac{1-2\xi}{\sqrt{1-\xi}}.
\label{eq:leef-ffh-mod}
\end{equation}

As mentioned these modifications assume a complete decoupling of the extra resonances belonging to the composite Higgs scenario while, in presence of a complete spectrum, extra effects with respect to eq.(\ref{eq:leef-vvh-vvhh-mod}) and eq.(\ref{eq:leef-ffh-mod}) can appear, due to the mixing between the SM and extra particles caused by the partial compositness mechanism.
Moreover in the case of loop induced couplings such as $hgg$ and $h\gamma\gamma$, relevant for the analysis of the Higgs sector, the presence of extra particles running into the loops can in principle bring another source of modification of the effective couplings, besides the one given by the two effects already mentioned.
All these effects have been taken into account in our analysis both at the partial compositness level, as already done in Chapter~\ref{chap-3}, and also at the level of extra particles running into the loops and, when necessary, a comparison between the low energy approach and the complete approach will be shown.

\section{Loop induced couplings}

Since the tool chosen to perform our analysis, CalcHEP, is by default a tree-level MC generator, in order to study the Higgs properties of the 4DCHM, it has been necessary to implement in it those couplings that don't arise at the tree-level such as $hgg$ and $h\gamma\gamma$.
To compute these effective couplings in the 4DCHM we need to consider both the tree-level modifications of the couplings of the SM particles to the Higgs boson, which thanks to our implementation are automatically taken into account, and also the contributions of extra states intervening in these  loop induced processes, which deserve a dedicated treatment.
Therefore we will now present the expressions for $hgg$ and $h\gamma\gamma$ within the 4DCHM, leaving however indicated in a general way the couplings of which they are functions that have been, of course, expressed in function of the model parameters in the CalcHEP implementation.
The 4DCHM with the addition of the $hgg$ and $h\gamma\gamma$ couplings has been uploaded onto the HEPMDB \cite{hepmdb} web site
under the name \emph{4DCHM (with HAA/HGG)}\footnote{That can be found at the following URL: 
\url{http://hepmdb.soton.ac.uk/hepmdb:0213.0123}.}.

\subsection{$hgg$ coupling}

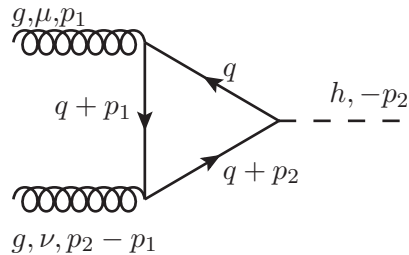
\begin{figure}[!h]
  \begin{picture}(292,118) (55,-47)
    \SetWidth{1.0}
    \SetColor{Black}
    \Text(194,60)[lb]{\Black{$g$,$\mu$,$p_1$}}        
    \Gluon(194,54)(243,54){4.5}{7}
    \Text(194,-25)[lb]{\Black{$g,\nu,p_2-p_1$}}       
    \Gluon(194,-5)(243,-5){4.5}{7}
    \Text(210,25)[lb]{\Black{$q+p_1$}}
    \Line[arrow,arrowpos=0.5,arrowlength=5,arrowwidth=2,arrowinset=0.2](243,54)(243,-5)
    \Text(273,0.5)[lb]{\Black{$q+p_2$}}    
    \Line[arrow,arrowpos=0.5,arrowlength=5,arrowwidth=2,arrowinset=0.2](243,-5)(293,24.5)
    \Text(273,39.5)[lb]{\Black{$q$}}     
    \Line[arrow,arrowpos=0.5,arrowlength=5,arrowwidth=2,arrowinset=0.2](293,24.5)(243,54)
    \Text(313,29.5)[lb]{\Black{$h,-p_2$}}      
    \Line[dash,dashsize=6](293,24.5)(338,24.5)  
  \end{picture}
  \caption[Feynman diagram for $hgg$ couplings]{Feynman diagram for the one loop $hgg$ coupling. With $q$ we indicate here any quark that can run in the loop. All external momenta are incoming.}
  \label{fig:ggh-prod}
\end{figure}

The coupling of the Higgs boson to a gluon pair is induced by the Feynman diagram of Fig.~\ref{fig:ggh-prod}, where with $q$ we indicate any possible quark that can run into the loop.
The structure of this coupling is well established in literature and the only modification that has to be taken into account for computing this amplitude within the 4DCHM is to sum over all the possible quarks that can intervene in the loop obtaining then
\begin{equation}
\mathcal M(g g\to h) = i \sum_{j=1}^{n_q} \frac{1}{2} \frac{N_c m_{q_j}}{2 \pi^2 m_H^2} g^2_s g_{hq_j \bar q_j}(-2-4 m^2_{q_j}C_0+m_H^2C_0)(g^{\mu\nu}p_1\cdot p_2-p_1^\nu p_2^\mu)
\end{equation}
where $m_{q_j}$ and $g_{h q_j \bar q_j}$ are the mass and the couplings of the $j$-th quark to the Higgs boson and 
\begin{equation}
C_0=C_0(0,0,m_H^2,m^2_{q_j},m^2_{q_j},m^2_{q_j})
\end{equation}
is the scalar three point Passarino Veltman function.
Finally we have included the next to leading order QCD $\kappa$ factor \cite{Spira:1995rr}, while for other production processes we have neglected their $k$ factors since their effect is expected to be smaller.

\subsection{$h\gamma\gamma$ coupling}

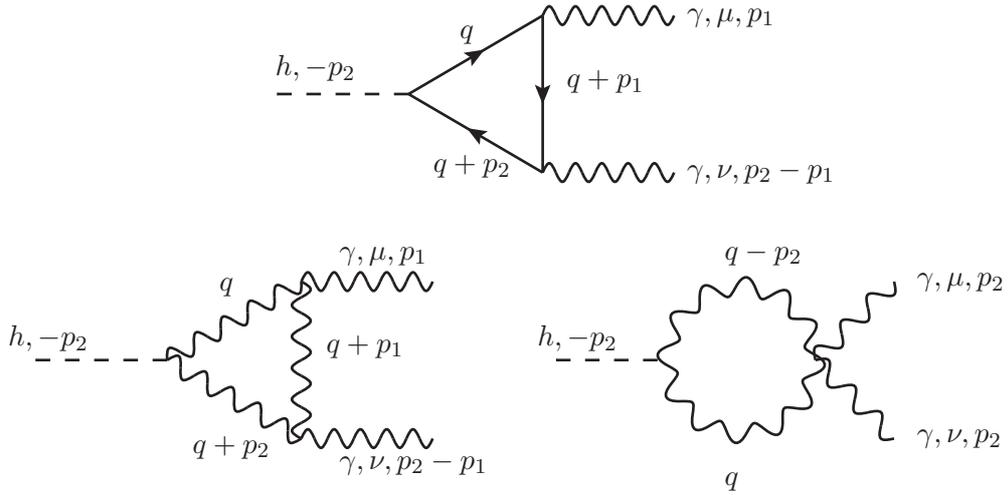
\begin{figure}[!h]
  \begin{picture}(292,218) (55,-147)
    \SetWidth{1.0}
    \SetColor{Black}
    \Text(184,29.5)[lb]{\Black{$h,-p_2$}}      
    \Line[dash,dashsize=6](184,24.5)(233,24.5)
    \Text(243,-7)[lb]{\Black{$q+p_2$}}       
    \Line[arrow,arrowpos=0.5,arrowlength=5,arrowwidth=2,arrowinset=0.2](283,-5)(233,24.5)
    \Text(253,43.5)[lb]{\Black{$q$}}      
    \Line[arrow,arrowpos=0.5,arrowlength=5,arrowwidth=2,arrowinset=0.2](233,24.5)(283,54)
    \Text(293,25)[lb]{\Black{$q+p_1$}}    
    \Line[arrow,arrowpos=0.5,arrowlength=5,arrowwidth=2,arrowinset=0.2](283,54)(283,-5)
    \Text(338,50)[lb]{\Black{$\gamma,\mu,p_1$}}      
    \Photon(283,54)(332,54){3.5}{5}
    \Text(338,-9)[lb]{\Black{$\gamma,\nu,p_2-p_1$}} 
    \Photon(283,-5)(332,-5){3.5}{5}   
    \Text(84,-72.5)[lb]{\Black{$h,-p_2$}}     
    \Line[dash,dashsize=6](94,-75.5)(143,-75.5)   
    \Text(153,-113)[lb]{\Black{$q+p_2$}}     
    \Photon(143,-75.5)(193,-105){3.5}{5}
    \Text(163,-51.5)[lb]{\Black{$q$}}     
    \Photon(193,-46)(143,-75.5){3.5}{5}
    \Text(203,-75)[lb]{\Black{$q+p_1$}}    
    \Photon(193,-105)(193,-46){3.5}{5}
    \Text(208,-40)[lb]{\Black{$\gamma,\mu,p_1$}}      
    \Photon(193,-46)(242,-46){3.5}{5}
    \Text(208,-119)[lb]{\Black{$\gamma,\nu,p_2-p_1$}}    
    \Photon(193,-105)(242,-105){3.5}{5}     
    \Text(282,-72.5)[lb]{\Black{$h,-p_2$}}        
    \Line[dash,dashsize=6](288,-75.5)(325.811,-75.5)
    \Text(352,-40)[lb]{\Black{$q-p_2$}}
    \Text(352,-125)[lb]{\Black{$q$}}      
    \PhotonArc(357,-75.5)(27.689,-56,304){-3.5}{12}
    \Text(424,-50)[lb]{\Black{$\gamma,\mu,p_2$}}          
    \Photon(384.689,-75.5)(414,-105){3.5}{3}
    \Text(424,-109)[lb]{\Black{$\gamma,\nu,p_2$}}    
    \Photon(384.689,-75.5)(414,-46){3.5}{3}      
   
  \end{picture}
  \caption[Feynman diagrams for $h\gamma\gamma$ couplings]{Feynman diagrams for the one loop $h\gamma\gamma$ coupling. With $f$ and $W$ we indicate here any fermion and charged vector that can run in the loop. All external momenta are incoming.}
  \label{fig:haa-decay}
\end{figure}

The coupling of the Higgs boson to two photons is mediated by both fermions and vector bosons and the Feynamn diagrams for this process are shown in Fig.~\ref{fig:haa-decay}.
Once again the structure of this coupling is well established in literature so that we only need to sum the amplitudes over all the possible fermions and charged vector bosons that can intervene in the loop obtaining then
\begin{equation}
\mathcal M_f(h \to \gamma\gamma) = i \sum_{j=1}^{n_f} \frac{N_c m_{f_j}}{2 \pi^2 m_H^2} e^2 Q^2_f g^2_{hq_j \bar q_j}(-2-4 m^2_{f_j}C_0+m_H^2C_0)(g^{\mu\nu}p_1\cdot p_2-p_1^\nu p_2^\mu)
\end{equation}
for the fermion loop amplitude and
\begin{equation}
\begin{split}
 \mathcal M_W(h \to \gamma\gamma) = & i \sum_{j=1}^{n_W} 
\frac{1}{8 \pi^2 m_H^2}e^2g_{hW_jW_j}
(-6-\frac{m^2_H}{m^2_{W_j}}-12 C_0 m^2_{W_j}+6 C_0 m^2_H)\cdot\\
& \cdot (g^{\mu\nu}p_1\cdot p_2-p_1^\nu p_2^\mu)
\end{split}
\end{equation}
for the charged vector boson loop, where $m_{f_j}$,$m_{W_j}$,$g_{h f_j \bar f_j}$,$g_{h W_j W_j}$ are the mass and the couplings of the $j$-th fermion or vector boson in the loop and 
\begin{equation}
C_0=C_0(0,0,m_H^2,m^2_{f_j/W_j},m^2_{f_j/W_j},m^2_{f_j/W_j})
\end{equation}

\section{LHC analysis}

We will present our results concerning the phenomenology of the 4DCHM Higgs in terms of the ratio of events predicted in the 4DCHM hypothesis with respect to the SM one for a given production process and in a given decay channel, by defining a \emph{signal strength} in the same way as done by the experimental collaborations
\begin{equation}
\mu_i=\frac{[\sigma(pp\to h X)_i Br(h\to YY)]_{{\textrm{4DCHM}}}}{[\sigma(pp\to h X)_i Br(h\to YY)]_{{\textrm{SM}}}}
\label{eq:mupar_theo}
\end{equation}
This theoretical prediction will be compared with the ATLAS and CMS measurements that are reported in Tab.~\ref{tab:higgs-mu-premor2013-postmor2013} (b), that is with the publicly available data after the Moriond 2013 winter conference\footnote{We mention however that more update results have been delivered from both the ATLAS \cite{ATLAS-CONF-2014-009} and CMS \cite{CMS:ril,Chatrchyan:2013mxa,Chatrchyan:2013iaa,Chatrchyan:2013zna} collaborations but that, on the other hand, these modifications will not change the conclusions of our analysis and we will then continue to refer to the values of Tab.~\ref{tab:higgs-mu-premor2013-postmor2013} (b).}.
For our analysis we will choose as unique production modes for the Higgs boson, gluon-gluon-fusion for the $\gamma\gamma$, $WW$ and $ZZ$ final states and Higgs-strahlung for the $b \bar b$ final state.
This choice is motivated by the fact that in the case of the decay of the Higgs in a pair of bottom quark the experimentalists, due to the large QCD background of this final state, tag it in combination with an associated production of a vector boson while for the other channels, cleaner than the former, gluon-fusion can be safely considered as the main production channel.

For the purpose of our analysis it is convenient to re-write eq.(\ref{eq:mupar_theo}) as follows
\begin{equation}
\mu_{YY}^{Y^\prime Y^\prime}=\frac{[\Gamma(h\to Y^\prime Y^\prime)\Gamma(h\to YY)]_{{\textrm{4DCHM}}}}{[\Gamma(h\to Y^\prime Y^\prime)\Gamma(h\to YY)]_{{\textrm{SM}}}}\frac{[\Gamma_{{\textrm{tot}}}(h)]_{{\textrm{SM}}}}{[\Gamma_{{\textrm{tot}}}(h)]_{{\textrm{4DCHM}}}}
\label{eq:mupar_theo2}
\end{equation}
where $Y^\prime Y^\prime$ denote the particles participating in the Higgs boson production, that are $gg$ for gluon-fusion and $VV=W^+W^-,ZZ$ for VBF and Higgs-strahlung, while $Y Y$ are again the particles arising from the Higgs boson decay.
In other words we trade a cross section for a width and this is possible as we will be carrying out our analysis at the lowest order without
the presence of radiative corrections due to either QCD or EW interactions, and this is possible since the latter are expected to be much smaller than the former which are the same both in the 4DCHM and the SM.
Moreover, following \cite{LHCHiggsCrossSectionWorkingGroup:2012nn}, we can also cast eq.(\ref{eq:mupar_theo2}) as
\begin{equation}
\mu_{YY}^{Y^\prime Y^\prime}=\frac{\kappa^2_{Y^\prime}\kappa^2_Y}{\kappa^2_H}
\label{eq:mupar_theo3}
\end{equation}
where
\begin{equation}
\kappa^2_{Y/Y^\prime}=\frac{\Gamma(h\to YY/Y^\prime Y^\prime)_{{\textrm{4DCHM}}}}{\Gamma(h\to YY/Y^\prime Y^\prime)_{{\textrm{SM}}}}, \qquad \kappa^2_H=\frac{\Gamma_{{\textrm{tot}}}(h)_{{\textrm{4DCHM}}}}{\Gamma_{{\textrm{tot}}}(h)_{{\textrm{SM}}}}
\label{eq:kappa-par}
\end{equation}
so that we can disentangle in the signal strengths the effects arising from the production, the partial width and the Higgs total width parts.

\begin{figure}[!h]
\centering
\epsfig{file=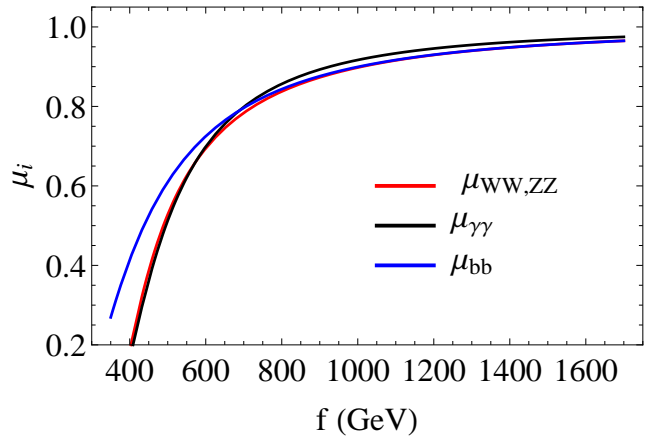, width=.58\textwidth}
\caption[Higgs signal strengths at the LHC in the decoupling limit as function of the model scale $f$]{Higgs signal strengths at the LHC in the decoupling limit as function of the model scale $f$. Red, black and blue lines correspond to $WW/ZZ$, $\gamma\gamma$ and $b\bar b$ final states respectively. For $WW/ZZ$ and $\gamma\gamma$ final states we have assumed unique gluon-fusion production, while for $b \bar b$ final state unique Higgs-strahlung production.}
\label{fig:LHC-mu-decoup}
\end{figure}

Before presenting our results for the analysis of the LHC phenomenology of the 4DCHM it is however instructive to see what  the  predictions are for the signal strengths for the final states of Tab.~\ref{tab:higgs-mu-premor2013-postmor2013} in the low energy assumptions, that is when the only source of modification of the Higgs couplings is the one given by its pNGB properties.
In this case the $\kappa$s parameters of eq.(\ref{eq:kappa-par}) can be computed straightforwardly from the expression for the Higgs-vector-vector and Higgs-fermion-fermion reduced couplings of eq.(\ref{eq:leef-vvh-vvhh-mod}) and eq.(\ref{eq:leef-ffh-mod}) assuming, for simplicity, that the total Higgs width, that is the $\kappa^2_H$ part, is made up only of the contribution of $\Gamma(h\to b \bar b)$, $\Gamma(h\to W^+ W^-)$, $\Gamma(h\to Z Z)$, $\Gamma(h\to g g)$ which are, as shown in Fig.~\ref{fig:higgs-br-xsbr} (a), the main decay modes of a SM Higgs with a mass $\sim$125 GeV.
From the expressions of eq.(\ref{eq:leef-vvh-vvhh-mod}) and eq.(\ref{eq:leef-ffh-mod}) we see that both the gluon-gluon-fusion and the Higgs-strahlung production rates will be decreased with respect to the SM predictions, that is $\kappa^2_{Y^\prime Y^\prime}<1$, and the same will happen for the partial widths of the Higgs in the various final state, $\kappa^2_{YY}<1$, causing the total Higgs width to decrease.
From a phenomenological point of view the reduction of the Higgs total width is of particular interest. In fact even if a measure of the latter is challenging at an hadron collider, we observe that we are in presence of contrasting effects intervening in eq.(\ref{eq:mupar_theo3}) that can induce an increase of the various signal strengths, pointing towards more stable values even though the production rates are generally decreased.
We show this in Fig.~\ref{fig:LHC-mu-decoup} where we plot the $\mu$ parameters as defined in eq.(\ref{eq:mupar_theo3}) for three different final states in function of the model scale $f$.

We now move onto present our results for the 4DCHM including all the possible effects due to its extra particle
content, focusing in particular on the importance of the extra fermionic resonances present in the spectrum.
The role of the fermionic partners in the LHC Higgs physics has been deeply investigated in literature in the context of CHMs.
For example the modification of the LHC phenomenology in presence of third generation quark partners has been discussed in \cite{Azatov:2011qy,Gillioz:2012se} while in \cite{Delaunay:2013iia} the composite Higgs phenomenology has been analysed in the case where the partial compositness mechanism is also implemented  for light quark generations.
In order to show our results for the analysis of the 4DCHM we adopt as a reference benchmark point the combination of $f$=1000 GeV and $g_\rho$=2 for which we have scanned on the parameters of the fermionic sector as explained in Section~\ref{subsec:2-implement} with the constraints due to the EWPT and direct search of extra gauge bosons and fermions illustrated in Section~\ref{subsec:2-bench} and Section~\ref{sec:ztt}.
However when presenting our final results for the compatibility with the LHC data we will do so for all the combinations of $f$ and $g_\rho$ of eq.(\ref{eq:bench-fvar}).

\begin{figure}[!h]
\centering
\epsfig{file=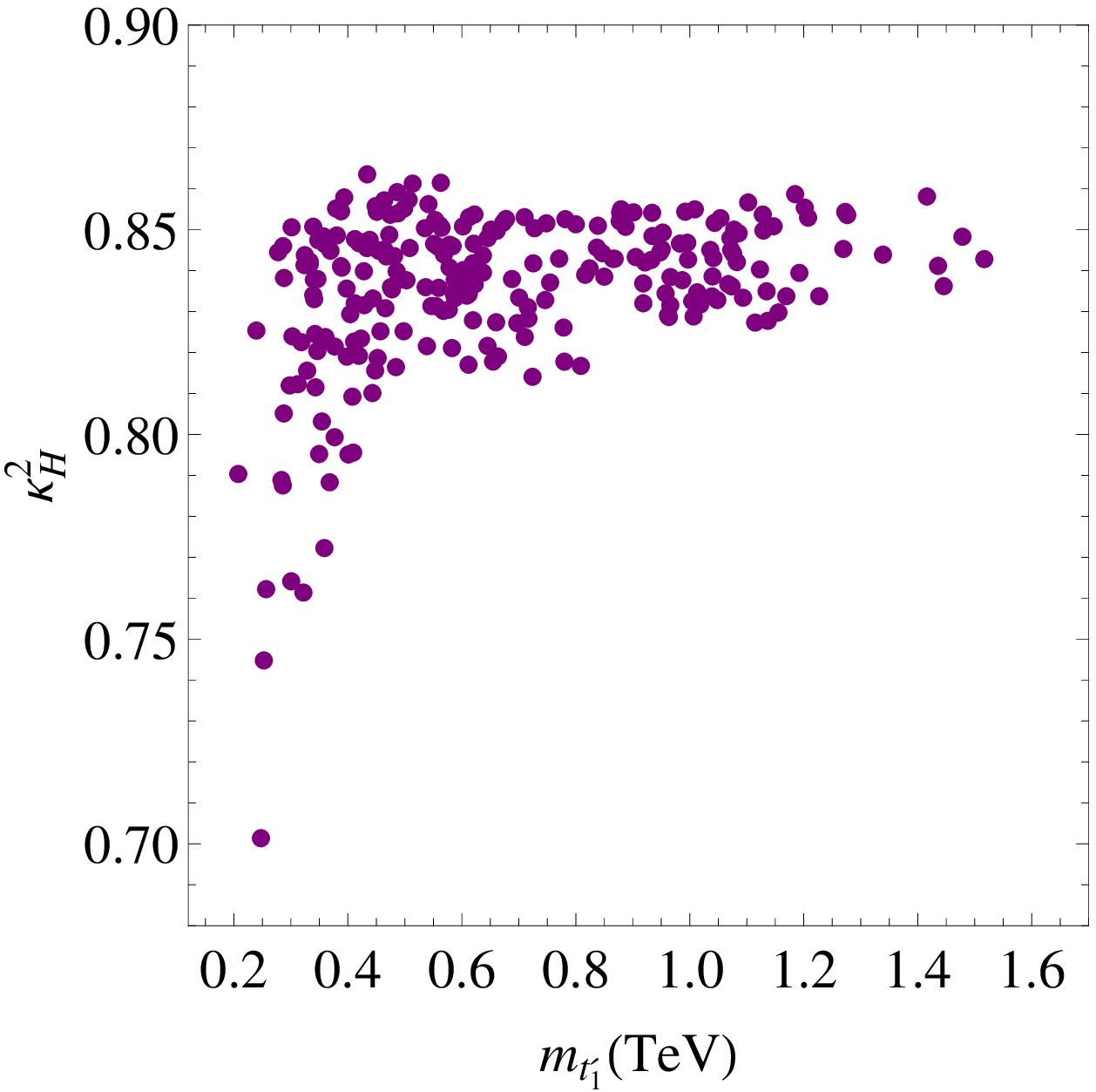, width=.46\textwidth}{(a)}\hfill
\epsfig{file=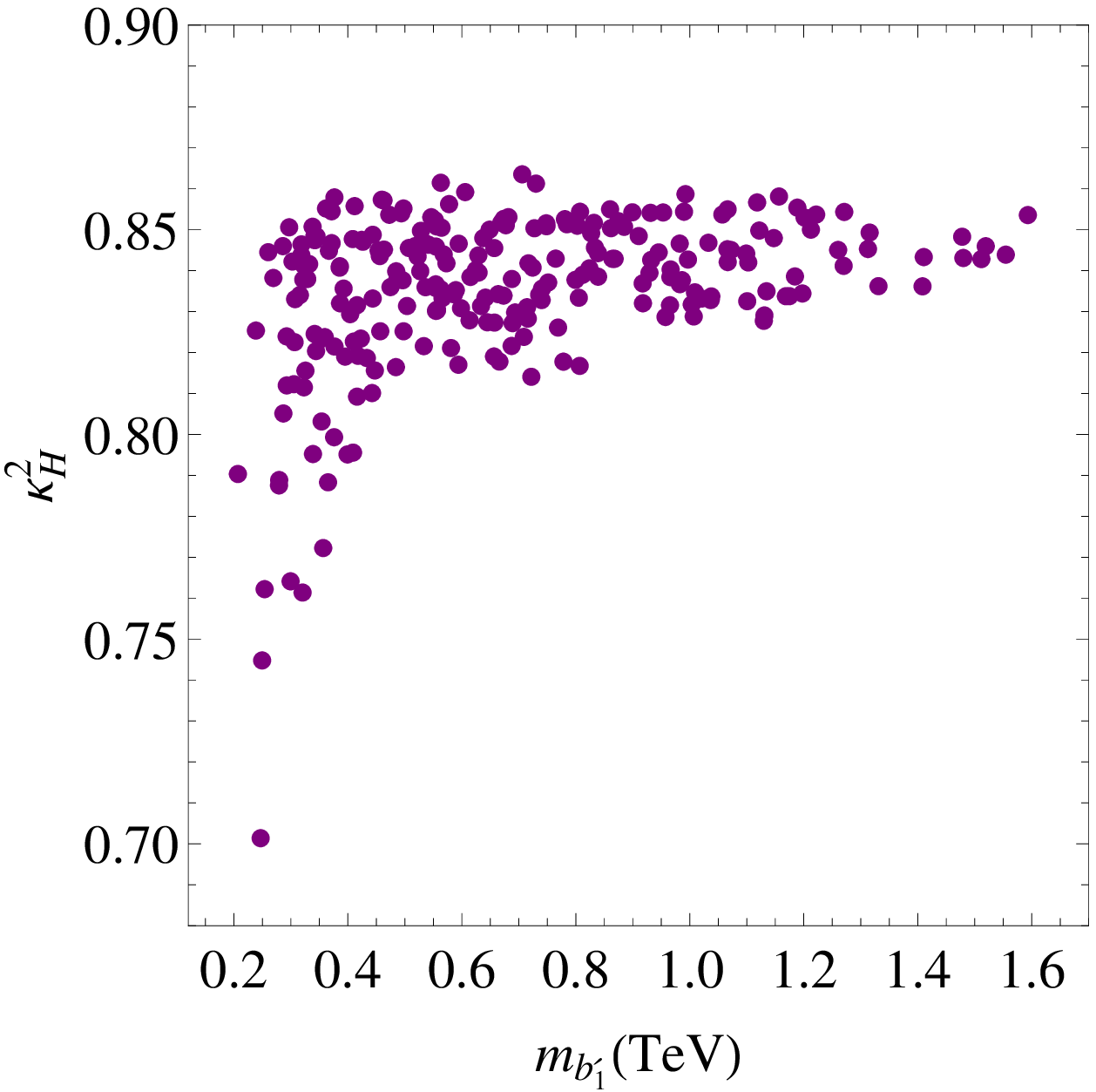, width=.46\textwidth}{(b)}
\caption[$\kappa^2_H$ in function of the lightest $t^\prime$ and $b^\prime$ for $f=$ 1000 GeV and $g_\rho$=2]{Distributions of $\kappa^2_H$ as a function of the lightest $t^\prime$ (a) and $b^\prime$ (b). }
\label{fig:k2h-tp-bp}
\end{figure}

We show the results for $\kappa^2_H$ in function of the mass of the lightest $t^\prime$ and $b^\prime$ in Fig.~\ref{fig:k2h-tp-bp} where, for illustrative purposes, we also keep in these plots points that do not comply with our reinterpretation of the limits on extra quarks for which, we recall, we have obtained a bound for their masses around 500 GeV.
Since the Higgs boson total width is dominated by the $b \bar b$ partial width, in the limit when the extra fermions are decoupled, and so the mixing effects between SM and extra quarks due to the partial compositness mechanism are negligible, $k^2_H$ tends towards the value of $(g_{ffh}/g^{SM}_{ffh})^2=(\frac{1-2\xi}{\sqrt{1-\xi}})^2\sim$ 0.82 for $f=1000$ GeV.
However even in regions of the parameter space where the masses of the extra fermions are consistent with our recast of the extra quark limits, deviations from this asymptotic limit, due to the mixing of the $b$ quark with extra fermions with charge $-1/3$, are still possible and can be quantified in $\sim$ 5\%.

\begin{figure}[!h]
\centering
\epsfig{file=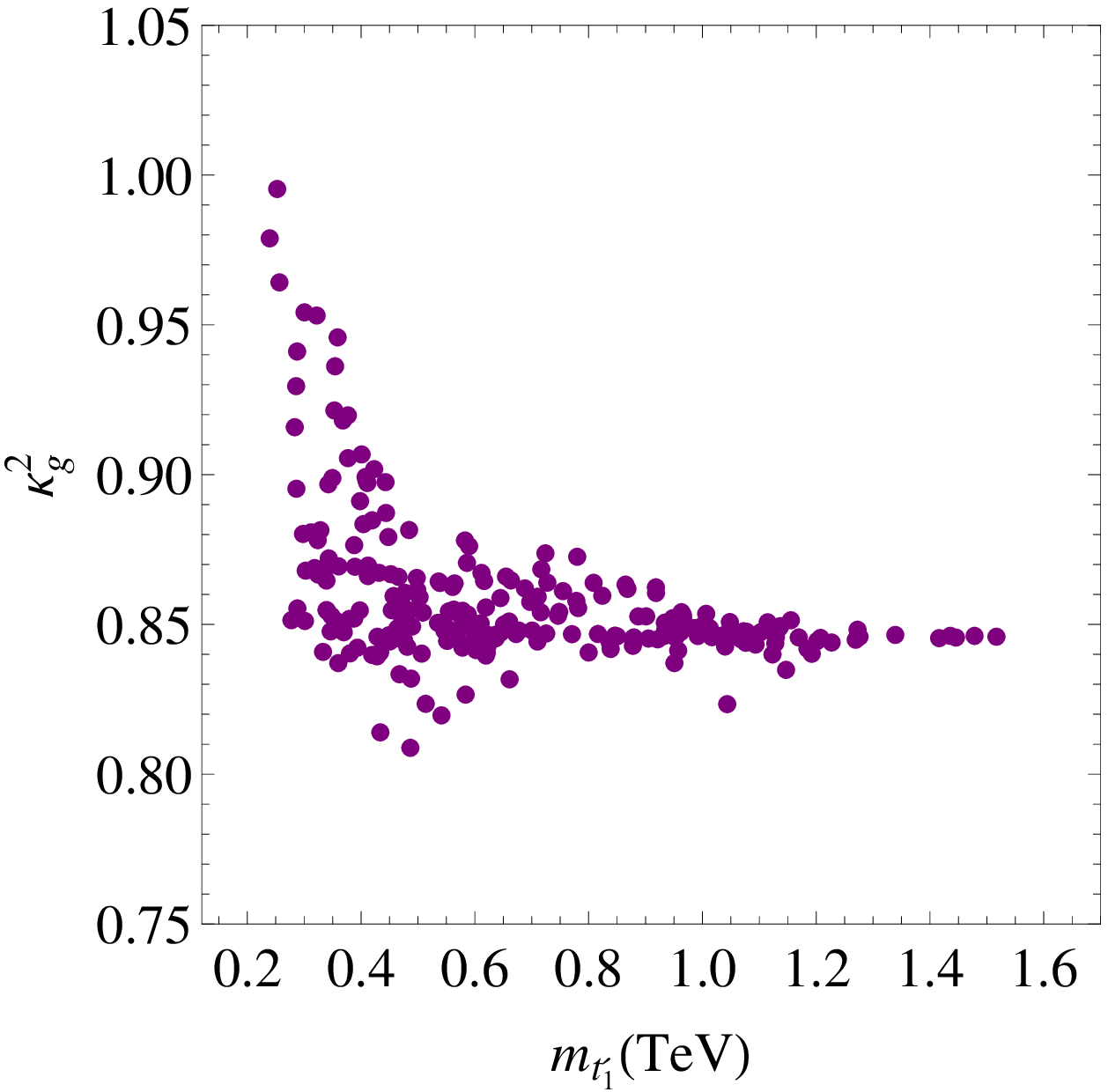, width=.46\textwidth}{(a)}\hfill
\epsfig{file=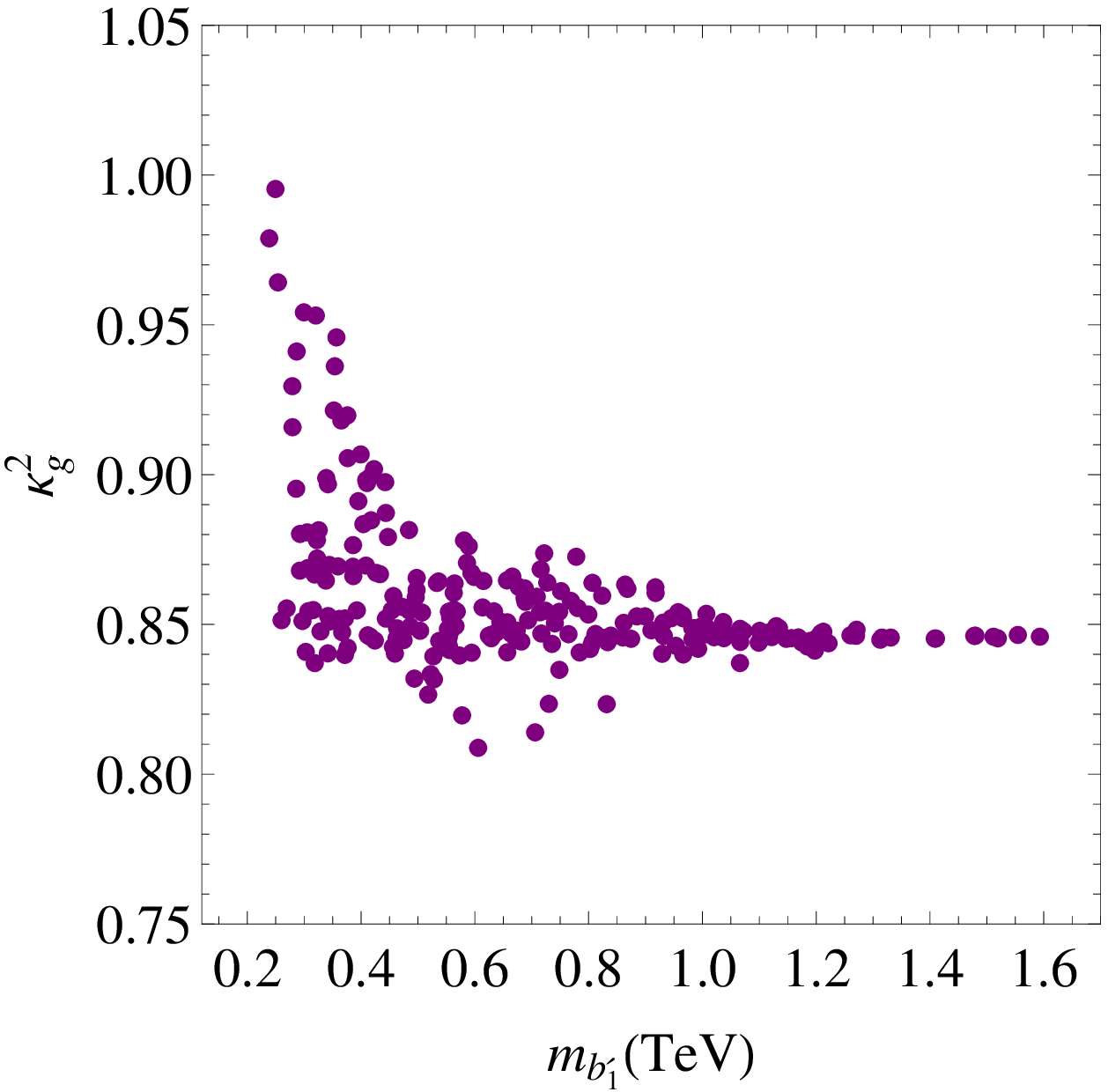, width=.46\textwidth}{(b)}
\caption[$\kappa^2_g$ in function of the lightest $t^\prime$ and $b^\prime$ for $f=$ 1000 GeV and $g_\rho$=2]{Distributions of $\kappa^2_g$ as a function of the lightest $t^\prime$ (a) and $b^\prime$ (b). }
\label{fig:k2g-tp-bp}
\end{figure}

This also happens  for the $\kappa$s related to production and decay processes and we show in Fig.~\ref{fig:k2g-tp-bp} the result for $\kappa^2_g$ that gives the measure of the variation of the gluon-fusion production rates with respect to the SM.
Again, if the extra states belonging to the 4DCHM are decoupled, this parameter reaches its asymptotic value $(\frac{g_{ffh}}{g^{SM}_{ffh}})^2\sim 0.82$ while, if smaller masses of the extra quarks are allowed, it will be brought towards higher values.
In this dynamics we recognise the effects of the $t^\prime$s and $b^\prime$s intervening in the loop such as the lighter these states are the bigger their loop contribution will be, which will balance and overcome the reduction of the $ht \bar t$ coupling, as the top quark contribution is the dominant in the loop induced $hgg$ coupling, due to the partial compositness mechanism.

\begin{figure}[!h]
\centering
\epsfig{file=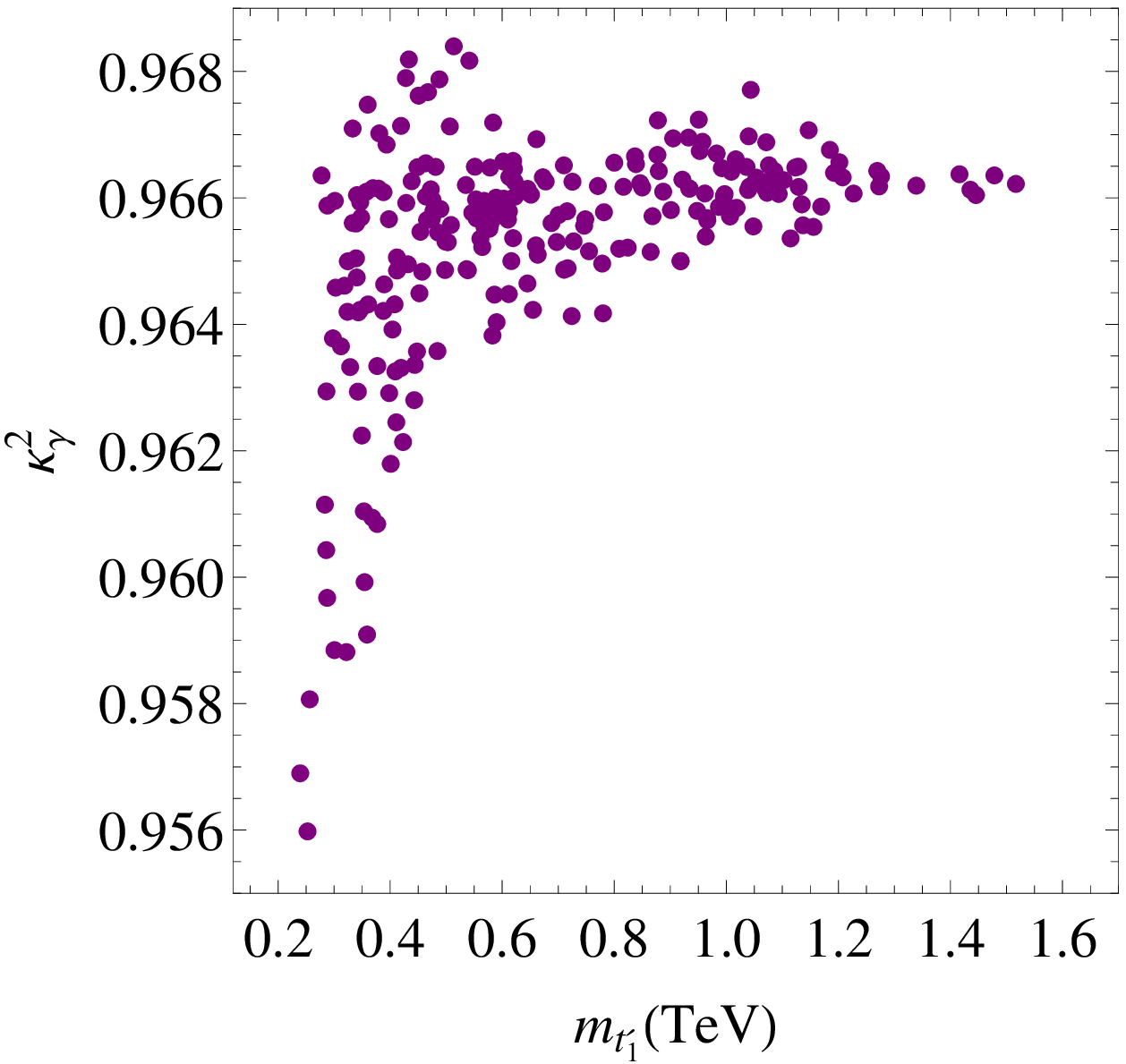, width=.46\textwidth}{(a)}\hfill
\epsfig{file=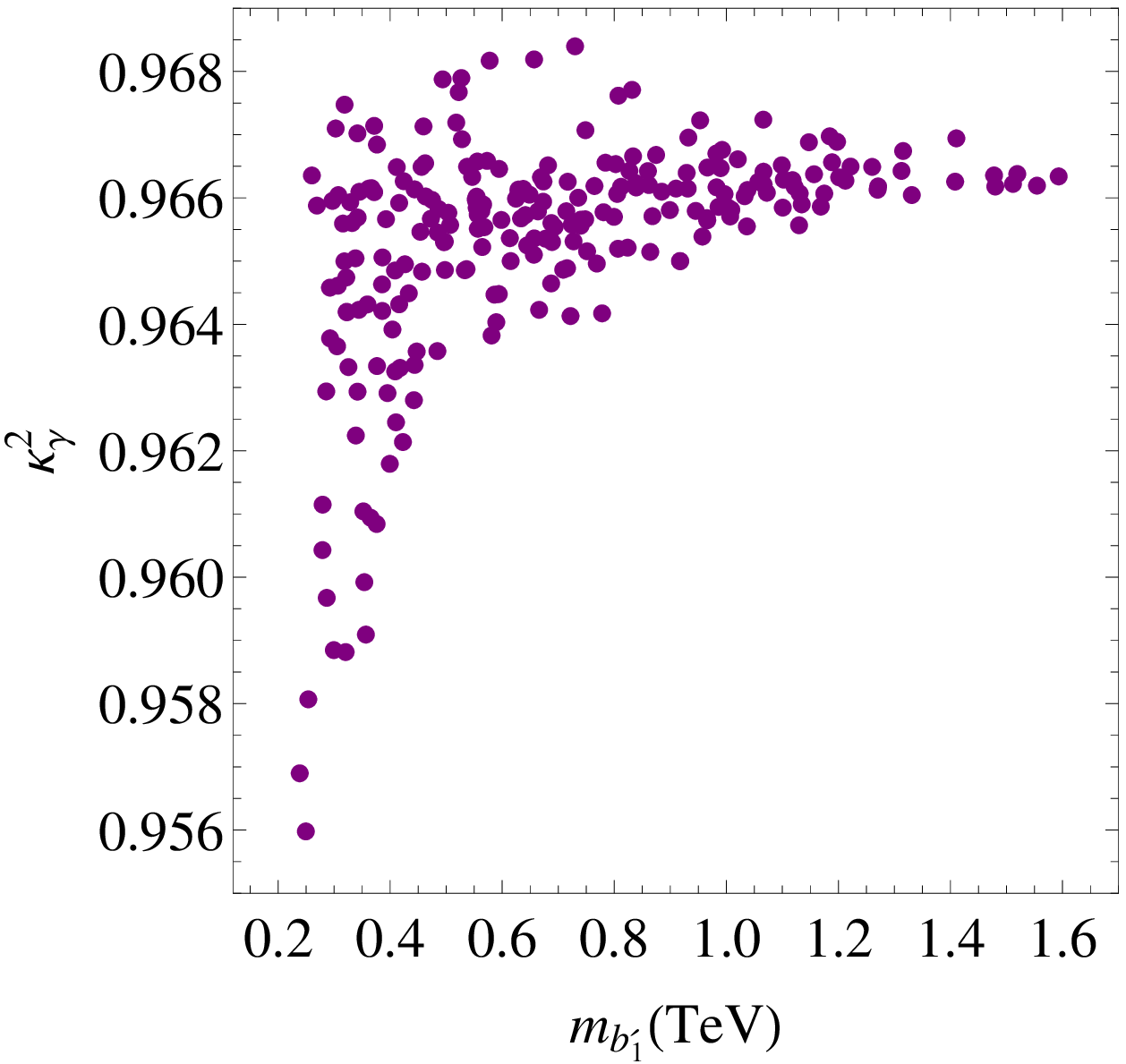, width=.46\textwidth}{(b)}
\caption[$\kappa^2_\gamma$ in function of the lightest $t^\prime$ and $b^\prime$ for $f=$ 1000 GeV and $g_\rho$=2]{Distributions of $\kappa^2_\gamma$ as a function of the lightest $t^\prime$ (a) and $b^\prime$ (b). }
\label{fig:k2a-tp-bp}
\end{figure}

Regarding the $h\gamma\gamma$ coupling we show in in Fig.~\ref{fig:k2a-tp-bp} the results for $\kappa^2_\gamma$ again in function of the masses of the lightest $t^\prime$ and $b^\prime$ in the spectrum.
Since the $W$ boson is the dominant contribution in this loop induced coupling, the value of this parameter points in the asymptotic limit towards higher values for $f$= 1000 GeV and, since the top and top partners loop contributions interfere destructively with the $W$ contribution, it is brought towards smaller values if smaller masses of these states are allowed.

\begin{figure}[!h]
\centering
\epsfig{file=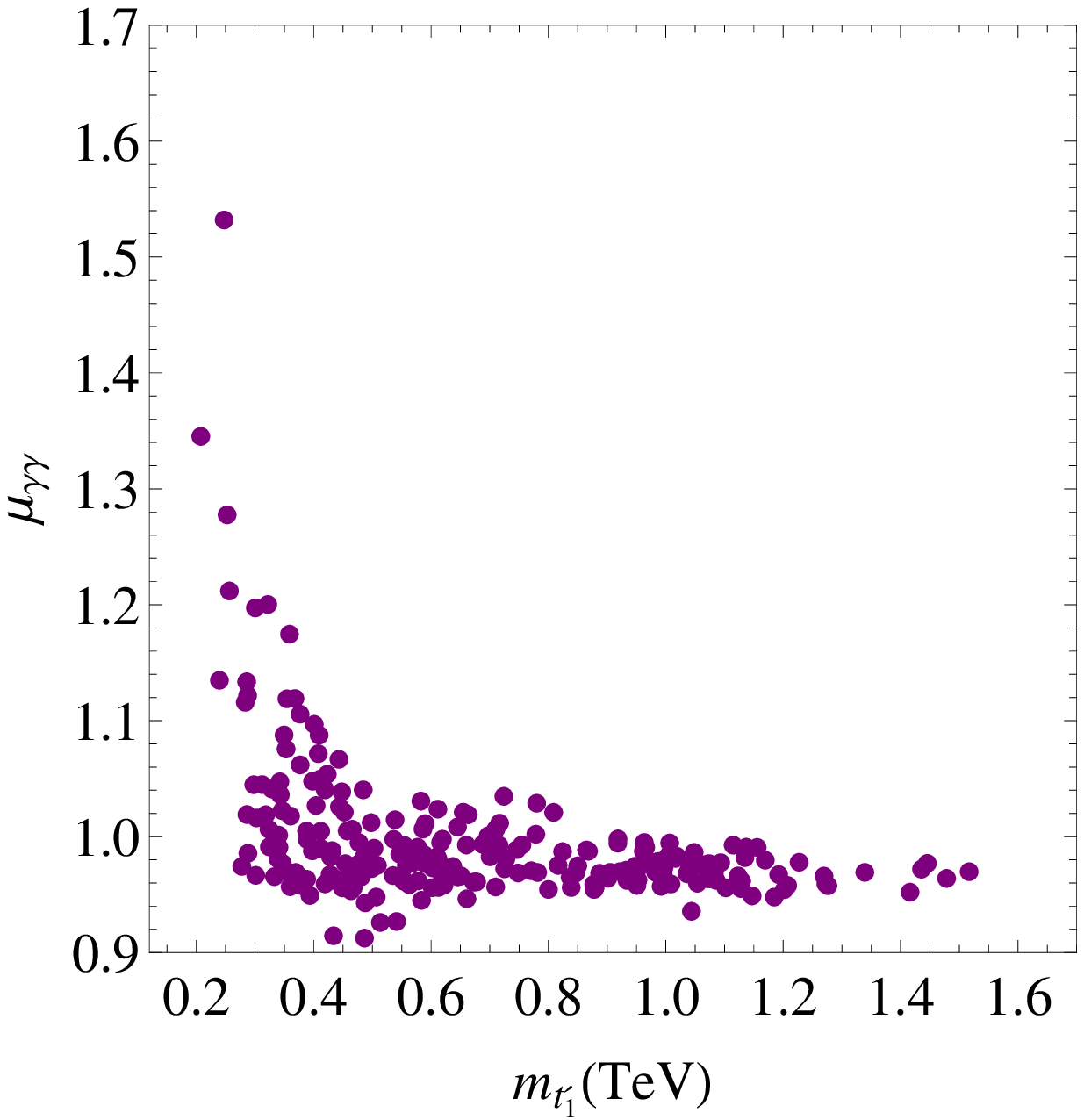, width=.46\textwidth}{(a)}\hfill
\epsfig{file=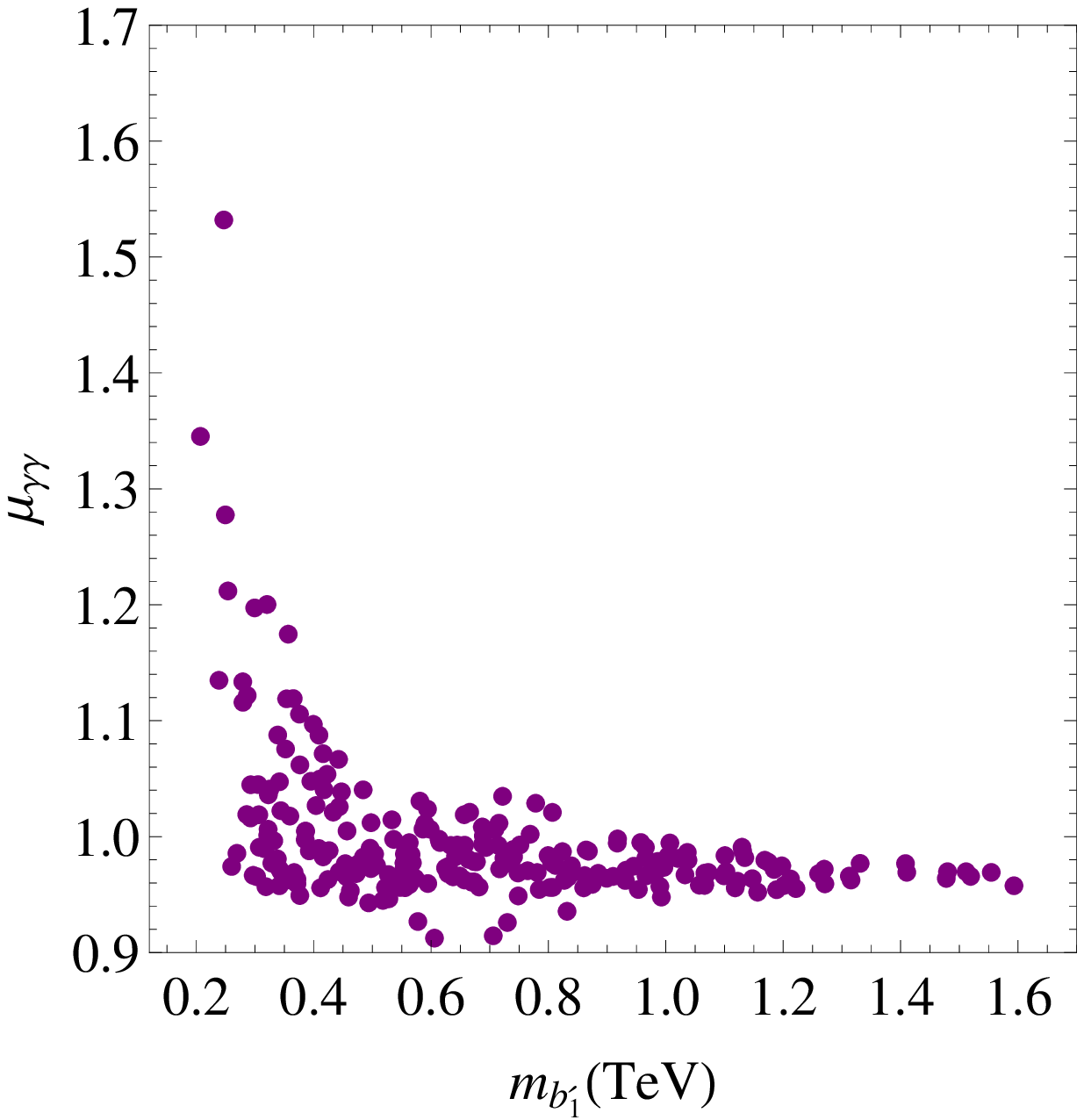, width=.46\textwidth}{(b)}
\caption[$\mu_{\gamma\gamma}$ in function of the lightest $t^\prime$ and $b^\prime$ for $f=$ 1000 GeV and $g_\rho$=2]{Distributions of $\mu_{\gamma\gamma}$ as a function of the lightest $t^\prime$ (a) and $b^\prime$ (b). }
\label{fig:muaa-tp-bp}
\end{figure}

We then combine these three parameters in order to have the results for the various signal strengths and, as an example, we plot the result for $\mu_{\gamma\gamma}$ in Fig.~\ref{fig:muaa-tp-bp}.
We observe that the net effect is again an overall balance pointing towards a value of the signal strength around one with a possible leakage of points towards higher values if smaller masses of the extra fermions are allowed, but still consistent with our recast bound of 500 GeV.

\begin{figure}[!h]
\centering
\epsfig{file=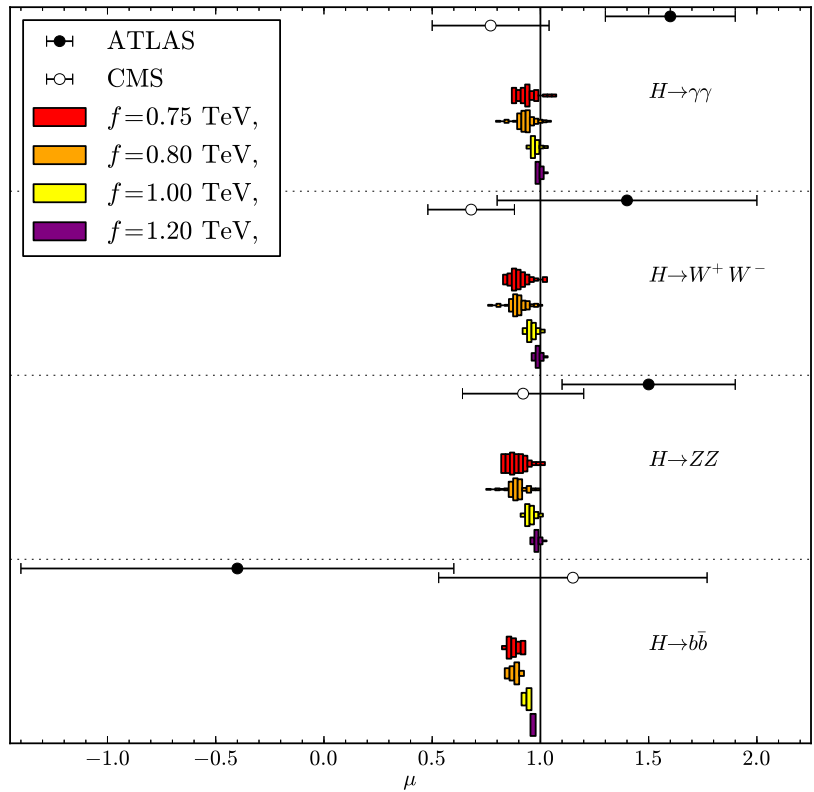, width=.76\textwidth}
\caption[Distributions, as a series of normalized histograms, of the $\mu$s for various choices of $f$ and $g_\rho$]{Distributions of the $\mu$s for the $b \bar b$, $ZZ$, $WW$ and $\gamma\gamma$ final states for the choices of $f$ and $g_\rho$ of eq.(\ref{eq:bench-fvar}).}
\label{fig:mutotal_fg}
\end{figure}

The final results for the analysis of the Higgs sector of the 4DCHM for the four final state chosen,
$b \bar b$, $WW$, $ZZ$ and $\gamma\gamma$, and for the benchmark choices of eq.(\ref{eq:bench-fvar}) are shown in Fig.~\ref{fig:mutotal_fg} where we plot them  as a series of normalized histograms, in order to demonstrate the relative number of points surviving the scan that take a given value of the $\mu$s; the ATLAS and CMS experimental results are represented by a black and white dot respectively, together with their 1$\sigma$ error bars, and we exclude from the presentation of these results the points where the masses of the extra fermions are below 500 GeV.
We observe the tendency of the various $\mu$s to point towards the SM value of $\mu_{ii}$=1 with the model scale $f$ increasing, as expected, together with the overall presence of a smaller number of points for a higher model scale $f$ which is essentially due to the higher fine-tuning of the model which further constraints its parameters.

\begin{figure}[!h]
\centering
\epsfig{file=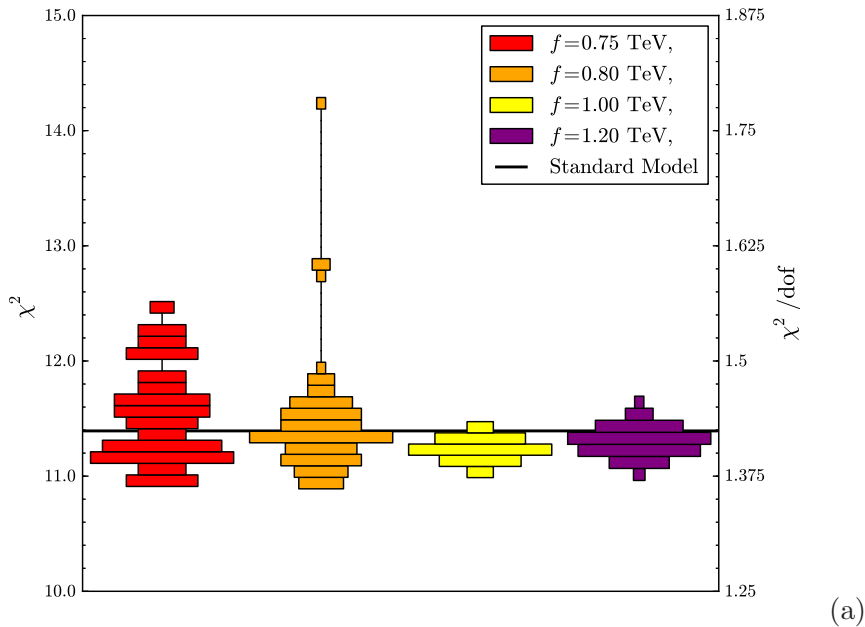, width=.76\textwidth}{(a)}\hfill
\caption[Distributions, as a series of normalized histograms, of the $\chi^2$s for various choices of $f$ and $g_\rho$]{Distributions of the $\chi^2$s for the choices of $f$ and $g_\rho$ of eq.(\ref{eq:bench-fvar}). The SM $\chi^2$ is represented as a black line.}
\label{fig:chi2_fg}
\end{figure}

In order to have a clear and quantitative picture on how the 4DCHM fares against the experimental data, and also in
relation to the SM, we have calculated the $\chi^2$ goodness assuming that all the channels are independent and simply summing the corresponding $\chi^2$s, obtaining then at the end eight degrees of freedom, as
\begin{equation}
\chi^2=\sum_{i=1}^8 \frac{(\mu^{{\textrm{4DCHM}}}_i-\hat \mu_i)}{\delta \hat \mu_i}
\end{equation}
where $\hat \mu_i$ and $\delta \hat \mu_i$ are the measured values and errors reported in Tab.~\ref{tab:higgs-mu-premor2013-postmor2013} (b).
The values of our $\chi^2$ function for each of the parameter scan points are shown in Fig.~\ref{fig:chi2_fg}
where we plot, again as normalized histograms, the results for the 4DCHM  and, as horizontal black line, the SM $\chi^2$ value (computed then with $\mu_{ii}=1$), from which we observe that for most of the points considered the fit of the 4DCHM to the 8 TeV LHC data is as good as, if not better than, that of the SM thus making clear the compatibility of our framework with the results arising from the CERN machine.

\begin{figure}[!h]
\centering
\epsfig{file=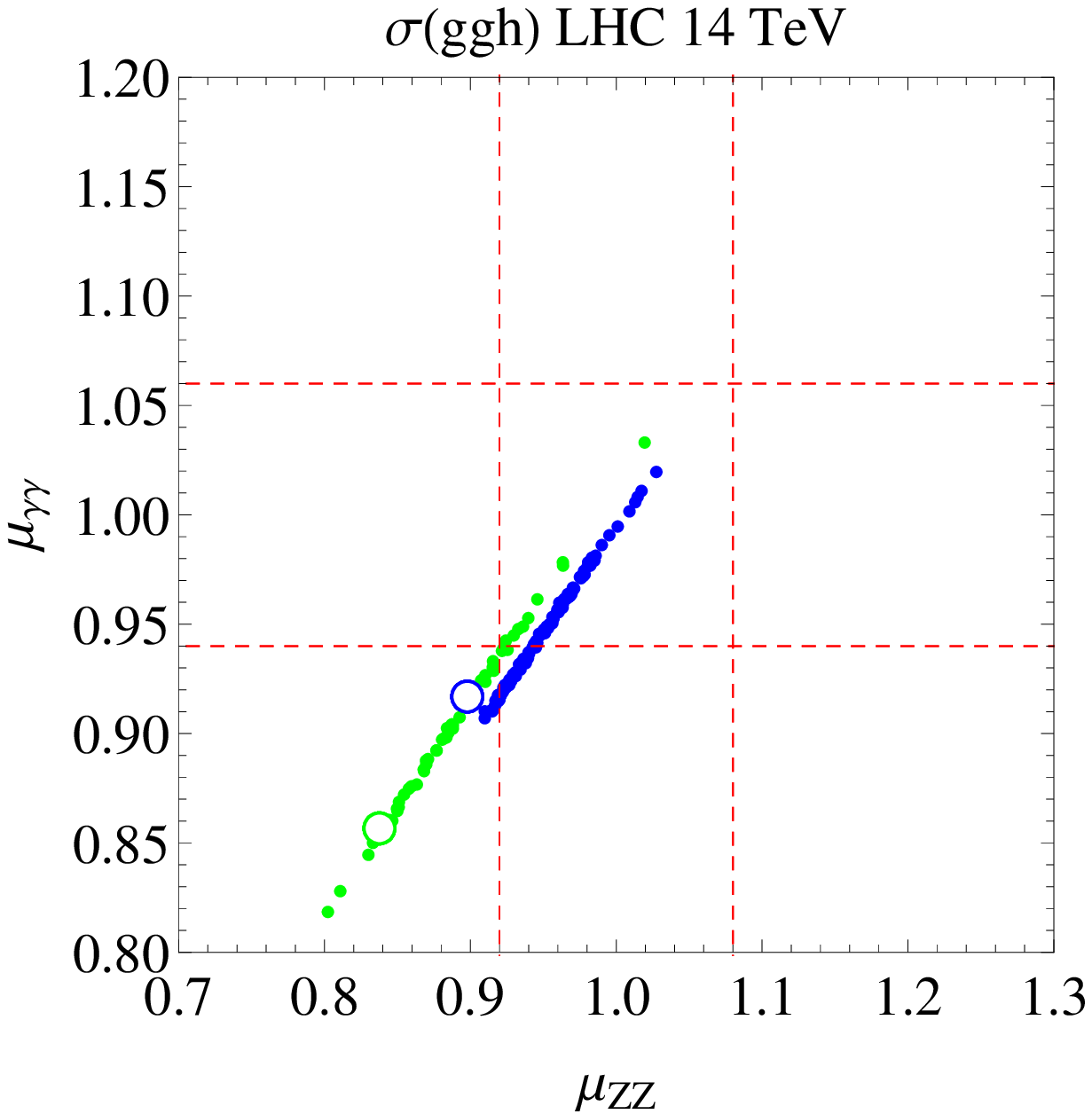, width=.46\textwidth}{(a)}\hfill
\epsfig{file=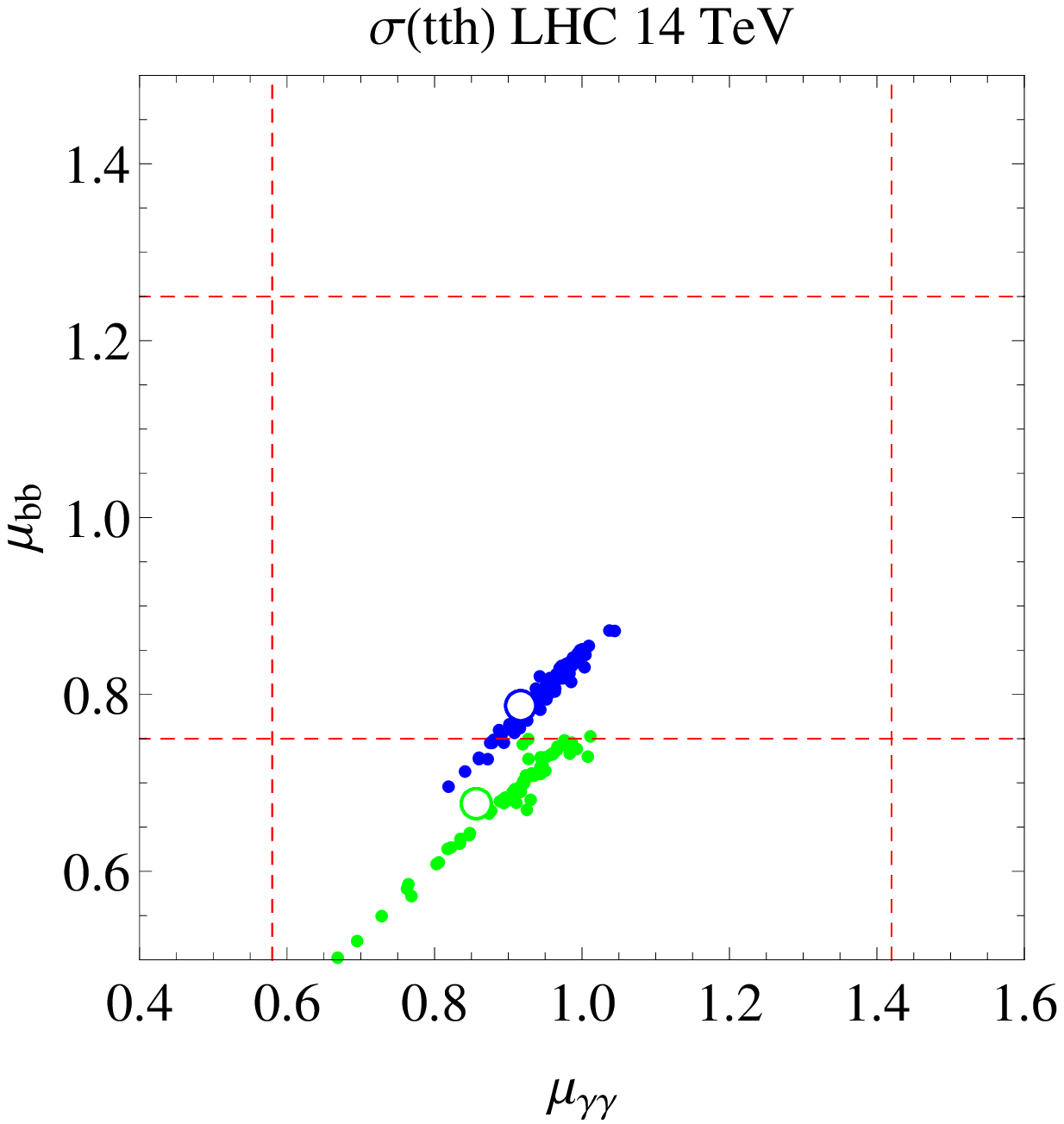, width=.442\textwidth}{(b)}
\caption[Correlations between $\mu$ parameters at the 14 TeV LHC]{Correlations between the relevant $\mu$ parameters at the 14 TeV LHC with 300 fb$^{-1}$ of integrated luminosity for $ggh$ (a) and $tth$ (b) production modes for $f$=800 GeV and $g_\rho$=2.5 (green points) and $f$=1000 GeV and $g_\rho$=2 (blue points).
The circles represent the signal strengths predictions in the limit where all the extra particle in the spectrum are decoupled.
The area between the dashed lines represents the experimental precision limits given in Tab.~\ref{tab:mu-err-lhc14}.
}
\label{fig:mu-lhc14}
\end{figure}

Before closing this Section we want to illustrate, although limited to just an example, what the capabilities of the next stage of the LHC could be in 
disentangling the 4DCHM from the SM by means of the measurements of the signal strengths, comparing our predictions with the expected 14 TeV LHC experimental accuracies\footnote{These accuracies, and the ones we will present in the next Section, are based on statistical errors. Even though systematic errors can degrade these predictions, especially those claiming below 1\% accuracies, we will rely on these numbers for our analysis.
} that can be found in \cite{Peskin:2012we,Baer:2013cma}, and that are given for an integrated luminosity of 300 fb$^{-1}$.
We do this for two different Higgs production processes, gluon-gluon-fusion and associated top production, and for different final states associated to these production mechanisms, for which our predicted values for the $\mu$s parameters lie outside the LHC expected experimental errors reported in Tab.~\ref{tab:mu-err-lhc14}. 
The results for two of the benchmark points of eq.(\ref{eq:bench-fvar}), $f$=800 GeV and $g_\rho$=2.5 and $f$=1000 GeV and $g_\rho$=2 are shown in Fig.\ref{fig:mu-lhc14} where we observe that in certain regions of the parameter space the 4DCHM predictions lie outside of the expected experimental accuracies, while
the deviations of the Higgs signal strengths with respect to the SM for the other possible combinations of production and decay modes are inaccessible even at this stage of the CERN machine.

\begin{table}[!h]
\begin{center}
\begin{tabular}{|l|l|}
\hline
                     & $ggh$ \\
\hline
\hline
$\mu_{\gamma\gamma}$ & 0.06  \\
$\mu_{ZZ}$           & 0.08  \\
\hline
\end{tabular}{(a)}
\hspace{2cm}
\begin{tabular}{|l|l|}
\hline
                     & $tth$ \\
\hline
\hline
$\mu_{\gamma\gamma}$ & 0.42 \\
$\mu_{b\bar b}$      & 0.25 \\
\hline
\end{tabular}{(b)}
\end{center}
\caption[Expected accuracies on the $\mu$ parameters at the 14 TeV LHC ]{Expected accuracies for the determination of the $\mu$ parameters at the 14 TeV LHC with $300$ fb$^{-1}$ of integrated luminosity for $ggh$ (a) and $tth$ (b) production mechanisms as reported in~\cite{Baer:2013cma}.}
\label{tab:mu-err-lhc14}
\end{table}

\section{Future collider analysis}

Even without the evidence of new physics at the next stage of the LHC the physics community needs to make an important and critical decision in the near future regarding the next generations of colliders.
In fact the precision measurements of the Higgs properties, for which the LHC has limited capabilities due to its hadron collider nature, and the understanding of its nature are essentials steps to shedding light on the mechanism of EWSB.
In this respect many types of colliders are at the moment in their 
design phase, among which we count both hadron collider, workshops studying a 100 TeV proton proton collider have in fact recently started, and also $e^+e^-$ colliders, as CLIC \cite{Aicheler:2012bya}, ILC \cite{Behnke:2013xla} and TLEP \cite{Gomez-Ceballos:2013zzn}, the first two being linear colliders and the latter a circular one.
For many aspects it is clear that the advantage of an electron-positron collider with respect to an hadron collider are multiple. The cleanliness of the environment, the precision of the measurements and the large number of Higgs bosons that could be produced therefore make an analysis of the prospects of these 
types of machines in disentangling the nature of the Higgs of primary importance nowadays and in this respect we will discuss in this Section the potentiality of a proposed $e^+ e^-$ collider in testing the 4DCHM as a framework for the composite Higgs hypothesis. Recent literature has also studied the Higgs state as a pNGB in the context of these future colliders (see for example \cite{Contino:2013gna} where the authors focus on specific production processes showing the reach of a future lepton collider onto the parameter $\xi$ ) but once again our goal will be to analyse the properties of a model in which the entire particle 
spectrum is taken into account also comparing them with the asymptotic limit, as done 
for the LHC analysis.

As is known a feature specific to $e^+e^-$ colliders is the presence of initial state radiation and Beamstrahlung and for our numerical computation with CalcHEP both of these aspects have been taken into account. For the former CalcHEP implements the Jadach, Skrzypek and Ward expressions of \cite{Jadach:1988gb,Skrzypek:1990qs} while for the latter we adopt the parametrisation specific for the ILC project \cite{Behnke:2013xla} that is
\begin{itemize}
\item[-] Beam size (x+y) = 645.7 nm
\item[-] Bunch length = 300 $\mu$m
\item[-] Bunch population = 2$\cdot10^{10}$
\end{itemize}
While fine details of the emerging electron and positron spectra may be different for other colliders, we can confirm that the main features of our phenomenological analysis are captured by our choice.

\begin{figure}[!h]
\centering
\epsfig{file=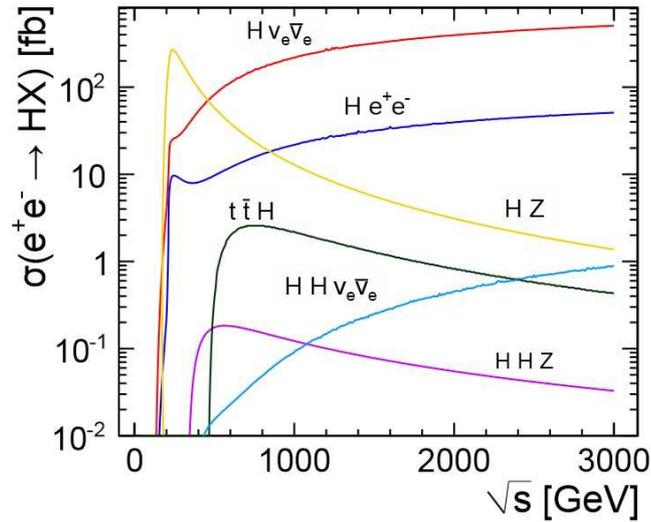, width=.68\textwidth}
\caption[Higgs boson production cross sections at the ILC]{125 GeV Higgs boson production cross sections at an $e^+e^-$ collider in function of the CM energy.}
\label{fig:ILC_prod_500}
\end{figure}

With the choice of the ILC prototype we will choose for our analysis three standard values of its CM energy for which this machine could run, that are 250, 500 and 1000 GeV.
The relevant Higgs production modes at the ILC can be seen from Fig.~\ref{fig:ILC_prod_500} where the production cross sections for a 125 GeV Higgs boson in an $e^+ e^-$ collider in function of $\sqrt{s}$ are shown.
For $\sqrt{s}$= 250 GeV the machine will be at the peak of the reaction $e^+e^-\to Zh$
with the Higgs-strahlung production mode, Fig.~\ref{fig:ilc-vh-vbf} (left), dominating the VBF production mode, Fig.~\ref{fig:ilc-vh-vbf} (right). This stage of the machine will allow a precise measurement of the Higgs mass and quantum numbers, the former being made particularly clean in the process $e^+e^-\to Zh$ with $Z\to e^+ e^-/\mu^+\mu^-$ from which a measurement of the invariant mass recoiling against the reconstructed $Z$ can provide a precise value of $m_H$.
At the energy of $\sqrt{s}$=500 GeV two important processes will become accessible, $e^+e^-\to t\bar t h$ and $e^+ e^-\to Z h h$, the former containing the 
top Yukawa coupling and the latter the triple Higgs self-coupling at the tree-level. Finally, the energy stage of $\sqrt{s}$=1000 GeV, for which the dominant Higgs production mode is VBF mediated by off-shell charged gauge bosons, will allow a great number of measurements sensitive to the Higgs boson 
coupling to the top quark and the Higgs boson self coupling, that could be relevant as probes of a strongly interacting or composite Higgs, as well as being an important CM energy for the search of new exotic particles.
For these three energy stages of the ILC we will choose in the following as an integrated luminosity the values of 250,500 and 1000 fb$^{-1}$ respectively, following the values reported in \cite{Baer:2013cma}.
We will present once again our results comparing our simulated data, presented in terms of Higgs boson signal strengths as defined in eq.(\ref{eq:mupar_exp}), with the experimental predicted accuracies that are given in \cite{Peskin:2012we,Asner:2013psa,Baer:2013cma}.

\subsection{Results}

\begin{figure}[!h]
\centering
\epsfig{file=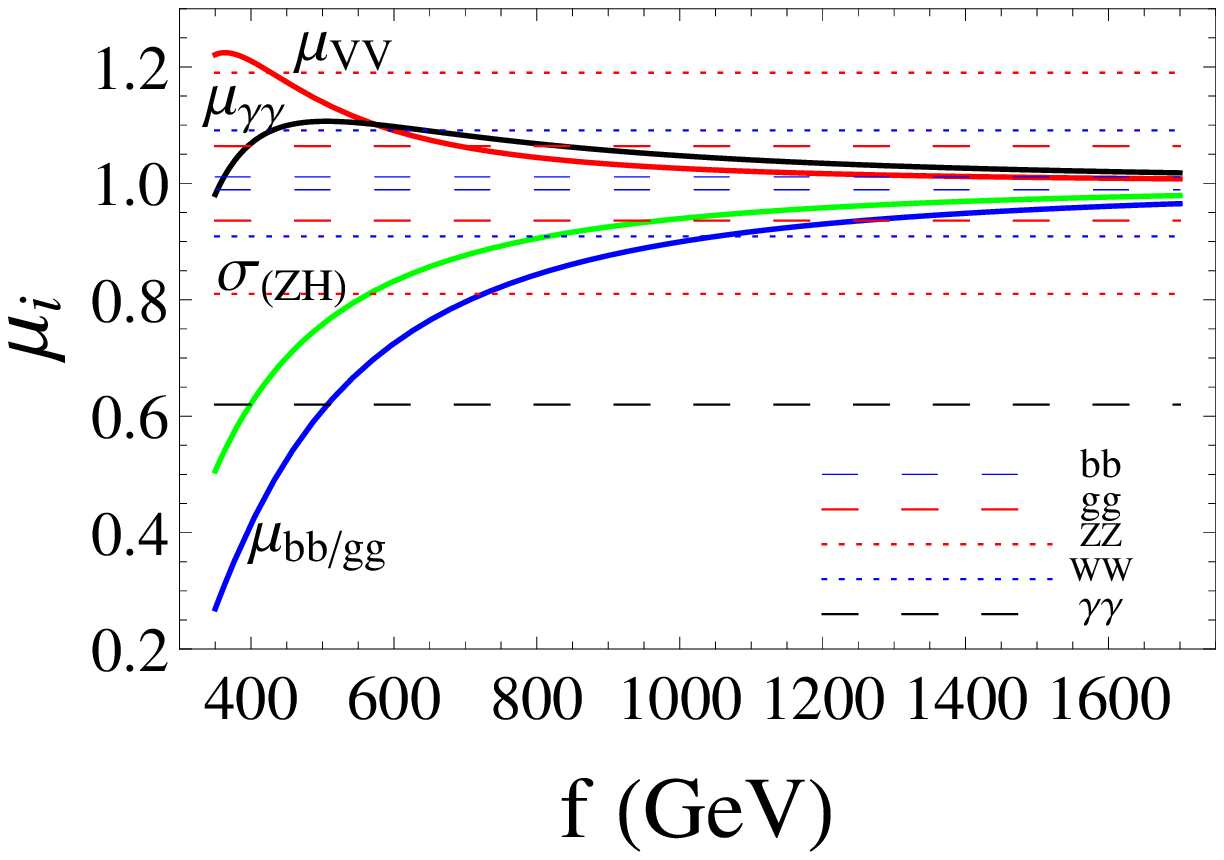, width=.44\textwidth}{(a)}\hfill
\epsfig{file=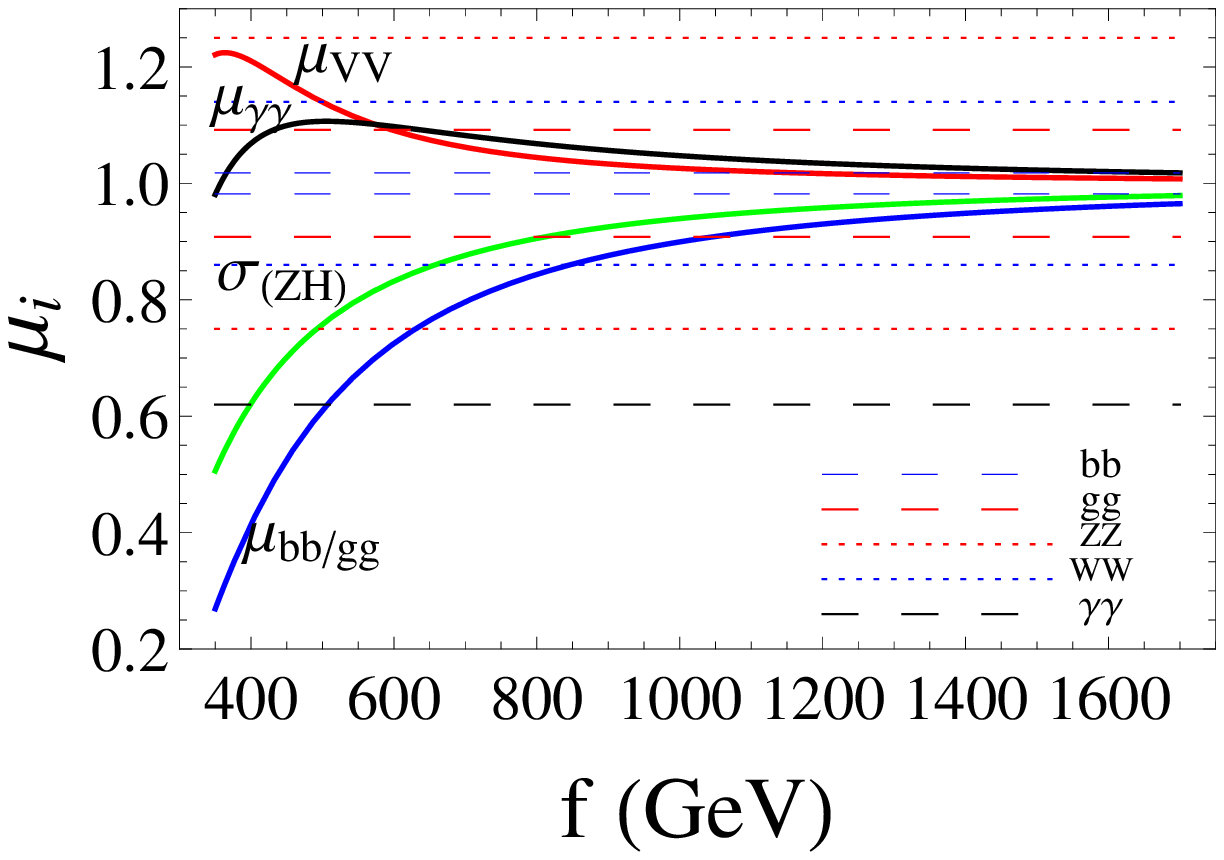, width=.44\textwidth}{(b)}
\caption[Higgs signal strengths at the ILC in the decoupling limit as function of the model scale $f$]{Higgs signal strengths at the ILC in the decoupling limit as function of the model scale $f$ for Higgs-strahlung production process at $\sqrt{s}$=250 GeV (a) and 500 GeV (b).
Red, black and blue lines correspond to $WW/ZZ$, $\gamma\gamma$ and $b\bar b/gg$ final state respectively, while the green line corresponds to the ratio of the inclusive production cross section. Horizontal dashed lines are the expected accuracies for the channels under consideration as reported in Tab.~\ref{tab:mu-err-ilc-stra}.}
\label{fig:ilc-mu-decoup}
\end{figure}

As for the analysis of the Higgs sector of the 4DCHM at the LHC, we will first present our results in the decoupling limit, that is assuming that all the extra particles in the spectrum are heavy enough so that the only modifications to the Higgs couplings are the ones given by eq.(\ref{eq:leef-vvh-vvhh-mod}) and eq.(\ref{eq:leef-ffh-mod}), and for which we can again compute the Higgs signal strengths just as a function of the model scale $f$.
We present these results just for the case of Higgs-strahlung production process at $\sqrt{s}=250,500$ GeV in Fig.~\ref{fig:ilc-mu-decoup} in which we also show, with horizontal dashed lines, the expected experimental accuracies for the channels under consideration, as reported in Tab.~\ref{tab:mu-err-ilc-stra}.
We also present the ratio of the inclusive cross section with respect to the SM prediction since this is a quantity that at an electron-positron collider is measurable via the invariant mass recoiling against the reconstructed $Z$ boson.
From the figures we observe that, limited to certain final states, there exists a sensitivity to the compositness scale $f$ up to $\sim$ 1200 GeV for the Higgs-strahlung production processes.

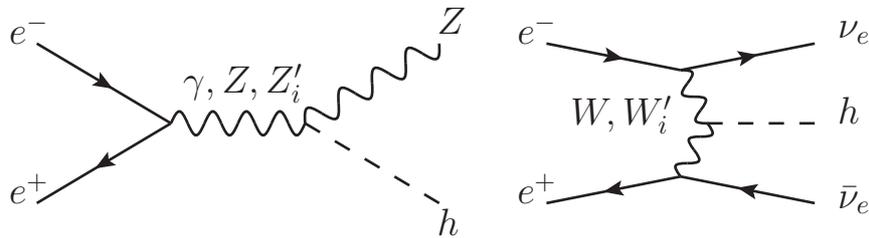
\begin{figure}[!h]
  \begin{picture}(292,118) (35,-47)
    \SetWidth{1.0}
    \SetColor{Black}
    \Text(63,54)[lb]{\Large{\Black{$e^-$}}}      
    \Line[arrow,arrowpos=0.5,arrowlength=5,arrowwidth=2,arrowinset=0.2](73,54)(123,24)
    \Text(63,-6)[lb]{\Large{\Black{$e^+$}}}      
    \Line[arrow,arrowpos=0.5,arrowlength=5,arrowwidth=2,arrowinset=0.2](123,24)(73,-6)
    \Text(128,32)[lb]{\Large{\Black{$\gamma,Z,Z^\prime_i$}}}      
    \Photon(123,24)(173,24){4.5}{4}
    \Text(223,58)[lb]{\Large{\Black{$Z$}}} 
    \Photon(173,24)(223,54){4.5}{4}
    \Text(223,-18)[lb]{\Large{\Black{$h$}}}      
    \Line[dash,dashsize=6](173,24)(223,-6)
    \Text(253,54)[lb]{\Large{\Black{$e^-$}}}     
    \Line[arrow,arrowpos=0.5,arrowlength=5,arrowwidth=2,arrowinset=0.2](263,54)(313,44)
    \Text(253,-6)[lb]{\Large{\Black{$e^+$}}}     
    \Line[arrow,arrowpos=0.5,arrowlength=5,arrowwidth=2,arrowinset=0.2](313,4)(263,-6)
    \Text(273,20)[lb]{\Large{\Black{$W,W^\prime_i$}}}     
    \Photon(313,44)(323,24){3.5}{2}
    \Photon(313,4)(323,24){3.5}{2}
    \Text(373,54)[lb]{\Large{\Black{$\nu_e$}}}       
    \Line[arrow,arrowpos=0.5,arrowlength=5,arrowwidth=2,arrowinset=0.2](313,44)(363,54)
    \Text(373,-10)[lb]{\Large{\Black{$\bar\nu_e$}}}       
    \Line[arrow,arrowpos=0.5,arrowlength=5,arrowwidth=2,arrowinset=0.2](363,-6)(313,4)
    \Text(373,24)[lb]{\Large{\Black{$h$}}}      
    \Line[dash,dashsize=6](323,24)(363,24)     
  \end{picture}
  \caption[Feynman diagrams for Higgs-strahlung and VBF Higgs production processes at an $e^+e^-$ collider]{Feynman diagrams for Higgs-strahlung (left) and VBF (right) Higgs production processes at an $e^+e^-$ collider.}
  \label{fig:ilc-vh-vbf}
\end{figure}

\begin{table}[!h]
\begin{center}
\begin{tabular}{|l|l|l|}
\hline
HS                   & 250 GeV & 500 GeV \\
\hline
\hline
$\sigma(ZH)$         & 0.025   & \\
$\mu_{b \bar b}$     & 0.011   & 0.018 \\
$\mu_{WW}$           & 0.064   & 0.092 \\
$\mu_{ZZ}$           & 0.19    & 0.25 \\
$\mu_{\gamma\gamma}$ & 0.38    & 0.38 \\
$\mu_{gg}$           & 0.091   & 0.14 \\
\hline
\end{tabular}{(a)}
\hfill
\begin{tabular}{|l|l|l|l|}
\hline
VBF                   & 250 GeV & 500 GeV & 1000 GeV\\
\hline
\hline
$\mu_{b \bar b}$     & 0.105   & 0.0066   & 0.0047  \\
$\mu_{WW}$           &         & 0.026    & 0.033   \\
$\mu_{ZZ}$           &         & 0.082    & 0.044   \\
$\mu_{\gamma\gamma}$ &         & 0.26     & 0.10    \\
$\mu_{gg}$           &         & 0.041    & 0.031   \\
\hline
\end{tabular}{(b)}
\end{center}
\caption[Expected accuracies on the $\mu$ parameters at the ILC for Higgs-strahlung and VBF production processes]{Expected accuracies for the determination of the $\mu$ parameters at the ILC for Higgs-strahlung (a) and VBF (b) production processes as reported in \cite{Baer:2013cma}. Blank entries correspond to values not provided by the ILC design report.}
\label{tab:mu-err-ilc-stra}
\end{table}

As done for the LHC study we now move onto the analysis of our concrete CHM realisation including all the other effects that can produce modifications to the physical observables, such as extra modifications of the couplings and also extra particles that can be exchanged in the processes of Fig.~\ref{fig:ilc-vh-vbf}.

\begin{figure}[!h]
\centering
\epsfig{file=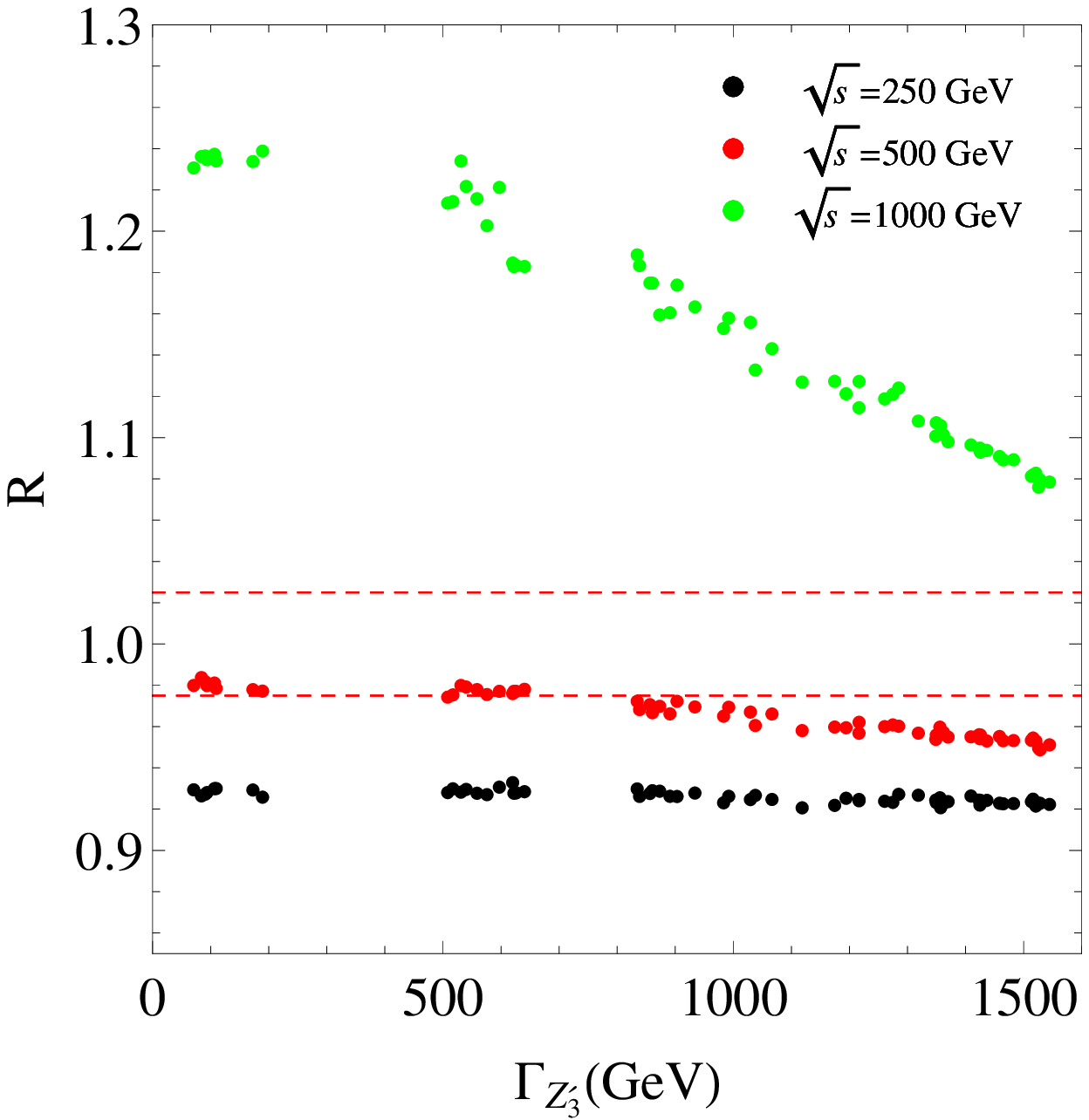, width=.48\textwidth}\hfill
\epsfig{file=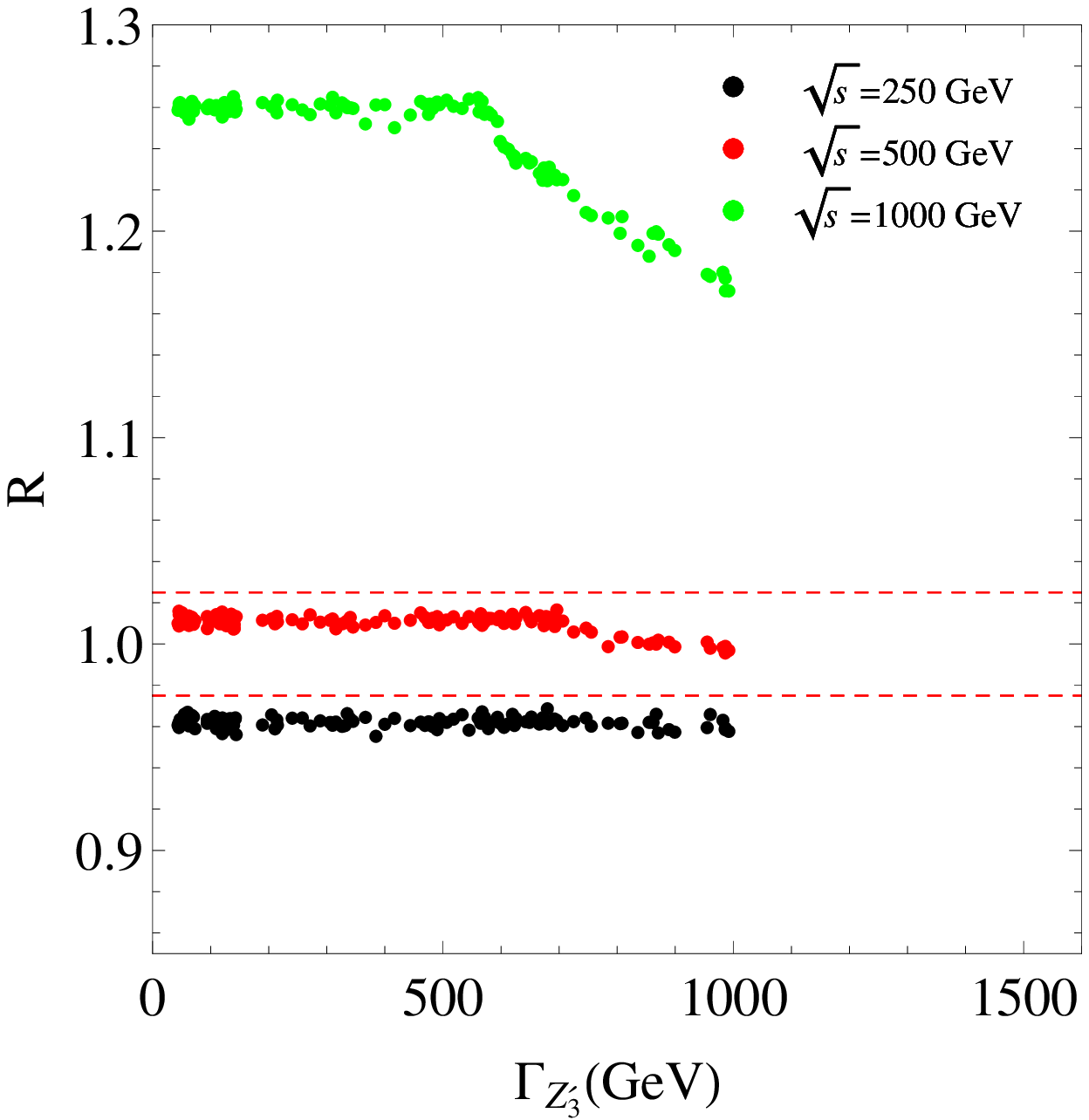, width=.48\textwidth}\\
\vskip 15pt
\epsfig{file=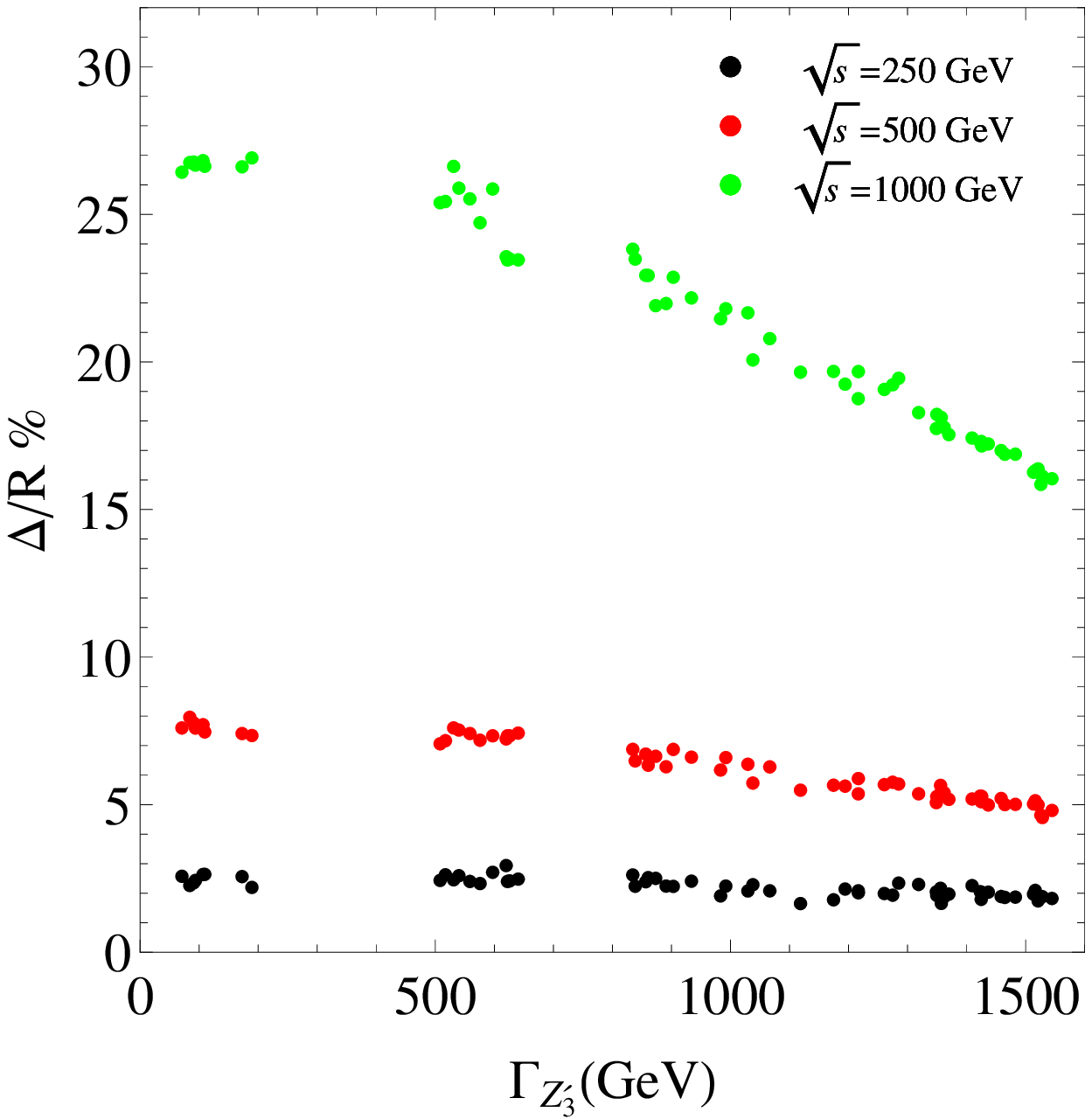, width=.48\textwidth}\hfill
\epsfig{file=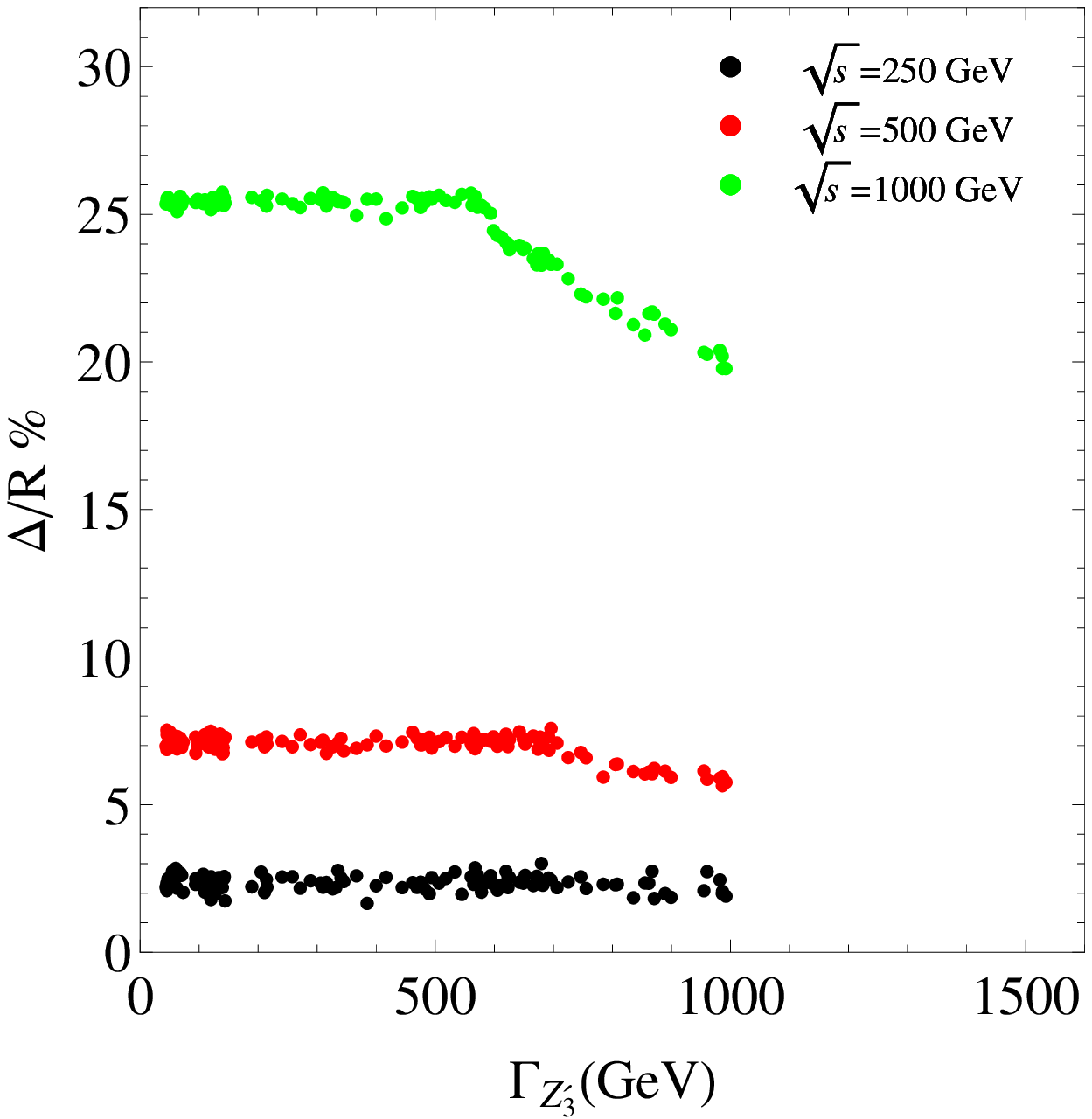, width=.48\textwidth}
\caption[$R$ and $\Delta/R$ quantities in function of the width of $Z^\prime_3$ for $f$=800 GeV, $g_\rho=2.5$ and $f$=1000 GeV, $g_\rho=2$]{$R$ and $\Delta/R$ quantities of eq.(\ref{eq:ilc-R-delta}) in function of the width of $Z^\prime_3$ for $f$=800 GeV, $g_\rho=2.5$ (left) and $f$=1000 GeV, $g_\rho=2$ (right) for 
three values of the CM energy, 250,500 and 1000 GeV. The horizontal red dashed line represents the expected experimental precision in determine the inclusive strahlung cross section for $\sqrt{s}$=250~GeV.}
\label{fig:ilc-R-delta}
\end{figure}

As a first step we introduce the following two parameters, $R$ and $\Delta$, for the inclusive Higgs-strahlung production cross section, $\sigma(Zh)$, as follows
\begin{equation}
R=\frac{\sigma(Zh)_{{\textrm{4DCHM}}}}{\sigma(Zh)_{{\textrm{SM}}}}, \quad \Delta=R-\kappa^2_{hZZ}, \quad \kappa_{hZZ}=\frac{g^{{\textrm{4DCHM}}}_{hZZ}}{g^{{\textrm{SM}}}_{hZZ}},
\label{eq:ilc-R-delta}
\end{equation}
where $\Delta$ then gives  a measure of the direct impact of the extra resonances intervening in this production processes and for which we have checked, by numerical computation,  that if the new classes of gauge bosons are completely stripped off the calculation, then $R\to\kappa^2_{hZZ}$ that is $\Delta\to 0$.
However, when we include the $Z^\prime$s in our computation, also for $\sqrt{s}$ smaller than the new physics scale $f$, where the new gauge boson 
spectrum lives, the Higgs-strahlung cross section is always affected by propagator effects (see also \cite{Conley:2005et} and \cite{Hartling:2012ss,Contino:2011np} for related studies in the littlest Higgs and CHM respectively).
This is well illustrated in Fig.~\ref{fig:ilc-R-delta} where we plot $R$ and $\Delta/R$ in function of the width of $Z^\prime_3$ for two of the benchmark points of eq.(\ref{eq:bench-fvar}) for which we have scanned over the parameters of the fermionic sector as described in Chapter~\ref{chap-2} and for which the 
rescaling factors are respectively $R\simeq0.91$ for $f$=800 GeV and $R$=0.94 for $f$=1000 GeV.
The slopes present in the plots make it clear that we are in presence of width dependent propagator effects and the deviations from the SM limit span from $\sim$ 2\% for $\sqrt{s}$=250 GeV up to 25\% for $\sqrt{s}$=1000 GeV.
It is interesting to notice that, contrary to the case where the extra gauge bosons are completely decoupled and so that, following eq.(\ref{eq:leef-vvh-vvhh-mod}),  it is always $R<$1, in case of the inclusion of the extra gauge boson the total result can be an enhancement of the SM cross section already for $
\sqrt{s}$=500 GeV, that is well below the mass of the extra gauge bosons. We have also analysed  the case of VBF were we have established that such an interference effect is taking place but with a contribution which is one order of magnitude lower than in the Higgs-strahlung case.

\begin{figure}[!h]
\centering
\epsfig{file=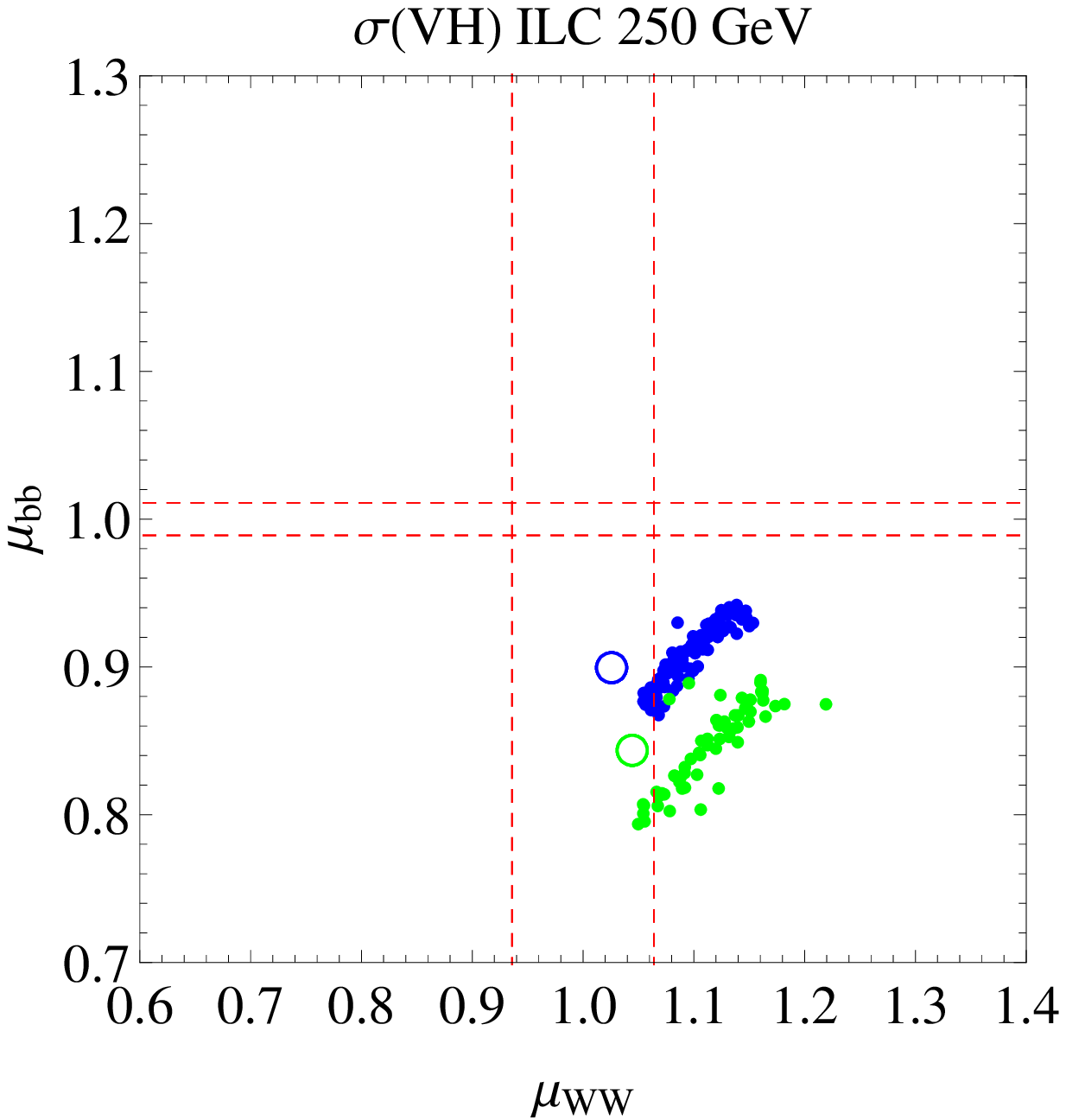, width=.48\textwidth}\hfill
\epsfig{file=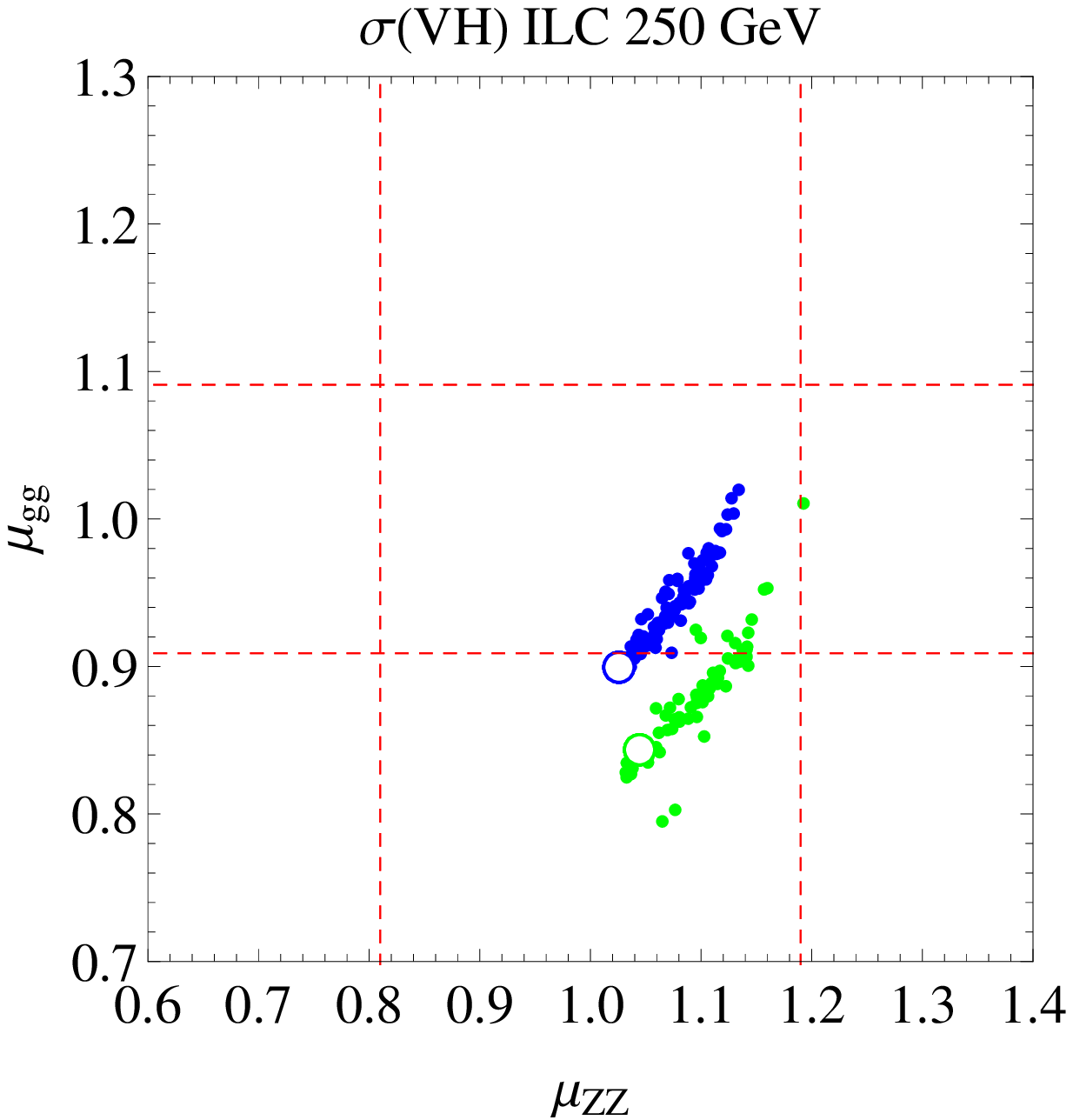, width=.48\textwidth}\\
\epsfig{file=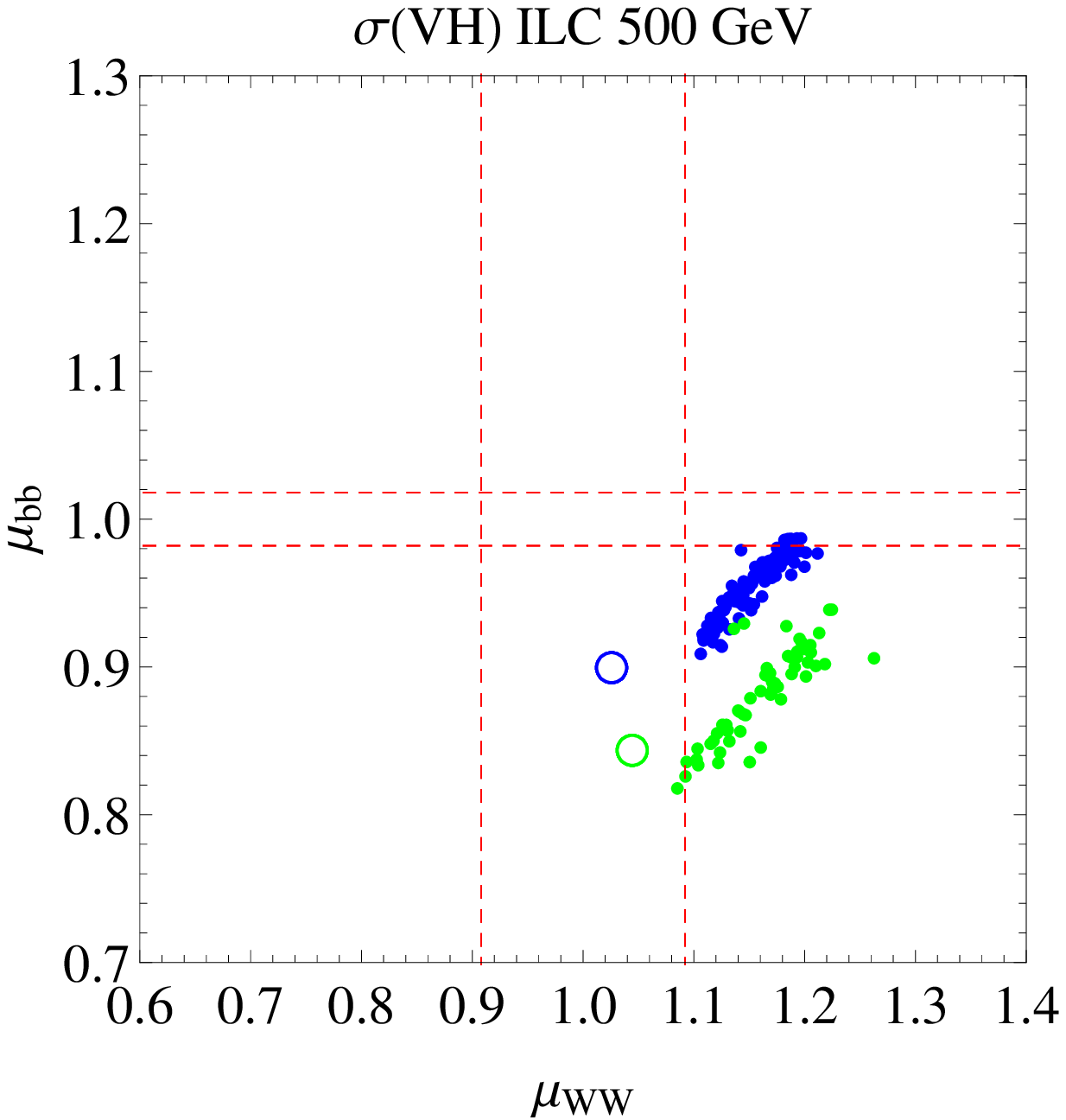, width=.48\textwidth}\hfill
\epsfig{file=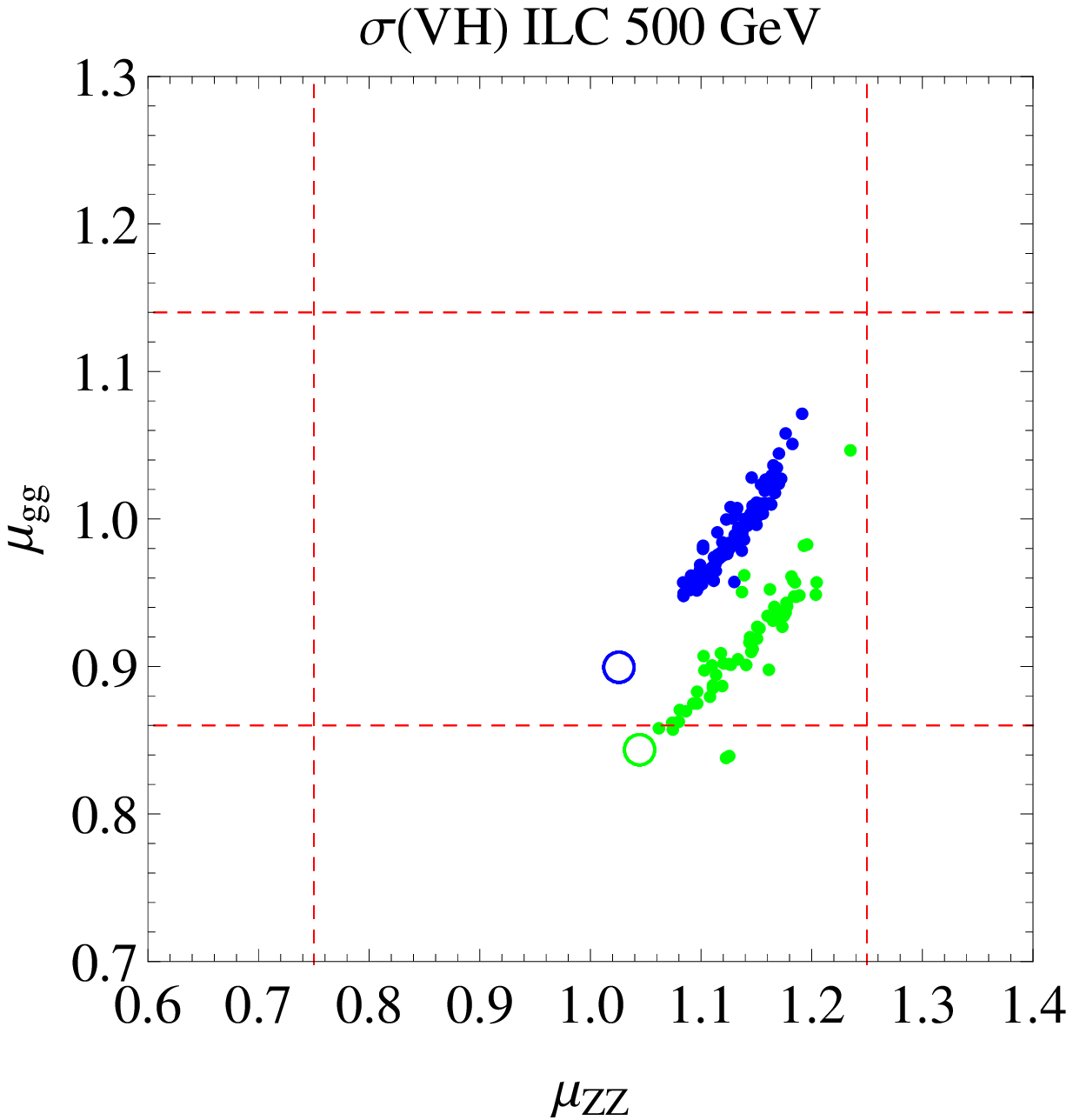, width=.48\textwidth}
\caption[Correlations between $\mu$ parameters at the ILC for $\sqrt{s}$=250,500 GeV]{Correlations between the relevant $\mu$ parameters at the 250 (upper row) and 500 (lower row) GeV ILC with 250 and 500 fb$^{-1}$ of integrated luminosity respectively for $f$=800 GeV and $g_\rho$=2.5 (green points) and $f$=1000 GeV and $g_\rho$=2 (blue points). The circles represent the signal strengths predictions in the limit where all the extra particle in the spectrum are decoupled.
The area between the dashed lines represents the experimental precision limits given in Tab.~\ref{tab:mu-err-ilc-stra}. }
\label{fig:ilc-vh}
\end{figure}

\begin{figure}[!h]
\centering
\epsfig{file=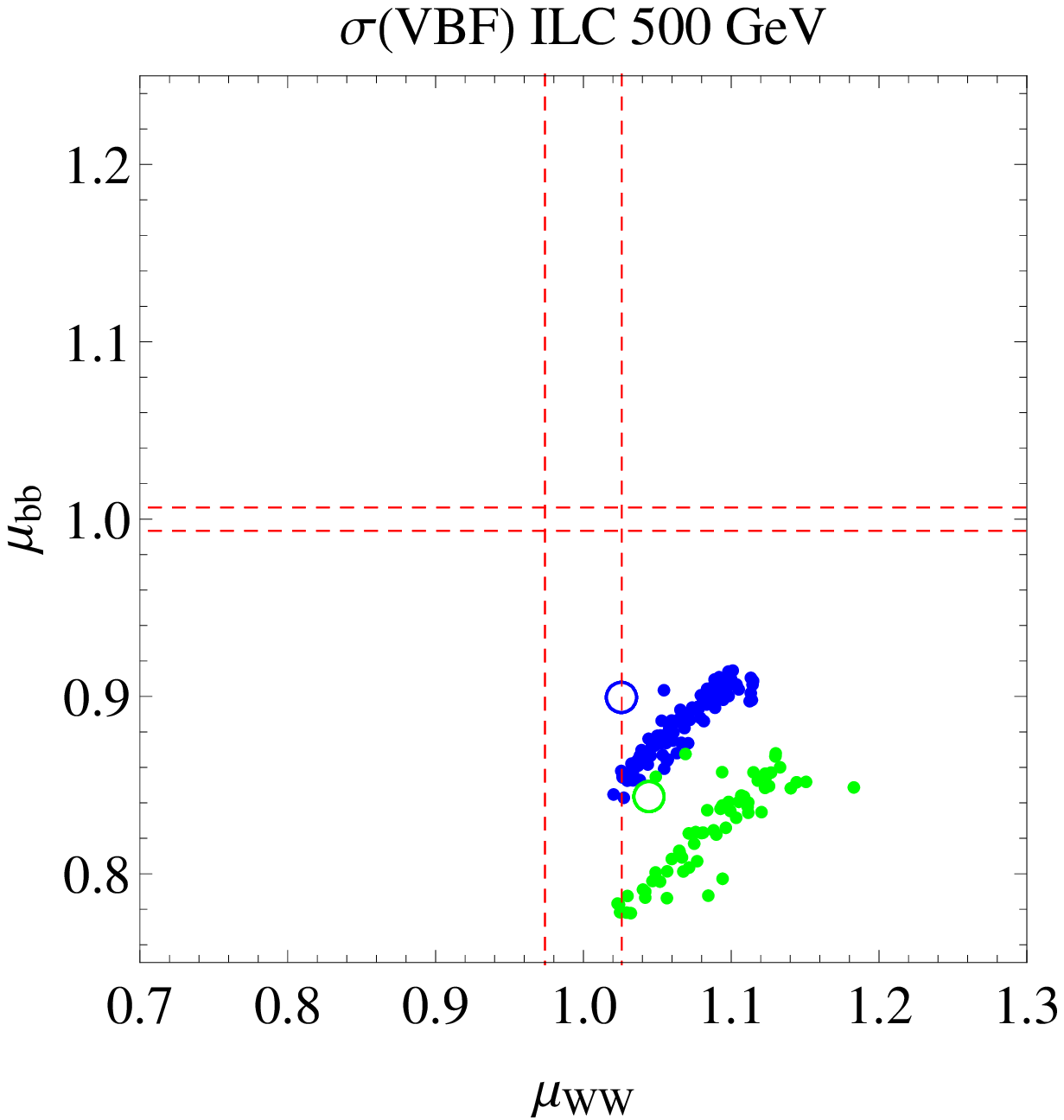, width=.48\textwidth}\hfill
\epsfig{file=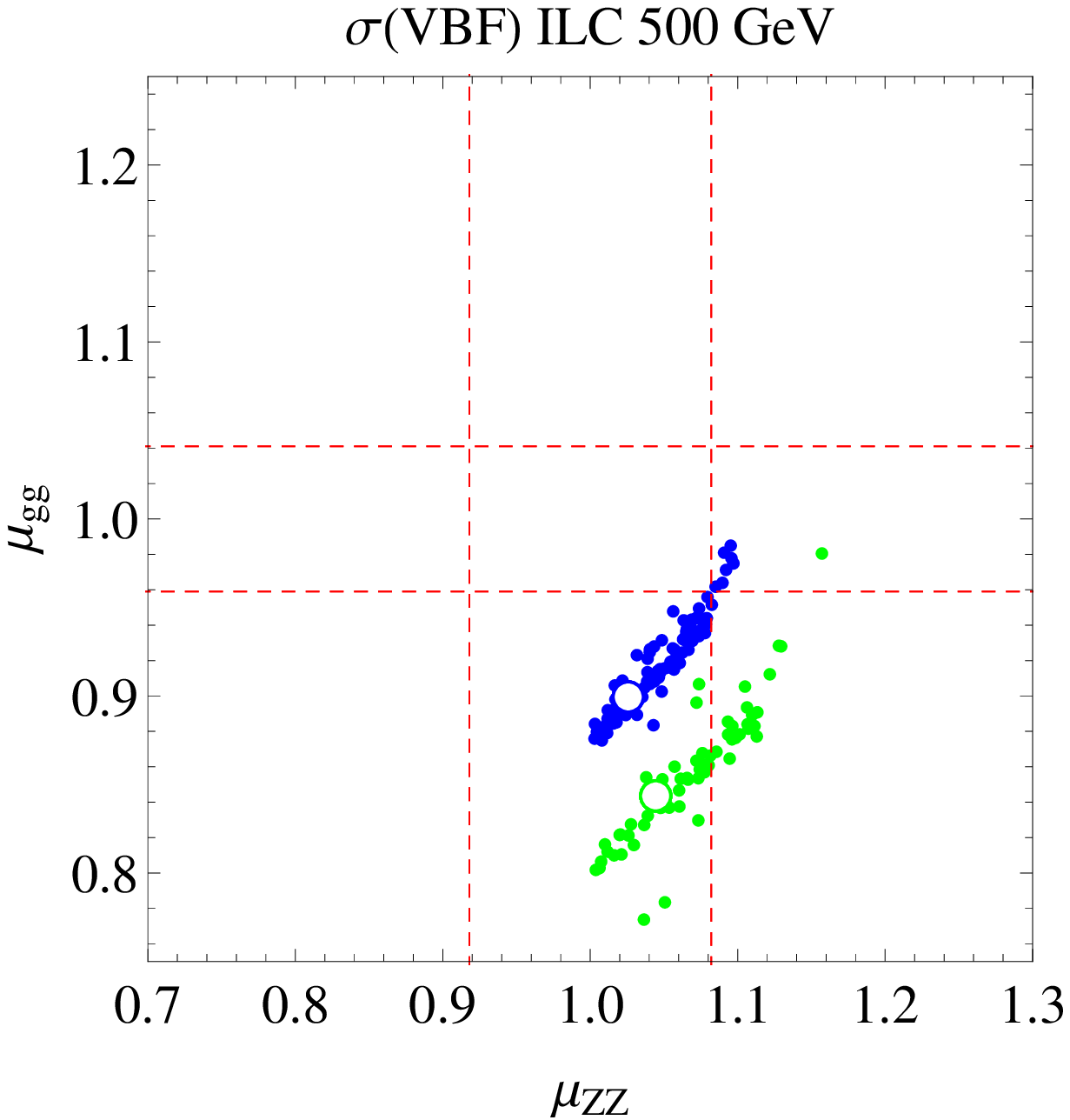, width=.48\textwidth}\\
\epsfig{file=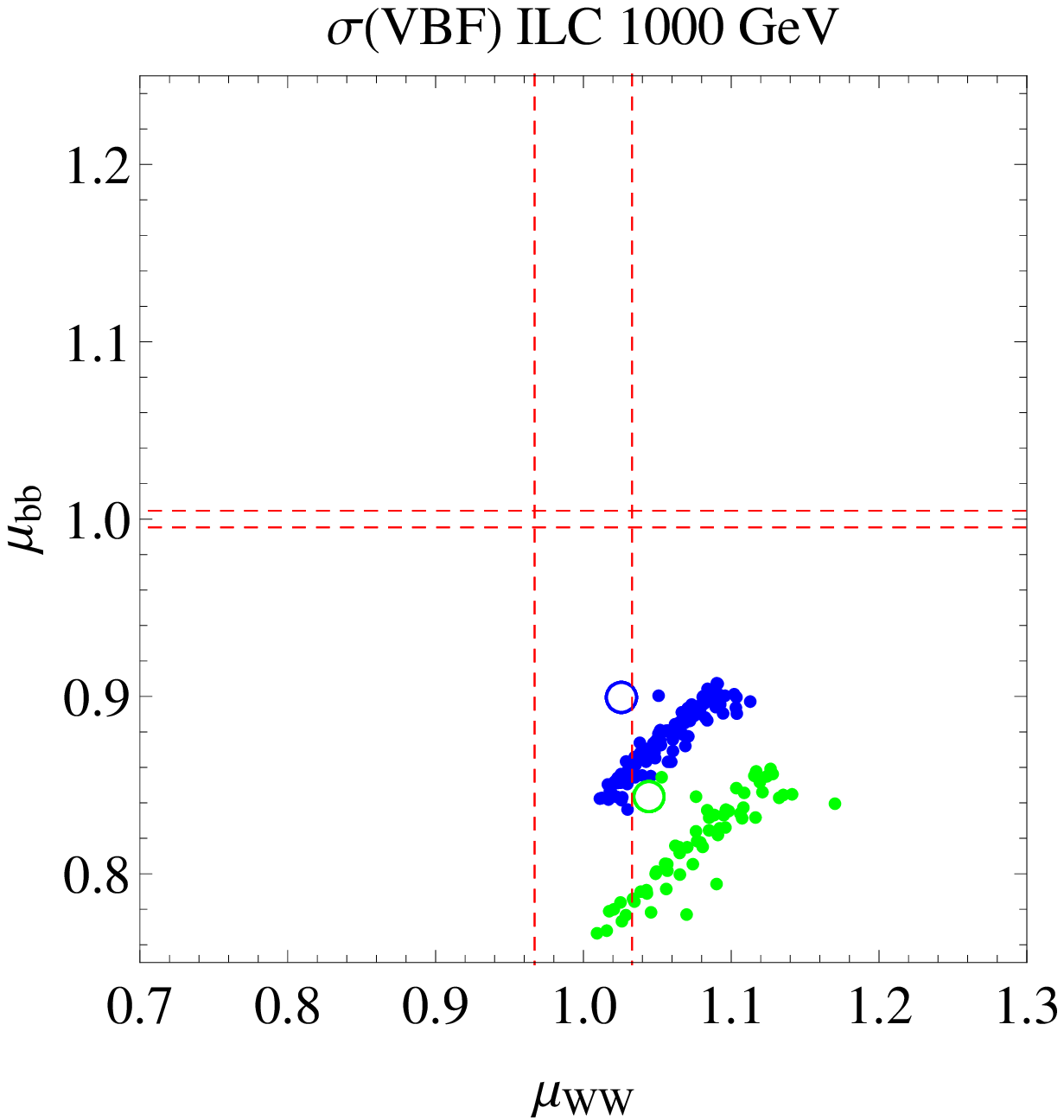, width=.48\textwidth}\hfill
\epsfig{file=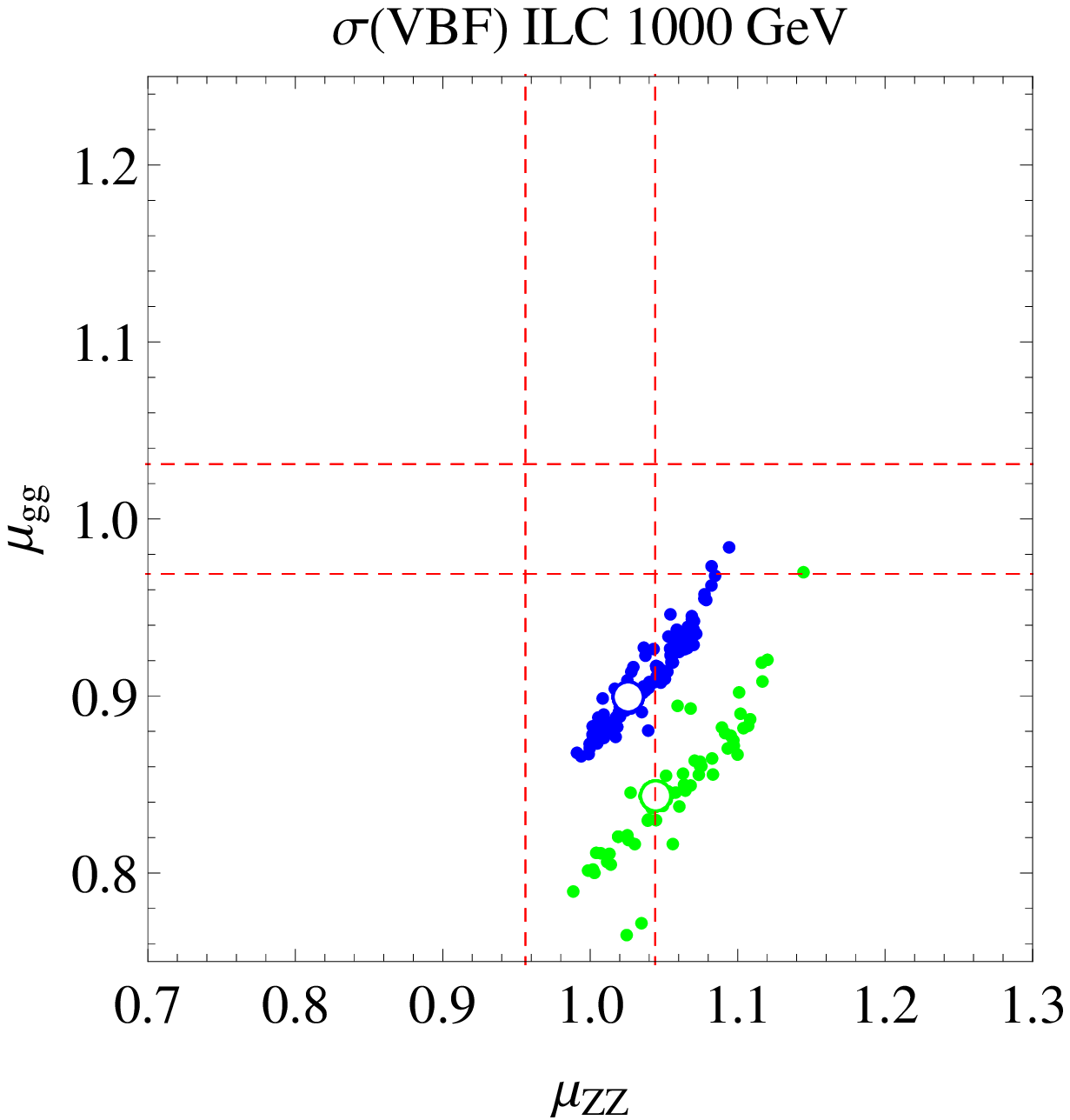, width=.48\textwidth}
\caption[Correlations between $\mu$ parameters at the ILC for $\sqrt{s}$=500,1000 GeV]{Correlations between the relevant $\mu$ parameters at the 500 (upper row) and 1000 (lower row) GeV ILC with 500 and 1000 fb$^{-1}$ of integrated luminosity respectively for $f$=800 GeV and $g_\rho$=2.5 (green points) and $f
$=1000 GeV and $g_\rho$=2 (blue points). The circles represent the signal strengths predictions in the limit where all the extra particle in the spectrum are decoupled.
The area between the dashed lines represents the experimental precision limits given in Tab.~\ref{tab:mu-err-ilc-stra}.}
\label{fig:ilc-vbf}
\end{figure}

After having determined, once again, the importance of keeping the finite spectrum effects in our computation we now present  our results in terms of the $\mu$ parameters relating them to the expected ILC experimental accuracies of Tab.~\ref{tab:mu-err-ilc-stra}. We do this for the choice $f=800$ GeV, $g_\rho=2.5$ and $f=1000$ GeV, $g_\rho=2$ both for both the Higgs-strahlung and VBF production processes in Fig.~\ref{fig:ilc-vh} and Fig.~\ref{fig:ilc-vbf}. 
From the results it is clear that, for both production processes, deviations from the case in which the full particle spectrum is not taken into account, represented in the plots by the circles, could modify the signal strengths for various channels is such a way that the modifications could be, depending on 
the Higgs boson decay mode, fully disentangleable, for some aspects more in the case of Higgs-strahlung than in the case of VBF due to the different topologies of Feynman diagrams.
Altogether, though, it is clear the potentiality of future leptonic machines to pin down the possible composite nature of the 125 GeV scalar boson already at the lowest CM energy here considered.

As mentioned, at the running energy stages of 500 and 1000 GeV another important process to be analysed at a future electron positron collider is $e^+ e^-\to t\bar t H$ for which the relevant Feynman diagram is reported in Fig.~\ref{fig:ilc-tth}\footnote{Note that at the energies considered the contribution from the Higgs off $Z$ diagram is negligible \cite{Baer:2013cma}.}.
Other than the effects already seen in the previously analysed production modes, due to the exchange of $s$ channel extra gauge bosons, in this channel effects due to the exchange of $t^\prime$s can also occur that could affect this production process significantly if these particles are light enough and produced resonantly. In order 
to illustrate the results for this process we have now restricted the allowed mass of the extra fermions to be heavier than 600 GeV which, once again, does not intend to be an accurate and precise exclusion, for which we refer to the next Chapter, but an indication of what realistic mass bounds on these states could 
be, which is an approach indeed sufficient for the purpose of our analysis.
Again following the guidance of \cite{Peskin:2012we,Baer:2013cma} we quote in Tab.~\ref{tab:mu-err-ilc-tth} the expected accuracies on the $\mu$ parameters for the 500 and 1000 GeV stage of the ILC considering however only the $b\bar b$ Higgs decay mode, that is the only decay mode for which the analysis of \cite{Baer:2013cma} gives the expected errors.
For the same two benchmarks previously analysed we present our results in Fig.~\ref{fig:ilc-tth} where on the left we present the results with the inclusion of the extra $t^\prime$s and on the right without.
From the comparison of the two panels it is clear that the enhancement of the $\mu_{b\bar b}$ parameter up to a factor two or so for $\sqrt{s}$=1000 GeV is due to the exchange of extra $t^\prime$s with a mass smaller than $\sqrt{s}-m_{top}$ that can indeed be resonant in the production of the Higgs top pair final 
state. Conversely at $\sqrt{s}$=500 GeV the deviations are due to non-resonant $Z^\prime$s and $t^\prime$s effects as, at this CM energy, no $t^\prime$ mass can be larger than $m_{top}+m_H$ and smaller than $\sqrt{s}-m_{top}$ due to our choice of restricting $m_{t^\prime s}>$ 600 GeV. Altogether we observe then 
that, also in this channel, and limited to the $b\bar b$ Higgs decay mode, the potential of a future $e^+ e^-$ machine in assessing the 4DCHM comprehensive of its finite mass spectrum is significant.

\begin{figure}[!h]
  \begin{picture}(292,118) (-45,-47)
    \SetWidth{1.0}
    \SetColor{Black}
    \Text(63,54)[lb]{\Large{\Black{$e^-$}}}      
    \Line[arrow,arrowpos=0.5,arrowlength=5,arrowwidth=2,arrowinset=0.2](73,54)(123,24)
    \Text(63,-6)[lb]{\Large{\Black{$e^+$}}}      
    \Line[arrow,arrowpos=0.5,arrowlength=5,arrowwidth=2,arrowinset=0.2](123,24)(73,-6)
    \Text(128,32)[lb]{\Large{\Black{$\gamma,Z,Z^\prime_i$}}}      
    \Photon(123,24)(173,24){4.5}{4}
    \Text(223,58)[lb]{\Large{\Black{$t$}}} 
    \Line[arrow,arrowpos=0.5,arrowlength=5,arrowwidth=2,arrowinset=0.2](173,24)(223,54)
    \Text(223,-18)[lb]{\Large{\Black{$\bar t$}}}      
    \Line[arrow,arrowpos=0.5,arrowlength=5,arrowwidth=2,arrowinset=0.2](198,9)(173,24)
    \Line[arrow,arrowpos=0.5,arrowlength=5,arrowwidth=2,arrowinset=0.2](223,-6)(198,9)
    \Text(233,25)[lb]{\Large{\Black{$h$}}}       
    \Line[dash,dashsize=6](198,9)(223,25)
  \end{picture}
  \caption[Feynman diagram for Higgs production in association with $t\bar t$ pair at an $e^+e^-$ collider]{Feynman diagram for Higgs production in association with $t\bar t$ pair at an $e^+e^-$ collider.}
  \label{fig:ilc-tth}
\end{figure}
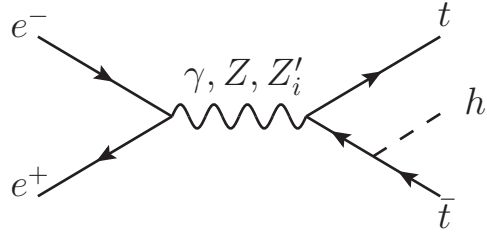

\begin{table}[!h]
\begin{center}
\begin{tabular}{|l|c|c|}
\hline
$tth$                   & 500 GeV & 1000 GeV \\
\hline
\hline
$\mu_{b\bar b}$              & 0.35    & 0.087 \\
\hline
\end{tabular}
\end{center}
\caption[Expected accuracies on the $\mu$ parameters at the ILC for $tth$]{Expected accuracies for the determination of the $\mu$ parameters at the ILC for the $tth$ production process as reported in \cite{Baer:2013cma}. Only the $b \bar b$ decay mode is considered.}
\label{tab:mu-err-ilc-tth}
\end{table}

\begin{figure}[!h]
\centering
\epsfig{file=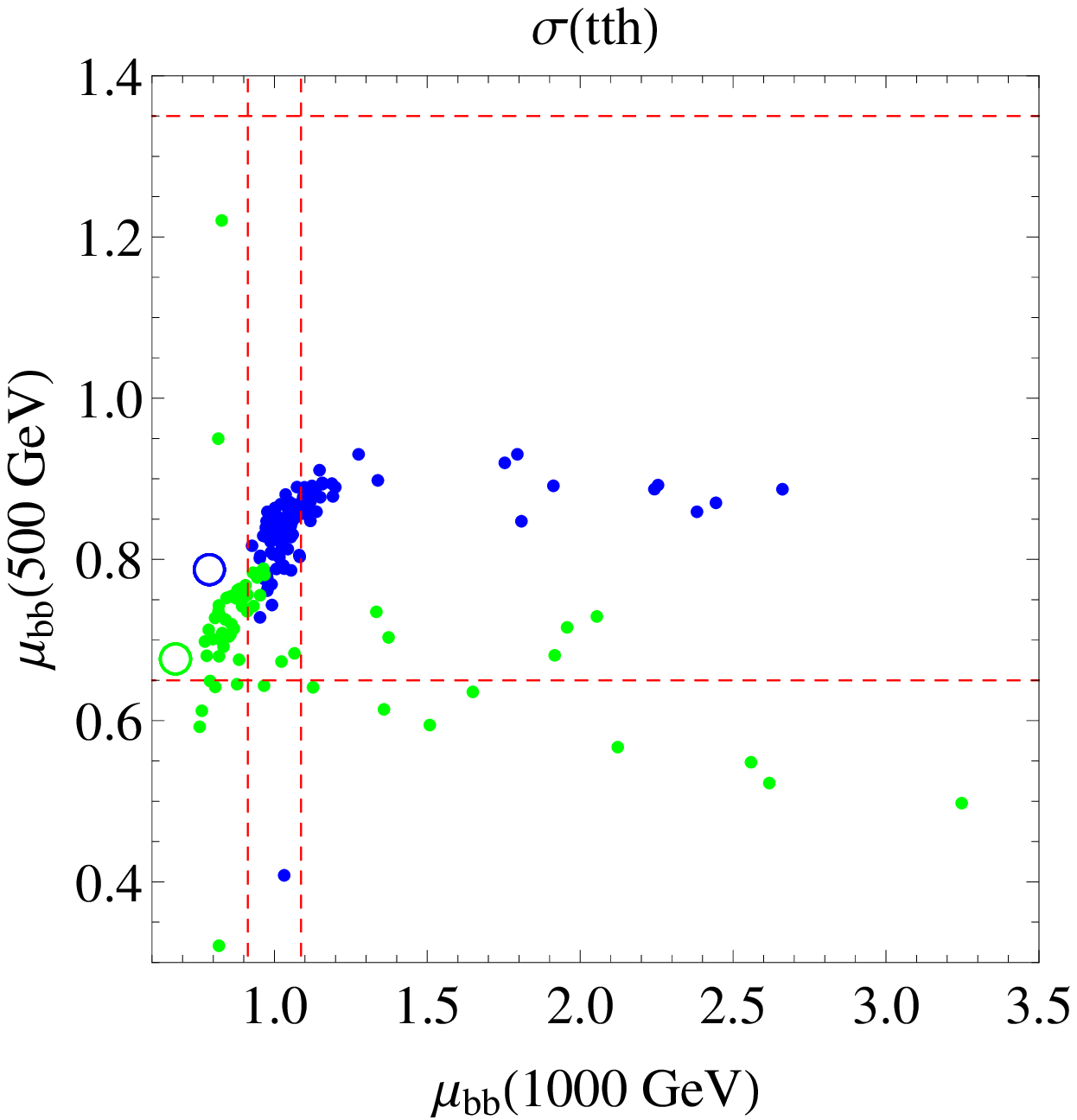, width=.44\textwidth}{(a)}
\epsfig{file=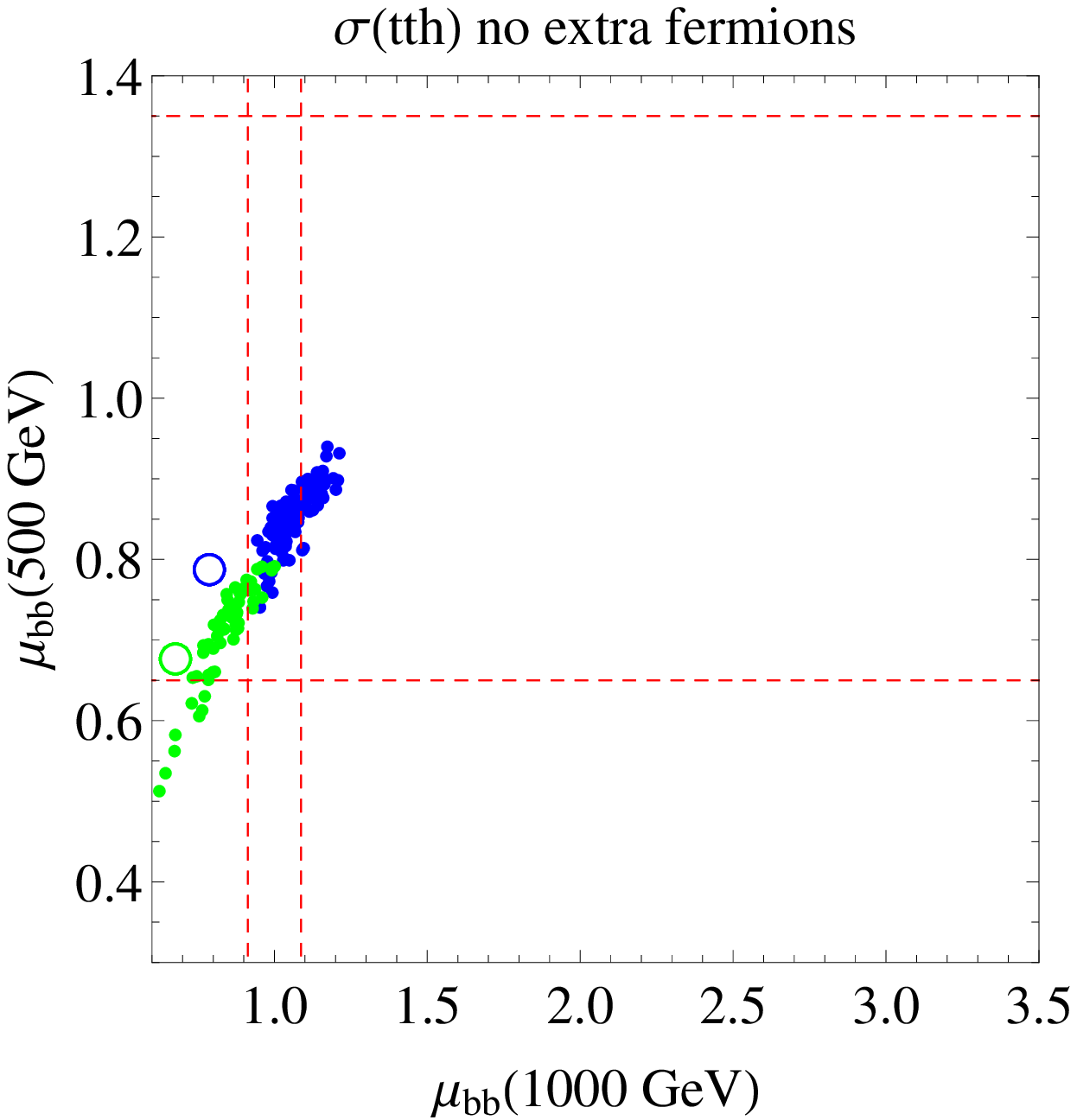, width=.44\textwidth}{(b)}
\caption[Correlations between $\mu$ parameters at the ILC for $\sqrt{s}$=500,1000 GeV for $tth$ production]{Correlations between the relevant $\mu$ parameters at the 500 and 1000 GeV ILC with 500 and 1000 fb$^{-1}$ of integrated luminosity respectively for $f$=800 GeV and $g_\rho$=2.5 (green points) and $f$=1000 GeV and $g_\rho$=2 (blue points) with the inclusion of extra fermions (a) and without (b).
The circles represent the signal strengths predictions in the limit where all the extra particle in the spectrum are decoupled.
The area between the dashed lines represents the experimental precision limits given in Tab.~\ref{tab:mu-err-ilc-stra}.}
\label{fig:ilc-tth}
\end{figure}

We conclude this Chapter by presenting results regarding a production process relevant at $\sqrt{s}=$~500 and 1000 GeV, that allows a precision measurement that is challenging at the LHC, but that can potentially be measured at an $e^+e^-$ collider such as the ILC, namely the one of the triple Higgs coupling.
As explained in Chapter~\ref{chap-2} in a CHM the Higgs potential is radiative generated and, with the choice of the 4DCHM fermionic sector, it turns out to be UV finite. 
It is possible then to extract the Higgs VEV and mass from the expression of the 4DCHM potential and its derivative \cite{DeCurtis:2011yx}, together with the triple self-coupling $\lambda$ for which, at the leading order in the contribution of the gauge and fermionic loops, we get
\begin{equation}
\lambda=\frac{3m^2_H}{v}\frac{1-2\frac{v^2}{f^2}}{\sqrt{1-\frac{v^2}{f^2}}}=\lambda_{SM}\frac{1-2\xi}{\sqrt{1-\xi}}.
\label{eq:ilc-hhh}
\end{equation}

\begin{figure}[!h]
  \begin{picture}(292,118) (35,-47)
    \SetWidth{1.0}
    \SetColor{Black}
    \Text(63,48)[lb]{\Black{$e^-$}}  
    \Line[arrow,arrowpos=0.5,arrowlength=5,arrowwidth=2,arrowinset=0.2](73,44)(103,24)
    \Text(63,-8)[lb]{\Black{$e^+$}}      
    \Line[arrow,arrowpos=0.5,arrowlength=5,arrowwidth=2,arrowinset=0.2](103,24)(73,4)
    \Photon(103,24)(133,24){3.5}{4}
    \Text(169,44)[lb]{\Black{$Z$}}     
    \Photon(133,24)(163,44){3.5}{4}
    \Text(169,24)[lb]{\Black{$h$}}       
    \Line[dash,dashsize=6](148,14)(133,24)
    \Text(169,4)[lb]{\Black{$h$}}     
    \Line[dash,dashsize=6](163,4)(148,14)
    \Line[dash,dashsize=6](148,14)(163,24)
    \Text(183,48)[lb]{\Black{$e^-$}}
    \Text(183,-8)[lb]{\Black{$e^+$}}
    \Text(289,44)[lb]{\Black{$h$}} 
    \Text(289,24)[lb]{\Black{$h$}}
    \Text(289,4)[lb]{\Black{$Z$}}      
    \Line[arrow,arrowpos=0.5,arrowlength=5,arrowwidth=2,arrowinset=0.2](193,44)(223,24)
    \Line[arrow,arrowpos=0.5,arrowlength=5,arrowwidth=2,arrowinset=0.2](223,24)(193,4)
    \Photon(223,24)(253,24){3.5}{4}
    \Line[dash,dashsize=6](253,24)(283,44)
    \Photon(268,14)(253,24){-3.5}{2}
    \Photon(282,4)(268,14){-3.5}{2}
    \Line[dash,dashsize=6](268,14)(282,24)
    \Text(302,48)[lb]{\Black{$e^-$}}
    \Text(302,-8)[lb]{\Black{$e^+$}}
    \Text(414,44)[lb]{\Black{$h$}} 
    \Text(414,24)[lb]{\Black{$h$}}
    \Text(414,4)[lb]{\Black{$Z$}}      
    \Line[arrow,arrowpos=0.5,arrowlength=5,arrowwidth=2,arrowinset=0.2](312,44)(342,24)
    \Line[arrow,arrowpos=0.5,arrowlength=5,arrowwidth=2,arrowinset=0.2](342,24)(312,4)
    \Photon(342,24)(372,24){3.5}{4}
    \Line[dash,dashsize=6](372,24)(402,44)
    \Photon(372,24)(402,4){-3.5}{4}
    \Line[dash,dashsize=6](372,24)(402,24)       
  \end{picture}
  \caption[Feynman diagrams for double Higgs production via Higgs-strahlung at an $e^+e^-$ collider]{Feynman diagrams for double Higgs production via Higgs-strahlung at an $e^+e^-$ collider. All neutral gauge bosons can be exchanged in the intermediate states.}
  \label{fig:ilc-hh-vh}
\end{figure}
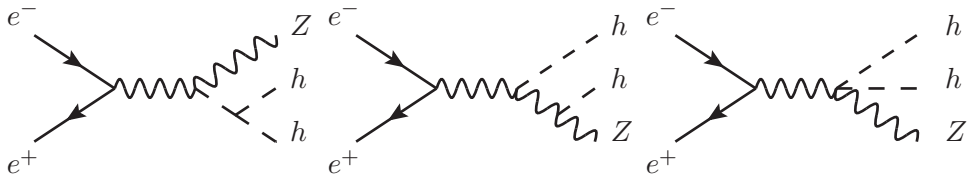

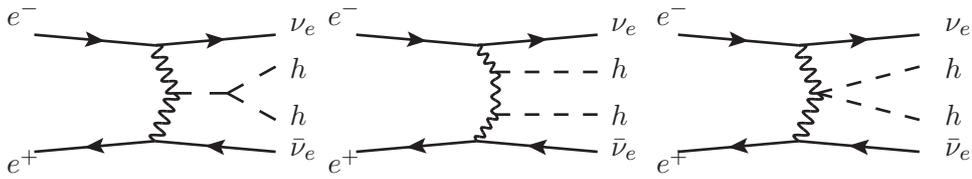
\begin{figure}[!h]
  \begin{picture}(292,118) (35,-47)
    \SetWidth{1.0}
    \SetColor{Black}
    \Text(63,48)[lb]{\Black{$e^-$}}  
    \Text(63,-8)[lb]{\Black{$e^+$}}      
    \Text(169,44)[lb]{\Black{$\nu_e$}}     
    \Text(169,28)[lb]{\Black{$h$}}       
    \Text(169,10)[lb]{\Black{$h$}}   
    \Text(169,-3)[lb]{\Black{$\bar \nu_e$}}     
    
    \Line[arrow,arrowpos=0.5,arrowlength=5,arrowwidth=2,arrowinset=0.2](73,44)(118,40)
    \Line[arrow,arrowpos=0.5,arrowlength=5,arrowwidth=2,arrowinset=0.2](118,4)(73,0)
    \Line[arrow,arrowpos=0.5,arrowlength=5,arrowwidth=2,arrowinset=0.2](118,40)(163,44)
    \Line[arrow,arrowpos=0.5,arrowlength=5,arrowwidth=2,arrowinset=0.2](163,0)(118,4)
    \Photon(118,40)(125,22){2.5}{4}
    \Photon(118,4)(125,22){2.5}{4}
    \Line[dash,dashsize=6](125,22)(145,22)
    \Line[dash,dashsize=6](145,22)(163,32)  
    \Line[dash,dashsize=6](145,22)(163,12)      
    \Text(183,48)[lb]{\Black{$e^-$}}  
    \Text(183,-8)[lb]{\Black{$e^+$}}      
    \Text(289,44)[lb]{\Black{$\nu_e$}}     
    \Text(289,28)[lb]{\Black{$h$}}       
    \Text(289,10)[lb]{\Black{$h$}}   
    \Text(289,-3)[lb]{\Black{$\bar \nu_e$}}   
    
    \Line[arrow,arrowpos=0.5,arrowlength=5,arrowwidth=2,arrowinset=0.2](193,44)(238,40)
    \Line[arrow,arrowpos=0.5,arrowlength=5,arrowwidth=2,arrowinset=0.2](238,4)(193,0)
    \Line[arrow,arrowpos=0.5,arrowlength=5,arrowwidth=2,arrowinset=0.2](238,40)(283,44)
    \Line[arrow,arrowpos=0.5,arrowlength=5,arrowwidth=2,arrowinset=0.2](283,0)(238,4)
    \Photon(238,40)(245,30){1.5}{3}
    \Photon(238,4)(245,14){1.5}{3}
    \Photon(245,30)(245,14){1.5}{3}
    \Line[dash,dashsize=6](245,30)(283,30)  
    \Line[dash,dashsize=6](245,14)(283,14)      
    \Text(306,48)[lb]{\Black{$e^-$}}  
    \Text(306,-8)[lb]{\Black{$e^+$}}      
    \Text(414,44)[lb]{\Black{$\nu_e$}}     
    \Text(414,28)[lb]{\Black{$h$}}       
    \Text(414,10)[lb]{\Black{$h$}}   
    \Text(414,-3)[lb]{\Black{$\bar \nu_e$}}       
    \Line[arrow,arrowpos=0.5,arrowlength=5,arrowwidth=2,arrowinset=0.2](313,44)(358,40)
    \Line[arrow,arrowpos=0.5,arrowlength=5,arrowwidth=2,arrowinset=0.2](358,4)(313,0)
    \Line[arrow,arrowpos=0.5,arrowlength=5,arrowwidth=2,arrowinset=0.2](358,40)(403,44)
    \Line[arrow,arrowpos=0.5,arrowlength=5,arrowwidth=2,arrowinset=0.2](403,0)(358,4)
    \Photon(358,40)(365,22){2.5}{4}
    \Photon(358,4)(365,22){2.5}{4}
    \Line[dash,dashsize=6](365,22)(403,32)  
    \Line[dash,dashsize=6](365,22)(403,12)    
  \end{picture}
  \caption[Feynman diagrams for double Higgs production via VBF at an $e^+e^-$ collider]{Feynman diagrams for double Higgs production via VBF at an $e^+e^-$ collider. All charged gauge bosons can be exchanged in the intermediate states.}
  \label{fig:ilc-hh-vbf}
\end{figure}

\begin{figure}[!h]
\centering
\epsfig{file=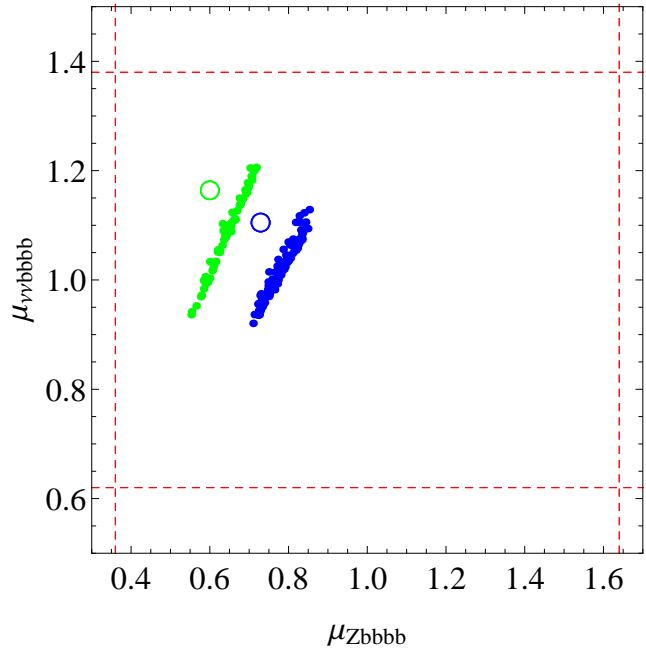, width=.58\textwidth}
\caption[Correlation between $\mu$ parameters at the ILC for $\sqrt{s}$=500,1000 GeV for double Higgs production]{Correlation between the relevant $\mu$ parameters at the 500 and 1000 GeV ILC with 500 and 1000 fb$^{-1}$ of integrated luminosity respectively for $f$=800 GeV and $g_\rho$=2.5 (green points) and $f$=1000 GeV and $g_\rho$=2 (blue points) for the double Higgs production processes.
The circles represent the signal strengths predictions in the limit where all the extra particle in the spectrum are decoupled.
The area between the dashed lines represents the experimental precision limits given in Tab.~\ref{tab:mu-err-ilc-stra}.}
\label{fig:ilc-hh}
\end{figure}

This modified coupling intervenes in the Feynman diagram of double Higgs production via Higgs-strahlung and VBF, Fig.~\ref{fig:ilc-hh-vh} and  Fig.~\ref{fig:ilc-hh-vbf}, for which the expected accuracies in determining the $\mu$ parameters for Higgs-strahlung at 500 GeV and VBF at 1000 GeV production processes 
\cite{Peskin:2012we,Baer:2013cma} are reported in Tab.~\ref{tab:mu-err-ilc-hh}, where again we assume the decays of both the Higgs bosons just into a $b\bar b$ final state.
We then present our results in Fig.~\ref{fig:ilc-hh} where we observe that sizeable deviations with respect to the SM, as well as to the decoupling limit, predictions are possible.
However due to the poor predicted accuracies for the corresponding measurements, the expected errors ought to be reduced by at least factor of two or so, in order to disentangle the 4DCHM from SM effects, and this is in line with the pursuit of the so called ILC(LumUP) high luminosity scenario 
considered in \cite{Asner:2013psa}, in the hope of decreasing the predicted uncertainties on the measurements of such an observable down to a level comparable to their expected departures from the SM value.

\begin{table}[!h]
\begin{center}
\begin{tabular}{|l|c|c|}
\hline
                        & ZH~~500 GeV & VBF~~1000 GeV \\
\hline
\hline
$b \bar b$              & 0.64    & 0.38 \\
\hline
\end{tabular}
\end{center}
\caption[Expected accuracies on the $\mu$ parameters at the ILC for double Higgs production]{Expected accuracies for the determination of the $\mu$ parameters at the ILC for the double Higgs production processes as reported in \cite{Baer:2013cma}. Only the $b \bar b$ decay mode is considered.}
\label{tab:mu-err-ilc-hh}
\end{table}

\section{Conclusions}

In this Chapter we have analysed the properties of the Higgs sector of the 4DCHM with respect firstly to the current available LHC data and then to proposed $e^+ e^-$ colliders.
We have shown that the 4DCHM compatibility with the 7 and 8 TeV Higgs data of the LHC is as good as the SM in pointing to a discovery of a neutral Higgs boson with a mass $\sim$~125 GeV.
In computing the various signal strengths $\mu$ we have analysed all the effects intervening in it and shown that, for values of the extra fermions around 500 GeV, enhancements in these values are possible, albeit moderate.
We have compared our results obtained with the complete particle spectrum of the model with those obtained in the asymptotic limit showing that possible observable deviations between the two approaches can arise, for both the 7,8 and 14 TeV LHC.
We have then tested our model against future proposed electron-positron colliders, borrowing the specific configurations of the ILC, and showing the potentiality of this machine in disentangling the nature of the composite Higgs boson, again by means of the signal strengths, in both the asymptotic limit as well as with the full particle spectrum of the model in various Higgs production modes.
In particular we have shown that the composite Higgs scenario can be tested already at a centre of mass energy of 250 GeV in the case of Higgs-strahlung production process, albeit limited to final states with a predicted error $<$10\%, while, under the same assumptions on the errors upper bound, VBF and top associated production can be used at a probe for $\sqrt{s}=500$ and 1000 GeV.

\chapter{Heavy extra quarks in BSM scenarios}
\label{chap-5}
\lhead{Chapter 5. \emph{Heavy extra quarks in BSM scenarios}}

This Chapter is devoted to the analysis of the phenomenology of extra heavy quarks that are present in many extensions of the SM, as in the case of CHMs, although here, differently from the rest of the Thesis, we will not directly study the case of the 4DCHM.
We will illustrate a framework and discuss a dedicated tool called \verb|XQCAT| (extra quark combined analysis tool) \cite{Barducci:2014ila} based on publicly available experimental data  which allows us to determine the exclusion confidence level (eCL) of a given scenario involving multiple heavy extra quarks with respect to the available direct searches for \emph{top partners}.
After describing the motivations for this study, we will explain our analysis strategy and apply our tool to some specific scenarios, including a recently proposed simplified CHM, indeed reproducible in the 4DCHM case, in order to show its potentiality.
Finally we will show that since the dedicated experimental searches for \emph{top partners} are not sensitive yet to the case of decay of these states into light quarks, the re-interpretation of SUSY searches  could be important so as to set bounds  in these scenarios as well.

\section{Introduction and motivations}

During our analysis we have so far set bounds on the masses of the extra quarks present in the 4DCHM by rescaling the pair production cross sections of the lightest extra quarks to take into account the non-100\% branching ratios of these states into the channels for which the experimentalists are searching for and we have then derived a bound on their masses around 500-600 GeV.
The motivations for this approach have been both the fact that for the purpose of our analysis the approximation that we adopted was sufficient in order to illustrate the relevant phenomenology of the gauge and Higgs sectors of the 4DCHM, and the fact that in presence of a large quark spectrum a detailed analysis of it has to take into account other aspects besides the simple rescaling of the rates of the lightest quarks.
The presence of a wider spectrum can in fact increase the final signal rate into a given search channel due to the possibility that two quarks can either have the same decay channels or different decay channels which however feed, even partially, the same final state.
This, together with the necessity to take into account the different branching ratios of the extra quarks with respect to the one assumed by the experimental collaborations, makes a re-interpretation of the experimental mass bounds not trivial.
However this re-interpretation is of primary importance nowadays since \emph{top partners} are expected to play an important role in many extensions of the SM that offer alternative explanations to the mechanisms of EWSB and might have a crucial role in softening the quadratic divergences contributing to the Higgs mass term, which is the origin of the hierarchy problem discussed in Chapter~\ref{chap:1}.
New heavy quarks can appear for example in models with extra dimensions \cite{Antoniadis:1990ew,Csaki:2003sh,Cacciapaglia:2009pa}, little Higgs models \cite{ArkaniHamed:2002qy,Schmaltz:2005ky}, models with gauging of the flavour group \cite{Davidson:1987tr,Babu:1989rb,Grinstein:2010ve,Guadagnoli:2011id}, 
non minimal SUSY extensions of the SM \cite{Moroi:1991mg,Moroi:1992zk,Babu:2008ge,Martin:2009bg,Graham:2009gy,Martin:2010dc}, grand unified theories \cite{Rosner:1985hx,Robinett:1985dz} as well as composite Higgs models \cite
{Dobrescu:1997nm,Chivukula:1998wd,He:2001fz,Hill:2002ap,Agashe:2004rs,Contino:2006qr,Barbieri:2007bh,Anastasiou:2009rv} and it is therefore important to be able to set bounds on the masses of these states in order to constrain or exclude portions of the parameters space of these scenarios, and up to now 
extra quarks coupled to third generation quarks have been thoroughly investigated from a phenomenological point of view \cite{DeSimone:2012fs,delAguila:2000rc,AguilarSaavedra:2005pv,AguilarSaavedra:2009es,Cacciapaglia:2010vn,Matsedonskyi:2014lla}, together with the possibility of having sizeable couplings to light generations \cite{Atre:2011ae,Cacciapaglia:2011fx}.

In principle, for cut based analyses, a MC simulation and a simplified detector emulation can allow the conversion of the results of the existing experimental searches and allow for the re-interpretation of the mass bounds for any given model.
However this is a very time-consuming task since for each model point, corresponding to different mass spectra and decay modes, the simulation has to be re-done.
Nevertheless we will show that it is possible, given the parameter space describing the extra quark spectrum and the experimental results from their searches, to find allowed and excluded regions in the parameter space of a model without performing any dedicated simulation and in this respect some
studies have attempted to tackle this problem proposing tools such as CheckMATE \cite{Drees:2013wra}, SModelS \cite{Kraml:2013mwa} or Fastlim \cite{Papucci:2014rja} where the former adopts a rather model independent approach while the others are mostly dedicated to SUSY scenarios.
Clearly in order to achieve the task of reinterpreting the LHC data a pre-loaded database of MC data and experimental searches has to be created and these two aspects, which are the most time-consuming ones, can be overcome by the use of the tool that we propose, called \verb|XQCAT|\footnote{Which is available for public use at the pages \url{http://www.hep.phys.soton.ac.uk/xqcat} and \url{https://launchpad.net/xqcat}.} which requires as input 
masses, charges and branching ratios of the extra quarks of a given model and gives as output the corresponding eCL calculated by reinterpreting the searches implemented into the program database.

To achieve a certain degree of model independence we restricted ourselves to considering only pair production processes induced by QCD interactions, so that the production rates of these new states are sensitive only to their mass, although a similar procedure could in principle also be applied  in the case of 
single production, as shown in \cite{Buchkremer:2013bha}. Moreover we considered that none of the new particles can decay into each other or other non-SM particles, such as $Z^\prime$s, $W^\prime$s, extra scalars and DM candidates.
Under these assumptions we will show that the cross sections for various final states can be decomposed in model independent subsets containing all the kinematic information from the decays and that it is then possible to reconstruct the signal coming from a general scenario by combining with appropriate 
weights the different topologies which generate the signal thanks to the fact that the efficiencies of the various signal channels, each of which in principle has a different kinematic, are included into the database of \verb|XQCAT| so that no further simulation is needed.

\section{Analysis strategy}

\subsection{General approach}

Given the assumption that the extra quarks can only decay into SM quarks and bosons and considering just dimension-four interactions, the electric charges for the extra quarks which are of interest to us are
restricted to be  $5/3,2/3,-1/3,-4/3$, which for example arise naturally in the context of CHMs as vector like quarks belonging to singlet, doublet or triplet representation of $SU(2)_L$, see \cite{Okada:2012gy} for a review.
These states, that we have called $X,t^\prime,b^\prime$ and $Y$ for the case of the 4DCHM, can therefore have just the following decays channels
\begin{equation}
\begin{split}
& X_i\to W^+ u_i, \\
& t^\prime_i \to W^+ d_i,Z u_i, H u_i,\\
& b^\prime_i \to W^- u_i,Z d_i, H d_i,\\
& Y_i\to W^- d_i,
\end{split}
\end{equation}
where with $i=1,2,3$ we label the SM families not restricting  ourselves to decays just within the third generation of quarks.

We now want to show that knowing the number of signal events surviving the selection cuts for any given possible decay channel combination of the extra quarks with a certain mass $m_Q$  and for any given experimental search (that is, knowing the efficiency of the corresponding search for each of the 
subprocesses that can contribute to the final state) it is possible to express the overall signal as a weighted sum of signals arising from each of the subprocesses, where the weights are given by a product of the branching ratios combination and efficiencies for any given final state.
To be more clear let's illustrate this with an example in the case of just two extra quarks, $X$ and $b^\prime$, for which the corresponding QCD pair production cross sections are $\sigma(m_X)$ and $\sigma(m_{b^\prime})$ and their decay rates are given by ${\textrm{BR}}(X\to W^+t)$, ${\textrm{BR}}(b^\prime\to W^-u)$ and ${\textrm{BR}}(b^\prime\to W^-t)$,
 that is, we assume here that the $b^\prime$ can decay just via charged current, but also into light generation quarks. Knowing the efficiencies for a 
given bin of a search for each production and sub sequent decay process\footnote{As an example here we are considering the $W^{\pm}$ as the final state of the processes, which of course has been decayed in performing our analysis.}, that are $\epsilon(m_{X^\prime},W^+tW^-\bar t)$, $\epsilon(m_{b^\prime},W^-uW^+\bar u)$, $\epsilon(m_{b^\prime},W^-uW^+\bar t)$, 
$\epsilon(m_{b^\prime},W^-tW^+\bar u)$, $\epsilon(m_{b^\prime},W^-tW^+\bar t)$, it is possible to compute the final signal rate for a given integrated luminosity $\mathcal L$ as
\begin{equation}
\begin{split}
&\#\textrm{events}=\mathcal{L}\cdot[\sigma(m_X){\textrm{BR}}(X\to W^+t)^2\epsilon(m_{X^\prime},W^+tW^-\bar t)+\\
&+\sigma(m_{b^\prime}){\textrm{BR}}(b^\prime\to W^-u){\textrm{BR}}(\bar b^\prime\to W^+\bar t)\epsilon(m_{b^\prime},W^-uW^+\bar t)+...]
\label{eq:XQCAT-ev}
\end{split}
\end{equation}
and, again, if the efficiencies, together with the QCD production cross sections, are stored in a database no simulations are needed in order to compute the final number of events.
From this event rate it is then possible, knowing the background and the observed number of events provided by the experimental collaborations in their analysis, to calculate the eCL of the chosen scenario using the CL method
\begin{equation}
eCL=1-\frac{CL(s+b)}{CL(b)}=1-\frac{1-p\textrm{-value}(s+b)}{1-p\textrm{-value}(b)}
\label{eq:XQCAT-ecl}
\end{equation}
that can be extended straightforwardly to the case of multiple bins by introducing the products of $p$-values.

For our purpose we have then simulated signals for different values of the masses of the extra quarks and for each possible decay combinations, distinguishing between quarks and antiquarks, and have then used a \verb|C++| based analysis code to apply the selection cuts of the experimental searches in order to find the respective efficiencies for each subprocess contributing to each individual signature.
These efficiencies have been stored into the database of \verb|XQCAT| that, for a given input, allows to automatically evaluate the corresponding final signal rate, eq.(\ref{eq:XQCAT-ev}), and eCLs, eq.(\ref{eq:XQCAT-ecl}), for each individual implemented search and, when possible, combination of searches.

\subsection{Generation of the efficiency database}

The number of processes that have been simulated in order to obtain a complete efficiency database is related to the number of considered masses, decay 
channels and chirality of the couplings\footnote{The tool assumes in fact a dominant chirality for the extra quarks couplings, which is a correct assumptions in case that these states are vector like \cite{AguilarSaavedra:2009es}, as in the case of CHMs.}.
Assuming that these states can decay in all three of the quark generations and neglecting the possibility of a charm tagging (so that 1$^{th}$ and 2$^{nd}$ generations are indistinguishable from an experimental point of view), both the $t^\prime$s and the $b^\prime$s can have 6 possible decay 
modes, 4 via neutral current and 2 via charged current, while they are just 2 in the case of the $X$ and $Y$ species, since they can only decay into the $W^{\pm}$ boson and a SM quark.
Considering that we are assuming QCD pair production and simulating both the case of left and right chiral couplings, the total number of channels for each mass is equal to $2\times 2\times(6\times 6 + 2 \times 2)=160$ and we have performed our simulations in the mass range of $400-2000$ GeV at steps of 100 GeV for a total of 2720 simulated processes for each LHC centre of mass energy.
The simulations have been done with MadGraph5, v.1.5.8 \cite{Alwall:2011uj}, with the decays of the extra quarks into SM quarks and gauge bosons performed by means of BRIDGE, v.2.24 \cite{Meade:2007js}. For the decay of SM quarks and bosons and subsequent hadronisation and parton showering we have used 
PYTHIA, v.6.4 \cite{Sjostrand:2006za}, and the detector simulation has been performed with Delphes2, v.2.0.2 \cite{Ovyn:2009tx}. Finally, since the pair production is a QCD process, jet matching up to two jets has been considered in simulating the processes in MadGraph5.
Limiting our study just to CMS searches we have implemented in our program two different kinds of experimental analyses.
\begin{itemize}
\item \underbar{\textit{Direct search of extra quarks}}:
we have implemented the CMS analysis B2G-12-015 \cite{Chatrchyan:2013uxa}, at $\sqrt{s}=8$ TeV with a 19.5 fb$^{-1}$ dataset, for pair produced $t^\prime$ quarks that mix only with third generation SM quarks and can therefore decay to $W^+b, Zt$ or $Ht$ with variable branching ratios.
The CMS collaboration presents the 95\% eCL lower limits on the $t^\prime$ quark mass for different combinations of its branching ratios, see Fig.\ref{fig:CMS-bounds}, using six mutually exclusive channels: two single lepton (single electron and single muon), three di-lepton (two opposite sign and one same sign, the former requiring different number of jets) and one tri-lepton. 
Since the sensitivity of the search is mostly driven by the multi-lepton channels, in the present version of the tool we have only implemented three bins:
one opposite sign (called OS1 in the CMS paper), the same sign (SS) and the tri-lepton channels. More details about this choice will be given in the next Section where we will describe the validation of our framework. The limits for the multi lepton channels only, can be found in the twiki page of the search \cite{twikiB2G12015} and the quoted observed bounds are in the range $592-794$ GeV depending on the assumed branching ratios.
\item \underbar{\textit{SUSY searches}}
we have implemented four searches inspired by SUSY scenarios, characterised by the presence of different numbers of leptons in the final state and large missing transverse energy: 0 lepton (called $\alpha_{T}$ in the CMS paper) \cite{Chatrchyan:2012wa}, mono-lepton ($L_{p}$) \cite{Chatrchyan:2012sca}, opposite sign di-lepton (OS) 
\cite{Chatrchyan:2012te} and same sign di-lepton (SS) \cite{Chatrchyan:2012sa}, considering the entire 4.98 fb$^{-1}$ 2011 dataset at $\sqrt{s}=7$ TeV. We have also included the updated $\alpha _{T}$ \cite{Chatrchyan:2013lya} and SS \cite{Chatrchyan:2012paa} searches at 8 TeV, with 11.7 fb$^{-1}$ and 
10.5 fb$^{-1}$, respectively. It has been verified that the selected searches are uncorrelated and, therefore, it is possible to statistically combine them without the need of a correlation matrix, yielding 95\% CL bounds at 7 TeV (combination of 4 searches), 8 TeV (combination of 2 searches) and 7+8 TeV (combination of 6 searches).
\end{itemize}

Before giving details on how our framework has been validated we need to mention that there are 
several effects, that the code currently doesn't take into account, which might affect the calculation of the final signal rates, and therefore of the eCLs, such as possible chain decays between the extra quarks, decays into states beside the SM ones, interference effects in the pair production of two extra quarks and loop corrections to masses and mixing between the extra states.
These issues can in principle reduce the final number of predicted signal events therefore providing, if not taken into account, a non-conservative bound on a given scenario. 
It is however also important to consider all the possible effects that might enhance the signal rate, since an over-conservative estimate would result in too weak a bound; a detailed discussion of these effects is given in \cite{Barducci:2014ila} and future upgrades of the code will also take into account these aspects of the phenomenology of extra quarks.

\section{Validation of the framework}

While for the SUSY inspired searches the framework for computing the efficiencies has been validated and used in previous works \cite{Buchmueller:2012hv,Buchmueller:2013exa}, for the case of the direct search for \emph{top partners} we need to validate it against the experimental results provided by the CMS collaboration.
The validation part also includes the test of the limit code with which we compute the eCLs from the final signal rates.

\subsection{Validation of the computation of the eCLs}

This validation has the scope to test any discrepancy between the statistical method used in our approach and the one used in the experimental analysis, and has been done by comparing the expected and observed limits computed using the signal, background and data information provided in the experimental search documentation \cite{Chatrchyan:2013uxa}, for which we show the results in Tab.\ref{tab:XQ-validationlimitcode}
where the signal rates have been computed for a $t^\prime$ with branching ratios of 50\%, 25\% and 25\% in $W^+b$, $Zt$ and $Ht$ final states respectively.
First of all, from the column where the eCLs have been computed using all the channels of the search, we observe that we are not able to reproduce the mass bounds considering the single lepton channels in combination with the multi-lepton ones, and this is due to a different analysis technique used by the experimental collaboration.
Considering however only the multi-lepton channels we obtain a mass bounds of 626 and 630 GeV for the expected and observed bounds, obtained by linearly interpolating the eCLs between the simulated mass points that are reported in the table.
We are therefore able, with our statistical technique, to reproduce in this case the experimental values of 683 and 668 GeV with a discrepancy of $-8\%$ and $-6\%$ and for this reason we will only consider the multi-lepton channels in the rest of our discussion.

\begin{table}
\scriptsize
\centering
\setlength{\tabcolsep}{2pt}
\begin{tabular}{|c|cc|cccc|cc|cc|}
\cmidrule{1-7}                                
& \multicolumn{2}{c|}{Single lepton channels} & \multicolumn{4}{c|}{Multi lepton channels} \\
\cmidrule{1-7}                                
& Muon & Electron & OS1 & OS2 & SS & 3l \\
\cmidrule{1-7}                                
Bg & $61900\pm13900$ & $61500\pm13700$ & $17.4\pm3.7$ & $84\pm12$ & $16.5\pm4.8$ & $3.7\pm1.3$ \\
Data & 58478 & 57743 & 20 & 86 & 18 & 2 \\
\midrule
\multicolumn{7}{|c|}{\multirow{2}{*}{Signal events for nominal point (BR$(Wb)=0.5$ and BR$(Zt)={\rm BR}(Ht)=0.25$)}} & \multicolumn{2}{c|}{${\rm eCL}_{\rm\tiny all}$} & \multicolumn{2}{c|}{${\rm eCL}_{\tiny\mbox{multilepton}}$} \\
\multicolumn{7}{|c|}{} & Exp & Obs & Exp & Obs \\
\midrule
500 GeV  & 850  & 840  & 16.7 & 35.1 & 21.3 & 19.1 & 0.29 & 0.36 & 1     & 1 \\
600 GeV  & 280  & 280  & 8.9  & 16.6 & 7.5  & 8.5  & 0.20 & 0.25 & 0.998 & 0.999 \\
700 GeV  & 97   & 98   & 4.0  & 6.6  & 2.8  & 3.1  & 0.10 & 0.13 & 0.831 & 0.851 \\
800 GeV  & 36   & 37   & 1.6  & 2.5  & 1.0  & 1.3  & 0.05 & 0.06 & 0.438 & 0.487 \\
\midrule                                
\midrule                                
\multicolumn{7}{|c|}{95\% exclusion limit computed by the limit code} & - & - & 626 GeV & 630 GeV \\
\bottomrule
\end{tabular}
\caption[95\% exclusion confidence level mass bounds obtained with the statistical combination of the search bins implemented in the limit code]{eCLs  and 95\% lower mass bounds obtained with the statistical combination of search bins implemented in the limit code. The quoted values for the expected (observed) lower mass bounds in \cite{Chatrchyan:2013uxa} are 773 GeV (696 GeV) considering all channels, and 683 GeV (668 GeV) considering only the multi-lepton channels (as reported in the corresponding twiki page \cite{twikiB2G12015}, and to be compared to the limit code results in the table).}
\label{tab:XQ-validationlimitcode}
\end{table}

\subsection{Validation of the efficiency extraction code}

The extraction of the efficiencies depends on the interplay of different factors among which the most important ones are the accuracy of the MC simulation, the correct reproduction of the detector effects and the correct implementation of the experimental selection cuts. 
Considering only the multi-lepton channels and again for the same choice of branching ratios of the $t^\prime$ as before, the number of events for a given mass of the quark and the difference between the ones reported by the CMS analysis in Tab.~\ref{tab:XQ-validationlimitcode} are reported in Tab.~\ref{tab:XQ-efficiencytable}.
We observe that our results present an offset between 20\% and 30\% in all the channels except the second opposite sign di-lepton, OS2, where we find a larger deviation which is around 50\%.
This discrepancy can however be  explained by the unavoidable differences in the reproduction of the detector effects and selection cuts and for this reason we have decided to omit the OS2 channel from our implementation of this experimental analysis, leaving us with just three channels, the first opposite sign di-lepton, OS1, the same sign di-lepton, SS, and the tri-lepton, 3l.

\begin{table}[!h]
\small
\centering
\setlength{\tabcolsep}{2pt}
\begin{tabular}{|c||c|c|c|c|}
\toprule
Mass & OS1 & OS2 & SS & 3l \\
\midrule
500 GeV  & 13.1 (-22\%) & 14.8 (-58\%) & 16.1 (-24\%) & 15.1 (-21\%) \\
600 GeV  & 6.6  (-26\%) & 7.5  (-55\%) & 5.6  (-25\%) & 6.1  (-28\%) \\
700 GeV  & 3.2  (-20\%) & 3.1  (-53\%) & 2.0  (-29\%) & 2.6  (-16\%) \\
800 GeV  & 1.4  (-13\%) & 1.2  (-52\%) & 0.7  (-30\%) & 1.0  (-23\%) \\
\bottomrule                                
\end{tabular}
\caption[Number of signal events in multi-lepton channels for various masses of a $t^\prime$ with BR$(Wb)=50\%$ and BR$(Zt)={\rm BR}(Ht)=25\%$]{Number of signal events in the multi-lepton channels for various masses of a $t^\prime$ with BR$(Wb)=50\%$ and BR$(Zt)={\rm BR}(Ht)=25\%$. In parenthesis, the relative discrepancy with the number of events quoted in \cite{Chatrchyan:2013uxa} and also reported in Tab.~\ref{tab:XQ-validationlimitcode}.}
\label{tab:XQ-efficiencytable}
\end{table}

\subsection{Comparison with the experimental results}

After having validated both the expression for the calculation of the eCLs and the extraction of the experimental efficiencies, we now need to compare our final results, obtained using the NLO QCD pair production cross section supplemented by the next-to-next-to-leading-logarithmic (NNLL) resummation \cite{Cacciari:2011hy}, with those presented by the CMS collaboration for different choices of the $t^\prime$ branching ratios into SM bosons and third generation quarks.

\begin{figure}[!h]
\centering
\epsfig{file=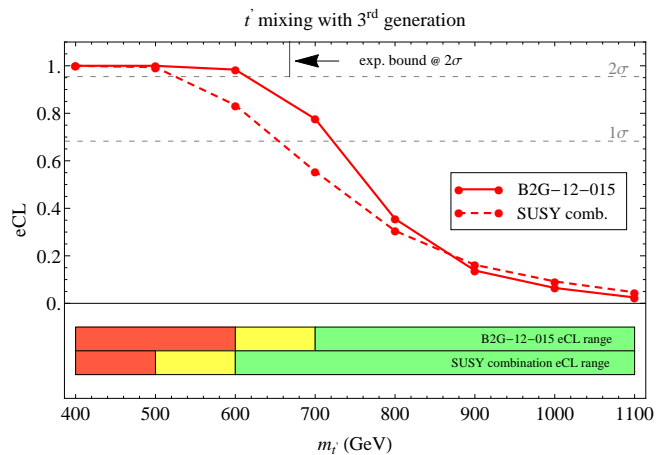, width=.58\textwidth}
\caption[Exclusion confidence levels for a $t^\prime$ mixing only with the third generation of quarks with BR$(Zt)={\rm BR}(Ht)=0.25$ and BR$(Wb)=0.5$]{Exclusion confidence levels for a $t^\prime$ mixing only with the third generation of quarks with BR$(Zt)={\rm BR}(Ht)=0.25$ and BR$(Wb)=0.5$. The dots 
correspond to the simulated points, while the lines are linear interpolations of the eCLs. The solid line corresponds to the eCLs obtained using the direct search \cite{Chatrchyan:2013uxa}, while the dashed line corresponds to the combination of the SUSY searches at $\sqrt{s}=7$ and 8 TeV. Below the plot the red region is excluded at 95\% CL, the yellow region is where the $2\sigma$ eCL can be found, the green region is not excluded at 95\% CL.}
\label{fig:XQ-validationTsingletthirdgen}
\end{figure}

Firstly we show in Fig. \ref{fig:XQ-validationTsingletthirdgen} the eCL for a $t^\prime$, in function of its mass, again for the nominal point with BR$(Zt)={\textrm{BR}}(Ht)=0.25$ and BR$(Wb)=0.5$, for which, linearly interpolating the eCLs between the simulated points represented by the dots, we obtain a $2\sigma$ mass bound of 614 GeV to be compared with the 668 GeV limit for the multi-lepton channel only that is quoted in \cite{twikiB2G12015}.
The colour code for the strips below the plot is the following: in red and green are the regions where both the configurations with the extreme masses are 
excluded or allowed at 95\% CL respectively, while in yellow are the regions where one configuration is allowed, the one with higher mass, and one excluded, the one with lower mass, allowing us to claim that the 2$\sigma$ bound can be found in the yellow range.
The motivation for the explicit presentation of the colour code below the plot is due to the fact that the most accurate results are those obtained for 
the simulated masses, which we remember are from 400 to 2000 GeV in steps of 100 GeV, while for intermediate masses we don't have direct information about the experimental efficiencies.
Besides linearly interpolating the eCLs between the simulated mass points another possibility exists in order to provide an eCL for a given scenario, which is to linearly interpolate the efficiencies and cross sections between the simulated masses. A detailed discussion of two methods, with the respective pros and cons, is reported in Appendix A of \cite{Barducci:2014ila}, to which we refer for further details.
In the plot we also show, with the dashed red line, the eCL obtained from the combination of the 7 and 8 TeV SUSY searches for which, again interpolating the eCLs between the simulated points, we can claim a 2$\sigma$ exclusion at a mass of 525 GeV.
We clearly see then that the direct search is, of course, more sensitive to this particular scenario but that the the mass bound obtained from the SUSY searches combination is not too far from it, which is remarkable since these analyses are not designed to be sensitive to this kind of final state and this 
could be important, as we will show, in scenarios were the extra quarks can also decay  in light generations for which direct searches, that usually requires $b$-jets into the final state, might not be sensitive. However we have to point out that progress on this front is currently underway in the 
experimental community in both designing searches and interpreting results in the context of models where the extra quarks can have couplings also to the light generations of SM quarks.

\begin{figure}[!h]
\centering
\epsfig{file=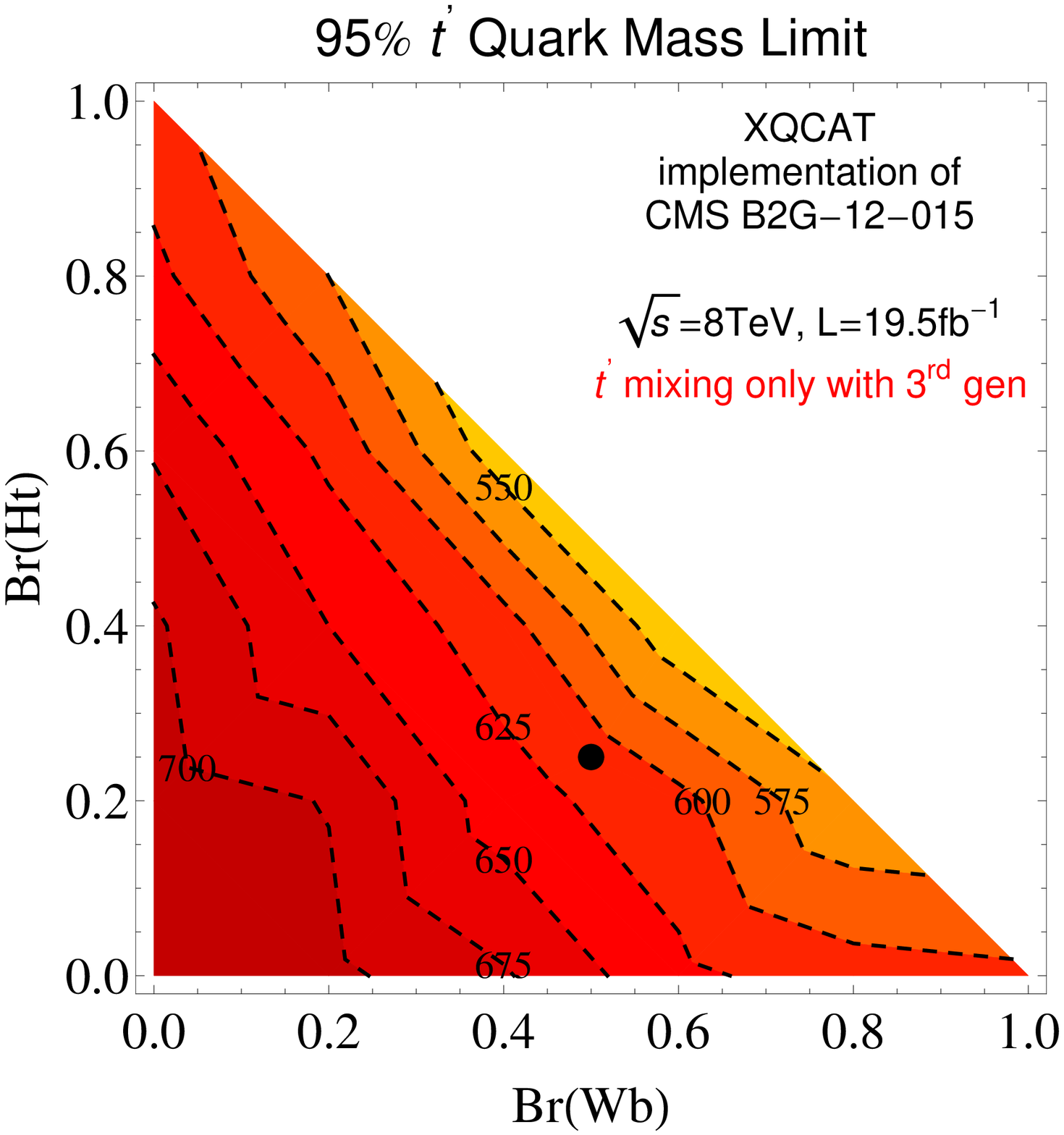, width=.44\textwidth}{(a)}\hfill
\epsfig{file=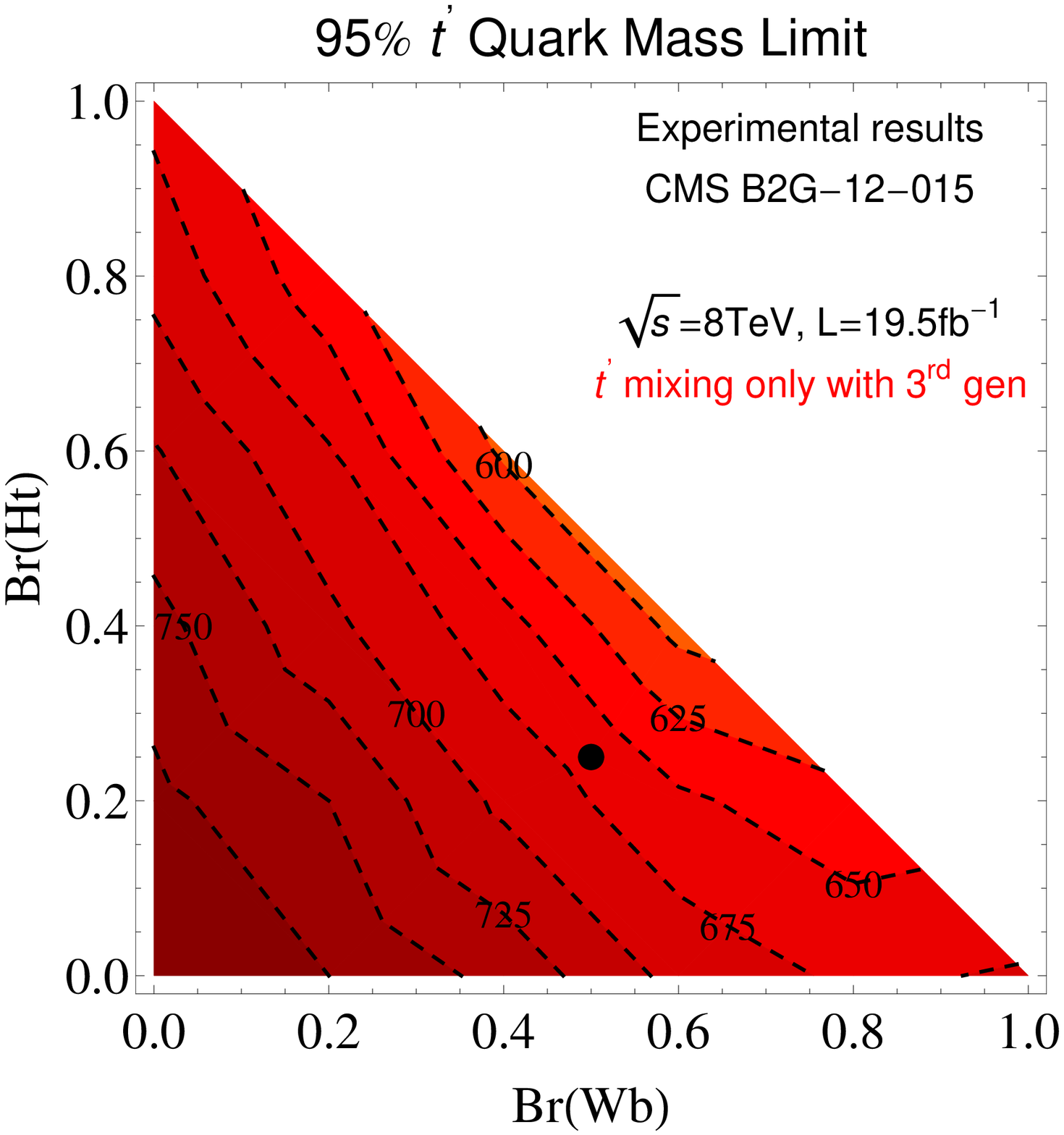, width=.44\textwidth}{(b)}\\
\epsfig{file=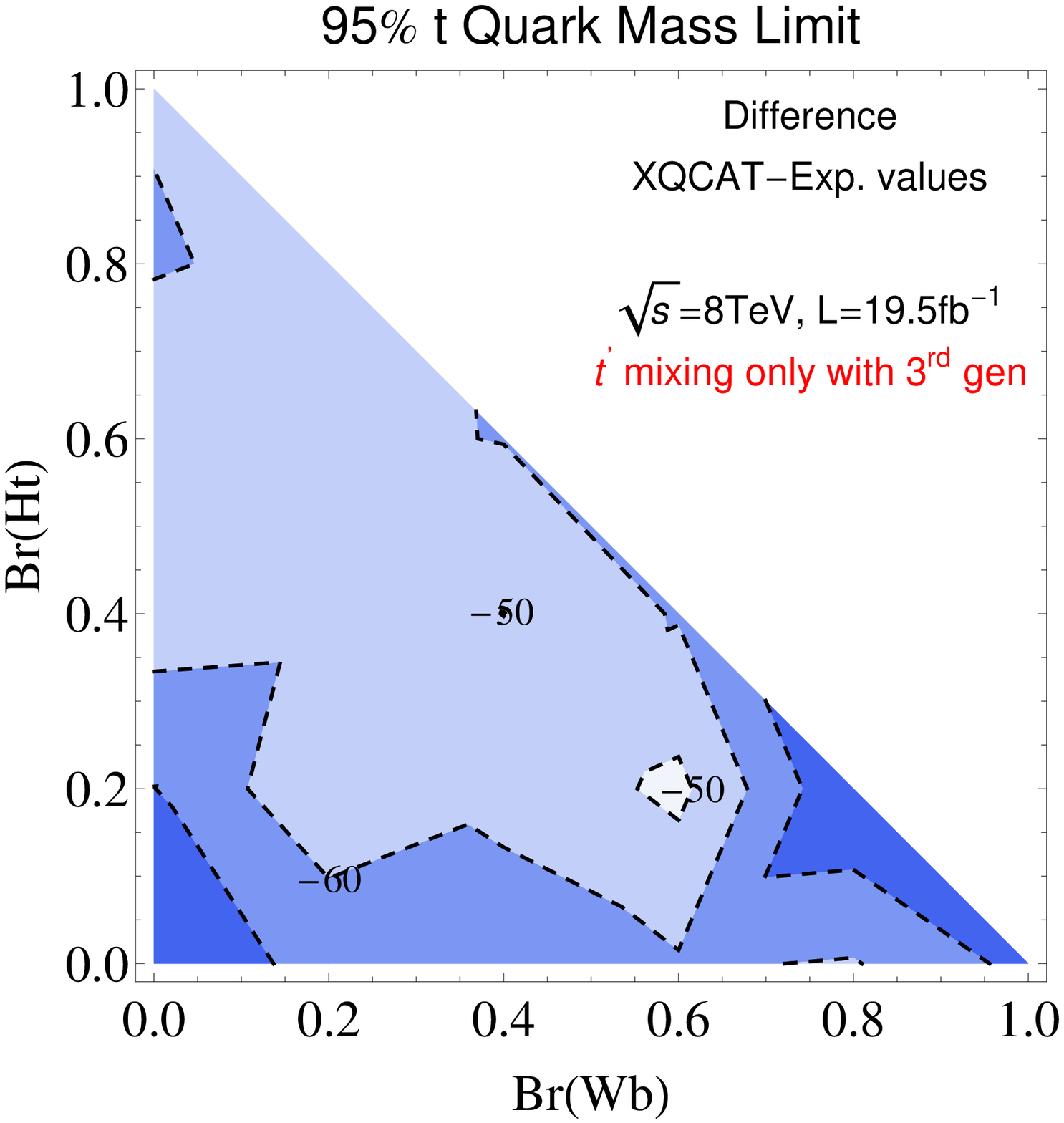, width=.44\textwidth}{(c)}\\
\caption[95\% eCL mass limit for a $t^\prime$ mixing only with the third generation of quarks with variable branching ratios.]{95\% eCL mass limit for a $t^\prime$ mixing only with the third generation of quarks with variable branching ratios for the results obtained with our code (a) and the experimental 
results of \cite{Chatrchyan:2013uxa} (b), while the difference between the two are reported in (c). The black dot represent the point with the combination of branching ratios BR$(Zt)={\rm BR}(Ht)=0.25$ and BR$(Wb)=0.5$.}
\label{fig:XQ-validationtriangle}
\end{figure}

We now allow two of the $t^\prime$ branching ratios to vary between 0 and 1 and we calculate the corresponding eCLs for which we show in Fig.~\ref{fig:XQ-validationtriangle} (a) the 95\% CL mass limit contours in the ${\rm BR}(t^\prime\to Wb)$ ${\rm BR}(t^\prime\to Ht)$ plane, again obtained by linearly 
interpolating the eCLs computed for the simulated mass points, together with the results provided by the CMS collaboration, Fig.~\ref{fig:XQ-validationtriangle} (b).
From the plots we can observe the remarkable similarity of the contour lines of our results with respect to the experimental ones in almost all the regions of branching ratios and, from Fig.~\ref{fig:XQ-validationtriangle} (c), that the mass bounds obtained from our analysis are consistent with the CMS ones within 60 GeV for most of the branching ratio configurations.

\section{Results}

\subsection{Direct searches of \emph{top partners}}

Having validated our framework and compared our results with the experimental ones in the case of a single extra quark, we now want to apply our tool to analyse scenarios where more than one extra state is present in the spectrum.
In this case a given bin of a search can receive contributions from different physical states in the following two cases
\begin{itemize}
\item[-] the model contains states with the same decay channels which therefore produce the same final state
\item[-] the model contains states with different decay channels, however, the selection cuts are (even partially) sensitive to the different final states
\end{itemize}
that are automatically taken into account by \verb|XQCAT| by using eq.(\ref{eq:XQCAT-ev}) generalised to an arbitrary number of extra quarks.

\begin{figure}[!h]
\centering
\epsfig{file=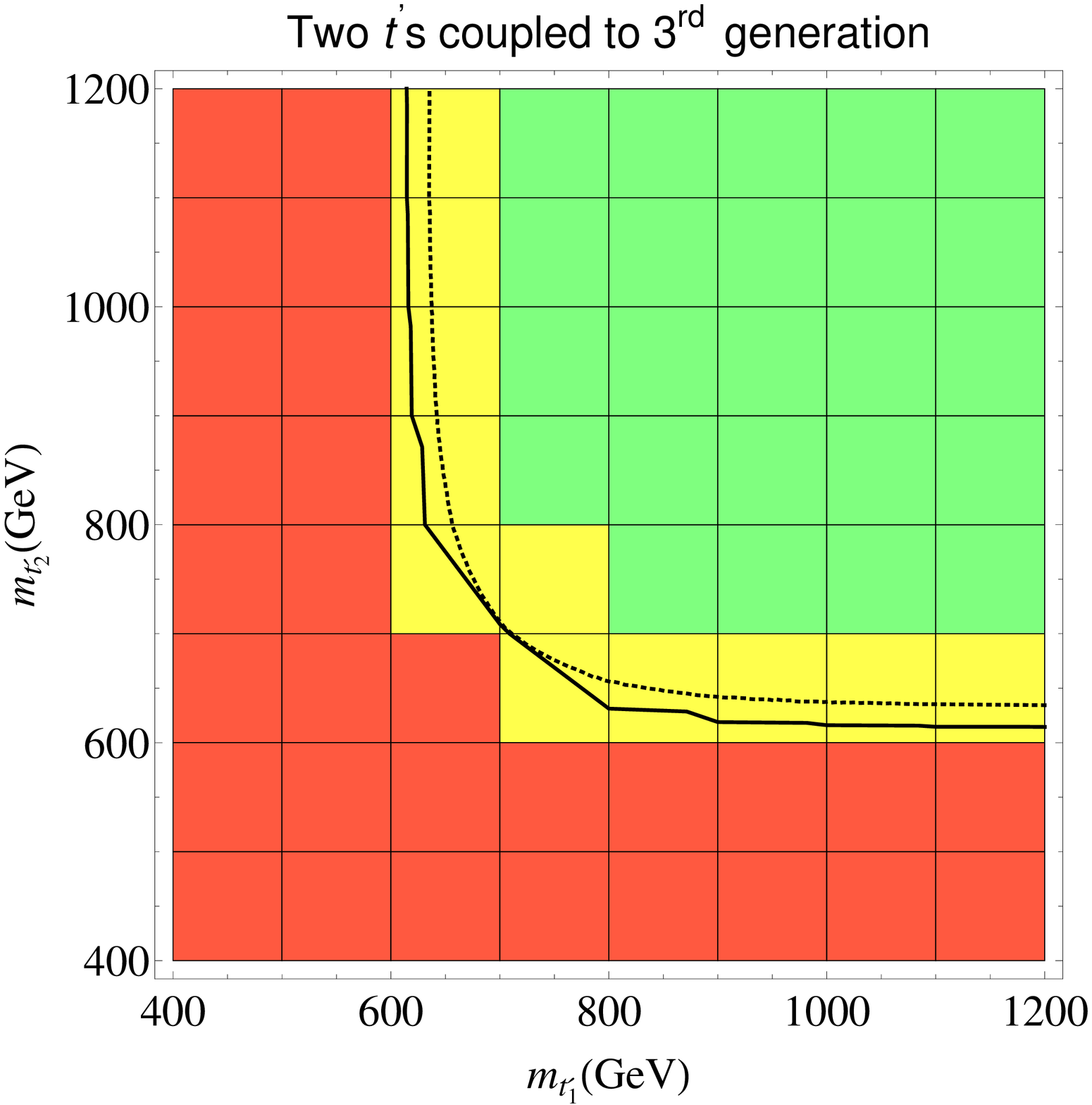, width=.44\textwidth}{(a)}\hfill
\epsfig{file=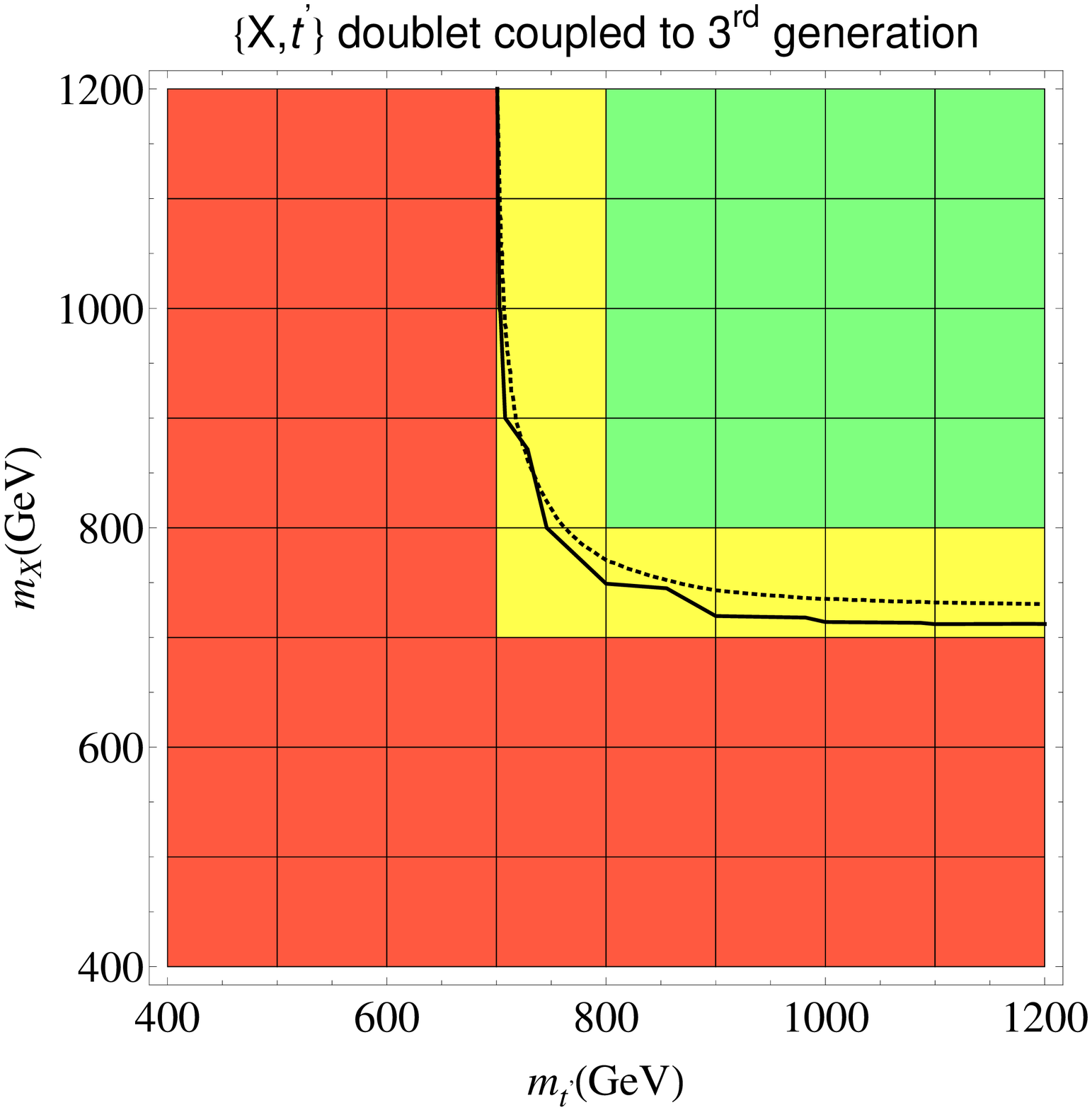, width=.44\textwidth}{(b)}\hfill
\caption[Exclusion confidence level for two simplified multi extra quark configurations]{Exclusion confidence level calculated using the CMS direct search for top partner \cite{Chatrchyan:2013uxa} for a configuration with two $t^\prime$s (a) with BR$(Zt)={\rm BR}(Ht)=0.25$ and BR$(Wb)=0.5$ or one $t^\prime$ with 
the same branching ratios and one $X$ quark with 100\% branching ratio into $W^+t$ (b). The excluded, red, non excluded, green, and boundary, yellow, regions at 95\% CL are shown. The solid black line corresponds to the 95\% CL obtained from a liner interpolation of the eCLs between the simulated points, while the dashed black line to the one obtained by linearly interpolating efficiencies and cross sections.}
\label{fig:XQ-multi-toy-3rd}
\end{figure}

As a first step we apply our tool to two simplified scenarios, one containing two $t^\prime$s, both with the same branching ratios used in the previous subsection, and one containing one $t^\prime$, again with the same decay rates as before, and a $X$ quark that decays 100\% into $W^+t$.
These configurations can arise, for example, in the case of the introduction of two singlet representations of $SU(2)_L$ with hypercharge 2/3 or a doublet 
with hypercharge 7/6, in which the extra quarks are embedded (see again \cite{Okada:2012gy} for a review) and the numerical implementation of these two scenarios can be found in \cite{FeynRules,FeynRulesVLQ} and \cite{hepmdb,hepmdbVLQ}.
In both cases we have calculated the eCLs given by the direct search for top partners by varying independently the extra quark masses \{$m_{t^\prime_1}$,$m_{t^\prime_2}$\} and \{$m_{t^\prime}$,$m_{X}$\} although the second case 
might not correspond to a physically realistic situation since the mass splitting between these two quarks that usually appear in the $SU(2)_L$ doublet can be only generated via mixing with the Higgs boson, therefore giving rise to a small mass difference.
For the first case the results are shown in Fig.~\ref{fig:XQ-multi-toy-3rd} (a) where the crossing points in the
grid correspond to the masses that we have simulated. Shown in the plot are the red squares, whose corners
are all excluded at the 95\% confidence level and the green ones, whose corners are all allowed, while in yellow
are the squares that present some corners allowed and some excluded so that, in the same spirit of the discussion in the case of just one $t^\prime$, we can then affirm that the exclusion limit should be a line crossing the yellow squares.
This is shown in the plot by the black solid line, which is obtained by interpolating the eCLs between simulated mass points, as is usually done in the experimental analysis, using the inverse distance weighted (IDW) interpolation \cite{IDW} described in Appendix A of \cite{Barducci:2014ila}, since we are dealing with a 2-dimensional function, while with the black dashed line we illustrate the results obtained interpolating efficiencies and cross sections.
The plot displays interesting physical results: we observe in fact that  when the mass gap between the two quarks is less than, say, 200 GeV, the obtained bound is more stringent than in the case of just one extra quark in the spectrum, up to the point that the eCL can be found in the 700-800 GeV mass range 
in this quasi degenerate case. In contrast, in the case where one of the two $t^\prime$s is assumed to be way heavier than the other, the mass bound coincides with the one previously obtained for the case of the single extra top.

We then study the case where a quark with a different charge with respect to the one for which the search is designed is present in the spectrum. The exotic quark with charge 5/3 has been searched for at the LHC from 
both ATLAS  and CMS in the SS di-lepton channel, which gives a bound of 670 \cite{ATLAS:2012hpa} and 800 \cite{Chatrchyan:2013wfa} GeV respectively, assuming 100\% of decay into $W^+t$. The plot in Fig.~\ref{fig:XQ-multi-toy-3rd} (b) again shows the interplay between the two quarks in giving rise to a higher bound where the mass splitting is relatively small and also the interesting result that the CMS analysis dedicated to the search of a $t^\prime$ state is actually sensitive to the $X$ state and able to set bounds on its mass.

We now move on to apply our tool to a physical motivated scenario and show the bound on its parameters space given by our code.
The framework that we consider is a model for a pNGB boson Higgs based on the $SO(5)/SO(4)$ breaking pattern recently presented in \cite{DeSimone:2012fs}.
In this work the authors assume that the \emph{top partners} belong to an $SO(4)$ bidoublet and singlet; since the former is typically expected to be lighter than the latter we will study the case of the bidoublet for illustrating our results.
This fourplet can be decomposed in the $SU(2)_L\otimes U(1)_Y$ doublets \{$T$,$B$\} and \{$X_{5/3}$,$X_{2/3}$ \}, where we are using, for consistency, the notation chosen by the authors of \cite{DeSimone:2012fs}, labelling then with $T$ and $B$ the 2/3 and $-1/3$ charged quarks belonging to the $SU(2)_L$ 
doublet with SM like hypercharge, 1/6, and with $X_{5/3}$ and $X_{2/3}$ the 5/3 and 2/3 charged quarks belonging to the $SU(2)_L$ doublet with exotic hypercharge, 7/6.
These states are assumed to decay exclusively into third generation SM quarks with the following branching ratios
\begin{equation}
\begin{split}
& {\rm{BR}}(X_{5/3} \rightarrow W^{+}t)={\rm BR}(B\rightarrow W^{-}t)=100\%,\\
& {\rm BR}(X_{2/3}\rightarrow Zt)={\rm BR}(X_{2/3}\rightarrow Ht)=50\%, \\ 
& {\rm BR}(T\rightarrow Zt)={\rm BR}(T\rightarrow Ht)=50\%.
\end{split}
\end{equation}

\begin{figure}[!h]
\centering
\epsfig{file=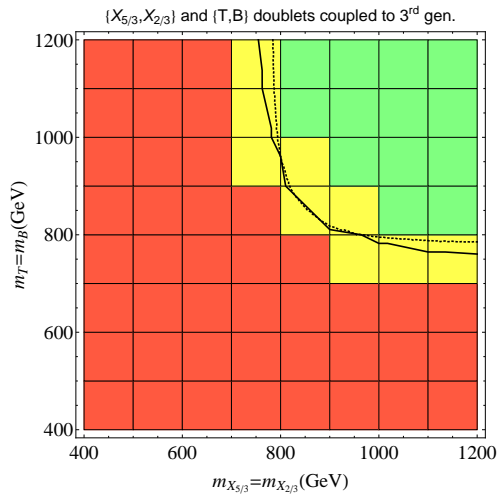, width=.44\textwidth}
\caption[Exclusion confidence level for the composite Higgs model with an $SO(4)$ bidoublet]{Exclusion confidence level calculated using the CMS direct search for top partner \cite{Chatrchyan:2013uxa} for the model with the the $SO(4)$ bidoublet of \cite{DeSimone:2012fs}. The excluded, red, non excluded, green, and 
boundary, yellow, regions at 95\% CL are shown. The solid black line corresponds to the 95\% CL obtained from a liner interpolation of the eCLs between the simulated points, while the dashed black line to the one obtained by linearly interpolating efficiencies and cross sections.}
\label{fig:XQ-multi-chm-3rd}
\end{figure}

For our study we assume a common mass for the states belonging to the two doublets which is indeed a natural approximation, as previously explained, since 
the mass splitting between the two quarks can only be generated by the Higgs VEV and is therefore relatively small with respect to the mass scales of the extra quarks, while the mass difference between the two doublets is generally large, as argued by the authors of \cite{DeSimone:2012fs}.
However for our purpose we have decided to independently vary the mass parameters of the two doublets in the 400$-$2000 GeV range and we show our results in Fig. \ref{fig:XQ-multi-chm-3rd}.
From the plot we observe that for this scenario with four extra quarks in the spectrum the bounds are more stringent as in the previous examples.
Moreover, in the region of small splitting between the two doublets the bound can reach the 900-1000 GeV mass range while it can be found between 700 and 800 GeV when one of the two doublets is much heavier than the other.

\subsection{Complementarity with other searches}

As already noticed in Fig.~\ref{fig:XQ-validationTsingletthirdgen}, the mass bound that can be obtained from the re-interpretation and combination of the SUSY analysis implemented in our tool is not too far from the one obtained from the dedicated experimental analysis and in this subsection we will show 
that SUSY inspired searches are able to set bounds on scenarios where the extra quarks can also decay  into the light quark generations.
From a phenomenological point of view this case has recently received a great deal of attention due to the fact that it can potentially give rise to large single production cross sections \cite{Atre:2011ae} and that it has been shown that flavour bounds might not disfavour cases where significant mixing with both the top and the 
up or charm quark are turned on \cite{Cacciapaglia:2011fx}. Nevertheless at the time being no specific searches focused on QCD pair production followed by decays to light jets are available and we will show the potentiality of SUSY inspired searches in constraining  these scenarios.

\begin{figure}[!h]
\centering
\epsfig{file=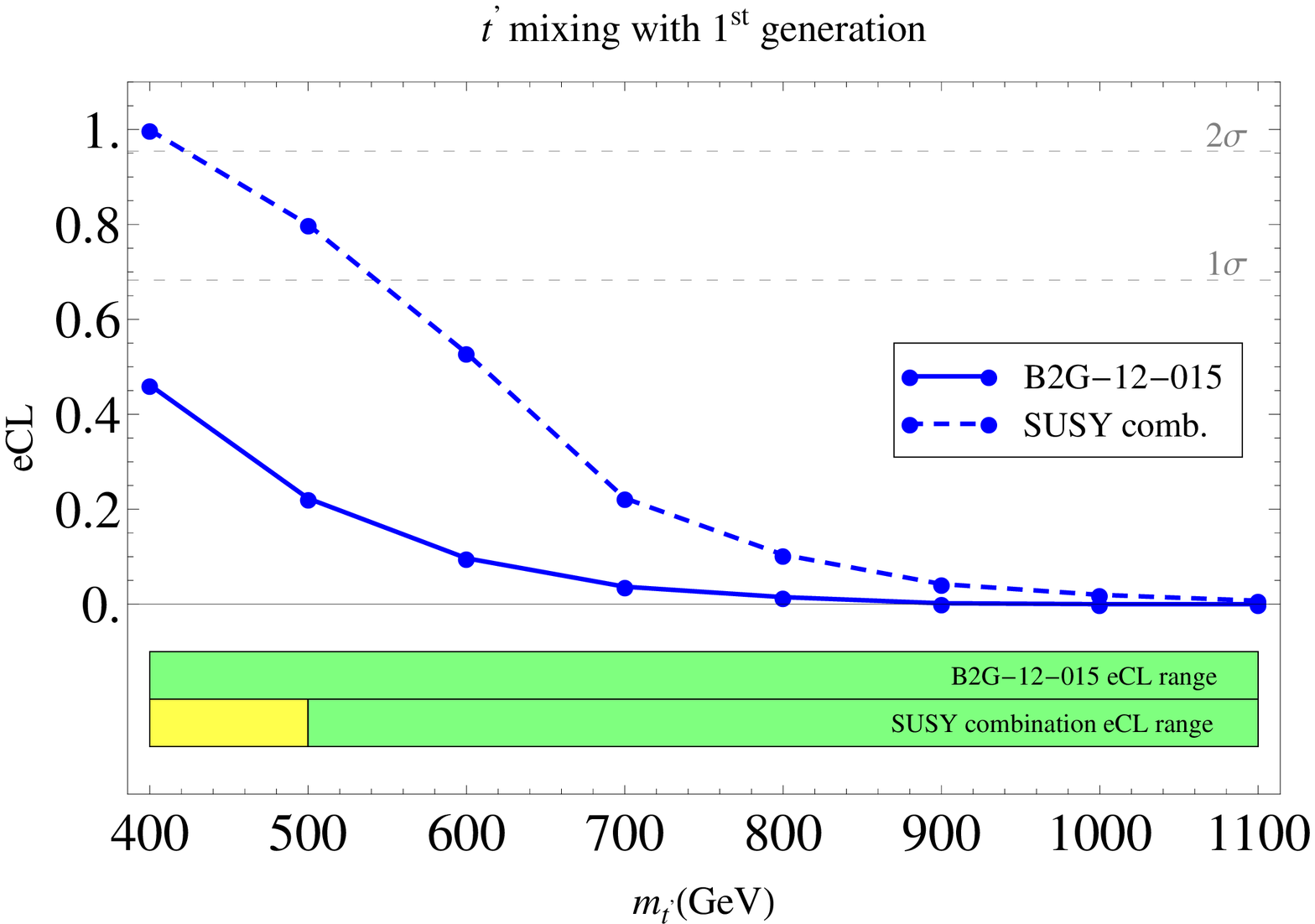, width=.48\textwidth}\hfill
\epsfig{file=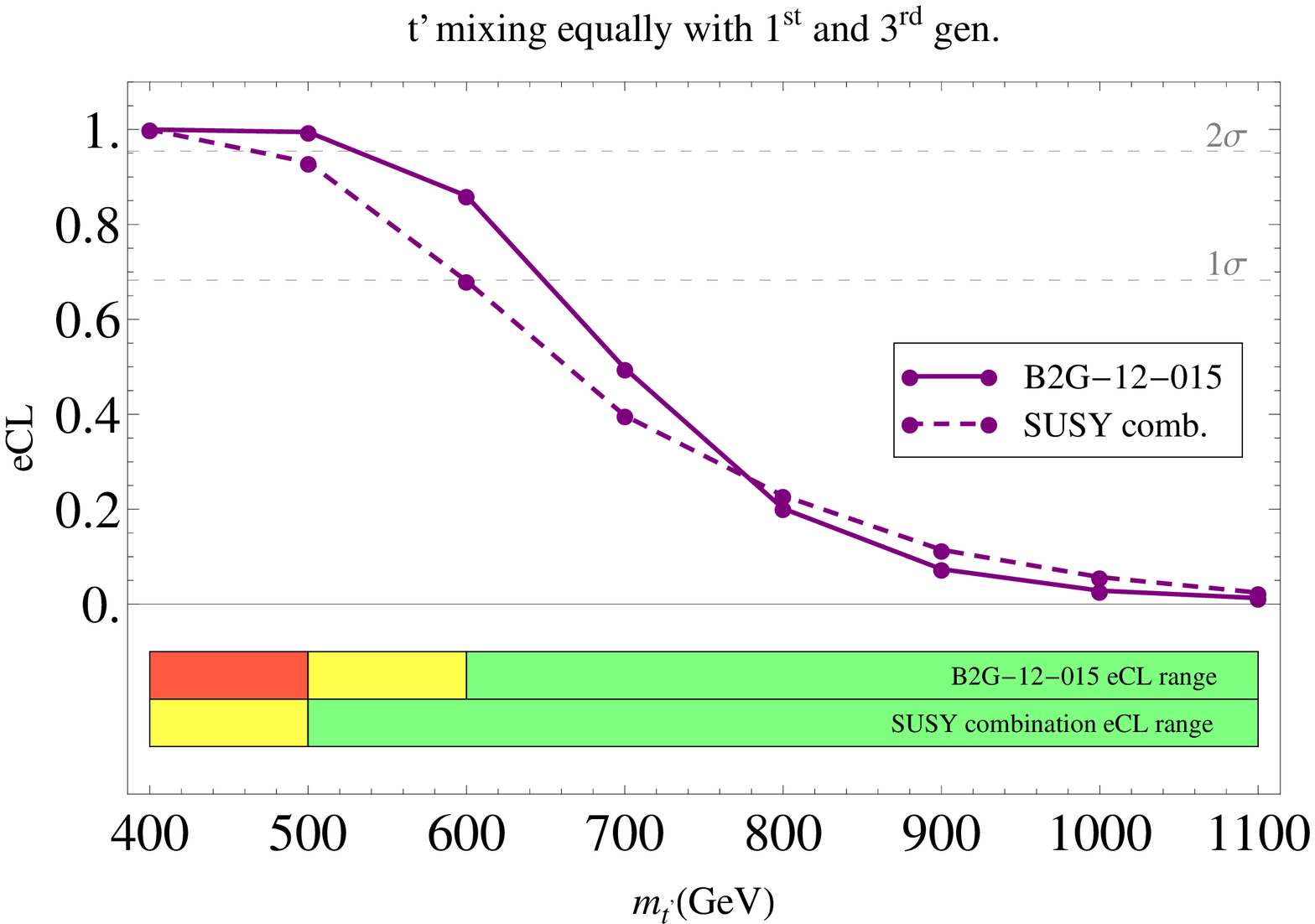, width=.48\textwidth}
\caption[Exclusion confidence levels for a $t^\prime$ mixing also with light quark generations]{Exclusion confidence levels for a $t^\prime$ mixing only with also with light quark generations with BR$(Zu)={\rm BR}(Hu)=0.25$, BR$(Wd)=0.5$ (a) and BR$(Zu)={\rm BR}(Zt)={\rm BR}(Hu)={\rm BR}(Ht)=0.125$ and BR$(Wd)={\rm BR}(Wb)=0.25$ (b).
The dots correspond to the simulated points, while the lines are linear interpolations of the 
eCLs. The solid line corresponds to the eCLs obtained using the direct search \cite{Chatrchyan:2013uxa}, while the dashed line corresponds to the combination of the SUSY searches at $\sqrt{s}=7$ and 8 TeV. Below the plots the red region is excluded at 95\% CL, the yellow region is where the $2\sigma$ 
eCL can be found, the green region is not excluded at 95\% CL.}
\label{fig:XQ-Tsingletlightgen}
\end{figure}

We start by showing this again for the case of just a $t^\prime$ quark mixing exclusively with the first generation of quarks, BR$(Zu)={\rm BR}(Hu)=0.25$, BR$(Wd)=0.5$, and equally with the first and third generation of quarks, BR$(Zu)={\rm BR}(Zt)={\rm BR}(Hu)={\rm BR}(Ht)=0.125$ and BR$(Wd)={\rm BR}(Wb)=0.25$, in Fig.~\ref{fig:XQ-Tsingletlightgen} (a) and (b) respectively.
Following the same method of interpolating the eCLs between the simulated points we observe that the sensitivity of the direct search is strongly reduced when the mixing with the light generations is turned on, which is
not a surprise since the lack 
of the top quark arising from the decay of the $t^\prime$s causes a reduction of $b$-jets and leptons to which the search is sensitive. It is remarkable however that the combinations of the SUSY searches can set a bound on the mass of the extra quark between 400 and 500 GeV in case of exclusive mixing with light quarks, and that this bound is stronger that the one obtained by the re-interpretation of the direct search for which no bound above 400 GeV is found.

Finally we apply our tool for the same physical motivated scenario based on the pNGB Higgs analysed before with the difference that the extra quarks are now  assumed to couple just to the light quark generations, that is, assuming the following branching ratios for the states belonging to the $SO(4)$ bi-doublet
\begin{equation}
\begin{split}
& {\rm{BR}}(X_{5/3} \rightarrow W^{+}u)={\rm BR}(B\rightarrow W^{-}u)=100\%,\\
& {\rm BR}(X_{2/3}\rightarrow Zu)={\rm BR}(X_{2/3}\rightarrow Hu)=50\%, \\ 
& {\rm BR}(T\rightarrow Zu)={\rm BR}(T\rightarrow Hu)=50\%.
\end{split}
\end{equation}

\begin{figure}[!h]
\centering
\epsfig{file=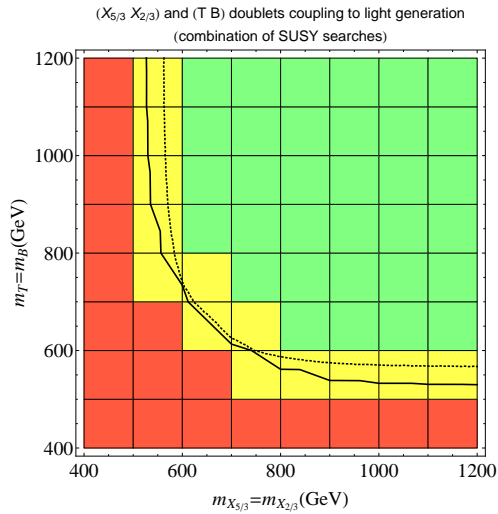, width=.44\textwidth}
\caption[Exclusion confidence level for the composite Higgs model with an $SO(4)$ bidoublet assuming mixing just with light quark generations]{Exclusion confidence level calculated using the combination of SUSY searches for the model with the the $SO(4)$ bidoublet of \cite{DeSimone:2012fs} assuming mixing just with the light quark generations. The excluded, red, non excluded, green, and boundary, yellow, regions at 95\% CL are shown. The solid black line corresponds to the 95\% CL obtained from a liner interpolation of the eCLs between the simulated points, while the dashed black line to the one obtained by linearly interpolating efficiencies and cross sections.}
\label{fig:XQ-multi-chm-1st}
\end{figure}

The re-interpretation of the direct search for the extra $t^\prime$ quarks doesn't allow us to set any bound
above 400 GeV and so we only show in Fig.~\ref{fig:XQ-multi-chm-1st} the limits arising from the combinations of the SUSY searches, from which we observe that in this scenario too it is possible to set a bound between 500 and 600 GeV, which reaches the 700$-$800 GeV region in case of quasi degenerate doublets.

\section{Conclusions}

In this last Chapter we have proposed and illustrated a framework for the re-interpretation of any given experimental search so as to set bounds on extra quarks that are assumed to be pair produced via QCD interactions.
Our analysis, based on a simple cut and count approach, is done by means of a code called \verb|XQCAT| that will soon be publicly available.
After having validated our framework we have shown how limits on scenarios with more than one extra quark in the spectrum can be set, using as examples both simplified and physically motivated models, without the need of any further simulation since all the physical information of the relevant experimental searches is encoded in the code database.
Moreover, in the case where the extra quarks can also mix  with the light quark generations, as for example can be the case in CHMs where the partial compositness mechanism is implemented for all three generations of quarks, we have shown how reinterpreting SUSY inspired searches also allows to set bounds on these scenarios for which, at the moment, no dedicated searches are available.

Even though our code has not yet been directly applied to the case of the 4DCHM, some comments on our analysis ought to be made.
We have shown that the presence of a higher number of extra quarks in the model can cause an increase on their mass bounds.
For the study of the gauge sector of the 4DCHM this effect has little impact, since our analyses have been performed in the \emph{small width regime} for which the extra quarks are heavy enough so as to escape more stringent bounds.
Also for the case of the Higgs sector we expect the more stringent bounds not to have a dramatic impact.
In fact the compatibility of the 4DCHM with the 7 and 8 TeV LHC data will not be spoiled by further constraining the parameter space of the model, and the modifications of the signal strengths in the case of the $e^+e^-$ study will remain accessible with respect to the expected experimental accuracies, albeit some moderate decrease could happen in the case of the $t\bar t$ Higgs associated production where the extra fermions play an important role in the signal enhancement.

\chapter{Conclusions}
\label{chap-6}
\lhead{Chapter 6. \emph{Conclusions}}

This Thesis has been devoted to the analysis of the phenomenology of a model with a composite Higgs arising as a pNGB bosons, which is a compelling alternative to SUSY in order to solve the hierarchy problem present in the SM and restore the naturalness of the EW scale.
The composite Higgs scenario that we have chosen for our investigation is the recently proposed 4DCHM which realises the Higgs state as a pNGB by means of the most economical symmetry breaking pattern also containing  a custodial symmetry: $SO(5)/SO(4)$.

In Chapter \ref{chap-2} we have explained the general idea of the (composite) Higgs boson as a pNGB and described the particle spectrum of the 4DCHM which, besides just one physical scalar, presents eight extra gauge bosons, five neutral and three charged, and twenty new fermions, both with SM and exotic electric charge.
Moreover we have illustrated how the 4DCHM has been implemented into automated tools that have been used for our phenomenological analysis.

The properties of the $Z^\prime$s and $W^\prime$s have been analysed in Chapter \ref{chap-3} where, after presenting 
the expressions for the masses and couplings of these states, we have investigated the capabilities of the 14 TeV stage of the LHC in 
discovering or excluding these states by means of DY and di-boson production processes, also paying  attention to specific features of the 4DCHM that are usually not captured by general analysis of CHMs, such as the presence of quasi degenerate neutral resonances that, in some regions of the parameter space, might be distinguishable and that can represent a hallmark signature of the model. Finally we have shown that also the production of top antitop pairs can be an important probe for this composite Higgs scenario both in cross section as well as in asymmetry studies, as already done in the case of the leptonic final states.

Chapter \ref{chap-4} has been devoted to the study of the properties of the Higgs boson of the 4DCHM.
We have summarised the status of the Higgs searches that are ongoing at the LHC and then reviewed the main production and decay mechanisms of the SM Higgs boson moving finally to the analysis of the 4DCHM Higgs. Assuming firstly a decoupling regime and then performing a phenomenological analysis considering all the effects arising from the complete model particle spectrum, again usually not captured by general analysis of CHMs, we have shown the compatibility of our framework with the 7 and 8 TeV LHC data by means of the Higgs signal strengths, giving also an example of the capabilities of the 14 TeV stage of the LHC in testing this scenario.
Motivated then by the critical decision that the physics community needs to make regarding the next generation of colliders, we have then tested the 4DCHM against future proposed $e^+ e^-$ colliders, using the ILC as a benchmark machine.
We have considered standard benchmark energies and luminosities proposed for this machine and we have compared our predictions for the Higgs signal rates with the expected precision in measuring them, for various types of Higgs production processes, highlighting the potentiality of this type of collider in testing the composite Higgs scenario.

We have then shown that the mixing and finite spectrum effects of the 4DCHM are sizeable in all chosen contexts, both in tree level as well as in loop induced processes, and that they can provide modifications of the physical observables, with respect to the case of the decoupling limit, relevant at the level of seen or expected experimental deviations.
Despite the fact that this point can only be made in a specific scenario, the 4DCHM in our case, we firmly believe that the kind of approach we have used in this Thesis is of great importance when studying CHMs with partial compositness.

Finally in Chapter \ref{chap-5} we have illustrated a framework which allows for the reinterpretation of the bounds that the experimental collaborations set on extra quarks present in many BSM models, usually obtained by assuming just one extra state besides the SM matter content, in scenarios where more than one heavy quark, with general branching ratios, is present, like in the 4DCHM (though not only).
By means of a code that we propose, called \verb|XQCAT|, we have applied our analysis both to simplified scenarios as well
as to a physical motivated model showing how the bounds on the extra quarks masses change with the presence of extra states.
Finally, allowing the extra quarks to have sizeable couplings also to the light SM quark generations, we have illustrated how the reinterpretation of SUSY searches can be used to constrain these scenarios for which no experimental results are yet available.

\section*{Acknowledgements}

I thank the NExT Institute for financial funding in the form of a NExT Ph.D. Scholarship split between the Southampton High Energy Physics group and the Particle Physics Department at the Rutherford Appleton Laboratory.

I thank the Particle Physics Department at the Rutherford Appleton Laboratory, in particular in the person of Claire Shepherd-Themistocleous, for the opportunity they gave me to spend three months at CERN collaborating with the CMS group.

\appendix
\chapter{The CCWZ prescription}
\lhead{Appendix}
\label{chap:App-A}

In this Appendix we will give some details about the most common approach used in literature for writing phenomenological lagrangians that encode a SSB described just in terms of the lightest degrees of freedom, the GBs, 
explaining the formalism due to Coleman, Callan, Wess and Zumino (CCWZ) for writing non-linear $\sigma$ models. 

Let's suppose to have a Lie Group $\mathcal{G}$ and a subgroup $\mathcal{H}$ both simple and compact.
We can divide the generator of $\mathcal{G}$ in those belonging to $\mathcal{H}$ and the remaining that we will call broken generators
\begin{equation}
T^A = \{T^a \in Lie \mathcal{H}, T^{\hat a} \in Lie \mathcal{G} - Lie \mathcal{H}\}.
\end{equation}

Every element of $g\in \mathcal{G}$ can be locally expressed as
\begin{equation}
g = \xi h, \quad  h \in \mathcal{H} 
\end{equation}
where $\xi=\xi(\pi)$ is a generic element that represents $\mathcal{G}/\mathcal{H}$ and that can be written as
\begin{equation}
\xi(\pi)=\exp(i \pi^{\hat a} T^{\hat a}/f).
\end{equation}

By multiplying both sides of the equation with an element $g\in \mathcal{G}$ we obtain the transformation law for
$\xi(\pi)$
\begin{equation}
\xi(\pi^\prime)=g\xi(\pi)h^\dag(\pi,g)
\end{equation}
that shows that the transformations properties of $\xi$ are non-linear.

We now introduce a useful quantity, the Maurer-Cartan form, defined as
\begin{equation}
\alpha_\mu=\xi^\dag(\pi)\partial_\mu\xi(\pi)
\end{equation}
that, since it has value in $\mathcal{G}$, can be decomposed in terms of the generators of $\mathcal{H}$ and
$\mathcal{G}/\mathcal{H}$
\begin{equation}
\alpha_\mu(\pi)=\alpha_\mu^\parallel(\pi)+\alpha_\mu^\perp(\pi)=\alpha_\mu^{\parallel,a}(\pi)T^a+
\alpha_\mu^{\perp,\hat a}(\pi)T^{\hat a}.
\end{equation}

The parallel and perpendicular components transform as
\begin{equation}
\begin{split}
& \alpha_\mu^\parallel(\pi^\prime)=h(\pi,g)\alpha_\mu^\parallel(\pi)h^\dag(\pi,g)+h(\pi,g)\partial_\mu h^\dag(\pi,g),\\
& \alpha_\mu^\perp(\pi^\prime)=h(\pi,g)\alpha_\mu^\perp(\pi)h^\dag(\pi,g),\\
\end{split}
\end{equation}
and it is then possible, using $\alpha_\mu^\perp(\pi)$, to write $\mathcal{G}$ invariant Lagrangian written in terms of 
just the elements of the coset that, in an effective field theory approach and at the lowest order derivative, reads
\begin{equation}
\mathcal{L}\propto f^2 Tr[\alpha_\mu^\perp(\pi)\alpha^{\mu^\perp}(\pi)].
\end{equation}

This Lagrangian depends only on the groups choice and can be specialised for the two cases relevant to our work.

In the first case we chose a $SO(N)/SO(N-1)$ coset and it is possible to prove that
\begin{equation}
f^2 Tr[\alpha_\mu^\perp(\pi)\alpha^{\mu^\perp}(\pi)]\propto f^2 (\partial_\mu \Phi)^\dag (\partial^\mu \Phi)
\end{equation}
with
\begin{equation}
\Phi=\xi(\pi)\phi_0=\exp(i \pi^{\hat a}T^{\hat a}/f)\phi_0 \quad \phi_0 = (0,0,\dots,1).
\end{equation}

In fact:
\begin{equation}
\begin{split}
  & (\partial_\mu \Phi)^T (\partial^\mu \Phi) = (\partial_\mu(\xi \phi_0))^\dag(\partial^\mu(\xi \phi_0))=\\
& =  \phi_0^T (\partial_\mu \xi)^\dag (\partial_\mu \xi) \phi_0 = \phi_0^T (\xi^\dag \partial_\mu \xi)^\dag (\xi^\dag \partial_\mu \xi)\phi_0 = \\
&= \phi_0^T \alpha_\mu^\dag \alpha^\mu \phi_0=\phi_0^T(\alpha_\mu^{\parallel,a}T^a+
\alpha_\mu^{\perp,\hat a}T^{\hat a})^\dag(\alpha_\mu^{\parallel,b}T^b+
\alpha_\mu^{\perp,\hat b}T^{\hat b})\phi_0=\\
&=\phi_0^T(\alpha_\mu^{\perp,\hat a}T^{\hat a})^\dag(\alpha_\mu^{\perp,\hat b}T^{\hat b})\phi_0=\\
&=\frac{1}{2}\alpha_\mu^{\perp,\hat a}\alpha_\mu^{\perp,\hat b}\delta^{\hat a \hat b}=\\
&=\frac{1}{2}Tr[\alpha_\mu^{\perp}\alpha_\mu^{\perp}]
\end{split}
\end{equation}
where we have used eq.(\ref{eq:so5-prop}) generalized to $SO(N)$.

In a similar way it is possible to show that for the breaking $SO(N)_L\otimes SO(N)_R/SO(N)_V$ we have
\begin{equation}
f^2 Tr[\alpha_\mu^\perp(\pi)\alpha^{\mu^\perp}(\pi)]\propto f^2 Tr[(\partial_\mu \Omega)^\dag(\partial_\mu \Omega)]
\end{equation}
where $\Omega$ is an element of $SO(N)_V$ that parametrises the GBs of this coset.

\chapter{Algebra of $SO(5)$}
\label{chap:App-B}

In a convenient basis the generators of $SO(5)$ are given by the following matrices
\begin{equation}
t^{ab}_{ij}=\delta^a_i\delta^b_j-\delta^a_j\delta^b_i \quad a,b,i,j=1,...,5
\end{equation}
for which the algebra is the following
\begin{equation}
[t^{ab},t^{cd}]=\delta^{ad}t^{bc}+\delta^{bc}t^{ad}-\delta^{ac}t^{bd}-\delta^{bd}t^{ac}.
\end{equation}

Among the 10 generators of $SO(5)$ we can identify the 6 generators of $SO(4)\simeq SU(2)_L\otimes SU(2)_R$ and in the remaining 4 of the coset $SO(5)/SO(4)$
\begin{equation}
T^a=\left\{ T^a_L,T^a_R \in SO(4),T^{\hat a} \in SO(5)/SO(4)\right\}
\end{equation}
with $a=1,2,3$ and $\hat a=1,2,3,4$ giving now the following commutations rules
\begin{equation}
\begin{alignedat}{3}
& [T^a_L,T^b_L]=i \epsilon^{a,b,c}T^c_L, &&\qquad  [T^a_R,T^b_R]=i \epsilon^{a,b,c}T^c_R, && \qquad [T^a_L,T^b_R]=0,\\
& [T^{\hat a},T^{\hat b}]=\frac{i}{2}\epsilon^{a \hat b \hat c}(T^c_L+T^c_R) , && \qquad [T^{\hat a}, T^{\hat 4}]=\frac{i}{2}(T^a_L-T^a_R), &&\\
& [T^a_{L,R},T^{\hat b}]=\frac{i}{2}(\epsilon^{a,b,c}T^{\hat c}\pm\delta^{ab}T^{\hat 4}), && \qquad [T^a_{L,R},T^{\hat 4}]=\mp \frac{i}{2}T^{\hat a}. &&
\end{alignedat}
\end{equation}

The explicit expressions of the generators with the normalization condition
\begin{equation}
Tr[T^aT^B]=\delta^{AB}
\label{eq:so5-normalizz}
\end{equation}
are
\begin{equation}
T^1_{L,R}=
-\frac{i}{2}
\left(
\begin{array}{ccccc}
0&0&0&\pm1&0\\
0&0&+1&0&0\\
0&-1&0&0&0\\
\mp1&0&0&0&0\\
0&0&0&0&0\\
\end{array}
\right),\quad
T^2_{L,R}=
-\frac{i}{2}
\left(
\begin{array}{ccccc}
0&0&-1&0&0\\
0&0&0&\pm1&0\\
+1&0&0&0&0\\
0&\mp1&0&0&0\\
0&0&0&0&0\\
\end{array}
\right),
\quad
\nonumber
\end{equation}

\begin{equation}
T^3_{L,R}=
-\frac{i}{2}
\left(
\begin{array}{ccccc}
0&+1&0&0&0\\
-1&0&0&0&0\\
0&0&0&\pm1&0\\
0&0&\mp1&0&0\\
0&0&0&0&0\\
\end{array}
\right),
\nonumber
\end{equation}
\begin{equation}
T^{\hat{1}}=
-\frac{i}{\sqrt{2}}
\left(
\begin{array}{ccccc}
0&0&0&0&+1\\
0&0&0&0&0\\
0&0&0&0&0\\
0&0&0&0&0\\
-1&0&0&0&0\\
\end{array}
\right),
\quad
T^{\hat{2}}=
-\frac{i}{\sqrt{2}}
\left(
\begin{array}{ccccc}
0&0&0&0&0\\
0&0&0&0&+1\\
0&0&0&0&0\\
0&0&0&0&0\\
0&-1&0&0&0\\
\end{array}
\right),
\nonumber
\end{equation}

\begin{equation}
T^{\hat{3}}=
-\frac{i}{\sqrt{2}}
\left(
\begin{array}{ccccc}
0&0&0&0&0\\
0&0&0&0&0\\
0&0&0&0&+1\\
0&0&0&0&0\\
0&0&-1&0&0\\
\end{array}
\right),
\quad
T^{\hat{2}}=
-\frac{i}{\sqrt{2}}
\left(
\begin{array}{ccccc}
0&0&0&0&0\\
0&0&0&0&0\\
0&0&0&0&0\\
0&0&0&0&+1\\
0&0&0&-1&0\\
\end{array}
\right).
\end{equation}

Finally, some useful properties of these generators are
\begin{equation}
\begin{split}
& \phi_0^T T^{\hat a}T^{\hat b} \phi_0=\frac{1}{2}\delta^{\hat a\hat b},\\
& \phi_0^T T^aT^{\hat b} \phi_0=\phi_0^T T^{\hat a}T^b \phi_0=\phi_0^T T^a T^b \phi_0 = 0.\\
\end{split}
\label{eq:so5-prop}
\end{equation}


\chapter{Benchmark points}
\label{chap:App-C}
In this Appendix we present the benchmark points of the model used for the analysis of the gauge sector in Chapter \ref{chap-3}, splitting them into benchmarks with fixed model scale $f$ and coupling constant $g_\rho$, for which we present various configurations or the fermionic parameters that correspond to various widths \emph{regimes}, and benchmarks with different values of $f$ and $g_\rho$, for which we present only points belonging to the \emph{small width regime}.

In presenting these benchmarks we refer to the parameters that have to be fed into CalcHEP as an input. Since the algorithms for the constrained numerical diagonalisation of the mass matrices and the computation of the Higgs potential have not been implemented into CalcHEP, we also need to give as an input to the program parameters that are in principle not free, such as $g_0$, $g_{0Y}$, $\langle h\rangle$ and $m_H$.
For this reason the list of the input parameters is the following
\begin{equation}
f,g_\rho,g_0,g_{0Y},\langle h \rangle, \Delta_{t,b/L,R},Y_{T,B},m_{Y_{T,B}}, m_H.
\end{equation}

\section{Benchmarks with $f$=1200 GeV and $g_\rho$=1.8}
\label{app-bench-fgfixed}

These input parameters are related to benchmark points with a fixed model scale $f$ and coupling constant $g_\rho$, chosen to be $f=1200$ GeV and $g_\rho$=1.8, that have been used in order to show the impact of the fermionic parameter of the model, and in particular of the mass of the lightest extra fermion, on the analysis of the gauge sector of the 4DCHM performed in Chapter~\ref{chap-3}.

\begin{table}[htb]
\begin{tabular}{l l||l| l|}
\hline
\multicolumn{1}{|c|}{f  }				& 1200		 & $m_*$  	     	&	 2219	 \\ 
\multicolumn{1}{|c|}{$g_\rho$} 				&	1.8  	 & $\Delta_{t_L}$  	&     	 2366	\\
\multicolumn{1}{|c|}{$g_0$} 				&0.69		 &  $\Delta_{t_R}$ 	& 2245		\\
\multicolumn{1}{|c|}{$g_{0Y}$} 			& 0.37 		 & $Y_T$  	     	& 2824		\\
\multicolumn{1}{|c|}{$\langle h \rangle$  } &   248  	 & $M_{Y_T}$  	&	 $-1043$	\\
\cline{1-2}
								&		 & $\Delta_{b_L}$   &	 202	\\
 								&		 & $\Delta_{b_R}$   & 284		\\
 								& 		 & $Y_B$ 		& 2543		\\
 								& 		 & $M_{Y_B}$ 		&   $-1378$    	\\
 								& 		 & $m_H$  		&   125   \\
 \cline{3-4}
\end{tabular}{(a)}
\hfill
\begin{tabular}{l l||l| l|}
\hline
\multicolumn{1}{|c|}{f  }				&1200		 & $m_*$  	     	&2014 		 \\ 
\multicolumn{1}{|c|}{$g_\rho$} 				&1.8	  	 & $\Delta_{t_L}$  	& 3303     		\\
\multicolumn{1}{|c|}{$g_0$} 				&0.69		 &  $\Delta_{t_R}$ 	&3158 		\\
\multicolumn{1}{|c|}{$g_{0Y}$} 			& 0.37		 & $Y_T$  	     	&1907		\\
\multicolumn{1}{|c|}{$\langle h \rangle$  } &  248     	 & $M_{Y_T}$  	& $-647$ 		\\
\cline{1-2}
								&		 & $\Delta_{b_L}$   &493 		\\
 								&		 & $\Delta_{b_R}$   &366 		\\
 								& 		 & $Y_B$ 		&1126 		\\
 								& 		 & $M_{Y_B}$ 		& $-1884$      	\\
 								& 		 & $m_H$  		& 126    \\
\cline{3-4}
\end{tabular}{(b)}\vspace{0.5cm}
\begin{tabular}{l l||l| l|}
\hline
\multicolumn{1}{|c|}{f  }				&1200		 & $m_*$  	     	& 1908		 \\ 
\multicolumn{1}{|c|}{$g_\rho$} 				&1.8	  	 & $\Delta_{t_L}$  	&    3328  		\\
\multicolumn{1}{|c|}{$g_0$} 				&0.69		 &  $\Delta_{t_R}$ 	& 4585		\\
\multicolumn{1}{|c|}{$g_{0Y}$} 			& 0.37		 & $Y_T$  	     	& 1762		\\
\multicolumn{1}{|c|}{$\langle h \rangle$  } &  248    	 & $M_{Y_T}$  	& $-715$ 		\\
\cline{1-2}
								&		 & $\Delta_{b_L}$   &	340 	\\
 								&		 & $\Delta_{b_R}$   &414		\\
 								& 		 & $Y_B$ 		& 999		\\
 								& 		 & $M_{Y_B}$ 		&  $-725$      	\\
 								& 		 & $m_H$  		&125      \\

\cline{3-4}
\end{tabular}{(c)}
\hfill
\begin{tabular}{l l||l| l|}
\hline
\multicolumn{1}{|c|}{f  }				&1200		 & $m_*$  	     	&2031		 \\ 
\multicolumn{1}{|c|}{$g_\rho$} 				&1.8	  	 & $\Delta_{t_L}$  	&  4423    		\\
\multicolumn{1}{|c|}{$g_0$} 				&0.69		 &  $\Delta_{t_R}$ 	& 4419		\\
\multicolumn{1}{|c|}{$g_{0Y}$} 			& 0.37		 & $Y_T$  	     	& 1636 		\\
\multicolumn{1}{|c|}{$\langle h \rangle$  } &   248    	 & $M_{Y_T}$  	&$-558$		\\
\cline{1-2}
								&		 & $\Delta_{b_L}$   & 127 		\\
 								&		 & $\Delta_{b_R}$   &286 		\\
 								& 		 & $Y_B$ 		&	4543	\\
 								& 		 & $M_{Y_B}$ 		&  $-1394 $    	\\
 								& 		 & $m_H$  		&   124   \\
\cline{3-4}
\end{tabular}{(d)}\vspace{0.5cm}
\begin{tabular}{l l||l| l|}
\hline
\multicolumn{1}{|c|}{f  } & 1200 & $m_*$   & 2216\\ 
\multicolumn{1}{|c|}{$g_\rho$} & 1.8& $\Delta_{t_L}$  &2434\\
\multicolumn{1}{|c|}{$g_0$} &0.70 &  $\Delta_{t_R}$ &2362\\
\multicolumn{1}{|c|}{$g_{0Y}$} &0.37 & $Y_T$  &2771\\
\multicolumn{1}{|c|}{$\langle h \rangle$  } & 248& $M_{Y_T}$  &$-1031$\\
\cline{1-2}
& & $\Delta_{b_L}$ &327\\
 & & $\Delta_{b_R}$ &299\\
 & & $Y_B$ &2815\\
 & & $M_{Y_B}$ &$-4093$\\
 & & $m_H$  & 124   \\
\cline{3-4}
\end{tabular}{(e)}
\hfill
\begin{tabular}{l l||l| l|}
\hline
\multicolumn{1}{|c|}{f  } & 1200 & $m_*$   &1293 \\ 
\multicolumn{1}{|c|}{$g_\rho$} & 1.8& $\Delta_{t_L}$  &4714\\
\multicolumn{1}{|c|}{$g_0$} &0.70 &  $\Delta_{t_R}$ &3402\\
\multicolumn{1}{|c|}{$g_{0Y}$} &0.37 & $Y_T$  &4165\\
\multicolumn{1}{|c|}{$\langle h \rangle$  } & 248& $M_{Y_T}$  &$-1503$\\
\cline{1-2}
& & $\Delta_{b_L}$ &224\\
 & & $\Delta_{b_R}$ &480\\
 & & $Y_B$ &4260\\
 & & $M_{Y_B}$ &$-2835$\\
 & & $m_H$  &  125   \\
\cline{3-4}
\end{tabular}{(f)}
\caption[Input parameters for the benchmark points with fixed $f$ and $g_\rho$]{Input parameters for the benchmark points with fixed $f$ and $g_\rho$. Benchmark (a), (e) and (f) are the ones corresponding to the small, medium and large width regime discussed in Section~\ref{subsec:2-bench}. All the parameters, except $g_{0}$, $g_{0Y}$ and $g_\rho$, are expressed in GeV.}
\label{tab:input_bench-fgfixed}
\end{table}

\section{Benchmarks with variable $f$ and $g_\rho$}
\label{app-bench-fgvar}

These input parameters are related to benchmarks points with variable model scale $f$ and coupling constant $g_\rho$, with $f g_\rho\simeq$ 2 TeV, that we have chosen in order to show the features of our model in different regions of the parameter space. However these parameters are chosen in such a way that the widths of the extra gauge bosons are sufficiently narrow, that is in order to be in the \emph{small width regime} described in Section~\ref{subsec:2-bench}.

\begin{table}
\begin{tabular}{l l||l| l|}
\hline
\multicolumn{1}{|c|}{f  }				&	750	 & $m_*$  	     	&1673		 \\ 
\multicolumn{1}{|c|}{$g_\rho$} 				&	  2  	 & $\Delta_{t_L}$  	&    974 		\\
\multicolumn{1}{|c|}{$g_0$} 				&	0.68	 &  $\Delta_{t_R}$ 	&1780		\\
\multicolumn{1}{|c|}{$g_{0Y}$} 			&	0.37	 & $Y_T$  	     	&2442		\\
\multicolumn{1}{|c|}{$\langle h \rangle$  } &      251	 & $M_{Y_T}$  	&	$-1231$	\\
\cline{1-2}
		&	 	 & $\Delta_{b_L}$   &77		\\
		&		 & $\Delta_{b_R}$   &238		\\
		& 		 & $Y_B$ 		&2884		\\
		& 		 & $M_{Y_B}$ 		&  $-1878 $   	\\
		& 		 & $m_H$  		&   126  \\
\cline{3-4}
\end{tabular}{(a)}\hfill
\begin{tabular}{l l||l| l|}
\hline
\multicolumn{1}{|c|}{f  }				&	800	 & $m_*$  	     	&	1700	 \\ 
\multicolumn{1}{|c|}{$g_\rho$} 				&	2.5     	 & $\Delta_{t_L}$  	&     1225		\\
\multicolumn{1}{|c|}{$g_0$} 				&	0.67	 &  $\Delta_{t_R}$ 	&1391		\\
\multicolumn{1}{|c|}{$g_{0Y}$} 			&0.36		 & $Y_T$  	     	&	2770	\\
\multicolumn{1}{|c|}{$\langle h \rangle$  } &	250	 & $M_{Y_T}$  	&	$-1339$	\\
\cline{1-2}
								&		 & $\Delta_{b_L}$   & 222		\\
 								&		 & $\Delta_{b_R}$   &	99	\\
 								& 		 & $Y_B$ 		&2485		\\
 								& 		 & $M_{Y_B}$ 		&   $-1185$   	\\
 								& 		 & $m_H$  		&   125  \\
\cline{3-4}
\end{tabular}{(b)}\vspace{0.5cm}
\begin{tabular}{l l||l| l|}
\hline
\multicolumn{1}{|c|}{f  }				&	1000	 & $m_*$  	     	&	1915	 \\ 
\multicolumn{1}{|c|}{$g_\rho$} 				&	   2 	 & $\Delta_{t_L}$  	&     1503		\\
\multicolumn{1}{|c|}{$g_0$} 				&	0.69 	 &  $\Delta_{t_R}$ 	&1972		\\
\multicolumn{1}{|c|}{$g_{0Y}$} 			&0.37		 & $Y_T$  	     	&2901		\\
\multicolumn{1}{|c|}{$\langle h \rangle$  } &	249	 & $M_{Y_T}$  	&	$-1303	$\\
\cline{1-2}
								&		 & $\Delta_{b_L}$   &196		\\
 								&		 & $\Delta_{b_R}$   &187		\\
 								& 		 & $Y_B$ 		&2662		\\
 								& 		 & $M_{Y_B}$ 		&$ -984$     	\\
 								& 		 & $m_H$  		& 126    \\
\cline{3-4}
\end{tabular}{(c)}\hfill
\begin{tabular}{l l||l| l|}
\hline
\multicolumn{1}{|c|}{f  }				&	1200	 & $m_*$  	     	&  	2219 	 \\ 
\multicolumn{1}{|c|}{$g_\rho$} 				&	1.8     	 & $\Delta_{t_L}$  	&  2366 		\\
\multicolumn{1}{|c|}{$g_0$} 				&  0.69		 &  $\Delta_{t_R}$ 	& 2245  		\\
\multicolumn{1}{|c|}{$g_{0Y}$} 			& 0.37 		 & $Y_T$  	     	&  2824		\\
\multicolumn{1}{|c|}{$\langle h \rangle$  } & 248 	 & $M_{Y_T}$  	& $-1043$	  	\\
\cline{1-2}
								&		 & $\Delta_{b_L}$   & 202 	\\
 								&		 & $\Delta_{b_R}$   & 284		\\
 								& 		 & $Y_B$ 		& 2543		\\
 								& 		 & $M_{Y_B}$ 		& $-1378$    	\\
 								& 		 & $m_H$  		& 125\\
\cline{3-4}
\end{tabular}{(d)}
\caption[Input parameters for the benchmark points with variable $f$ and $g_\rho$]{Input parameters for the benchmark points with variable $f$ and $g_\rho$, that correspond to a \emph{small width regime}.
Note that benchmark (d) correspond to benchmark (a) of Tab.~\ref{tab:input_bench-fgfixed}.
All the parameters, except $g_{0}$, $g_{0Y}$ and $g_\rho$, are expressed in GeV.}
\label{tab:input_bench-fgvar}
\end{table}

\newpage
\clearpage
\newpage\null\thispagestyle{empty}\newpage
\section*{List of publications}

This Thesis is based on the following published papers and preprints produced during the course of my Ph.D. studies
\begin{enumerate}
\item[-] \emph{Exploring Drell-Yan signals from the 4D Composite Higgs Model at the LHC}\\
D. Barducci, A. Belyaev, S. De Curtis, S. Moretti and G.M. Pruna\\
JHEP 1304 (2013) 152, arXiv:1210.2927 [hep-ph]
\item[-] \emph{Leptonic final states from di-boson production at the LHC in the 4-Dimensional Composite Higgs Model}\\
D. Barducci, L. Fedeli, S. De Curtis, S. Moretti and G.M. Pruna\\
JHEP 1304 (2013) 038, arXiv:1212.4875 [hep-ph]
\item[-] \emph{Multiple $Z^\prime \to t\bar t$ signals in a 4D Composite Higgs Model}\\
D. Barducci, A. Belyaev, M.S. Brown, S. De Curtis, S. Moretti and G.M. Pruna\\
Phys.Rev. D88 (2013) 074024, arXiv:1212.5948 [hep-ph]
\item[-] \emph{The 4-Dimensional Composite Higgs Model (4DCHM) and the 125 GeV Higgs-like signals at the LHC}\\
D. Barducci, S. De Curtis, K. Mimasu and S. Moretti\\
JHEP 1309 (2013) 047, arXiv:1302.2371 [hep-ph]
\item[-] \emph{Future Electron-Positron Colliders and the 4-Dimensional Composite Higgs Model}\\
D. Barducci, S. De Curtis, S. Moretti and G.M. Pruna\\
JHEP 1402 (2014) 005, arXiv:1311.3305 [hep-ph]
\item[-] \emph{Model Independent Framework for Analysis of Scenarios with Multiple Heavy Extra Quarks}\\
D. Barducci, A. Belyaev, M. Buchkremer, G. Cacciapaglia, A. Deandrea, S. De Curtis, J. Marrouche, S. Moretti and L. Panizzi\\
arXiv:1405.0737 [hep-ph]
\end{enumerate}


\addtocontents{toc}{\vspace{2em}} 

\appendix 


\addtocontents{toc}{\vspace{2em}}  
\backmatter

\label{Bibliography}
\lhead{\emph{Bibliography}}  
\bibliographystyle{Sources/JHEP}
\bibliography{Thesis}  

\end{document}